\documentclass[aps,prd,preprint,groupedaddress,superscriptaddress,nofootinbib,showpacs,eqsecnum]{revtex4-2}
\usepackage[T1]{fontenc}

\usepackage{amssymb}
\usepackage{amsmath}
\usepackage{cancel}
\usepackage{epstopdf}
\usepackage{mathtools}
\usepackage{bbm}
\usepackage{faktor}
\usepackage{tensor}
\usepackage{bbm}
\usepackage{shuffle}
\usepackage{slashed}

\usepackage{graphicx}
\usepackage[export]{adjustbox}
\usepackage[usenames,dvipsnames]{xcolor}
\usepackage{tikz}
\usepackage[compat=1.1.0]{tikz-feynman}
\tikzfeynmanset{/tikzfeynman/warn luatex = false} 
\usetikzlibrary{calc}

\usepackage{relsize}
\usepackage{titlesec}
\usepackage[markifdraft]{gitinfo2}
\usepackage{setspace} 
\usepackage{placeins}
\usepackage{subcaption} 
\usepackage{enumitem}
\usepackage{diagbox}

\usepackage[textsize=tiny]{todonotes}

\makeatletter
\pretocmd{\@affil@script}{\vspace*{2mm}}{}{}{}
\makeatother

\tikzset{invclip/.style={clip,insert path={{[reset cm]
        (-16383.99999pt,-1c6383.99999pt) rectangle (16383.99999pt,16383.99999pt)
      }}}}

\definecolor{allOrderBlue}{rgb}{0.4,0.5,1}
\definecolor{patternBlue}{rgb}{0,0,1}
\definecolor{photonRed}{rgb}{1,0.2,0.2}

\def\Ord{\mathcal{O}}

\def\backdexpatch latexntA{\hspace*{-6mm}}
\def\nsand#1.#2.#3{%
  \left\langle\smash{#1}{\vphantom1}\right|{#2}%
  \left|\smash{#3}{\vphantom1}\right]}
\def\nsandaa#1.#2.#3{%
  \left\langle\smash{#1}{\vphantom1}\right|{#2}%
  \left|\smash{#3}{\vphantom1}\right\rangle}
\def\nsandbb#1.#2.#3{%
  \left[\smash{#1}{\vphantom1}\right|{#2}%
  \left|\smash{#3}{\vphantom1}\right]}
\def\nsandba#1.#2.#3{%
  \left[\smash{#1}{\vphantom1}\right|{#2}%
  \left|\smash{#3}{\vphantom1}\right\rangle}

\def\spa#1.#2{\left\langle#1\,#2\right\rangle}
\def\spb#1.#2{\left[#1\,#2\right]}

\def\spash#1.#2{\spa{\smash{#1}}.{\smash{#2}}}
\def\spbsh#1.#2{\spb{\smash{#1}}.{\smash{#2}}}

\def\sect#1{Sect.~{\ref{#1}}}

\def\app#1{App.~{\ref{#1}}}

\def\fig#1{Fig.~{\ref{#1}}}
\def\Fig#1{Fig.~{\ref{#1}}}
\def\figs#1#2{Figs.~{\ref{#1}} and {\ref{#2}}}

\def\tab#1{Table~{\ref{#1}}}

\def\eq#1{eq.~(\ref{#1})}
\def\eqn#1{eq.~(\ref{#1})}
\def\eqns#1#2{eqs.~(\ref{#1}) and (\ref{#2})}

\def\<{\langle}
\def\>{\rangl\noaffiliatione}

\newbox\charbox
\newbox\slabox
\def\s#1{{      
    \setbox\charbox=\hbox{$#1$}
    \setbox\slabox=\hbox{$/$}
    \dimen\charbox=\ht\slabox
    \advance\dimen\charbox by -\dp\slabox
    \advance\dimen\charbox by -\ht\charbox
    \advance\dimen\charbox by \dp\charbox
    \divide\dimen\charbox by 2
    \raise-\dimen\charbox\hbox to \wd\charbox{\hss/\hss}
    \llap{$#1$}
  }}
\def\cut#1{{      
    \setbox\charbox=\hbox{$#1$}
    \setbox\slabox=\hbox{$|$}
    \dimen\charbox=\ht\slabox
    \advance\dimen\charbox by -\dp\slabox
    \advance\dimen\charbox by -\ht\charbox
    \advance\dimen\charbox by \dp\charbox
    \divide\dimen\charbox by 2
    \raise-\dimen\charbox\hbox to \wd\charbox{\hss$|$\hss}
    \llap{$#1$}
  }}

\def\pol{\varepsilon}

\def\kb{\bar k}

\def\kbperpi#1{\kb\_{i\perp}}

\def\LIPS{\textrm{LIPS}}

\usepackage{./packages/spinors}
\usepackage{./packages/auxiliary_definitions}

\def\tree{{(0)}}

\def\twoloop{{(2)}}

\newcommand{\cyclicgroup}{Z}
\def\ctop{{\textrm{T}}}
\def\cbottom{{\textrm{B}}}
\def\ESet{E}
\DeclareMathOperator{\Exch}{Exch}
\def\CutClass#1{{\{#1\}}}
\def\GeneratingCuts{G}
\def\inner{\textrm{I}}
\def\central{\textrm{C}}
\def\ibar{\bar\imath}
\def\jbar{\bar\jmath}
\newcommand{\cycperm}{\rho}
\newcommand{\CutSet}{\mathrm{AllLabels}}
\newcommand{\CutSetUnique}{\mathrm{UniqueCuts}}
\newcommand{\LProd}{L}
\newcommand{\ssix}{s^{(6)}}
\newcommand{\sfour}{s}
\newcommand{\Rota}{\text{Rot}}

\makeatletter
\def\CutSetClass#1{\def\CSC@thearg{#1}\CutSet\@ifnextchar_{\CSC@insertsub}{(\CSC@thearg)}}
\def\CSC@insertsub_#1{_{#1}(\CSC@thearg)}

\def\GeneratingCutsClass#1{\def\GSC@thearg{#1}\GeneratingCuts\@ifnextchar_{\GSC@insertsub}{(\GSC@thearg)}}
\def\GSC@insertsub_#1{_{#1}(\GSC@thearg)}
\makeatother

\newcommand{\six}{L}

\usepackage{orcidlink}

\usepackage[]{hyperref}

\graphicspath{{./figures/}} 

\begin{document}

\newcommand{\Ampl}{\mathcal{A}}
\newcommand{\AmplB}{\mathcal{\bar A}}
\def\Int{\textrm{Int}}

\hfuzz=15 pt

\title{A Unitarity Approach to Two-Loop All-Plus Rational Terms}

\def\squeezev{\vspace*{-2.5mm}}

\author{David~A.~Kosower\orcidlink{0000-0002-9087-0071}}
\affiliation{\looseness=-1%
  \linespread{1}\selectfont%
  Institut de Physique Th\'eorique, CEA, CNRS, Universit\'e Paris--Saclay,
  F--91191 Gif-sur-Yvette cedex, France
  \\ {\sf David.Kosower@ipht.fr}
}

\author{Sebastian P{\"o}gel\orcidlink{0000-0003-4323-9743}}
\affiliation{\looseness=-1%
  \linespread{1}\selectfont%
  PRISMA Cluster of Excellence, Institut f{\"u}r Physik,\\
Johannes Gutenberg-Universit{\"a}t Mainz, 
D--55099 Mainz, Germany\\ 
{\sf Poegel@uni-mainz.de}
}
\date\today

\begin{abstract}
  We present a calculation of the rational terms in
two-loop all-plus gluon amplitudes using $D$-dimensional
unitarity.  We use a conjecture of separability of the
two loops, and then a simple generalization of one-loop
$D$-dimensional unitarity to perform calculations.  We compute
the four- and five-point rational terms analytically, and
the six- and seven-point ones numerically.  We find agreement
with previous calculations of Dalgleish, Dunbar, Godwin, Jehu, Perkins, and Strong.  
For a special subleading-color amplitude, we compute the
eight- and nine-point results numerically, and find agreement
with an all-$n$ conjecture of Dunbar, Perkins, and Strong.
\end{abstract}

\pacs{\hspace{1cm}}

\preprint{1}{SAGEX-22-26-E}

\preprint{1}{MITP-22-040}

\maketitle
\newpage
\thispagestyle{empty}

\section{Introduction}
\label{sec:introductionsection}

Increasing integrated luminosity at the Large Hadron Collider (LHC) in the
coming decade will drive experimenters' search for physics beyond the Standard
Model (SM).
The LHC will not be surging into a new energy domain, but will be
sensitive to ever-fainter discrepancies from SM predictions.
The greater
sensitivity will emerge both from increased statistics and from a better
understanding of systematic uncertainties.
Greater experimental sensitivity
does not suffice, however, in order to find small deviations from theoretical
expectations.
We also need higher-precision calculations, in particular in
perturbative QCD, in order to reduce theoretical uncertainties.

The current frontier for perturbative QCD calculations is at
next-to-next-to-leading order (NNLO), where one may broadly hope that results
will reduce these latter uncertainties to below a few percent.
Because of
renormalization-scale sensitivity in leading-order (LO) calculations,
next-to-leading order (NLO) calculations provide the first truly quantitative
predictions, and NNLO is then the first order at which theoretical
uncertainties can be assessed quantitatively.

Many results, primarily for $2\rightarrow1$ and $2\rightarrow2$ processes, are
already available at NNLO (and sometimes beyond).
The generalization of NNLO
calculations to multijet processes requires developments of several aspects,
most notably of two-loop amplitudes and of infrared-regulation techniques.
In
this article, we explore a technique for computing certain contributions to a
simple class of two-loop Yang--Mills amplitudes, the so-called ``all-plus''
amplitudes, with all external gluons of identical helicity.

These amplitudes are simpler than general two-loop amplitudes.
At tree level, they vanish.
This vanishing can be proven diagrammatically, or understood as
the consequence of a supersymmetry
identity~\cite{ManganoParkeReview}.

As a result, the one-loop amplitudes are free of ultraviolet and infrared
divergences, and indeed are purely rational in the external
spinors~\cite{Ellis:1985er,Bern:1993mq,Bern:1993qk,Mahlon:1992fs,Mahlon:1993si}.
In turn, two-loop amplitudes have 
singularities in dimensional regularization of
the same degree as general one-loop amplitudes.
The polylogarithmic weights in their finite terms are the same as those in
one-loop amplitudes.
The two-loop all-plus amplitudes are in a sense intermediate in complexity
between one-loop amplitudes and general two-loop amplitudes.
In this respect, they are a good laboratory for exploring aspects of
Yang--Mills amplitudes beyond what is probed in ${\cal N}=4$ supersymmetric
amplitudes.
These amplitudes may also be a portal to exploring hidden connections between
different theories: at one loop, there is an intriguing
connection~\cite{Bern:1996ja,Britto:2020crg} between the
all-plus amplitude and the simplest ${\cal N}=4$ super-Yang--Mills amplitude,
scattering all gluons but two of like helicity (MHV).
Furthermore, for four and five gluons in the planar limit, 
the leading 
transcendental weight parts of all-plus amplitudes 
at two loops (and three loops for four gluons) have been 
shown to be dual to 
those of ${\cal N}=4$ Wilson loops with Lagrangian 
insertions at one lower loop order~\cite{AllPlusAndWilsonLoops}. 
This duality is conjectured to hold for any loop order.

The four-point all-plus amplitude at two loops
was computed long ago by Bern, Dixon, and
one of the authors~\cite{Bern:2000dn}.
Much later, Badger, Frellesvig,
and Zhang (BFZ) 
computed~\cite{Badger:2013gxa} the leading-color part 
of the
five-point all-plus amplitude (partly analytically and partly 
numerically);
Badger, Mogull, Ochirov, and O'Connell computed the full
integrand~\cite{TwoLoopAllPlus5ptFull} for the five-point amplitude; Gehrmann,
Henn, and Lo~Presti gave an analytic form of the leading-color
part~\cite{Gehrmann:2015bfy}.
More recently, Abreu, Dormans,
Febres~Cordero, Ita, Page, and~Zeng gave an analytic form for the planar
amplitude~\cite{Abreu:2017hqn} by
reconstructing
an
expression from a
numerical calculation, while Badger, Chicherin, Gehrmann, Heinrich, Henn, Peraro, Wasser, Zhang and Zoia
gave an analytic form of the full amplitude by integrating the earlier
integrand~\cite{Badger:2019djh}.

There are also results at three loops: the leading-color
four-gluon amplitude was computed by 
Jin and Luo~\cite{Jin:2019nya}, 
while the full, non-planar QCD result was recently 
computed by Caola, Chakraborty, Gambuti, 
von~Manteuffel, and Tancredi~\cite{Caola:2021izf}.

The structure of the all-plus amplitude at two loops, as we shall review in the next section, has made it possible for Dunbar,
Jehu, and Perkins (DJP)~\cite{TwoLoopAllPlusPolylogs} to compute the polylogarithmic terms for an
arbitrary number of external gluons at leading color.
In addition, Dalgleish, Dunbar, Godwin, Jehu, Perkins, and 
Strong~\cite{Dunbar:2016gjb,Dunbar:2016aux,Dunbar:2016cxp,%
Dunbar:2017nfy, Dunbar:2017azf,Dalgleish:2020mof,Dunbar:2019fcq} 
have also computed the
rational terms in the five- and six-point amplitudes at 
leading and subleading
color, as well as the leading-color seven-point amplitude.
They
made use of recursive techniques to do so.
Dunbar, Perkins, and Strong~(DPS) presented an all-$n$
conjecture for a special subleading-color 
amplitude~\cite{Dunbar:2020wdh}.
Badger, Mogull and Peraro (BMP)~\cite{Badger:2016ozq} 
have also computed the
leading-color five- and six-point amplitudes through a reconstruction of the
integrand.
In addition, they presented a conjecture for the all-$n$
integrand on which we shall rely in our calculations.

This article is organized as follows.  In the next section, 
we review 
general aspects of all-plus amplitudes and present
the separability conjecture along with
the dimensional decompositions we use.  
In \sect{ColorStructureSection}, we discuss the color structure and generating sets of unitarity cuts.
We look at the required tree-level amplitudes in
\sect{IngredientAmplitudesSection}, and
discuss the \textsl{Mathematica\/} implementation of 
our calculations in \sect{MathematicaImplementationSection}.
We present the four-point calculation in detail
in \sect{FourGluonSection}, and an overview of higher-point 
calculations in \sect{HigherPointSection}.
We make concluding remarks
in \sect{ConclusionSection}.
In the appendices, we detail our 
conventions~(\app{ConventionsAppendix});
list the massive-scalar tree amplitudes we 
use~(\app{ScalarAmplitudesAppendix});
present a current-based alternative calculation
of contact contributions to four-scalar 
amplitudes~(\app{OffShellCurrentAppendix});
present an alternate form of the
seven-point subleading-color single-trace 
amplitude~(\app{SevenGluonSubleadingAppendix}); 
give an $n$-point momentum-twistor 
parametrization~(\app{MomentumTwistorAppendix});
and list the one-loop expressions for integral
coefficients we use in our 
calculations~(\app{GeneralizedUnitarityAppendix}).
We also attach a set of auxiliary files, containing the \mm\ packages 
implementing the methods presented in this paper;
lists of the unitarity cuts required for the rational parts of all partial 
amplitudes with up to nine gluons;
and analytic expressions for the five-point rational parts derived 
using our automated code.

\section{Review, Separability, and Dimensional Reconstruction}
\label{BackgroundSection}
\label{sec:background-and-review}

We know from considerations of infrared and ultraviolet divergences that
two-loop all-plus amplitudes have no singularities stronger than $1/\eps^2$ in
dimensional regularization, and that the singular terms are proportional to
the one-loop amplitude.
What about the finite, polylogaritmic terms?
We can
imagine looking at four-dimensional cuts to extract information about them.
For example, localizing both loop momenta in a double box in four dimensions,
following the prescriptions of ref.~\cite{TwoLoopProjectors}, we see that
there is no possible helicity assignment of internal legs that yields a
non-vanishing coefficient for the integral.
While the analogous procedure is
not known for other integrals, one would
presumably be drawn to a similar conclusion from
examining integrands~\cite{Badger:2016ozq}.

\begin{figure}
  \includegraphics[height=12em,valign=c]
  {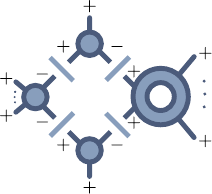}
  \caption{
    \label{fig:OneLoopQuadCut}
    A four-dimensional quadruple cut of
    one of the loops in the leading-color two-loop all-plus amplitude.
    The all-plus configuration requires the cut to be either of the one-mass or easy
    two-mass box type.
    In addition, the second loop can only appear as a one-loop all-plus
    amplitude.
  }
\end{figure}

Even with ordinary unitarity, once we cut one of the two loops, we are left
with too simple a function to cut another loop.
If we fully localize one of
the loop integrals, performing a quadruple cut on it, we will find a product
of rational functions, as illustrated in \fig{fig:OneLoopQuadCut}.
One of these
functions will be the one-loop all-plus amplitude, a rational functions of the
spinor variables just like the tree-level amplitudes at the other three
corners.

This implies that the calculation of the polylogarithmic terms follows a
one-loop recipe faithfully: one computes the coefficients of one-loop box,
triangle, and bubble integrals following the prescriptions of
refs.~\cite{BCF,Forde:2007mi,Badger:2008cm}, or alternatively, the integrand
approach of ref.~\cite{Ossola:2006us}.
This is the procedure that DJP carried out 
in ref.~\cite{TwoLoopAllPlusPolylogs},
and Dunbar, Godwin, Jehu, and Perkins in
refs.~\cite{Dunbar:2016}.

Four-dimensional cuts do not suffice, however, in order to compute the
rational terms.
Dunbar, Godwin, Jehu, Perkins, and Strong
made use of recursion relations to compute them through the seven-point
amplitude~\cite{Dunbar:2016,Dunbar:2020wdh,DalgleishDunbar:2020}.
We will explore a different approach,
using $D$-dimensional unitarity, to do so.

\subsection{General Structure of All-Plus Amplitudes}

At tree level, all-plus amplitudes vanish in Yang--Mills theories.  We can see
this in a number of ways.  One approach relies on the BCFW on-shell recursion
relations~\cite{BCF,BCFWrecursion}.  The three-point all-plus amplitude vanishes
in pure Yang--Mills with a dimensionless coupling.  The factorization
channel(s) in the four-point amplitude necessarily involve this three-point
amplitude, and hence likewise vanish.  At higher points, each factorization
channel necessarily involves a lower-point all-plus amplitude, and hence this
argument continues inductively to all multiplicity.

\begin{figure}[ht]
  \begin{equation}
    \includegraphics[height=6em,valign=c]{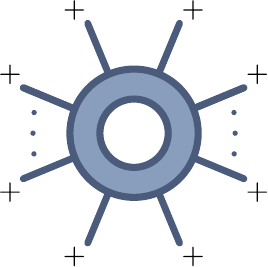}
    \qquad \scalebox{1.5}{$\rightarrow$} \qquad
    \includegraphics[height=6em,valign=c]{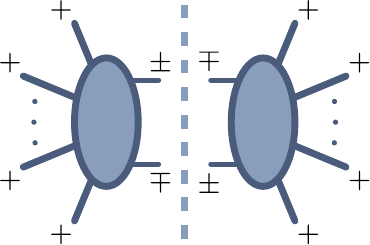}
    \qquad \scalebox{1.5}{$+$} \qquad
    \includegraphics[height=6em,valign=c]{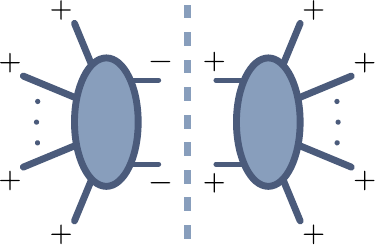}
  \end{equation}
  \caption{An ordinary cut of a one-loop all-plus amplitude}
  \label{OneLoopAllPlus4DCut}
\end{figure}

The tree-level vanishing simplifies loop amplitudes.
At one-loop, amplitudes can be written in a form exposing their universal
singular structure~\cite{Giele:1991vf,BDDK,Kunszt:1994np,Catani:1996vz},
\begin{equation}\label{eq:one-loop_catani}
  \AOne = \ATree I^{(1)} + F^{(1)} + \mathcal{O}(\epsilon)\,.
\end{equation}
Here, $I^{(1)}$ is a universal function of the Lorentz invariants, which
through double and single poles in $\eps$ summarizes the singular structure of
the amplitudes.
The next term, $F^{(1)}$, is of $\Ord(\eps^0)$.
The vanishing
of $\ATree$ immediately implies the finiteness of the one-loop amplitudes.
In general, $F^{(1)}$ contains logarithms and polylogarithms in addition to
rational functions of the spinor variables.
In the all-plus amplitudes, it contains only rational terms.
The absence of logarithms and polylogarithms
can be understood in ordinary unitarity in four
dimensions~\cite{BDDK,Cutkosky:1960sp}.
Perform an ordinary cut in any channel of a color-ordered amplitude.
In the $K_{1\ldots j}$
channel, for example, we find for the discontinuities,
\begin{equation}
  \begin{aligned}
    \AOne_n&(1^+,\ldots,n^+)\big|_{s_{1\ldots j}\textrm{\ cut}} =
    \\ &
    \sum_{h_{a,b}=\pm} \int d^4\LIPS(k_a,k_b)\;\ATree(1^+,\ldots,j^+,k_a^{h_a},k_b^{h_b})
    \\ &\hspace{43mm}\times 
    \ATree((j+1)^+,\ldots,n^+,(-k_b)^{-h_b},(-k_a)^{h_a})\,,
  \end{aligned}
\end{equation}
where $d\LIPS$ is the Lorentz-invariant phase-space measure.
This cut is illustrated in \fig{OneLoopAllPlus4DCut}.
All terms contain either a single-minus or all-plus tree amplitude, both of
which vanish.
Thus, one-loop all-plus amplitudes are free of discontinuities,
and have to be entirely rational.
For the leading-color partial amplitude,
general expressions for these rational terms were conjectured in
ref.~\cite{Bern:1993qk} by demanding correct collinear factorization,
\begin{equation}\label{eq:A1_n:1_all-plus}
  \AOne (1^+\mathellipsis n^+)=
  -\frac{1}{3}\frac{\sum_{1\le i < j < k < l \le n}\spab{i|jkl|i}}
  {\spaa{12}\spaa{23}\mathellipsis\spaa{(n-1)n}\spaa{n1}}+\mathcal{O}(\epsilon)\,,
\end{equation}
a form which was later proven in ref.~\cite{Mahlon:1993si}.
The subleading-color amplitudes at one-loop can always be obtained from the
leading-color ones through color relations~\cite{Bern:1990ux}.
Compact forms for them are also known~\cite{DalgleishDunbar:2020},
\begin{align}
  \label{eq:A1_n:2_all-plus}
  \AOne_{n:2}(1^+;2^+\mathellipsis n^+) &=
                                              -\frac{\sum_{2\le i < j \le n}\spbb{1|ij|1}}
    {\spaa{23}\spaa{34}\mathellipsis\spaa{(n-1)n}\spaa{n2}}\,,\\
  \label{eq:A1_n:r_all-plus}
  \AOne_{n:r}(1^+\mathellipsis (r-1)^+;r^+\mathellipsis n^+) &=
                                                                   -2\frac{s_{1\mathellipsis (r-1)}^2}
                                                                   {\spaa{12}\spaa{23}\mathellipsis\spaa{(r-1)1}\spaa{r(r+1)}\mathellipsis\spaa{nr}},
\end{align}
which respectively multiply the subleading color structures
$\Tr(T^{a_1})\Tr(T^{a_2}\mathellipsis T^{a_n})$ and
$\Tr(T^{a_1}\mathellipsis T^{a_{r-1}})
\Tr(T^{a_r}\mathellipsis T^{a_n})$, 
where $r\ge3$.
The amplitude \(\AOne_{n:2}(1^+;2^+\mathellipsis n^+)\) is equivalent to the
one-photon amplitude \(\AOne(1^\gamma\, 2^+\mathellipsis n^+)\), for which a
compact all-\(n\) form was provided earlier in ref.~\cite{Bern:1993qk}.
Given their finiteness and absence of branch-cuts
these expressions seem more like tree-level amplitudes rather than
one-loop ones.

The first computation of a two-loop all-plus amplitude was 
presented in
refs.~\cite{Bern:2000dn,Glover:2001af,Bern:2002tk}, which 
provided analytic
expressions for the full-color four-gluon amplitude.
For five gluons, the planar and non-planar integrands were 
derived in
Badger, Frellesvig and Zhang~(BHZ)~\cite{Badger:2013gxa} and
by Badger, Mogull, Ochirov and 
O'Connell (BMOOC)~\cite{Badger:2015lda}
respectively, together with results from numerical integration.
Analytic results for the planar amplitude were first given in
ref.~\cite{Gehrmann:2015bfy}.
The full-color two-loop all-plus amplitude was then given
in ref.~\cite{Badger:2019djh}.

As in the one-loop case, two-loop amplitudes can be decomposed with
respect to their singularity structure~\cite{Catani:1998bh}, which
leads us to a
relation similar to that of eq.~\eqref{eq:one-loop_catani},
\begin{equation}\label{eq:IR_Catani}
  \ATwo= \ATree I^{(2)} + \AOne I^{(1)} + F^{(2)} + \mathcal{O}(\epsilon).
\end{equation}
Here, $I^{(2)}$ is (like $I^{(1)}$) a universal function of the Lorentz invariants
with divergences up to $\eps^{-4}$;  $I^{(1)}$ is the same function given 
in eq.~\eqref{eq:one-loop_catani}.
The remainder $F^{(2)}$ is finite in dimensional
regularization, and in general contains both polylogarithmic and rational terms.  
For all-plus amplitudes, we again find significant simplifications:
the vanishing of $\ATree$ and finiteness of
$\AOne$ allows only for divergences up to $\eps^{-2}$.
This is the same degree
of divergence that we would ordinarily expect in a \emph{one}-loop amplitude.  
The universal behavior allows us to obtain all
divergent terms in $\ATwo$, leaving us to compute only
the finite part $F^{(2)}$.
\begin{figure}[th]
  \hspace*{10mm}
  \includegraphics[height=6em,valign=c]{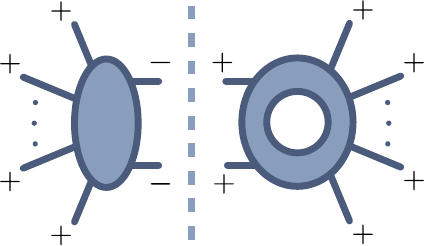}
  \caption{Ordinary cuts in the two-loop all-plus amplitude.}
  \label{TwoLoopBranchCuts}
\end{figure}

We may split $F^{(2)}$ into polylogarithmic and
rational parts $P^{(2)}$ and $R^{(2)}$,
\begin{equation}
  F^{(2)}=P^{(2)}+R^{(2)}\,.
\end{equation}
The polylogarithmic part $P^{(2)}$ has ordinary
branch cuts, which are shown schematically for the all-plus 
amplitude in \fig{TwoLoopBranchCuts},
and may therefore be computed using four-dimensional generalized unitarity.
In contrast, the rational part $R^{(2)}$ does not contain such
discontinuities and requires separate treatment.

\subsection{Dimensional Reconstruction}
\label{subsec:dim_reconstruction}

In dimensional regularization, scattering amplitudes depend on the spacetime
dimension $D$.
Dimensional regulators come in a number of different
flavors, which differ in their treatment of observed and unobserved particles,
and also in how the number of states for particles with spin are continued
away from four dimensions.
Over the years, different variants of dimensional regularization have been used.
In the conventional scheme (CDR), all particles are treated in $D$ dimensions,
as are the number of states.
Because the number of states for bosons and fermions
continues differently, this scheme is incompatible with maintaining manifest
supersymmetry;
and because it requires extra polarization states even for observed
particles, it is incompatible with efficient use of helicity methods.

In the so-called four-dimensional helicity scheme (FDH) and in the 't~Hooft--Veltman
scheme, observed particles are treated strictly in four dimensions, while unobserved
particles --- those inside loops, or soft or collinear real emissions --- are treated
in $D$ dimensions.
The dimensional reduction~(DR) scheme compactifies $2\eps$ dimensions,
thereby maintaining manifest compatibility with supersymmetry.
Its use of dimensions
smaller than four, however, is awkward for helicity methods beyond one loop.

In this article, we follow a modified approach originally introduced in
ref.~\cite{Bern:2000dn}, and later exploited at one loop in
ref.~\cite{Giele:2008ve} and at two loops by 
BMP~\cite{Badger:2016ozq} to help isolate rational
contributions.
It introduces two distinct dimensional parameters: $D$, which
controls the dimension of loop-momentum integrations, and $D_s$, which controls
the number of states, with $D_s$ taken to be greater than $D$.
Integrands of loop amplitudes then depend \emph{polynomially} on $D_s$, while
integrals depend in a general analytic fashion on $D$.

The value of $D_s$ is of course not integral.  Because of the polynomial
dependence, however, it is possible to reconstruct the full dependence on this
parameter by sampling the integrand at a number of integer values for it.
Ref.~\cite{Giele:2008ve} showed how to do this at one loop, where the dependence
is linear.  At two loops, the dependence in quadratic, but we can employ
the same strategy.

This allows us to avoid evaluating the tree amplitudes arising in generalized-unitarity
cuts in non-integer dimensions.  Instead, we are
free to perform all evaluations in integer dimensions.
This makes
dimensional reconstruction a useful technique for both analytic computations as
well as automated semi-numerical codes.
It has found application in determining various one-
and two-loop amplitudes in recent
years~\cite{Badger:2013gxa,Badger:2015lda,Badger:2016ozq,Badger:2018enw,Badger:2019djh,Abreu:2017xsl,Abreu:2017hqn,Abreu:2018jgq,Abreu:2018zmy,Abreu:2019rpt,Abreu:2019odu,Abreu:2020lyk,Abreu:2020cwb,Abreu:2021oya}.

BMP connect~\cite{Badger:2016ozq}
the rational terms in the two-loop
all-plus amplitude to the integrand's dependence on the state dimension $D_s$.
We review their construction briefly.  We also rely on the one-loop discussion~\cite{Giele:2008ve},
as well as on the
$L$-loop generalization presented in ref.~\cite{AccettulliHuber:2019abj}.

\def\hA{\hat A}
\def\vphj{{\vphantom{j}}}
In pure Yang--Mills amplitudes,
the $D_s$ dependence of loop amplitudes arises from contractions of the
metric tensor $\tensor{\eta}{^{\mu\nu}}$ along loops.  Each contraction
that ultimately closes on itself generates a yields a trace
$\tensor{\eta}{^\mu_\mu}=D_s$.
Vector bosons carry a single index, so that an $L$-loop scattering amplitude
can be written as a degree-$L$ polynomial in in $D_s$,
\begin{equation}
  A_{D_s}^{(L)}=\sum_{j=0}^{L}\hA_j^\vphj D_s^j.
\end{equation}
We can determine the coefficients $\hA_i$
by sampling the amplitude at $L+1$ different values of $D_s$.

Let us now discuss the constraints on the values which we can use.
We begin with one-loop amplitudes, the simplest case to
consider.  At one loop, we can have at most a single contraction
$\tensor{\eta}{^\mu_\mu}$, so amplitudes are linear in $D_s$,
\begin{equation}
  A_{D_s}^{(1)}=\hA_0^{(1)}+\hA_1^{(1)} D_s\,.
\end{equation}
A similar form holds for the integrand.
If we evaluate the amplitude in two different integer dimensions $D_0$ and $D_1$,
we can solve for the coefficients $\hA_{0,1}^{(1)}$, 
and then write,
\begin{equation}
  \label{eq:ADs_sampling_linear}
  A_{D_s}^{(1)}=\frac{D_s-D_0}{D_1-D_0}A_{1}^{(1)}+
  \frac{D_s-D_1}{D_0-D_1}A_{0}^{(1)}\,,
\end{equation}
in terms of the $D_{0,1}$-dimensional amplitudes $A_{0,1}^{(1)}$.
A similar result holds for integrands.
Choosing $D_1=D_0+1$, this expression simplifies,
\begin{equation}
  \label{eq:ADs_sampling_linear_simplified}
  A^{(1)}_{D_s}=(D_s-D_0)A_{1}^{(1)}-(D_s-D_1)A_{0}^{(1)}\,.
\end{equation}

As we take the external momenta to be four-dimensional, we must choose $D_0>4$
in order to capture the entire amplitude including rational contributions.
It suffices to choose $D_0$ large enough so that we can fully embed the
loop momentum $\ell$ in a $D_0$-dimensional space.
kinematics.  At one loop, we have only one $D$-dimensional vector, and so the
components beyond four dimensions can only appear in $\ell^2$.  It
can thus be embedded in five dimensions, and it suffices to choose $D_0=5$.

\def\ellb{\bar\ell}
\def\pol{\varepsilon}
With this choice, we then have $D_1=6$.  In the $D_1$-dimensional evaluation, we can
use Lorentz invariance to rotate away the sixth-dimensional components of the loop momentum.
We denote the four-dimensional components of the loop momentum by $\ellb$, so that
in both the $D_0$- and $D_1$-dimensional evaluations, the loop momentum has a nontrivial
fifth-dimensional component.  
Higher-dimensional components vanish.

In $D_s$ dimensions, gluons will have $D_s-2$ physical polarizations.  When we cut propagators,
we make use of a completeness relation to separate the $D_s$-dimensional metric tensor in its
state sum,
\begin{equation}
  \eta^{\mu\nu} = -\sum_{j=1}^{D_s-2} \pol_j^{\mu}(\ell) \pol_j^{\nu*}(\ell)\,.
  \label{eq:dim_reconstruction_pol_sum_1L}
\end{equation}
Each polarization vector is attached to an amplitude on one or the other side of the cut.
In generalized unitarity, all amplitudes arising from cuts will be tree amplitudes, now
taken to have the cut external legs with $D_s$-dimensional state counting.

In the five-dimensional evaluation, we have three polarization vectors satisfying the usual
conditions,
\begin{equation}
  \begin{aligned}
    \eta_{\mu\nu} \pol_i^{\mu}\pol_j^{\nu*} &= -\delta_{ij}\,,\quad
    \ell_\mu \pol_i^\mu &= 0\,.
  \end{aligned}
  \label{eq:pol_vectors_conditions}
\end{equation}

\def\scalar{\varphi}
In the six-dimensional evaluation,
we have four polarization states.  We can choose three of them to be the same as the
five-dimensional polarization states.  Because we have rotated the loop momentum to have
vanishing sixth component, we are free to choose the fourth and last polarization vector
to point in the sixth direction, $\pol_4^\mu = -\eta^{5\mu}$.  (The directions are as usual
labeled $0\ldots (D-1)$ in $D$ dimensions.)  This vector satisfies the conditions in
\eqn{eq:pol_vectors_conditions}.  Its only nonvanishing dot product is then with itself,
$\pol_4\cdot\pol_4$.  That is, it can only appear contracted to itself by a sequence of
metrics, so that it acts as though it is a scalar field.  (Of course, it transforms under
the color $SU(N)$ as an adjoint.)  We will call it $\scalar$.
\begin{figure}[thb]
  \captionsetup[subfigure]{justification=centering}
  \hspace*{1mm}
  \begin{subfigure}[]{0.3\textwidth}
    \includegraphics[height=3em,valign=c]{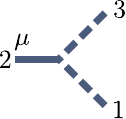}
    \caption{}
  \end{subfigure}
  \begin{subfigure}[]{0.3\textwidth}
    \includegraphics[height=3em,valign=c]
    {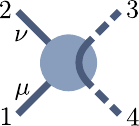}
    \caption{}
  \end{subfigure}
  \begin{subfigure}[]{0.3\textwidth}
    \includegraphics[height=3em,valign=c]
    {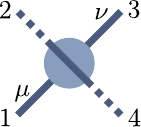}
    \caption{}
  \end{subfigure}
  \hspace*{10mm}
  \caption{Feynman vertices for an adjoint scalar field: 
  (a) the trilinear vertex
    $V_3(1_\scalar 2_g 3_\scalar)$ (b) the quadrilinear vertex $V_4(1_g 2_g 3_{\scalar} 4_{\scalar})$
    (c) the quadrilinear vertex $V_4^{\mu \nu}(1_g 2_{\scalar} 3_g 4_{\scalar})$.
    Solid lines represent gluons, while the dashed lines are scalars.
    Quartic vertices involving scalars are represented 
    as shaded disks,
    where the internal lines in the disk show the flow of 
    the scalar flavor.
    This will become more relevant in the multi-loop case, where there is more
    than one scalar.
    }
  \label{ScalarVertices}
\end{figure}

We can derive Feynman rules for $\scalar$ from the three- and
four-point gluon vertices simply by setting the Lorentz indices associated to $\scalar$
to $5$,
\begin{equation}
  \label{eq:Feyn_rules_scalars_one_loop}
  \begin{gathered}
    V_3^{\mu}(1_{\scalar} 2_g 3_{\scalar})=
    \frac{i}{\sqrt{2}}(k_1-k_3)^{\mu}\,, \qquad
    V_4^{\mu \nu}(1_g 2_g 3_{\scalar} 4_{\scalar})=
    \frac{i}{2}\eta^{\mu\nu}\,, \\
    V_4^{\mu \nu}(1_g 2_{\scalar} 3_g 4_{\scalar})=
    -i \eta^{\mu \nu}\,,
  \end{gathered}
\end{equation}
with all other configurations yielding vanishing vertices.
We also have the propagator,
\begin{equation}
  \label{eq:Feyn_rules_scalars_one_loop_propagators}
  \begin{gathered}
    \frac{i}{p^2}\,.
  \end{gathered}
\end{equation}
The construction above is a Kaluza--Klein reduction~\cite{KaluzaKlein} of the original six-dimensional gluon.
The Feynman rules accordingly match those of an
adjoint scalar field $\scalar^a$ with
Lagrangian density,
\begin{equation}
  \mathcal{L}_{\scalar}= D_\mu\scalar^a D^\mu\scalar^a,
\end{equation}
where $\scalar$ only couples to gluons through the covariant derivative.
The vertices are depicted%
\footnote{To represent the scalar flavor conservation in the quartic vertices we use the graphic design of ref.~\cite{Badger:2013gxa}.}
and explained
in \fig{ScalarVertices}.  At one-loop, a four-scalar
coupling plays no role.

Separating the last polarization vector $\pol_4$ from the others
allows us to decompose the $D_1$-dimensional amplitude $A_{1}^{(1)}$ into the sum,
\begin{equation}
  A_{1}^{(1)}=A_{0}^{(1)}+A_{s}^{(1)}\,,
\end{equation}
where $A_{s}^{(1)}$ is the contribution to a $D_0$-dimensional amplitude
arising from the scalar $\scalar$ circulating in the loop.
Substituting this decomposition back
into eq.~\eqref{eq:ADs_sampling_linear_simplified}, we obtain,
\begin{equation}
  \label{eq:dim_reconstruction_1L}
  \begin{aligned}
    A_{D_s}^{(1)}&=(D_s-D_0)\bigl(A_{0}^{(1)}+A_{s}^{(1)}\bigr)
    -(D_s-D_1)A_{0}^{(1)}
    \\
    &=A_{0}^{(1)}+(D_s-D_0)A_{s}^{(1)}
  \end{aligned}
\end{equation}
The amplitude can thus be reconstructed using a single value of
$D_s$, at the price of evaluating separately the
contributions from vectors and scalars circulating in the
loop.

This construction of amplitudes generalizes to arbitrary loop
order~\cite{AccettulliHuber:2019abj}.  At two loops, there can be up to two
contractions of the metric tensor, and so the amplitude is quadratic in
$D_s$,
\begin{equation}
  A^{(2)}= \hA_0 +\hA_1 D_s + \hA_2 D_s^2\,.
\end{equation}
We must evaluate it for three distinct values of $D_s$ in order to fix all terms.

Choosing dimensions,
\begin{equation}
  D_0\,,\quad D_1=D_0+1\,,\quad D_2 = D_0+2\,,
\end{equation}
we find for the $D_s$-dimensional amplitude,
\begin{equation}
  \label{eq:A_DS_sampling_2L}
  A_{D_s}^{(2)}=\frac{(D_s-D_2)}{2}(D_s-D_1)A_{0}^{(2)}-
  (D_s-D_0)(D_s-D_2)A_{1}^{(2)}+\frac{(D_s-D_0)}{2}(D_s-D_1)A_{2}^{(2)}\,,
\end{equation}
with $A_j^{(2)}$ the $D_j$-dimensional two-loop amplitude.
As a consistency check we can verify that for $D_s=D_0,D_1,D_2$ we obtain
$A_{D_0}^{(2)}$, $A_{D_1}^{(2)}$ and $A_{D_2}^{(2)}$.  We now have
two loop momenta $\ell_1,\,\ell_2$; we can rotate the extra-dimensional
components of the first into a fifth dimension, and the additional components of
the second into a sixth dimension.  We thus need $D_0\ge 6$; we choose $D_0=6$.

In the three consecutive dimensions, we have four, five, and six physical
polarizations, respectively.  We can choose the first four polarization vectors
to be the same in all dimensions.  We also choose the additional polarizations --- one for
$D_1$ and two for $D_2$ --- to have components only in the last two spatial
dimensions, and hence orthogonal to the loop momenta.  For the fifth state,
appearing in $D_{1,2}$, we choose $\pol_5^\mu = -\eta^{5\mu}$, and for the sixth
state, appearing only in $D_2$, we choose $\pol_6^\mu = -\eta^{6\mu}$.
As at one loop, the extra states in $D_1$ and $D_2$ behave like scalars.

\def\onescalar{1s}
\def\twoscalar{2s}
\def\withcontact{\times}
Thanks to the orthogonality of the additional polarization vectors
$\pol_{5,6}$ with respect to both loop momenta, we can express the
amplitudes $A_{1,2}^{(2)}$ in terms of $D_0$-dimensional amplitudes,
where scalars circulate in one or both of the loops.
Such a decomposition would rewrite,
\begin{equation}
  \label{eq:decomp_pseudo_two_loop}
  \begin{aligned}
    A_{1}^{(2)}&\to A_{0}^{(2)}\,, A_{\onescalar}^{(2)}\,, A_{\twoscalar}^{(2)}\,,
    \\
    A_{2}^{(2)}&\to A_{0}^{(2)}\,, A_{\onescalar}^{(2)}\,, A_{\twoscalar}^{(2)}\,,
    A_{\withcontact}^{(2)}\,.
  \end{aligned}
\end{equation}
Here, $A_{\onescalar}^{(2)}$ has one of the vector loops replaced by a scalar;
$A_{\twoscalar}^{(2)}$ has both vector loops replaced by scalars, with the two scalar loops
connected by an exchanged vector; and $A_{\withcontact}^{(2)}$ has both vector loops
replaced by scalars, with the two scalars of different flavors and
the two loops joined by
a four-scalar contact term.

\begin{figure}[thb]
  \captionsetup[subfigure]{justification=centering}
  \hspace*{1mm}
  \begin{subfigure}[]{0.4\textwidth}
    \includegraphics[height=3em,valign=c]{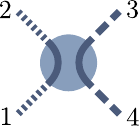}
    \caption{}
  \end{subfigure}
  \begin{subfigure}[]{0.4\textwidth}
    \includegraphics[height=3em,valign=c]{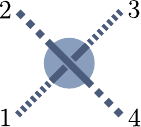}
    \caption{}
  \end{subfigure}
  \hspace*{10mm}
  \caption{Four-scalar Feynman vertices for two adjoint scalar fields: (a) the quadrilinear vertex
    $V_4(1_{\scalar} 2_{\scalar} 3_{\scalar'} 4_{\scalar'})$
    (b) the quadrilinear vertex $V_4(1_{\scalar} 2_{\scalar'} 3_{\scalar} 4_{\scalar'})$.
    The two types of dashing represent two distinct scalar flavors.}
  \label{FourScalarVertices}
\end{figure}

The mixed gluon--scalar Feynman rules for the
two scalars $\scalar\leftarrow\pol_5$ and $\scalar'\leftarrow\pol_6$ are the same as
given in \eqn{eq:Feyn_rules_scalars_one_loop} and shown in \fig{ScalarVertices}.
The four-scalar contact terms arise from the four-gluon vertex, and come in two
types,
\begin{equation}
  \label{eq:Feyn_rules_scalars}
  V_4(1_{\scalar} 2_{\scalar} 3_{\scalar'} 4_{\scalar'}) =
  -\frac{i}{2}\,,
  \qquad
  V_4(1_{\scalar} 2_{\scalar'} 3_{\scalar} 4_{\scalar'}) =
  i\,,
\end{equation}
shown in \fig{FourScalarVertices}.
We represent the two scalar flavors diagrammatically 
via two dashing styles.
As the scalars are interchangeable, the exact correspondence is
irrelevant.

\begin{figure}[thb]
  \begin{equation*}
    \includegraphics[height=4em,valign=c]
    {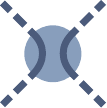}
    +
    \includegraphics[height=4em,valign=c]
    {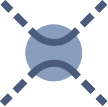}
    +
    \includegraphics[height=4em,valign=c]
    {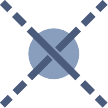}
    =0
  \end{equation*}
  \caption{Cancellation of different routings for four identical
    scalars.}
  \label{ScalarCancellation}
\end{figure}

Pure-scalar cubic vertices are absent, because at least one of their
momenta would have to be contracted with either $\pol_{5}$ or
$\pol_{6}$, and all such dot products vanish.
We can also omit contact terms for four identical scalar lines, as in
our application, all three configurations would add up and cancel.
The Feynman rules given here are then in agreement with those of
ref.~\cite{Badger:2013gxa}%
\footnote{In ref.~\cite{Badger:2013gxa}, the four-scalar
  contact-terms are defined respecting internal-scalar flavor
  conservation.  Each contact term would be present even for
  identical scalars.
  However, one would also have to sum over
  the three internal flavor routings: %
  $V_{4a}(1_{\scalar} 2_{\scalar} 3_{\scalar} 4_{\scalar}) +
  V_{4a}(2_{\scalar} 3_{\scalar} 4_{\scalar} 1_{\scalar}) +
  V_{4b}(1_{\scalar} 2_{\scalar} 3_{\scalar} 4_{\scalar}) = 0$.
  This cancellation is depicted in \fig{ScalarCancellation},
  such contributions thus vanishing as in our discussion.
}.

\begin{figure}[tbh]
  \centering
  \includegraphics[width=0.3\textwidth]{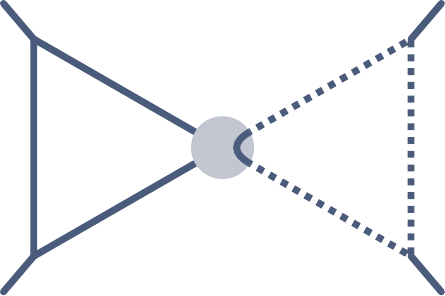}
  \hspace{1em}
  \includegraphics[width=0.3\textwidth]{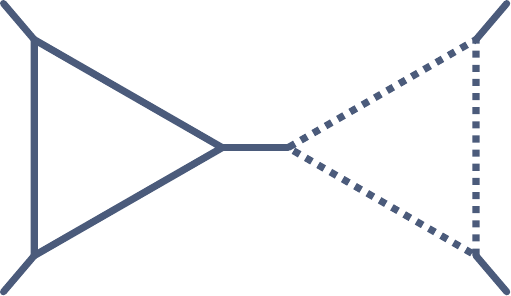}
  \\[1em]
  \includegraphics[width=0.3\textwidth]{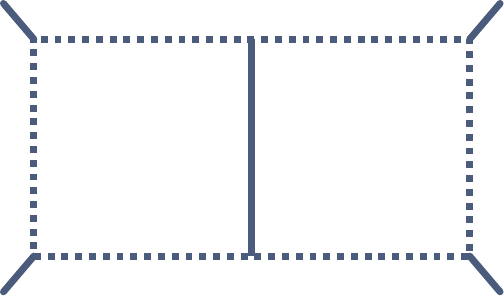}
  \hspace{1em}
  \includegraphics[width=0.3\textwidth]{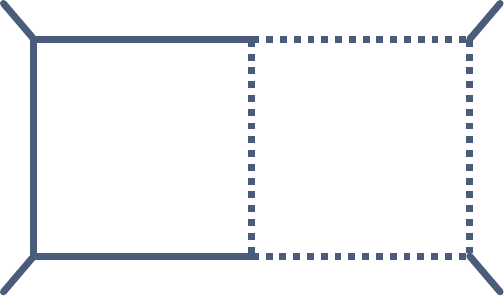}
  \caption{Representative Feynman diagrams contributing to
    $A_{\onescalar}^{(2)}$.
    The grey circle symbolizes a quartic gluon--scalar vertex.}
  \label{fig:A2_10_graphs}
\end{figure}

\begin{figure}[tbh]
  \centering
  \captionsetup[subfigure]{justification=centering}
  \begin{subfigure}[]{0.34\textwidth}
    \includegraphics[valign=c]{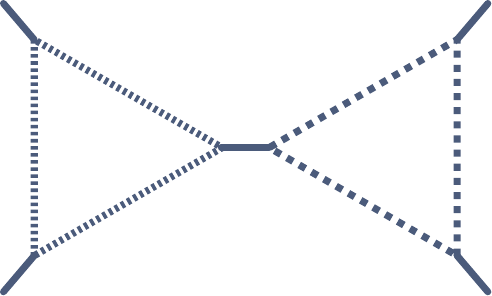}
    \caption{\label{fig:A2_20_graphs}}
  \end{subfigure}
  \begin{subfigure}[]{0.34\textwidth}
    \includegraphics[valign=c]{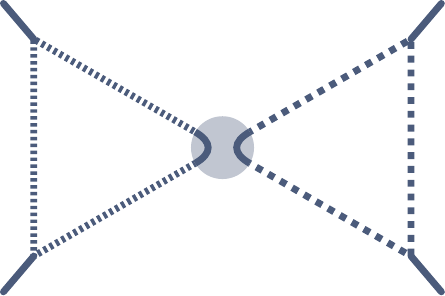}
    \caption{\label{fig:A2_11_graphs}}
  \end{subfigure}
  \caption{Representative of graphs contributing to $A_{D_0,2,0}^{(2)}$ (a),
    and $A_{D_0,1,1}^{(2)}$ (b).  In (a), the scalar flavors in the two loops
    can be the same, while the quartic scalar interaction represented in (b)
    requires the flavors to be different. The quartic interaction with all
    scalars having the same flavor vanishes. After summing over scalar flavors
    the diagrams in (a) and (b) therefore appear with prefactors of
    $(D_2-D_0)^2$ and $(D_2-D_0)(D_2-D_0-1)$ respectively.}
\end{figure}

Examples of Feynman diagrams contributing to $A_{\onescalar}^{(2)}$,
$A_{\twoscalar}^{(2)}$, and $A_{\withcontact}^{(2)}$
are shown respectively in figs.~\ref{fig:A2_10_graphs},%
~\ref{fig:A2_20_graphs} and~\ref{fig:A2_11_graphs}.
Separating the gluon-exchange and contact-term
contributions in $A^{(2)}_{2}$ will simplify expressions, as
$A^{(2)}_{\twoscalar}$ contributes to both the $A^{(2)}_{1}$ and
$A^{(2)}_{2}$.
We use a contact term which requires the
scalar flavors in two loops to be different, such that in one loop
$\pol_5$ circulates while $\pol_6$ circulates in the other.  In
the gluon-exchange contribution, there is no such restriction.

Given the specific Feynman rules above and the choices for
the different dimensions, we can now make the decomposition in
\eqn{eq:decomp_pseudo_two_loop} explicit,
\begin{equation}
  \begin{aligned}
    A_{1}^{(2)}&= A_{0}^{(2)}+(D_1-D_0)A_{\onescalar}^{(2)}+(D_1-D_0)^2
    A_{\twoscalar}^{(2)}\\
    &= A_{0}^{(2)}+A_{\onescalar}^{(2)}+A_{\twoscalar}^{(2)}\,,\\[1em]
    A_{2}^{(2)}&= A_{0}^{(2)}+(D_2-D_0)A_{\onescalar}^{(2)}+(D_2-D_0)^2
    A_{\twoscalar}^{(2)}+(D_2-D_0)(D_2-D_0-1)A_{\withcontact}^{(2)}\\
    &= A_{0}^{(2)}+2A_{\onescalar}^{(2)}+4A_{\twoscalar}^{(2)}+2A_{\withcontact}^{(2)}\,.\\
  \end{aligned}
  \label{eq:two_loop_AD1_AD2_decomposition}
\end{equation}
When summing over the scalar-like degrees of freedom, each scalar loop in
leads to a prefactor of
$(D_i-D_0)$ in the decomposition of $A_i^{(2)}$.
In $A_2^{(2)}$, we find a contribution from the scalar contact term,
which requires different scalars in the two loops.  There are
$(D_2-D_0)(D_2-D_0-1)$ such combinations.

Substituting \eqn{eq:two_loop_AD1_AD2_decomposition} into
\eqn{eq:A_DS_sampling_2L}, we finally obtain
\begin{equation}
  \label{eq:A2L_dim_reconstruction}
  A_{D_s}^{(2)}=A_{0}^{(2)}+(D_s-D_0)A_{\onescalar}^{(2)}+(D_s-D_0)^2
  A_{\twoscalar}^{(2)}+(D_s-D_0)(D_s-D_0-1)A_{\withcontact}^{(2)}.
\end{equation}

\def\hspA{\hspace*{5mm}}
\begin{figure}[tbh]
  \centering
  \includegraphics[height=.08\linewidth,valign=c]{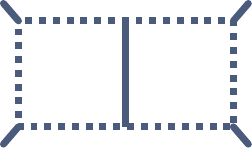}\hspA
  \includegraphics[height=.08\linewidth,valign=c]{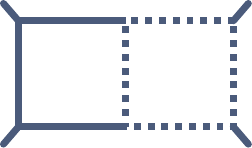}\hspA
  \includegraphics[height=.08\linewidth,valign=c]{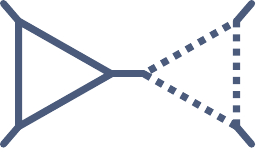}\hspA
  \includegraphics[height=.08\linewidth,valign=c]{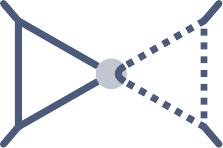}\hspA
  \includegraphics[height=.08\linewidth,valign=c]{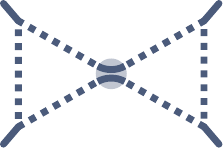}\hspA
  \includegraphics[height=.08\linewidth,valign=c]{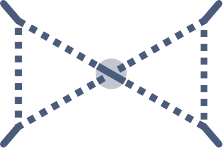}
  \caption{Diagrams with a single scalar loop contributing to the four-point amplitude.}
  \label{FourPointOneScalarLoop}
\end{figure}

\begin{figure}[tbh]
  \centering
  \includegraphics[height=.08\linewidth,valign=c]{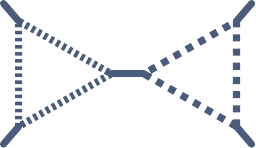}\hspA
  \includegraphics[height=.08\linewidth,valign=c]{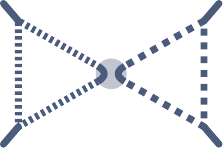}
  \caption{Diagrams with a two scalar loops contributing to the four-point amplitude.}
  \label{FourPointTwoScalarLoops}
\end{figure}

We can compare this result to
similar approaches to two-loop
computations in the literature.  As an example, consider the BFZ derivation~\cite{Badger:2013gxa}
of the four- and five-gluon all-plus amplitudes.
BFZ obtained these amplitudes by decomposing the integrand,
\begin{equation}
  \label{eq:A2L_integrand_reconstruction}
  \Delta_{D_s}^{(2)}=\Delta_{6}^{(2)}+(D_s-6)\Delta_{6,s}^{(2)}+(D_s-6)^2\Delta_{6,ss}^{(2)}\,,
\end{equation}
where $\Delta_{6}^{(2)}$ is the full integrand of the six-dimensional
amplitude, while $\Delta_{6,s}^{(2)}$ and $\Delta_{6,ss}^{(2)}$ are
integrands from diagrams involving the extra-dimensional scalars.  In the
four-point amplitude, these scalar integrands correspond respectively to
contributions from diagrams with a single scalar loop shown
in~\fig{FourPointOneScalarLoop}
and diagrams with two distinct scalar loops shown in \fig{FourPointTwoScalarLoops}.

We must be careful in separating the contributions containing
a four-scalar vertex, which necessarily have two scalar loops.
We assign the contributions with identical scalars in 
loops to $\Delta_{6,s}^{(2)}$, while
putting the contributions with different scalars in $\Delta_{6,ss}^{(2)}$.

For consistency between eq.~\eqref{eq:A2L_dim_reconstruction} and
eq.~\eqref{eq:A2L_integrand_reconstruction} with $D_0=6$ we need 
to have
the correspondence~\cite{AccettulliHuber:2019abj},
\begin{equation}
  \label{eq:dimReconstruction_example_integrand_to_amplitude}
  \Delta_{6,s}^{(2)}\to \bigl[ A_{\onescalar}^{(2)}-A_{\withcontact}^{(2)}\bigr]\big|_{D_0=6}\,,\quad
  \Delta_{6,ss}^{(2)} \to  \bigl[A_{\twoscalar}^{(2)}+A_{\withcontact}^{(2)}\bigr]\big|_{D_0=6}\,,
\end{equation}
after integration.
For $\Delta_{6,ss}^{(2)}$ we see that this is indeed the case, as the
diagrams shown in \fig{FourPointTwoScalarLoops} match the definitions
of $A_{\twoscalar}^{(2)}$ and $A_{\withcontact}^{(2)}$.
For $\Delta_{6,s}^{(2)}$,
we can obtain agreement using the relation,
\begin{equation}
  A_{6,1,1}^{(2)}\simeq \includegraphics[height=1cm,valign=c]{dim_reconstruction/A2_11_1_picto} =
  -\includegraphics[height=1cm,valign=c]{dim_reconstruction/A2_11_single_scalar_1_picto}-
  \includegraphics[height=1cm,valign=c]{dim_reconstruction/A2_11_single_scalar_2_picto},
\end{equation}
which follows from the quartic scalar Feynman rules of
eq.~\eqref{eq:Feyn_rules_scalars}.

\FloatBarrier
Adding and subtracting terms to \eqn{eq:A2L_dim_reconstruction},
we can introduce an additional sampling dimension $D_0'$,
which may be smaller than four,
\begin{equation}
  \begin{aligned}
    A_{D_s}^{(2)}= \,&A_{0}^{(2)}+(D_0'-D_0)A_{\onescalar}^{(2)}
    +(D_0'-D_0)^2A_{\twoscalar}^{(2)}
    +(D_0'-D_0)(D_0'-D_0-1)A_{\withcontact}^{(2)}\\
    &+(D_s-D_0')\bigl[A_{\onescalar}^{(2)}
    +2(D_0'-D_0)(A_{\twoscalar}^{(2)}+A_{\withcontact}^{(2)})\bigr]\\
    &+(D_s-D_0')^2A_{\twoscalar}^{(2)}+(D_s-D_0')(D_s-D_0'-1)A_{\withcontact}^{(2)}.
  \end{aligned}
  \label{eq:A2_D0D0P_shift}
\end{equation}

\FloatBarrier

\subsection{Separability and Two-Loop Rational Terms}
\label{subsec:separability}

\def\Rational{R^{(2)}}
\def\onehalf{{\textstyle\frac{\textrm{\small1}}{\lower2pt\hbox{\textrm{\small2}}}}}
\def\onequarter{{\textstyle\frac{\textrm{\small1}}{\lower2pt\hbox{\textrm{\small4}}}}}
BMP showed~\cite{Badger:2016ozq} how to decompose the two-loop
all-plus amplitudes into polylogarithmic terms $P_n^{(2)}$
and rational parts $\Rational_n$ associated to
different powers of the state dimension $D_s$.  More precisely, through $\Ord(\eps^0)$,
BMP conjectured that these
terms are associated to different powers of $D_s-2$,
\newcommand{\PFinite}{P}
\newcommand{\Rat}{R}
\begin{equation}
  F^{(2)}_{n:1}(1^+\mathellipsis n^+)=
  \onehalf (D_s-2)\PFinite^{(2)}(1^+\mathellipsis n^+)
  +\onequarter (D_s-2)^2\Rat^{(2)}(1^+\mathellipsis n^+)+\mathcal{O}(\eps)\,.
  \label{eq:two-loop_finite_statement}
\end{equation}
BMP verified this decomposition for the five- and six-gluon leading
color partial amplitudes by integrating the terms in the two-loop integrands proportional
to $(D_s-2)^2$.  They then checked the resulting analytic expressions numerically
against the known results of refs.~\cite{Dunbar:2016}.

\def\ExtractFinite{\mathcal{F}}
Let us now connect the BMP decomposition to the dimensional reconstruction picture
explained in the previous subsection.   Choose the base-dimension
$D_0=6$, and use the rearrangement of \eqn{eq:A2_D0D0P_shift} with
$D_0'=2$.
We then find,
\begin{equation}
  \begin{aligned}
    \PFinite^{(2)}&=2\bigl[\PFinite_{\onescalar}^{(2)}-2\times 4\,(\PFinite_{\twoscalar}^{(2)}+\PFinite_{\withcontact}^{(2)})-
    \PFinite_{\withcontact}^{(2)}\bigr]\,,\\
    \RT &= 4\bigl[\Rat_{\twoscalar}^{(2)}+\Rat_{\withcontact}^{(2)}\bigr]\,,
  \end{aligned}
  \label{eq:back_sep_PR_dimensional_reconstruction}
\end{equation}
where $\PFinite_{\onescalar}$, $\PFinite_{\twoscalar}$ and $\PFinite_{\withcontact}$
are the finite polylogarithmic pieces of $A_{\onescalar}$, $A_{\twoscalar}$ and
$A_{\withcontact}$, while $\Rat_{\twoscalar}$, $\Rat_{\withcontact}$ are the
corresponding rational parts.
As we will often consider the sum of $\Rat_{\twoscalar}$ and
$\Rat_{\withcontact}$ for the rational parts of the all-plus, we will
also use the shorthand,
\newcommand{\twoscalarcontact}{ss}
\begin{equation}
  \RT_{\twoscalarcontact}=\Rat_{\twoscalar}+\Rat_{\withcontact}\,,
\end{equation}
following ref.~\cite{Badger:2013gxa} and
\eqn{eq:dimReconstruction_example_integrand_to_amplitude}.

Rephrasing the problem of computing $\Rat$ in terms
of $\Rat_{\twoscalar}^{(2)}$ and~$\Rat_{\withcontact}^{(2)}$ reduces its
computational complexity.
For $A_{\twoscalar}^{(2)}$, the two loops are connected through the
exchange of a single gluon carrying no loop momentum; in
$A_{\withcontact}^{(2)}$, the two scalar
lines of the two loops meet directly, but only in a four-scalar
contact term.
In neither case do both loop momenta appear in a single propogator, so that
all required two-loop integrals
factorize into a product of one-loop integrals.
Because of the manner of
simplification, we call the BMP conjecture the 
\emph{separability\/}
conjecture.

We can pick a basis of integrals which factorize as well.
To determine
the coefficients of these integrals in $A_{\twoscalar}^{(2)}$ and
$A_{\withcontact}^{(2)}$ using generalized unitarity cuts in
each loop.  We treat the loops sequentially.  The first loop is
computed from tree amplitudes using the standard
one-loop unitarity technique; the second loop is computed
from tree amplitudes and the coefficient computed at
the first step, with the latter playing the role of
a tree amplitude.  In the second step,
we again use the one-loop unitarity technique.
We can pick either order for the two
loops.
The coefficient of the required two-loop integral is then 
the result of this two-stage computation.

\begin{figure}
  \centering
  \includegraphics[width=0.35\textwidth]{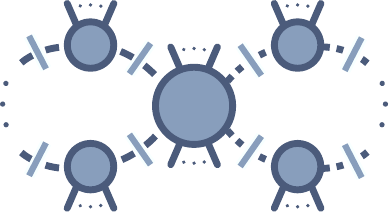}
  \caption{
    A generic two-loop cut needed for $A_{\twoscalar}^{(2)}$ or
    $A_{\withcontact}^{(2)}$.
    The two loops are connected only by a central
    tree amplitude, which contains no propagator dependent on both
    loop momenta.
    As we will see later, the shown attachment of the external gluons
    is associated to the leading-color trace structure
    $N_c^2 \Tr$.
  }
  \label{fig:general_cut}
\end{figure}

BMP made the separability conjecture for the leading-color partial amplitudes~\cite{Badger:2016ozq}.
We extend the conjecture to rational contributions
in subleading-color all-plus partial amplitudes as well, 
more specifically to those in
nonplanar partial amplitudes.  In later sections,
we will discuss the
construction of the non-planar versions of $A_{\twoscalar}^{(2)}$ and
$A_{\withcontact}^{(2)}$ via color-dressed unitarity, similar to the procedure for
polylogarithmic contributions of refs.~\cite{DalgleishDunbar:2020,Dunbar:2020wdh}.
We provide evidence for this extended conjecture
through comparison of direct unitary-based
computations of $A_{\twoscalar}^{(2)}$ and $A_{\withcontact}^{(2)}$ for several
subleading-color amplitudes to results in the
literature~\cite{DalgleishDunbar:2020}.  These comparisons suggest
that \eqn{eq:back_sep_PR_dimensional_reconstruction} 
does indeed hold
for all partial amplitudes in the color decomposition of two-loop all-plus
amplitudes.

\def\ellb{\bar\ell}
\def\elle{\breve{\ell}}
\subsection{\texorpdfstring{$D$}{D}-Dimensional Generalized Unitarity}
\label{subsec:d-dim_unitarity}
The dimensional reconstruction procedure described earlier makes use of various
integer dimensions for state-counting and matching in numerators of integrands.  For
the full integrands and for integrals, we of course must consider general dimensions
$D=4-2\eps$ for the loop momenta.

As usual in modern applications of dimensional regularization, we take the external momenta
to be strictly four-dimensional.  We split the loop momenta into their four- and extra-dimensional
components. The former we have already labeled $\ellb$;
the latter we label $\elle$,
\begin{equation}
  \ell_i = \ellb_i + \elle_i\,.
\end{equation}
Because the external kinematics is four-dimensional, the $\elle_i$ are
conserved within each loop, and they can only appear in the Lorentz invariants,
\begin{equation}
  \mu_{i}^2\equiv -(\elle_i\cdot\elle_i)\,,\qquad \mu_{12}\equiv - (\elle_1\cdot\elle_2)\,.
\end{equation}
Separability ensures that the second invariant can only appear in the numerator; and
in the numerator, performing either loop integral will make odd powers of it vanish, and
will replace even powers with the result of integrating over powers of $\mu_1^2\mu_2^2$.

At this point we can follow the standard approach~\cite{Badger:2008cm} for using
one-loop $D$-dimensional
generalized unitarity to compute rational terms.  We separate the loop integrals
into four- and extra-dimensional integrals,
\begin{equation}
  \int \frac{d^D\ell}{(2\pi)^D} =
  \int\frac{d^{-\eps} \mu^2}{(2\pi)^{-2\eps}}\,\int
  \frac{\diff^4 \ellb}{(2\pi)^4}\,.
\end{equation}
In cutting $D$-dimensional propagators, the massless on-shell condition for
$\ell_i$ is then equivalent to a massive on-shell condition for $\ellb_i$,
\begin{equation}
  \ell_i^2=0\ \Rightarrow \ \ellb_i^2=\mu_i^2\,,
\end{equation}
where the latter can be thought of as constant for the inner integral.  It corresponds
to the scalar mass in the dimensional reconstruction approach discussed earlier.
Were we to proceed and use generalized unitarity to obtain coefficients with respect
to the inner integration, we would obtain expressions with rational dependence
on the external spinor variables, and polynomial dependence on the $\mu_i^2$.
Terms independent of $\mu_i^2$ correspond to four-dimensional
cuts, and will lead to coefficients of the usual integrals.  Terms with nontrivial
powers of $\mu_i^2$ contain extra suppression, and will lead to either purely
rational terms, or to terms of $\Ord(\eps)$.
The maximal power of
$\mu_i^2$ is limited by power counting: for boxes in Yang--Mills theories
we get at most $\mu_i^4$, while for triangles and bubbles, at most
$\mu_i^2$ is allowed.  The box $\mu_i^2$ integral is of $\Ord(\eps)$,
and therefore does not contribute.

A generic one-loop amplitude may therefore be written as follows~\cite{Badger:2008cm},
\begin{equation}
  \begin{aligned}
    A^{(1)}&=\sum_{\textrm{boxes}} \CBox_{,[0]} I^D_4[1]
    +\sum_{\textrm{triangles}} \CTriangle_{,[0]}I_3[1]+
    \sum_{\textrm{bubbles}} \CBubble_{,[0]}I_2[1]\\
    &+\sum_{\textrm{boxes}} \CBox_{,[4]}I^D_4[\mu^4]
    +\sum_{\textrm{triangles}} \CTriangle_{,[2]}I_3[\mu^2]
    +\sum_{\textrm{bubbles}} \CBubble_{,[2]} I^D_2[\mu^2]+\Ord(\eps)\,,
  \end{aligned}
  \label{eq:D-Dim_one-loop_basis}
\end{equation}
where $I^D_n[\mathcal{N}]$ denotes an $n$-point $D$-dimensional
integral with $\mathcal{N}$ inserted in its numerator.
The coefficients $C_{\textrm{X},[j]}^{(1)}$ (corresponding to the
$j^{\textrm{th}}$ power of $\mu$ in integral X), are rational functions of the
external spinors.
The integrals on the first line, without powers of $\mu^2$, can be obtained using
four-dimensional generalized unitarity.
They correspond to the parts of the amplitude that have branch cuts;
in the case of the two-loop all-plus amplitude, these terms are known from
refs.~\cite{Gehrmann:2015bfy,Dunbar:2016aux,Dunbar:2016gjb,Dunbar:2016cxp,Badger:2019djh,
  Dunbar:2019fcq,Dalgleish:2020mof,Dunbar:2020wdh}.
The integrals on the second line of \eqn{eq:D-Dim_one-loop_basis} give rise to
the rational parts of the amplitudes, using their values~\cite{Badger:2008cm},
\begin{equation}
  \begin{aligned}
    I^D_4[\mu^4] &= -\frac{1}{6}+\mathcal{O}(\epsilon)\,,\\
    I^D_3[\mu^2] &= -\frac{1}{2}+\mathcal{O}(\epsilon)\,,\\
    I^D_2[\mu^2] &= -\frac{s}{6}+\mathcal{O}(\epsilon)\,.
  \end{aligned}
  \label{eq:D-dim_mu2_integrals}
\end{equation}
Here, $s$ is the square of the momentum flowing through the bubble.
The coefficients are given in terms of tree amplitudes,
\begin{equation}
  \begin{aligned}
    \CBox_{} &\equiv \CBox_{,[4]} \equiv \CBox_{,[4]}(A^\tree_1,A^\tree_2,A^\tree_3,A^\tree_4)\,,
    \\
    \CTriangle_{} &\equiv \CTriangle_{,[2]} \equiv \CTriangle_{,[2]}(A^\tree_1,A^\tree_2,A^\tree_3)\,,
    \\
    \CBubble_{} &\equiv \CBubble_{,[2]} \equiv \CBubble_{,[4]}(A^\tree_1,A^\tree_2)\,,
  \end{aligned}
\end{equation}
where the tree-amplitude arguments are understood to be 
functions of the
parametrizations required to extract appropriate terms.
In order to simplify the notation, we shall omit the 
subscript for the power of $\mu$ in
coefficients contributing to the rational term.

The one-loop rational terms then have the form,
\begin{equation}
  \ROne = -\frac16\sum_{\textrm{boxes}} \CBox_{}
  -\frac12\sum_{\textrm{triangles}} \CTriangle_{}
  -\frac16\sum_{\textrm{bubbles}} \CBubble_{}s^{\vphantom{(1)}}_{\textrm{Bub}}+\Ord(\eps)\,.
\end{equation}

\def\lloop{\textrm{L}}
\def\rloop{\textrm{R}}
\def\BoxBox{{\textrm{Box},\textrm{Box}}}
\def\BoxTri{{\textrm{Box},\textrm{Tri}}}
\def\BoxBub{{\textrm{Box},\textrm{Bub}}}
\def\TriBox{{\textrm{Tri},\textrm{Box}}}
\def\TriTri{{\textrm{Tri},\textrm{Tri}}}
\def\TriBub{{\textrm{Tri},\textrm{Bub}}}
\def\BubBox{{\textrm{Bub},\textrm{Box}}}
\def\BubTri{{\textrm{Bub},\textrm{Tri}}}
\def\BubBub{{\textrm{Bub},\textrm{Bub}}}
\def\CTwo{\coeff^\twoloop}

In the two-loop amplitudes $\ATwo_{\twoscalar}$ and $\ATwo_{\withcontact}$,
separability allows us to treat the loops consecutively.
Their integral bases include all two-loop integrals which
factorize into a product of one-loop integrals.
We call such two-loop integrals ``one-loop squared''.

As box, triangle and bubble integrals form a basis at one-loop,
a basis of one-loop squared integrals is given by their unique products.
There are six unique classes of such integrals,
\begin{equation}
  \label{eq:IntegralClasses}
  \begin{alignedat}{2}
    & I^{(2),D}_{\BoxBox}&&=\ I^{(1),D}_{4}[\mu^4]\times I^{(1),D}_{4}[\mu^4]\quad \sim\quad \includegraphics[valign=c,height=1.5em]{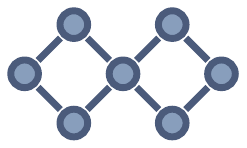},\\
    & I^{(2),D}_{\BoxTri}&&=\ I^{(1),D}_{4}[\mu^4]\times I^{(1),D}_{3}[\mu^2]\quad\sim\quad \includegraphics[valign=c,height=1.5em]{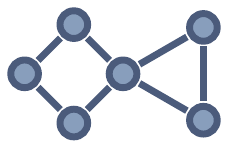},\\
    & I^{(2),D}_{\BoxBub}&&=\ I^{(1),D}_{4}[\mu^4]\times I^{(1),D}_{2}[\mu^2]\quad\sim\quad \includegraphics[valign=c,height=1.5em]{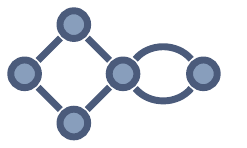},\\
    & I^{(2),D}_{\TriTri}&&=\ I^{(1),D}_{3}[\mu^2]\times I^{(1),D}_{3}[\mu^2]\quad\sim\quad \includegraphics[valign=c,height=1.5em]{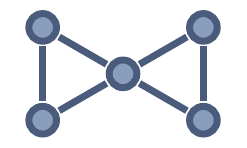},\\
    & I^{(2),D}_{\TriBub}&&=\ I^{(1),D}_{3}[\mu^2]\times I^{(1),D}_{2}[\mu^2]\quad\sim\quad \includegraphics[valign=c,height=1.5em]{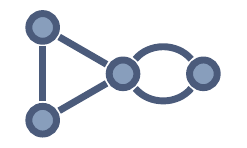},\\
    & I^{(2),D}_{\BubBub}&&=\ I^{(1),D}_{2}[\mu^2]\times I^{(1),D}_{2}[\mu^2]\quad\sim\quad \includegraphics[valign=c,height=1.5em]{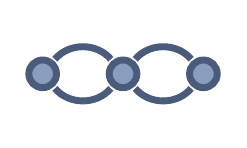},\\
  \end{alignedat}
\end{equation}
As we are only interested in the rational 
parts $\RT_{\twoscalar}$ and
$\RT_{\withcontact}$, we take these integrals to always 
include $\mu^2$ or $\mu^4$ numerators for both loops.
We will sometimes refer to $I^{(2),D}_{\TriTri}$
and $I^{(2),D}_{\BubBub}$ as `bow-tie' and `spectacles' integrals.
In a complete basis of one-loop squared integrals we 
include integrals in these classes for all unique 
arrangements of the external momenta.

Each unique one-loop squared integral is accompanied by a coefficient.
As for the integrals, there are six unique classes of such coefficients,
which we label $\CTwo_{\BoxBox}$, $\CTwo_{\BoxTri}$, $\CTwo_{\BoxBub}$,
$\CTwo_{\TriTri}$, $\CTwo_{\TriBub}$ and $\CTwo_{\BubBub}$.
As mentioned in the previous section, the coefficients can be
determined by computing a one-loop coefficient with the other
loop's one-loop coefficient nested inside alongside tree
amplitudes.
As an example, 
in the one-loop unitarity language introduced above, the coefficient
$\CTwo_{\TriTri}$ of an integral $I_{\TriTri}^{(2),D}$ would
be computed as follows,
\begin{equation}
  \CTwo_\TriTri =
  \CTriangle_{,[2]}(A_{1}^\tree, A_{2}^\tree ,
  \CTriangle_{,[2]}(A_{3}^\tree , A_{4}^\tree , A_{\central}^\tree))\,.
\end{equation}
Here, the amplitudes $\ATree_{1,2}$ and $\ATree_{3,4}$ 
are those of the
two loops, which are themselves connected by $\ATree_{\central}$.
Note that in the computation of the ``outer'' coefficient, the
``inner'' coefficient takes the role of an amplitude.
As the two loops are equivalent, the choice of which loop's
coefficient to compute first is arbitrary.
We could therefore just as well determine the same coefficient via
\begin{equation}
  \CTwo_\TriTri =
  \CTriangle_{,[2]}(A_{3}^\tree, A_{4}^\tree ,
  \CTriangle_{,[2]}(A_{1}^\tree , A_{2}^\tree , A_{\central}^\tree))\,.
\end{equation}

\def\vph{\vphantom{(1)}}
To summarize, the rational parts $\RT_{\twoscalar}$ and $\RT_{\withcontact}$
can be expressed as
\begin{equation}
  \begin{aligned}
    \RTwo_{\twoscalar,\withcontact} &= \frac{1}{36}\sum_{\textrm{box--boxes}} \CTwo_\BoxBox+
    \frac{1}{12}\sum_{\textrm{box--triangles}} \CTwo_\BoxTri +
    \frac{1}{36}\sum_{\textrm{box--bubbles}} \CTwo_\BoxBub s_{\textrm{Bub}}^{\vph} \\
    &\hphantom{{}={}}+\frac{1}{4}\sum_{\textrm{triangle--triangles}} \CTwo_\TriTri  +
    \frac{1}{12}\sum_{\textrm{triangle--bubbles}} \CTwo_\TriBub s_{\textrm{Bub}}^{\vph} \\
    &\hphantom{{}={}}+\frac{1}{36}\sum_{\textrm{bubble--bubbles}} \CTwo_\BubBub s_{\textrm{Bub}}^{\vph} s_{\textrm{Bub}}^{\vph}
    +\Ord(\eps)\,.
  \end{aligned}
\end{equation}

\FloatBarrier

\section{Color Structure and Generating Sets of Cuts}

\label{ColorStructureSection}
\label{sec:color_structure}

\def\TTr{\mathop{\rm ITr}\nolimits}
\def\modout#1#2{{#1}/{#2}}
\def\sufrac#1#2{{#1}/{#2}}

We can write a complete 
two-loop $\text{SU}(N_c)$ Yang--Mills amplitude as
follows~\cite{DalgleishDunbar:2020},
\begin{equation} 
  \begin{aligned}
    \mathcal{A}^{(2)}_n &=
    N_c^2 \sum_{\sigma\in \modout{S_n}{Z_n}}
    \TTr\bigl(\sigma(1\mathellipsis n)\bigr)
    \,A^{(2)}_{n:1}\bigl(\sigma(1\mathellipsis n)\bigr)
    \\
    & +N_c \sum_{r=3}^{\lfloor \sufrac{n}{2}\rfloor +1} \sum_{\sigma\in \modout{S_{n}}{P_{n:r}}}
    \TTr\bigl(\sigma(1\mathellipsis(r-1))\bigr)\TTr\bigl(\sigma(r\mathellipsis n)\bigr)
    \\[-2mm]
    &\hspace{9em}\times A^{(2)}_{n:r}\bigl(\sigma(1\mathellipsis(r-1);r\mathellipsis n)\bigr)
    \\
    &+\sum_{r=2}^{\lfloor \sufrac{n}{2}\rfloor}\sum_{k=r}^{\lfloor \sufrac{(n-r)}{2}\rfloor}
    \sum_{\sigma\in\modout{S_{n}}{P_{n:r,k}}}\TTr\bigl(\sigma(1\mathellipsis r)\bigr)
    \TTr\bigl(\sigma((r+1)\mathellipsis(r+k))\bigr)
    \\ &\hspace*{11em}\times
    \TTr\bigl(\sigma((r+k+1)\mathellipsis n)\bigr)
    \\
    &\hspace{9em}\times A^{(2)}_{n:r,k}\bigl(\sigma(1\mathellipsis r;(r+1)\mathellipsis(r+k);
    (r+k+1)\mathellipsis n)\bigr)\\
    &+\sum_{\sigma\in \modout{S_n}{Z_n}}\TTr\bigl(\sigma(1\mathellipsis n)\bigr)
    A^{(2)}_{n:1\mathrm{B}}\bigl(\sigma(1\mathellipsis n)\bigr)\,,
  \end{aligned}
  \label{eq:two-loop_color_decomposition}
\end{equation}
where the traces are,
\begin{equation}
  \begin{aligned}
    \TTr(i_1\mathellipsis i_n) = \Tr(T^{a_{i_1}}\mathellipsis T^{a_{i_n}})\,.
  \end{aligned}
\end{equation}
For the double- and triple-trace color-ordered
amplitudes, the sums over $r$ and $k$
are chosen such that the traces are 
ordered in increasing length.
By $S_n$ and $Z_n$ we denote as usual the symmetric and 
cyclic groups on $n$ objects respectively.
The permutation sets appearing above are,
\begin{equation}
  \begin{aligned}
    P_{n:r}&=\begin{cases}
      Z_{r-1}\times Z_{r-1}\times S_2,& n=2(r-1)\,,\\
      Z_{r-1}\times Z_{n-r-1},& \text{otherwise}\,;
    \end{cases},\\
    P_{n:r,k}&=\begin{cases}
      Z_{r}\times Z_{r}\times Z_{r}\times S_3,& r=k,n=3r\,,\\
      Z_{r}\times Z_{k}\times Z_{k}\times S_2,& n=r+2k\,,\\
      Z_{r}\times Z_{r}\times Z_{n-2r}\times S_2,& r=k\,,\\
      Z_{r}\times Z_{k}\times Z_{n-r-k},& \text{otherwise}\,.
    \end{cases}
  \end{aligned}
\end{equation}
These permutation sets account for the cyclic symmetry of the traces
($C_i$), as well as the exchanges of equal-length traces ($S_i$).
The factors of $N_c$ in the decomposition may be interpreted as
traces containing the identity $\mathbbm{1}_{N_c}$.

Our first task is to identify the two-loop cut 
topologies that contribute to the
different trace structures in full-color all-plus amplitudes.
It is useful to consider the corresponding
string-theory amplitude,
and use its color structure there as a guide for those in 
Yang--Mills theory~\cite{Bern:1990ux}.
This analysis was carried out at two loops in ref.~\cite{Dunbar:2020wdh}.

\begin{figure}
  \makebox[\textwidth][c]{
    \includegraphics[width=0.4\textwidth]
    {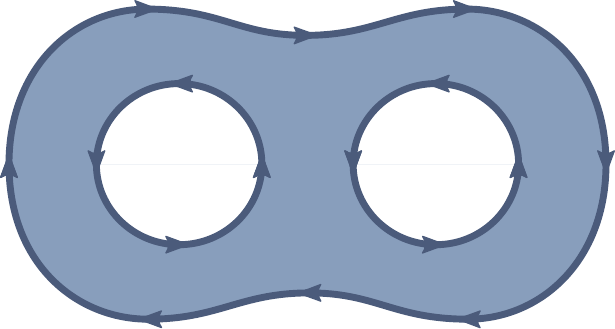}
  }
  \caption{The Yang--Mills partial amplitudes $A^{\twoloop}_{n:1}$, $A^{\twoloop}_{n:r}$ and
    $A^{\twoloop}_{n:r,k}$ are associated to one-, two-, and three-trace color structures
    also present in three-boundary orientable genus-two surfaces in open string theory depicted
    here.}
  \label{fig:two_puncture_disc_surface}
\end{figure}

Two-loop amplitudes in gauge theory can be obtained from taking the
infinite-tension limit of the corresponding
amplitudes in open string theory.  The latter amplitudes are
given as integrals over orientable
world sheets of genus two with at least one boundary.  There are two such
surfaces: the disc with two punctures, which has three boundaries, and the
punctured torus, which has only a single boundary.
We show representations of these two surfaces
in~\figs{fig:two_puncture_disc_surface}{fig:punctured_torus_surface}.

\begin{figure}
  \centering
  \captionsetup[subfigure]{justification=centering}
  \begin{subfigure}[]{0.45\textwidth}
    \includegraphics[width=\textwidth]{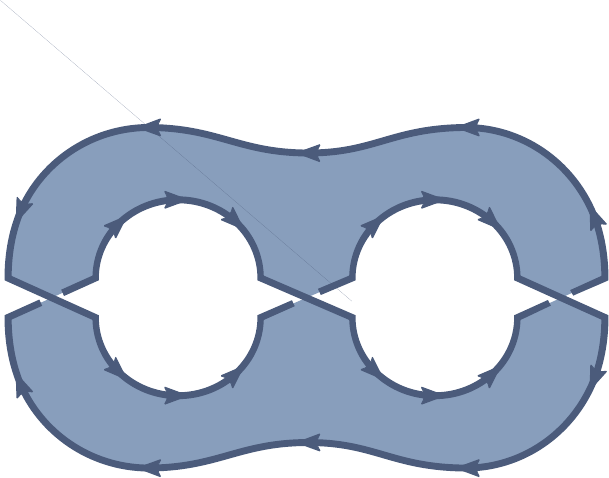}
    \caption{\label{subfig:1B_topo_twisted}}
  \end{subfigure}\\[3em]
  \begin{subfigure}[]{0.3\textwidth}
    \includegraphics[width=\textwidth]{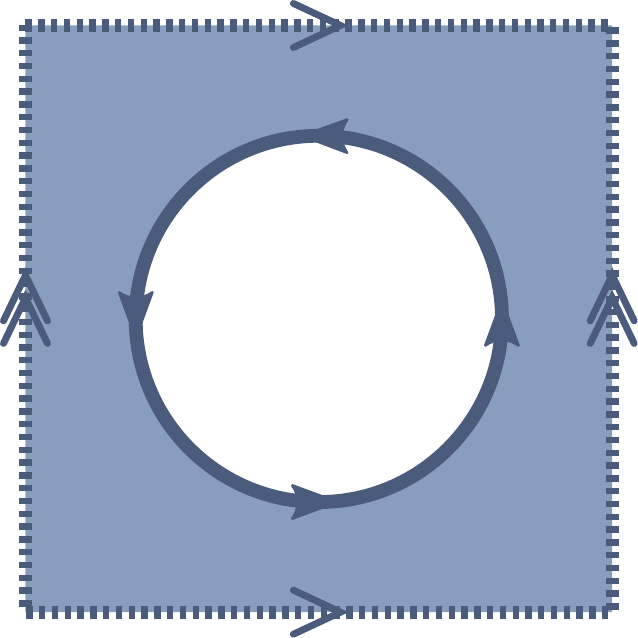}
    \caption{\label{subfig:1B_topo_punctured_torus}}
  \end{subfigure}
  \hspace{5em}
  \begin{subfigure}[]{0.3\textwidth}
    \includegraphics[width=\textwidth]{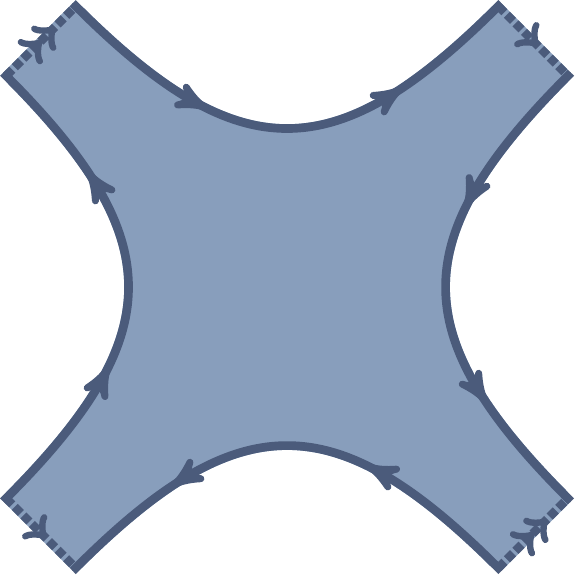}
    \caption{\label{subfig:1B_topo_untwisted}}
  \end{subfigure}
  \caption{The Yang--Mills partial amplitudes $A^{(2)}_{n:1\mathrm{B}}$ are associated
    to single-trace color structures with no accompanying 
    factors of $N_c$.  These
    structures are also present in contributions to open-string amplitudes arising
    from orientable single-boundary genus-two surfaces.
    Three equivalent
    representations are shown: (a) as obtained
    from the doubly punctured disk of
    \fig{fig:two_puncture_disc_surface}
    (b) as equivalent to a punctured torus
    (c) the most useful representation for constructing the
    required unitarity cuts for the rational contributions to
    $A^{(2)}_{n:\oneB}$.
    In (b) and (c) the dotted
    lines must be sewn together with orientation given by
    their arrows.}
  \label{fig:punctured_torus_surface}
\end{figure}

Open strings are dressed with color factors $T^a_{ij}$ at their ends.  Such factors
are also inserted along world-sheet boundaries by the vertex operators coupling
external states. The color-factor indices are contracted along each boundary,
giving rise to color or Chan--Paton~\cite{Paton:1969je} factors, one trace per boundary.
Traces with no vertex-operator insertion give rise to factors of $N_c$.

World sheets with the topology of \fig{fig:two_puncture_disc_surface}
thus give rise to the color structures $N_c^2\Tr$,
$N_c\Tr\Tr$, and $\Tr\Tr\Tr$, while world
sheets with the topology of \fig{fig:punctured_torus_surface} give rise
to single-trace structures with no accompanying factors of $N_c$.
These are exactly the color structures we expect in Yang--Mills theories,
and present in \eqn{eq:two-loop_color_decomposition}.

\begin{figure}[tbh]
  \captionsetup[subfigure]{justification=centering}
  \begin{subfigure}[]{0.49\linewidth}
    \includegraphics[valign=c]
    {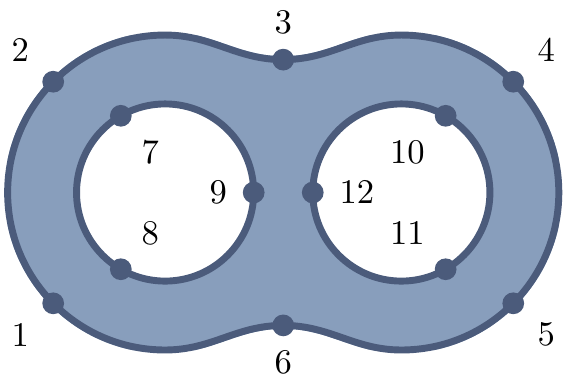}
    \caption{}
    \label{fig:cut_punctured_disk_topo}
  \end{subfigure}
  \begin{subfigure}[]{0.49\linewidth}
    \includegraphics[valign=c]
    {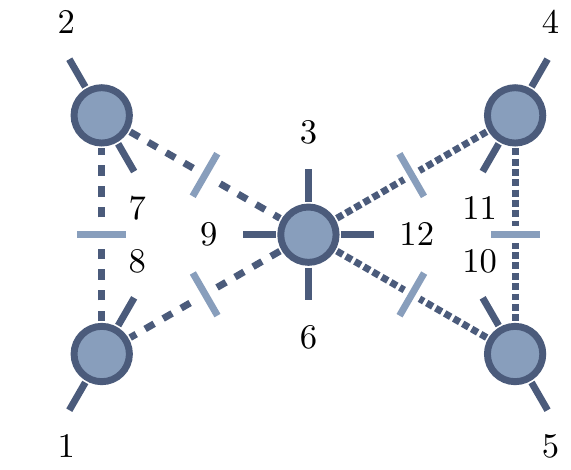}
    \caption{}
    \label{fig:cut_punctured_disk_cut}
  \end{subfigure}
  \caption{An example of a one-loop-squared cut contributing to the $A_{12;6,9}$ partial
    amplitude.
    As explained in the text, it contributions to the coefficient of the triple-trace
    structure $\TTr(1, 2, 3 ,4, 5, 6) \TTr(7, 8, 9)
    \TTr(10,11,12)$.}
  \label{fig:color_routing_example}
\end{figure}

We can reverse the sewing implicit in the Chan--Paton factors to obtain
the sets of cuts needed for the full color-dressed amplitude, and
thence the color-ordered cuts needed for each of the partial amplitudes.
For example, consider the contributions from the cut shown in
\fig{fig:color_routing_example}.
All tree amplitudes appearing in the cut are color-ordered.
Dressing the tree amplitudes with the
color factors imposed by the cyclic ordering of their legs, the
entire cut is associated to the color structure,
\begin{equation}
  \begin{aligned}
    &\TTr(1,a,8,f)\TTr(a,2,b,7)
    \TTr(b,3,c,12,e,6,f) \TTr(c,4,d,11)\Tr(d,5,e,10)\,.
  \end{aligned}
  \label{eq:sample_color_routing}
\end{equation}
Here, the indices $a,\ldots,f$ are implicitly summed over, as
they are sewn across cuts.
Using $U(N_c)$ Fierz identities, we can carry out this sewing,
ending up with a product of three traces,
\begin{equation}
  \label{eq:sample_color_routing_solution}
  \begin{aligned}
    &\TTr(1,2,3,4,5,6)\TTr(7,8,9)
    \TTr(10,11,12)\,.
  \end{aligned}
\end{equation}
recovering the three color traces we expected from
\fig{fig:two_puncture_disc_surface}.
Color flows clockwise in the outer trace,
and counter-clockwise in the two inner traces.

\begin{figure}[tbh]
  \captionsetup[subfigure]{justification=centering}
  \centering
  \begin{subfigure}{0.35\linewidth}
    \centering
    \includegraphics[height=12em]
    {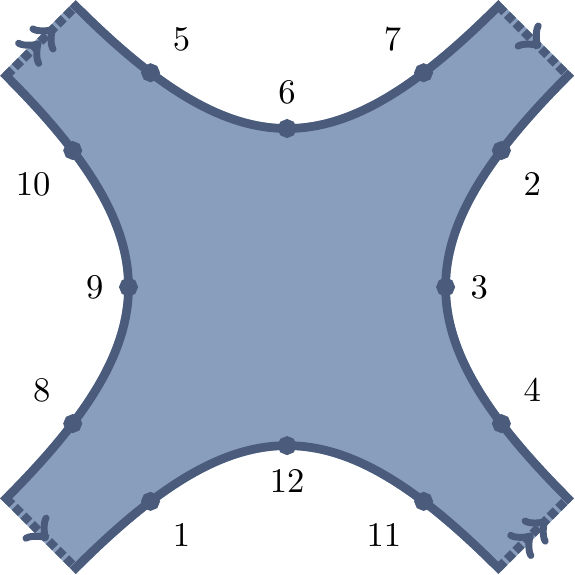}
    \caption{}
    \label{fig:1B_example_color_a}
  \end{subfigure}
  \hspace{3em}
  \begin{subfigure}{0.35\linewidth}
    \centering
    \includegraphics[height=12em]
    {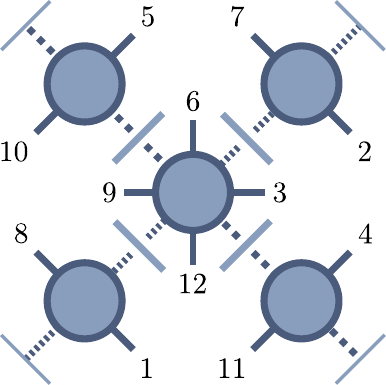}
    \caption{}
    \label{fig:1B_example_color_b}
  \end{subfigure}
  \caption{
    A graphical representation showing the heuristic relation between
    the string world-sheet and unitarity cuts of subleading single-trace
    amplitudes.
    The shown configuration belong to the color structure
    $\TTr(1,2,\ldots,11,12)$.
    In both cases, the dotted lines need to be sewn together
    according to the arrows shown.}
  \label{fig:1B_example_color}
\end{figure}

The partial amplitudes $A_{n:1\mathrm{B}}^{(2)}$,
on the other hand, arise
from configurations which follow the structure of the
punctured-torus worldsheet shown in~\fig{fig:punctured_torus_surface}.
\Fig{subfig:1B_topo_twisted} shows the relation of this world-sheet
to the doubly punctured disk, while
\fig{subfig:1B_topo_punctured_torus} shows that this surface is indeed
equivalent to a punctured torus.
Finally, the representation of \fig{subfig:1B_topo_untwisted} is the
one we will use to build color-ordered cuts.
The corresponding sewing of tree amplitudes is illustrated
at the one-loop squared cut shown in~\fig{fig:1B_example_color}.
The dotted lines at opposite corners are connected, similar
to the orientation-preserving sewing of segments in
\fig{subfig:1B_topo_untwisted}.

The main difference in one-loop squared cuts originating from the
doubly-punctured disk and punctured torus topologies lies in the
connection of the scalar lines at the central vertex.
In the former case, the two scalar lines are separated, 
while in the the latter they are required to cross.
It is this crossing that produces the expected single trace color
structure.
We can verify this fact by carrying out the color algebra explicitly.
For the particular cut shown in~\fig{fig:1B_example_color} we obtain,
\begin{equation}
  \label{eq:sample_color_routing_solution_1B}
  \begin{aligned}
    &\hphantom{={}}\TTr(1 ,a ,8 ,f)\TTr(a ,2 ,b ,7)
    \TTr(6 ,b ,3 ,c ,12 ,f ,9 ,e)\TTr(c ,4 ,d ,11)\TTr(d ,5 ,e ,10)\\
    &=\TTr(1 ,2 ,3 ,4 ,5 ,6 ,7 ,8 ,9
    ,10,11,12)\,,
  \end{aligned}
\end{equation}
as promised.
\FloatBarrier
\subsection{Leading-Color Partial Amplitudes}
\label{subsec:leading_color_cuts}
\begin{figure}
  \centering
  \includegraphics[width=0.6\textwidth,valign=c]
  {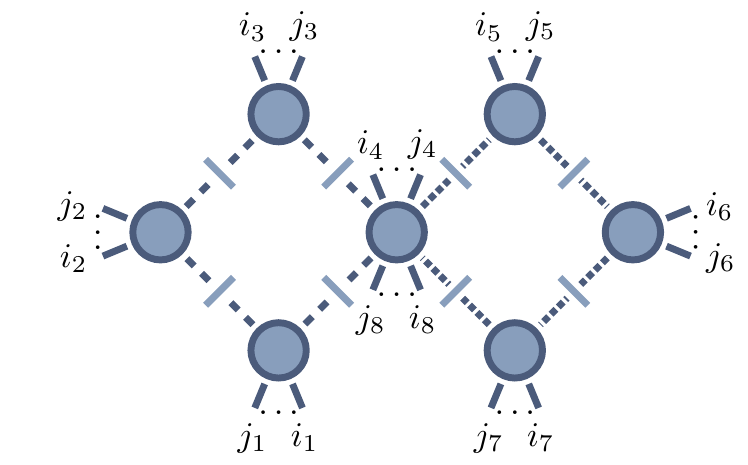}
  \caption{A generic cut for the leading-color rational part $\RTwo_{n:1}$}
  \label{fig:R2n1_generic_cut}
\end{figure}

Let us now describe the required sets of cuts.
We begin with the leading-color amplitude.
A generic cut of this amplitude is shown in \fig{fig:R2n1_generic_cut}.
It sews a product of tree amplitudes along the left and right loops,
joined by a four-scalar central amplitude.
For the latter, we must include both gluon-exchange and four-scalar
contact contributions.

To list all cuts systematically, we introduce labels for them.
The label of a leading-color cut consists of four sets of sequences,
\begin{equation}
  (\ESet_\lloop ; \ESet_\ctop ; \ESet_\rloop; \ESet_\cbottom)\,.
  \label{CutLabel}
\end{equation}
The four sets correspond, in order, to amplitudes on the left
loop ($\ESet_\lloop$); legs attached to the `top' of the central amplitude ($\ESet_\ctop$);
amplitudes on the right loop ($\ESet_\rloop$); and legs attached to the `bottom'
of the central amplitude ($\ESet_\cbottom$).
Each of these sets contains a number of sequences, which list the
external legs attached to a tree amplitude.
As the left and right loop are made up of either bubbles, triangles or boxes,
the sets $\ESet_\lloop$ and $\ESet_\rloop$ have either one, two or three of these sequences,
\begin{equation}
  \ESet_{\lloop} =
  \left\{\begin{aligned}
      &{(i_1\ldots j_1)}\,,
      \\ &{(i_1\ldots j_1), (i_2\ldots j_2)}\,,
      \\ &{(i_1\ldots j_1), (i_2\ldots j_2), (i_3\ldots j_3)}\,,
    \end{aligned}\right.
\end{equation}
along with a similar form for $\ESet_\rloop$.
The sets $\ESet_\ctop$ and $\ESet_\cbottom$ only contain a single
sequence,
\begin{equation}
  \begin{aligned}
    \ESet_{\ctop,\cbottom} &= {(i\ldots j)}\,,
  \end{aligned}
\end{equation}
In each sequence, the external leg labels are shown according to their clockwise order around the outside
of the diagram.
As an example, the cut of the $n$-point leading-color partial
amplitude shown in \fig{fig:R2n1_generic_cut} carries the label,
\begin{multline}
  \label{eq:cut_label_n:1}
  ({(i_1,\ldots,j_1),(i_2,\ldots,j_2),(i_3,\ldots,j_3)};
  {(i_4,\ldots,j_4)};\\
  {(i_5,\ldots,j_5),(i_6,\ldots,j_6),(i_7,\ldots,j_7)};
  {(i_8,\ldots,j_8)}).
\end{multline}
Any such label corresponds to a unitarity cut, provided that it
satisfies two conditions:
\begin{enumerate}
\item When $\ESet_\lloop$ or $\ESet_\rloop$  contain only a single
  sequence (\ie\ the corresponding loop is a bubble), this sequence has
  to contain at least two elements, to avoid scaleless bubbles.
\item When $\ESet_\lloop$ or $\ESet_\rloop$ contain
  two or three sequences (corresponding to a triangle or a box), each
  sequence needs to contain at least one element.
  In other words, all tree amplitudes are at least three-point amplitudes.
\end{enumerate}
We call the set of all labels \eqn{eq:cut_label_n:1} for $n$ momenta
that satisfy these conditions $\CutSet_{n:1}.$

In order to compute the rational contribution
to the leading-color partial amplitude $R^\twoloop_{n:1}(1,\ldots,n)$,
we need to sum over all unique cuts.
However, a cut does not have a single unique label.
We still have to
account for the equivalence of the two loops, giving each cut a $\mathbb{Z}_2$
symmetry.
This symmetry manifests itself in the cuts through an equivalence
relation $\Exch$, whose action is given by,
\begin{equation}
  \Exch\bigl[(\ESet_\lloop ; \ESet_\ctop ; \ESet_\rloop; \ESet_\cbottom)\bigr]
  = (\ESet_\rloop; \ESet_\cbottom; \ESet_\lloop ; \ESet_\ctop)\,.
  \label{ExchDefinition}
\end{equation}
This can be thought of as rotating a cut diagram by
180$^\circ$ in the plane.

We associate each label to
one of the six integral classes of shown in \eqn{eq:IntegralClasses}.
For each such class $C$, we define $\CutSetClass{C}_{n:1}$ to be the set of
all cut labels belonging to that class.
Due to the equivalence of the loops, there are always two labels
associated to a unitarity cut.
For example, the labels
\begin{equation}
  ((1),(2),(3);();(4),(5);()),\quad ((4),(5);(); (1),(2),(3);())
\end{equation}
are both elements of $\CutSet(\CutClass{\BoxTri})$, while describing
the same cut, as they are related by $\Exch$.
We therefore introduce the set $\CutSetUnique_{n:1}$, which contains
one label per unique cut.
As $\Exch$ only relates labels of the same class, we can construct
$\CutSetUnique_{n:1}$ in terms of equivalence
classes of  $\CutSetClass{C}_{n:1}$,
\begin{equation}
  \CutSetUnique_{n:1}=\bigcup\limits_{\text{classes }C}
  \left(
  \raisebox{3pt}{$\CutSetClass{C}_{n:1}$}
  \bigg/
  \raisebox{-3pt}{$\Exch$}\right)\,.
\end{equation}

For the rational part of the leading-color partial amplitude we then
sum over the contributions associated to the labels in $\CutSetUnique_{n:1}$,
\begin{equation}
  \RT_{n:1}(1,\ldots,n) = \sum_{c\in\CutSetUnique_{n:1}}\RT_{c}(1,\ldots,n).
\end{equation}
Each index $c$ can be given in the form described in \eqn{CutLabel}, with $\RT_{c}$ the rational contribution from that cut.
We give the precise definition of $\RT_{c}$ in
section~\ref{sec:rational_contributions}.

\begin{figure}[h]  
  \centering
  \begin{subfigure}[b]{\textwidth}
    \centering
    \begin{tikzpicture}
      \node[] (a) at (-6,0) {\(\includegraphics[]{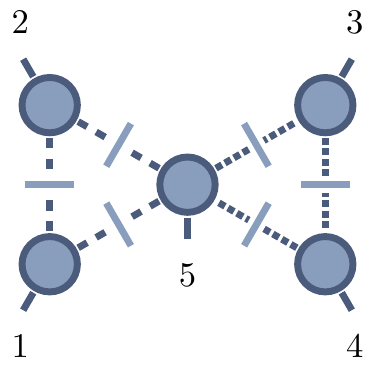}\)};
      \node[] (b) at  (0,0) {\includegraphics[width=0.15\linewidth]{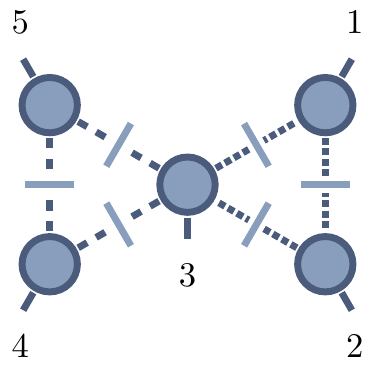}};
      \node[] (c) at  (6,0) {\(\includegraphics{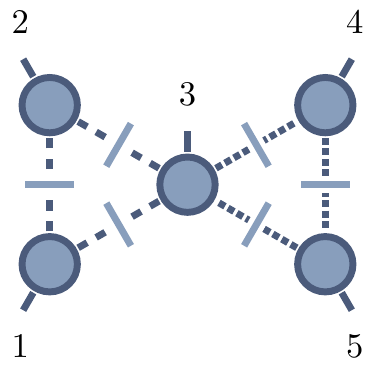}\)};
      \draw[<->,thick] (a) -- (b) node[midway,sloped,above] {$\cycperm_2$} ;
      \draw[<->,thick] (b) -- (c) node[midway,sloped,above] {$\Exch$} ;
    \end{tikzpicture}
    \caption{Example of a cut that is related to another
      cut via $\cyclicgroup_5\times\Exch$.
      We obtain the cut on the right from the one one the left
      by switching the two loops and cycling the external labels by two
      positions.
      We therefore include only one of these cuts in
      $\GeneratingCuts_{n:1}$.}
    \label{fig:symmetricCutExampleAsymmetric}
  \end{subfigure}
  \begin{subfigure}[b]{\textwidth}
    \centering
    \begin{tikzpicture}
      \node[] (a) at (-6,0) {\(\includegraphics[]{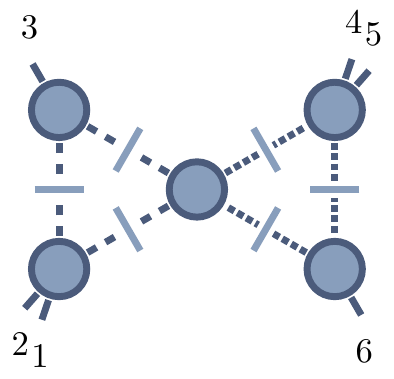}\)};
      \node[] (b) at  (0,0) {\includegraphics[width=0.15\linewidth]{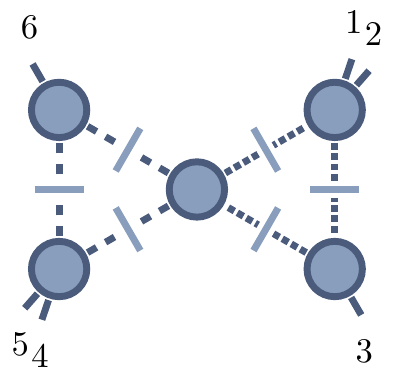}};
      \node[] (c) at  (6,0) {\(\includegraphics[]{generating_sets/leading_color_self_equivalence/graph_4}\)};
      \draw[<->,thick] (a) -- (b) node[midway,sloped,above] {$\cycperm_3$} ;
      \draw[<->,thick] (b) -- (c) node[midway,sloped,above] {$\Exch$} ;
    \end{tikzpicture}
    \caption{Example of a cut that is invariant under $\cyclicgroup_6\times\Exch$.
      Swapping the loops and cycling the external labels by three
      places leaves the cut invariant.
      We must therefore include this cut in $\GeneratingCuts_{n:1}$
      and associate to it a symmetry factor
      of $\tfrac{1}{2}$.}
    \label{fig:symmetricCutExampleSymmetric}
  \end{subfigure}
  \caption{Examples of relations between cuts for the leading-color
    rational parts $\RT_{n:1}$.}
  \label{fig:symmetricCutExample}
\end{figure}

We can go one step further and make the cyclic symmetry of the
all-plus configuration manifest.
We define $\TildeRT$ through the property that,
\begin{equation}
  R^\twoloop_{n:1}(1,\ldots,n) = \sum_{\cycperm\in \cyclicgroup_n}
  \TildeRT(\cycperm[1,\ldots,n])\,.
\end{equation}
To determine such an $\TildeRT$ we require only a generating subset of
all unitarity cuts $\GeneratingCuts_{n:1}\subset \CutSetUnique_{n:1}$.
We obtain such a subset by identifying cut labels that are related by a
cyclic permutation of the external particles,
\begin{equation}
  \GeneratingCuts_{n:1}
  = \left(\raisebox{3pt}{$\CutSetUnique_{n:1}$}\bigg/\raisebox{-3pt}{$\cyclicgroup_n$}\right)
  = \bigcup\limits_{\text{classes }C}
  \left(
  \raisebox{3pt}{$\CutSetClass{C}_{n:1}$}
  \bigg/
  \raisebox{-3pt}{$\Exch\times \cyclicgroup_n$}\right)\,.
    \label{eq:G_n1}    
\end{equation}
We can realize the identification of cut labels under $\cyclicgroup_n$
by fixing the position of one of the momenta.
We always make the choice that for each label in
$\GeneratingCuts_{n:1}$ we have
\begin{equation}
  \ESet_\lloop={(1,\mathellipsis),\mathellipsis}\,.
\end{equation}
In the generic cut of \fig{fig:R2n1_generic_cut} this corresponds to
setting $i_1=1$.
As a more concrete example, consider the following two
labels belonging to cuts of
$\RT_{5:1}$,
\begin{equation}
  c=({(1),(2)};();{(3),(4)};(5)),\quad c'=({(1),(2)};(3);{(4),(5)};())\,.
\end{equation}
Their associated cuts are shown in \Fig{fig:symmetricCutExampleAsymmetric}.
Of these two labels, we only include one 
in $\GeneratingCuts_{n:1}$, as,
\begin{equation}
  \begin{aligned}
    &\Exch\circ\, \cycperm_2
    \left[({(1),(2)};();{(3),(4)};(5))\right]\\
    =&\Exch[({(4),(5)};();{(1),(2)};(3))]\\
    =&({(1),(2)};(3);{(4),(5)};())\,,
  \end{aligned}
\end{equation}
with $\rho_2\in\cyclicgroup_5$ representing the clockwise 
shift of all labels by two positions.

\def\symfac#1{S^{\vphantom{(1)}}_{#1}}
The generator function $\TildeRT$ is determined by summing over all
labels within $\GeneratingCuts_{n:1}$, accompanied by a
symmetry factor $\symfac{c}$,
\begin{equation}
  \label{eq:R2n1_from_generating_set}
  \TildeRT(1,\ldots,n) =
  \sum_{c\in \GeneratingCuts_{n:1}} \symfac{c}\,\TildeRT_{c}(1,\ldots,n)\,.
\end{equation}
The symmetry factors are necessary to compensate for the overcounting
of cuts that are invariant under the
combined action of $\Exch$ and $\cyclicgroup_n$.
An example would be the label
\begin{equation}
  c=({(1,2),(3)};();{(4,5),(6)};())
\end{equation}
of $\RT_{6:1}$.
This cut label is invariant under shifting the momenta 
three places via
$\rho_3\in\cyclicgroup_6$ and exchanging the loops,
\begin{equation}
  \begin{aligned}
    &\Exch\circ\, \cycperm_3
    \left[({(1,2),(3)};();{(4,5),(6)};())\right] \\
    =&\Exch[({(4,5),(6)};();{(1,2),(3)};())]\\
    =&({(1,2),(3)};();{(4,5),(6)};())
  \end{aligned}
\end{equation}
This relation is shown diagrammatically in \Fig{fig:symmetricCutExampleSymmetric}.

Such an invariance can only occur for cut labels of the classes
$\CutClass{\BoxBox}$, $\CutClass{\TriTri}$ and $\CutClass{\BubBub}$
with an even number of external particles.
It causes the associated cut to appear
twice in the sum over cyclic permutations, and therefore requires a
symmetry factor of $S_c=\tfrac{1}{2}$.
For all remaining cuts we have $S_c=1$.

At one loop, computing the bubble coefficient using the approach of
refs.\cite{Forde:2007mi,Badger:2008cm} requires not
only the bubble cut itself, but also computing
three-propagator cuts from parent triangles.
These must then be accounted for according to 
\eq{eq:bubble_coefficient_total}.

\begin{figure}
  \centering
  \includegraphics[valign=c]
  {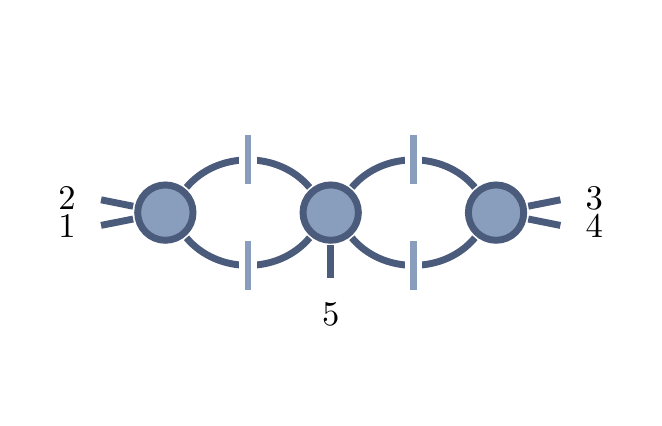}\\[1em]%
  \tikz[baseline] \draw [<-, line width=0.5pt, thick] (0,0) -- (0,1);\\
  \includegraphics[width=0.32\linewidth,valign=c]
  {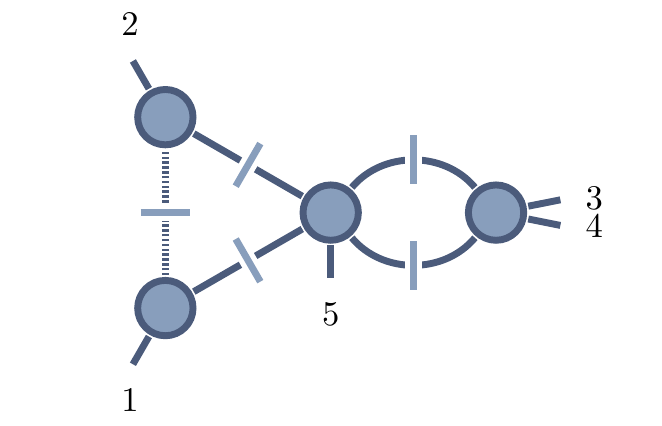}
  \includegraphics[width=0.32\linewidth,valign=c]
  {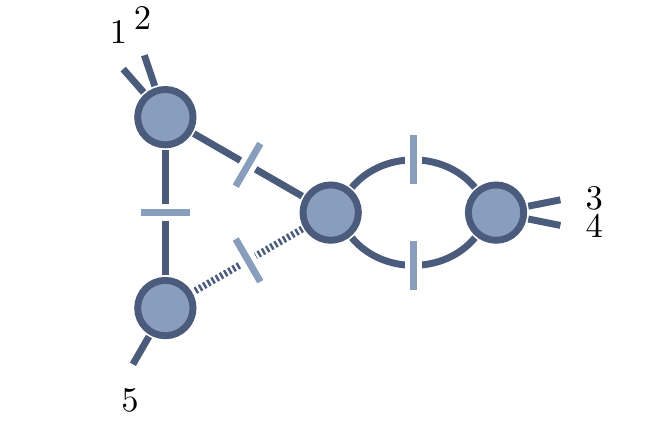}
  \includegraphics[width=0.32\linewidth,valign=c]
  {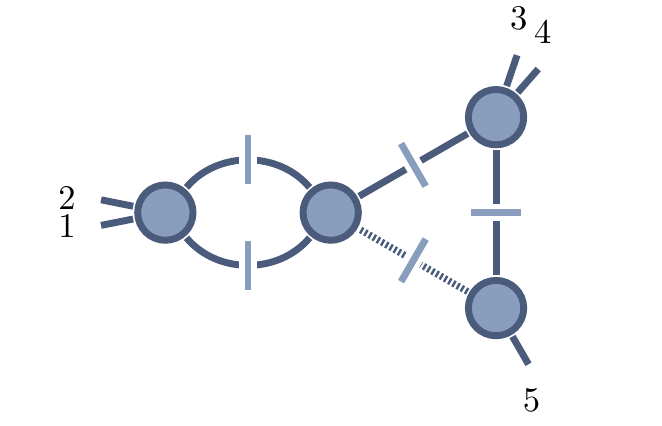}
  \includegraphics[width=0.32\linewidth,valign=c]
  {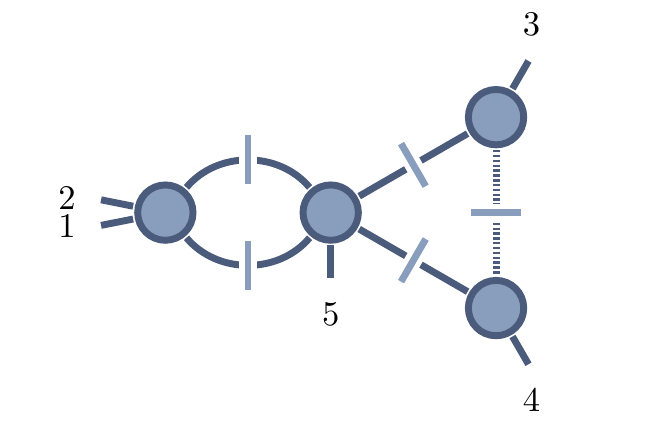}
  \includegraphics[width=0.32\linewidth,valign=c]
  {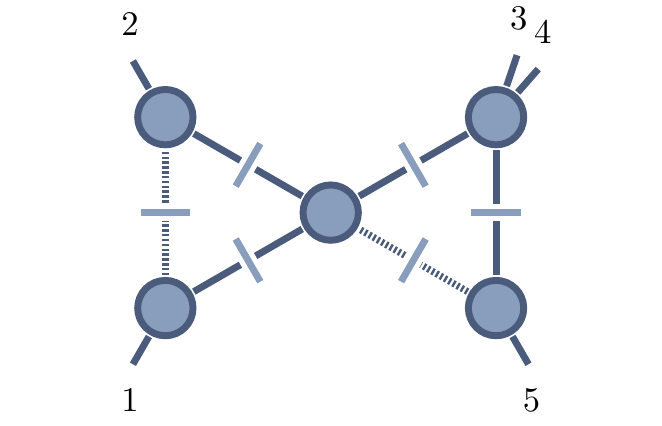}
  \includegraphics[width=0.32\linewidth,valign=c]
  {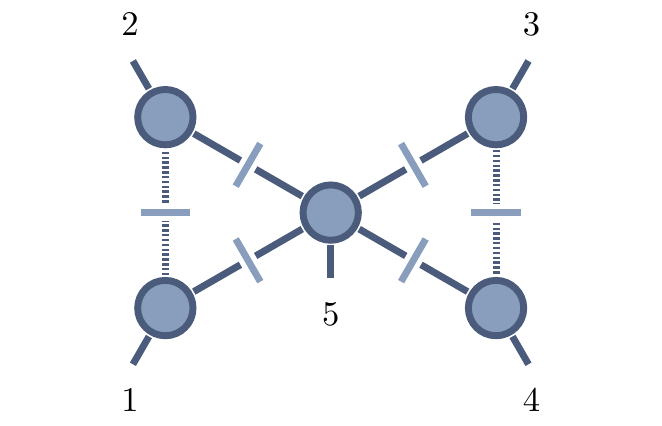}
  \includegraphics[width=0.32\linewidth,valign=c]
  {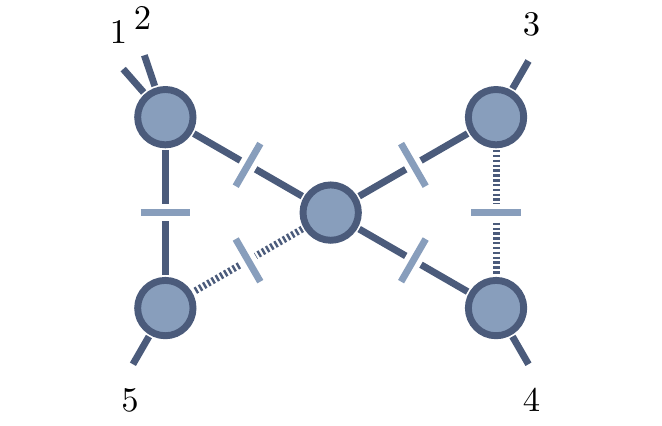}
  \caption{Examples of parent-triangle contributions that together with
    the bubble cut make up the bubble coefficient.
    For visual clarity, the cut propagators of the bubbles are shown as solid
    lines, while the additionally cut propagators are shown as
    dashed lines.}
  \label{fig:parentTriangleCutsExample}
\end{figure}

In computing the coefficients of one-loop squared integrals we must of 
course also include such contributions.
Specifically, for every cut-label in the generating set $\GeneratingCuts_{n:1}$ involving a bubble cut, we must also include cuts where the bubble
cut is replaced with a triangle cut.
Consider for example the cut shown in \fig{fig:parentTriangleCutsExample}
belonging to a $\GeneratingCuts_{5:1}$.
In addition to the two bubble cuts seen at the top, we also in general
need to compute the seven additional parent-triangle cuts shown below
it.
Here the short-dashed lines represent the extra cut introduced in addition 
to the original bubble cut.
In this procedure, only those cuts are allowed that belong to a one-loop squared topology.
Any topologies with a propagator depending on both loop momenta are still forbidden by the scalar Feynman rules.
\FloatBarrier

\subsection{Two- and Three-Trace Partial Amplitudes}
We turn next to the more general case of subleading-color
double- and triple-trace partial amplitudes.
In the following we focus on the coefficients of
$\Tr(\cdots)\Tr(\cdots)\Tr(\cdots)$.
The coefficients of $N_c\Tr(\cdots)\Tr(\cdots)$ 
can be obtained
using the same strategy by leaving one of the traces empty.

\begin{figure}
  \centering
  \includegraphics[width=0.6\textwidth,valign=c]
  {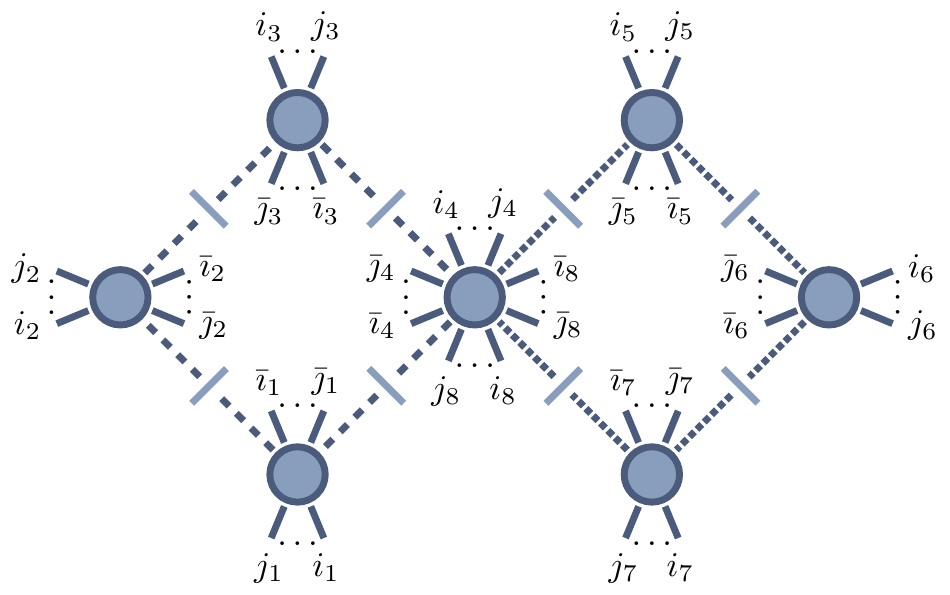}
  \caption{A generic cut for the sub-leading color rational
    part $\RTwo_{n:kl}$.}
  \label{fig:R2nkl_generic_cut}
\end{figure}

A generic three-trace cut has the form shown in
\fig{fig:R2nkl_generic_cut}.
To attach labels to such cuts, we extend the notation of
\eqn{CutLabel} by two further sets, separated by vertical bars.
From the point of view of \fig{fig:color_routing_example}, these sets
describe the attachment of legs to the inner edges of the left
and right loops.
We denote them respectively by a subscript $\lloop$ or $\rloop$ on the
vertical bar.
For simplicity, let us first consider the case where external
particles are only attached to the inside of the left loop.
A corresponding cut label takes the form,
\begin{equation}
  (\ESet_\lloop ; \ESet_\ctop ; \ESet_\rloop; \ESet_\cbottom |_\lloop
  \ESet_{\central,\lloop}; \ESet_{\inner,\lloop}) \,.
  \label{TwoTraceCutLabel}
\end{equation}
Here, $\ESet_{\lloop,\ctop,\rloop,\cbottom}$ denote the legs on the
`outside' of the diagram, $\ESet_{\central,\lloop}$ is the sequence of
legs attached to central amplitude `inside' the (left) loop,
while $\ESet_{\inner,\lloop}$ collects the
legs attached to the `inside' of tree amplitudes along the (left)
loop.
As $\ESet_\lloop$ and $\ESet_{\inner,\lloop}$ describe the attachments
to the same amplitudes, just 'inside' and 'outside' of the loop, they
need to have the same number of sequences---one for each amplitude in
the loop.
The legs within each sequence of $\ESet_{\inner,\lloop}$ are attached to
the same tree amplitude as those of the same-ordinal sequence with
$\ESet_\lloop$.
However, due to the reversed color-flow on the inside of the loop,
they are attached in reverse order.
That is, a label with,
\begin{equation}
  \begin{aligned}
    \ESet_\lloop &=
    (i_1\cdots j_1) \cdots (i_3\cdots j_3)\,,\\
    \ESet_{\inner,\lloop} &=
    (\jbar_1\cdots \ibar_1) \cdots (\jbar_3\cdots \ibar_3)\,,
  \end{aligned}
\end{equation}
corresponds to a cut containing the tree amplitudes
\begin{equation}
  \ATree(i_1,\ldots j_1,-\ell_1,\ibar_1,\ldots,\jbar_1,\ell_2)
  \times\cdots\times
  \ATree(i_3,\ldots j_3,-\ell_3,\ibar_3,\ldots,\jbar_3,\ell_4)\,,
\end{equation}
attached to a color structure of the form
\begin{equation}
  \TTr(i_1\cdots j_1\cdots i_3\cdots j_3\cdots)
  \TTr(\ibar_3\cdots \jbar_3\cdots \ibar_1\cdots\jbar_1\cdots)\cdots\,.
\end{equation}

As particles can now also be attached to the inside of a loop, we
need to extend the restrictions introduced for labels of leading-color
labels, which guarantee massive bubbles and tree amplitudes being at
least of three-point type.
They are respectively:
\begin{enumerate}
\item If $\ESet_{\lloop}$ and $\ESet_{\inner,\lloop}$ each contain
only one sequence, they together need to contain at
least two elements in total.
\item Sequences from $\ESet_{\lloop}$ and $\ESet_{\inner,\lloop}$
  belonging to the same amplitude have to contain at least one element
  in total.
\end{enumerate}
As long as these conditions are satisfied, sequences may contain any
number of elements.
In particular, they may now also be empty.

The extension to general cuts with the three-trace color structure is
now simple: when attaching external particles to the inner edges of
both the left and right loop, we add two sets to a cut label,
\begin{equation}
  (\ESet_\lloop ; \ESet_\ctop ; \ESet_\rloop; \ESet_\cbottom
  |_\lloop  \ESet_{\central,\lloop};\ESet_{\inner,\lloop}
  |_\rloop \ESet_{\central,\rloop}; \ESet_{\inner,\rloop})\,,
  \label{ThreeTraceCutLabel}
\end{equation}
where the number of sequences in $\ESet_{\inner,\lloop}$
($\ESet_{\inner,\rloop}$) must match that
of $\ESet_\lloop$ ($\ESet_\rloop$).
The conditions required by every amplitude having a particle attached
and the vanishing of massless bubbles now need to be satisfied by
both loops, \ie\ for both pairs  $\ESet_\lloop$,$\ESet_{\inner,\lloop}$ and
$\ESet_\rloop$,$\ESet_{\inner,\rloop}$.
The legs in $\ESet_{\inner,\rloop}$ are now also given in
\emph{reverse\/} order compared
to how they appear in the color-ordering,
and just as in the left loop, the legs of sequences in
$\ESet_{\inner,\rloop}$ are attached to the same tree amplitude as those
of the same-ordinal sequence of $\ESet_{\rloop}$.
The generic cut shown in \fig{fig:R2nkl_generic_cut} for example
carries the label
\begin{equation}
  \label{eq:triple-trace_generic_label}
  \begin{aligned}
    (&{(i_1,\ldots,j_1),(i_2,\ldots,j_2),(i_3,\ldots,j_3)}
    ;{(i_4,\ldots,j_4)};\\
    &{(i_5,\ldots,j_5),(i_6,\ldots,j_6),(i_7,\ldots,j_7)};
    {(i_8,\ldots,j_8)}\\
    &|_\lloop(\jbar_4,\ldots,\ibar_4);(\jbar_1,\ldots,\ibar_1),(\jbar_2,\ldots,\ibar_2),
    (\jbar_3,\ldots,\ibar_3)\\
    &|_\rloop(\jbar_8,\ldots,\ibar_8);(\jbar_5,\ldots,\ibar_5),(\jbar_6,\ldots,\ibar_6),
    (\jbar_7,\ldots,\ibar_7)).
  \end{aligned}
\end{equation}

As in the leading-color case, we have an equivalence
relation $\Exch$, due to equivalence of the two scalar loops.
Its action on the label of a two- or three-trace cut is given by,
\begin{equation}
  \begin{aligned}
    &\Exch\bigl[(\ESet_\lloop ; \ESet_\ctop ; \ESet_\rloop; \ESet_\cbottom
    |_X \ESet_{\central,X};\ESet_{\inner,X})\bigr]
    = (\ESet_\rloop; \ESet_\cbottom; \ESet_\lloop ; \ESet_\ctop
    |_{\Exch(X)} \ESet_{\central,X};\ESet_{\inner,X})\,,\\
    &\Exch\bigl[(\ESet_\lloop ; \ESet_\ctop ; \ESet_\rloop; \ESet_\cbottom
    |_\lloop \ESet_{\central,\lloop};\ESet_{\inner,\lloop}
    |_\rloop \ESet_{\central,\rloop};\ESet_{\inner,\rloop})\bigr]
    = 
    \\&\hspace*{35mm}
    (\ESet_\rloop; \ESet_\cbottom; \ESet_\lloop ; \ESet_\ctop
    |_\lloop; \ESet_{\central,\rloop}\ESet_{\inner,\rloop}
    |_\rloop \ESet_{\central,\lloop};\ESet_{\inner,\lloop})\,,\\
  \end{aligned}
\end{equation}
where in the former case we can have $X\in \{L,R\}$, while
$\Exch(\lloop) = \rloop$ and $\Exch(\rloop) = \lloop$.

\begin{figure}
  \centering
  \begin{subfigure}[b]{\textwidth}
    \centering
    \includegraphics[width=0.35\linewidth,valign=c]
    {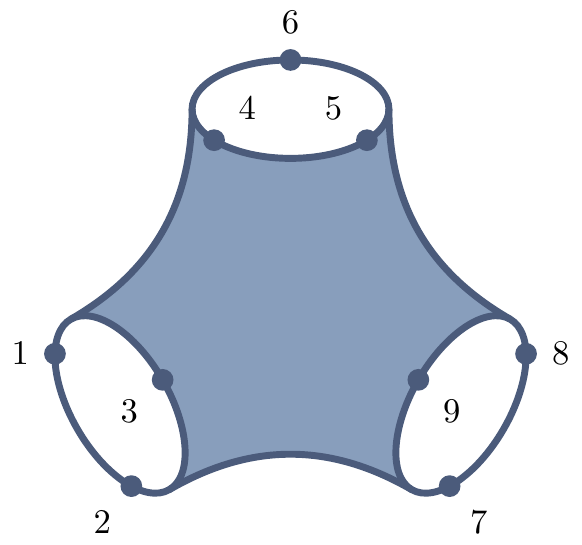}
    \caption{Representation of the world-sheet that makes the
      equivalence of the three edges evident.}
    \label{fig:string_amplitude_pants_representation}
  \end{subfigure}
  \begin{subfigure}[b]{\textwidth}
    \includegraphics[width=0.3\linewidth,valign=c]
    {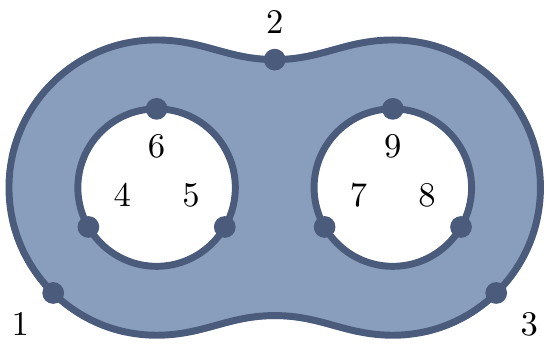}
    \hspace{1em}
    \includegraphics[width=0.3\linewidth,valign=c]
    {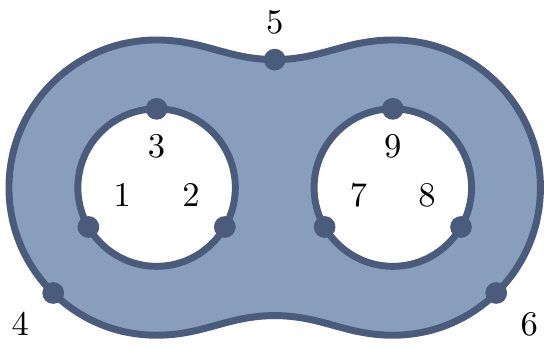}
    \hspace{1em}
    \includegraphics[width=0.3\linewidth,valign=c]
    {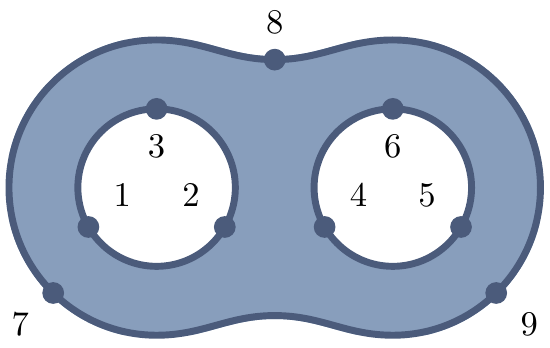}
    \includegraphics[width=0.3\linewidth,valign=c]
    {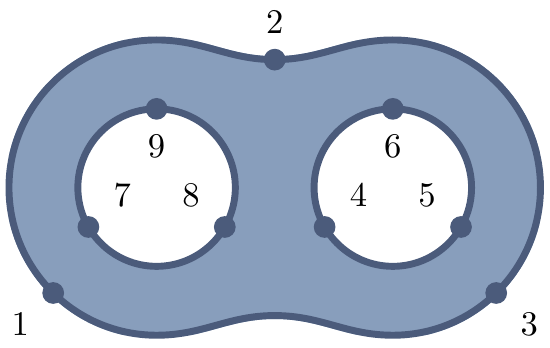}
    \hspace{1em}
    \includegraphics[width=0.3\linewidth,valign=c]
    {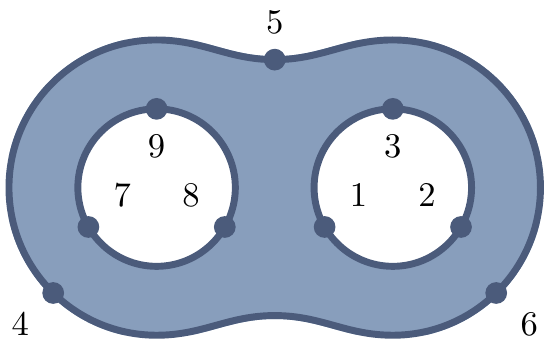}
    \hspace{1em}
    \includegraphics[width=0.3\linewidth,valign=c]
    {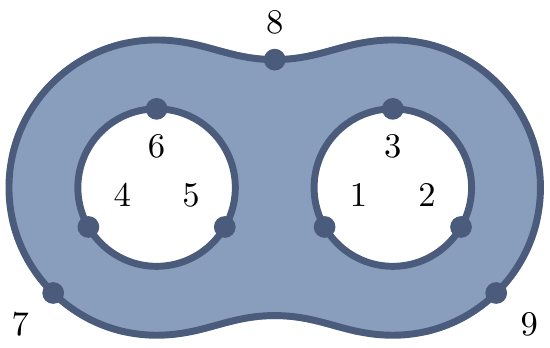}
    \caption{The six representations of the same world-sheet used to
      construct gauge-theory cuts.}
    \label{fig:string_amplitude_cut_representations}
  \end{subfigure}
  \caption{
    An example of a two-puncture disk open-string world-sheet.
    Fig.~(a) shows a three-dimensional embedding
    making the equivalence of the three edges manifest.
    Fig.~(b) shows the six different representations, for which we
    construct unitarity cuts.
    The insertions on the edges correspond to a color
    structure $\TTr(1,2,3)\TTr(4,5,6)\TTr(7,8,9)$ }
  \label{fig:string_amplitude_representations}
\end{figure}

We use the string picture as a
guide to construct the required cuts as shown
in~\fig{fig:color_routing_example}.
In the case of the string amplitude, the three boundaries of the
world-sheet---and
therefore the color traces---are on equal
footing.
We can smoothly deform the worldsheet, such that any two edges switch
places.
The form of the worldsheet shown in
\fig{fig:string_amplitude_pants_representation} makes this property
manifest.
However, when considering corresponding cuts as shown in
\fig{fig:R2nkl_generic_cut}, this symmetry between
the different edges is no longer present.
In particular, we cannot deform the cut, such that the particles on
the `outside' appear on the `inside' of one of the loops;
the `outer' edge of the a cut is therefore distinct%
\footnote{The role of the two inner edges can however be swapped using $\Exch$.}.
The resolution of this discrepancy is that the symmetry is not present at the level of individual cuts.
Rather, to restore it for the full amplitude we need sum over cuts for
each permutation of the edges, as exemplified in 
\fig{fig:string_amplitude_cut_representations}.

To be more explicit, let $\Tr_1$, $\Tr_2$ and $\Tr_3 $ be the three color
traces, and let $\sigma$ be a permutation of the set
$\{ \Tr_1,\Tr_2,\Tr_3\}$.
Define $\CutSet^\sigma\equiv\CutSet^{\{\sigma(1),\sigma(2),\sigma(3)\}}$
to be the set of all cut
labels of the
form given in \eqn{eq:triple-trace_generic_label}, in which the traces
$\sigma(1)$, $\sigma(2)$ and $\sigma(3)$ are associated to the outer,
left inner and right inner edge.
The momenta associated to $\sigma(1)$ therefore make up the sets
$\ESet_\lloop$,
$\ESet_\ctop$, $\ESet_\rloop$,
$\ESet_\cbottom$, while those of $\sigma(2)$ and $\sigma(3)$
respectively determine
$\ESet_{\inner,\lloop}$, $\ESet_{\central,\lloop}$ and 
$\ESet_{\inner,\rloop}$, $\ESet_{\central,\rloop}$.

\begin{figure}[h]
  \centering
  \includegraphics[width=0.49\textwidth]
  {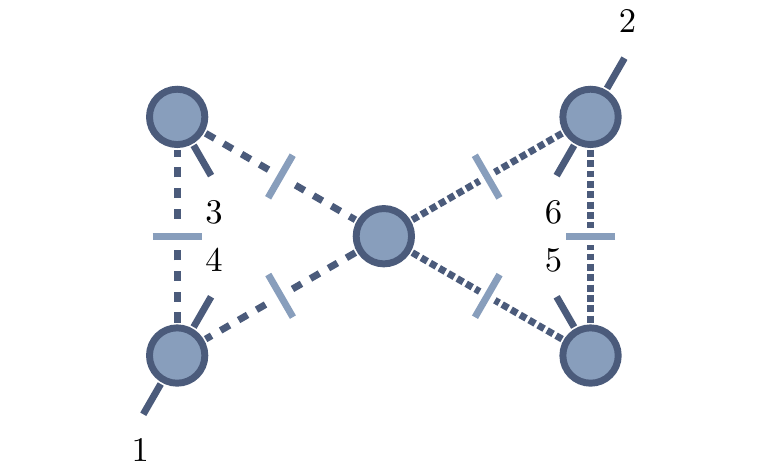}
  \includegraphics[width=0.49\textwidth]
  {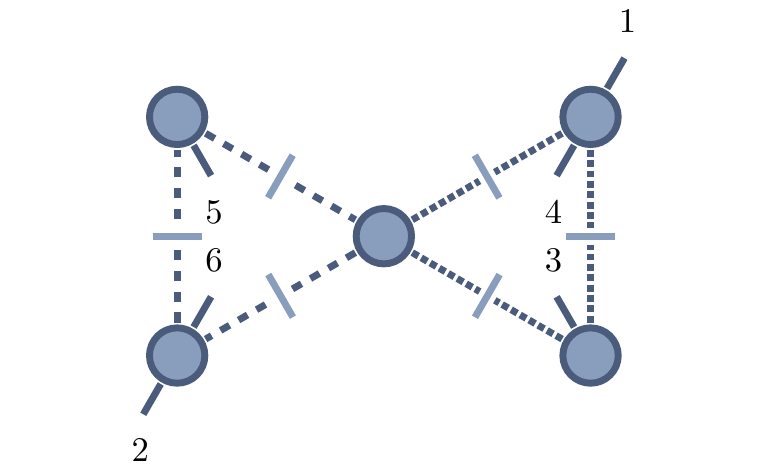}
  \caption{Graphic representation of two labels belonging to the same
    unitarity cut of $\RT_{6:2,2}$.}
\label{fig:R622_cut_label_example}
\end{figure}

The set of all labels compatible with a triple trace color structure
is then
\begin{equation}
  \CutSet_{n:kl}=\bigcup\limits_{\sigma\in S_3(\{\Tr_1,\Tr_2,\Tr_3\})}
    \CutSet^\sigma
\end{equation}
For the rational part $\RT_{n:kl}$ we again require only unique cuts.
Just as in the leading-color case, $\CutSet_{n:kl}$ contains two labels
for each cut, related by $\Exch$.
Consider for example, the labels
\begin{equation}
  \begin{aligned}
    &((1),();();(2),()|_\lloop();(4),(3)|_\rloop();(6),(5))\,,\\
    &((2),();();(1),()|_\lloop();(6),(5)|_\rloop();(4),(3))\,,
  \end{aligned}
\end{equation}
of $\RT_{6:2,2}$, whose visual representation is given in \Fig{fig:R622_cut_label_example}.
For $\Tr_1=\TTr(1,2)$, $\Tr_2=\TTr(3,4)$, $\Tr_3=\TTr(5,6)$, 
these labels belong respectively to $\CutSet^{\{\Tr_1,\Tr_2,\Tr_3\}}$ and
$\CutSet^{\{\Tr_1,\Tr_3,\Tr_2\}}$.
Both therefore appear in $\CutSet_{6:2,2}$.
However, from \Fig{fig:R622_cut_label_example} it is easy to see that
these labels describe the same unitarity cut.
Accordingly we find that they are related by $\Exch$,
\begin{equation}
  \begin{aligned} &\hphantom{{}={}}\Exch\bigl[((1),();();(2),()|_\lloop();(4),(3)|_\rloop();(6),(5))\bigr]\\
  &=((2,();();(1),()|_\lloop();(6),(5)|_\rloop();(4),(3))
  \end{aligned}
\end{equation}
Modding out such relations, we obtain the set of unique cuts,
\begin{equation}
  \label{eq:cutsetunique_nkl}
  \CutSetUnique_{n:kl}=
  \raisebox{5pt}{$\Biggl(\,\CutSet_{n:kl}\Biggr)$}
  \bigg/\raisebox{-3pt}{$\Exch$}\,.
\end{equation}
We can give a more concise form of this set.
As $\Exch$ switches the trace assignments of the
inner edges, we find that,
\begin{equation}
  \Exch\left[\CutSet^{\{\Tr_1,\Tr_2,\Tr_3\}}\right] =
  \CutSet^{\{\Tr_1,\Tr_3,\Tr_2\}}
\end{equation}
It is therefore simple to realize the action of $\Exch$ in \eqn{eq:cutsetunique_nkl}:
for every assignment of a trace to the outer edge, we include \emph{only one}
of the two possible left and right inner edge trace assignments.
We therefore only sum over cyclic permutations of $\{\Tr_1,\Tr_2,\Tr_3\}$, so that
\begin{equation}
  \label{eq:cutsetunique_nkl_simplified}
  \CutSetUnique_{n:kl}=\bigcup_{\sigma\in\cyclicgroup_3(\{\Tr_1,\Tr_2,\Tr_3\})}\CutSet^{\sigma},
\end{equation}
The rational part for the color structure 
$\TTr(1,\ldots,k)\TTr(k+1,\ldots,k+l)\TTr(k+l+1,\ldots,n)$ is
then,
\begin{multline}
  \RT_{n:kl}(1,\ldots,k;k+1,\ldots,k+l;k+l+1,\ldots,n) \\
  =\sum_{c\in \CutSetUnique_{n:kl}}
  \RT_{c}(1,\ldots,k;k+1,\ldots,k+l;k+l+1,\ldots,n).
\end{multline}

As in the leading-color case we can make the cyclic symmetry of the
partial amplitudes manifest.
We define a function $\TildeRT_{n:kl}$, which reproduces the full
rational part after summing over cyclic
permutations of particles in each trace,
\begin{multline}
  \RT_{n:kl}(1,\ldots,k;k+1,\ldots,k+l;k+l+1,\ldots,n) \\
  =\sum_{\cycperm\in\cyclicgroup_{n:kl}}
  \TildeRT_{n:kl}(\cycperm[1,\ldots,k;k+1,\ldots,k+l;k+l+1,\ldots,n]).
\end{multline}
Here, $\cyclicgroup_{n:kl}=\cyclicgroup_{k}\times \cyclicgroup_{l}
\times\cyclicgroup_{n-k-l}$ is the group of cyclic permutations of all
three traces.
We again obtain $\TildeRT_{n:kl}$ from a generating set of cut labels
$\GeneratingCuts_{n:kl}\subset\CutSetUnique_{n:kl}$, together with
potential symmetry factors $S_c$, \ie
\begin{multline}
  \TildeRT_{n:kl}(1,\ldots,k;k+1,\ldots,k+l;k+l+1,\ldots,n) \\
  =\sum_{c\in\GeneratingCuts_{n:kl}}
  S_c\RT_{c}(1,\ldots,k;k+1,\ldots,k+l;k+l+1,\ldots,n).
\end{multline}
We can again find such a generating set by identifying any unique cuts
that are related by a cyclic permutation of the traces,
\begin{equation}
  \GeneratingCuts_{n:kl}=\raisebox{3pt}{$\CutSet_{n:kl}$}\bigg/\raisebox{-3pt}{$\Exch\times\cyclicgroup_{n:kl}$}
  =\bigcup_{\sigma\in\cyclicgroup_3(\{\Tr_1,\Tr_2,\Tr_3\})}\raisebox{3pt}{$\CutSet^{\sigma}$}\bigg/\raisebox{-3pt}{$\cyclicgroup_{n:kl}$}.
  \label{eq:G_nkl}
\end{equation}

To find an explicit form of $\GeneratingCuts_{n:kl}$, we fix for every edge the position of one particle.
Given cut labels $\CutSet^{\sigma}$ with
\begin{equation}
  \sigma=\{\TTr(1,\ldots,k), \TTr(k+1,\ldots,k+l), \TTr(k+l+1,\ldots,n)\}\}\,,
\end{equation}
we choose to keep those labels in which the particles $1$, $k+1$ and
$k+l+1$ are at the first position after
the central vertex in the color-ordering.
For each such label we have,
\begin{equation}
  \begin{aligned}
    \ESet_\lloop\cup\ESet_{\ctop}\cup \ESet_\rloop\cup \ESet_{\cbottom} &=(1,\ldots),\ldots\,, \\
    \ESet_{\central,\lloop}\cup\ESet_{\inner,\lloop} &=\ldots,(\ldots,k+1)\,,\\
    \ESet_{\central,\rloop}\cup\ESet_{\inner,\rloop} &=\ldots,(\ldots,k+l+1)\,.
  \end{aligned}
\end{equation}
In the cut of \Fig{fig:R2nkl_generic_cut} this corresponds to
\begin{equation}
  i_1=1,\quad \ibar_3=k+1,\quad \ibar_7=k+l+1\,.
\end{equation}
The sequence containing $1$ in $\ESet_\lloop\cup\ESet_{\ctop}\cup \ESet_\rloop\cup \ESet_{\cbottom} $ can
potentially be preceded by empty sequences.
Similarly, the sequences containing $k+1$ and $k+l+1$ in
$\ESet_{\central,\lloop}\cup\ESet_{\inner,\lloop}$ and $ \ESet_{\central,\rloop}\cup\ESet_{\inner,\rloop} $ can be followed
by empty sequences.

In contrast to the leading-color case, we do not require symmetry
factors here, so that $S_c=1$ for all labels
$c\in\GeneratingCuts_{n:kl}$.
This absence of symmetry factors is due to the three traces being
unique.
At leading color, the symmetry factors were necessary due to an
overcounting of cuts, where the associated labels are
invariant under $\Exch$ and cyclic permutation of the
external legs.
From the point of view of the present discussion, this type of
equivalence relies on the two traces assigned to the insides of the
loops being identical---in the
leading-color case they are both empty.
If all traces are unique,
such invariant labels do not exist, and there is no
overcounting of cuts.

We again have to include for every bubble cut in $\GeneratingCuts_{n:kl}$ the corresponding set
of parent triangle cuts, which are required to obtain the full bubble coefficient.
The procedure is the same as in the leading-color case, and we refer to the 
discussion in section~\ref{subsec:leading_color_cuts}.

As mentioned at the beginning of the section, by taking one of the
traces to be empty, we immediately obtain the procedure for 
double-trace
color structures.
In this case, the traces are still unique, and we also require no
symmetry factors.
To some extent, we can also recover the leading-color case with two
empty traces.
The only trace assignment that contributes is then
the one where every particle is attached to the outer edge
of the cut.
Any other assignment leads to a loop without external
particles, for which no (non-vanishing) cuts exist.
Denoting empty traces by $N_c$, we have
\begin{equation}
  \begin{aligned}
    \CutSet_{n:1}&=\CutSet^{\{\Tr,N_c,N_c\}},\\
    \CutSetUnique_{n:1}&=
  \raisebox{5pt}{$\Biggl(\,\CutSet^{\{\Tr,N_c,N_c\}}\Biggr)$}
  \bigg/\raisebox{-3pt}{$\Exch$}\,,
  \end{aligned}
\end{equation}
with every label taking the form
\begin{equation}
  (\ESet_\lloop ; \ESet_\ctop ; \ESet_\rloop; \ESet_\cbottom
  |_\lloop  ();()\ldots
  |_\rloop ();()\ldots)\,.
\end{equation}
Note however that the construction of \eqn{eq:cutsetunique_nkl_simplified}
does \emph{not} translate to the leading-color case, as it relies on
traces being indistinguishable.
We also require symmetry factors, as previously discussed.

\subsection{Subleading-Color Single-Trace Partial Amplitudes}
\label{sec:sub_single_traces_generating_sets}
\begin{figure}[h]
  \centering
  \includegraphics[width=0.5\textwidth]
  {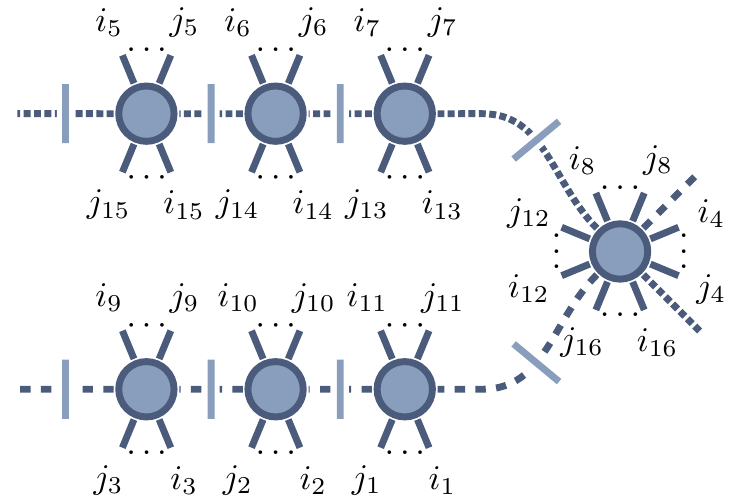}
  \caption{A generic cut contributing to $\RT_{n:\oneB}$. The scalar lines are assumed to be connected according to the dashing style.}
  \label{fig:1B_generic_cut}
\end{figure}

\def\outer{\textrm{O}}
\def\central{\textrm{C}}
The final class of subleading-color partial amplitudes are the
coefficients of the single-trace color structure with no accompanying
powers of $N_c$.
In the string picture, these correspond to the oriented
single-boundary worldsheets shown
in~\fig{fig:punctured_torus_surface}.
In \fig{fig:1B_example_color} we already gave an example of a unitarity
cut corresponding to such a color structure.
The most general cut we can encounter is shown in \fig{fig:1B_generic_cut},
belonging to a double-box integral.
We again define a notation to label such cuts.

A label will have four sections, separated by
vertical bars.
The general form is,
\begin{equation}
  (\ESet_{1}; \ESet_{\central,1}
  | \ESet_{2}; \ESet_{\central,2}
  | \ESet_{3}; \ESet_{\central,3}
  | \ESet_{4}; \ESet_{\central,4} )
  \,,
\end{equation}
Each section consists of two sets of sequences, $\ESet_{i}$  and
$\ESet_{\central,i}$.
The sets $\ESet_{i}$ are made up of one to three sequences, describing
the attachment of particles along the loops.
The set $\ESet_{\central_i}$ contains a single sequence, describing the
particles attached to the central vertex following the particles of
$\ESet_{i}$ in the color ordering.

The sets $\ESet_{1}$ and $\ESet_{3}$ together describe the amplitudes
along one scalar line.
They therefore have to contain the same number of sequences.
The association of the sequences in $\ESet_{3}$  to the amplitudes is
reversed from that of $\ESet_{1}$.
For example, the first sequence of $\ESet_{1}$ and last sequence of
$\ESet_{3}$ list legs attached to the same tree amplitude.
The same holds for $\ESet_{2}$ and $\ESet_{4}$, which describe the other
scalar line.

The label of the generic cut of \fig{fig:1B_generic_cut} is
\begin{equation}
  \begin{aligned}
    (&(i_1,\ldots,j_1),(i_2,\ldots,j_2),(i_3,\ldots,j_3);(i_4,\ldots,j_4)|\\
    &(i_5,\ldots,j_5),(i_6,\ldots,j_6),(i_7,\ldots,j_7);(i_8,\ldots,j_8)|\\
    &(i_9,\ldots,j_9),(i_{10},\ldots,j_{10}),(i_{11},\ldots,j_{11});(i_{12},\ldots,j_{12})|\\
    &(i_{13},\ldots,j_{13}),(i_{14},\ldots,j_{14}),(i_{15},\ldots,j_{15});(i_{16},\ldots,j_{16}))\,.
  \end{aligned}
\end{equation}
From the form of the label we can also immediately read off the associated
color trace,
\begin{equation}
  \TTr(i_1,\ldots,j_1,i_2,\ldots,j_2,\ldots,i_{15},\ldots,j_{15},i_{16},\ldots,j_{16})
\end{equation}

Just as in the triple-trace case, requiring that every bubble be
massive and that every tree amplitude have
at least three legs introduces restrictions on the cut labels:
\begin{enumerate}
\item If $\ESet_{1}$ ($\ESet_2$) and $\ESet_{3}$ ($\ESet_4$) each contain
only one sequence, they together need to contain at
least two elements in total.
\item Sequences from $\ESet_{1}$ ($\ESet_2$) and $\ESet_{3}$ ($\ESet_4$)
  belonging to the same amplitude have to contain at least one element
  in total.
\end{enumerate}
We define $\CutSet_{n:\oneB}$ to be the set of cut labels
associated to the color structure $\TTr(1,\ldots,n)$ that fulfill these
restrictions.

\begin{figure}
    \centering
    \begin{tikzpicture}
      \node[] (a) at (-6.4,0) {\(\includegraphics[]{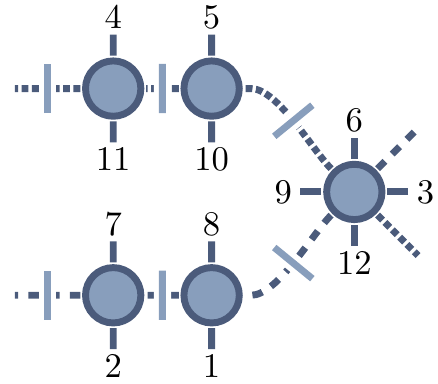}\)};
      \node[] (b) at  (0,0) {\includegraphics[]{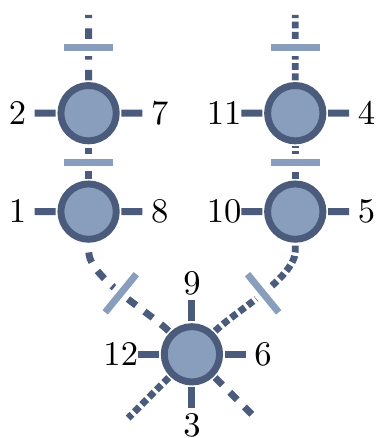}};
      \node[] (c) at  (5.4,0) {\(\includegraphics{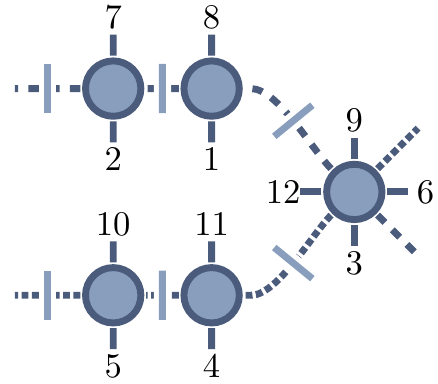}\)};
      \draw[->,thick] (a) -- (b) node[midway,sloped,above] {$\Rota$} ;
      \path (b) -- (c) node[midway,sloped] {\scalebox{1.3}{$\sim$}} ;
    \end{tikzpicture}
    \caption{
        Graphical representation of the action of $\Rota$ on the cut label $((1),(2);(3)|(4),(5);(6)|(7),(8);(9)|(10),(11);(12))$. The entire cut is rotated clockwise by $45^\circ$.
        Changing the attachment of the short-dashed scalar line then brings the cut back into the form shown in~\fig{fig:1B_generic_cut}.
    }
    \label{fig:1B_ROT_example}
\end{figure}

We again require only the set of cut labels where each label is
associated to a unique cut.
The cuts of $\RT_{n:\oneB}$ have an extended $\mathbb{Z}_4$ symmetry
beyond the $\mathbb{Z}_2$ symmetry of exchanging the loops seen previously.
This is becomes evident from the world-sheet origin of these cuts, as shown for
example in \fig{fig:1B_example_color}:
As there is only a single continuous edge, 
the representation of the 
world-sheet in \fig{fig:1B_example_color_a} is invariant under 
quarter-turn rotations.
This rotational invariance directly
translates to an invariance of the cut in \fig{fig:1B_example_color_b} under 
rotations of $45^\circ$.
On the level of cut labels, we can realize this symmetry using an operator
$\Rota$, which is defined by 
\begin{multline}
    \Rota[\{\ESet_1;\ESet_{\central,1},\ESet_2;\ESet_{\central,2},\ESet_3;\ESet_{\central,3},\ESet_4;\ESet_{\central,4}\}]
    =\{\ESet_2;\ESet_{\central,2},\ESet_3;\ESet_{\central,3},\ESet_4;\ESet_{\central,4},\ESet_1;\ESet_{\central,1}\}\,,
\end{multline}
In other words, under this symmetry, cut labels are cyclic in the
$\ESet_{i};\ESet_{\central,i}$ pairs.
As there are four such pairs, we have $\Rota^4=\mathbbm{1}$.
We can also identify the exchange of the two loops with two 
rotations, so 
that $\Rota^2=\Exch$.
As the pairs $\ESet_{1}$, $\ESet_{3}$ and $\ESet_{2}$, $\ESet_{4}$ 
each describe 
one of the loops, $\Rota^2$ can be thought of an ``inversion'' 
of each loop, 
while $\Rota$ and $\Rota^3$ then generate the two possible labels in 
which the two loops switch places.
\Fig{fig:1B_ROT_example} graphically shows the action of $\Rota$ on 
a cut label. 

As a consequence of this four-fold symmetry, the set $\CutSet_{n:\oneB}$
contains four labels for each unique unitarity cut:
For any $c\in\CutSet_{n:\oneB}$, the labels $c$, $\Rota[c]$, $\Rota^2[c]$ 
and $\Rota^3[c]$ all describe the same cut.
The set of unique cut labels is therefore given by,
\begin{equation}
    \CutSetUnique_{n:\oneB}=\raisebox{5pt}{$\Biggl(\,\CutSet_{n:\oneB}\Biggr)$}
  \bigg/\raisebox{-3pt}{$\Rota$}\,,
\end{equation}
such that the associated subleading single-trace rational part is,
\begin{equation}
    \RT_{n:\oneB}(1^+,\ldots,n^+)=\sum_{c\in\CutSetUnique_{n:\oneB}}\RT_c(1,\ldots,n)\,.
\end{equation}

We can again find a generating set of cut labels $\GeneratingCuts_{n:\oneB}$, which we 
use to define a function $\TildeRT_{n:\oneB}$ via,
\begin{equation}
    \TildeRT_{n:\oneB}(1,\ldots,n)=\sum_{c\in\GeneratingCuts_{n:\oneB}}S_c\RT_c(1,\ldots,n)\,,
    \label{eq:TildeRT_n1B}
\end{equation}
with the property that
\begin{equation}
    \RT_{n:\oneB}(1^+,\ldots,n^+)=\sum_{\sigma\in\cyclicgroup_n}\TildeRT_{n:\oneB}(\sigma[1,\ldots,n])\,.
    \label{eq:RT_n1B}
\end{equation}
This generating set can be obtained from $\CutSet_{n:\oneB}$ by modding out the combined
action of $\Rota$ and cyclic permutations of the external particles,
\begin{equation}
    \GeneratingCuts_{n:\oneB}=\raisebox{5pt}{$\Biggl(\,\CutSet_{n:\oneB}\Biggr)$}
  \bigg/\raisebox{-3pt}{$\cyclicgroup_n\times\Rota$}\,.
  \label{eq:G_n1B}
\end{equation}
For the subleading-color single-trace amplitude, 
however, we cannot construct $\CutSetUnique_{n:\oneB}$ by 
first picking an element of each $\Rota$ equivalence class, 
and then modding out $\cyclicgroup_n$.
In particular, 
\begin{equation}
    \GeneratingCuts_{n:\oneB}\ne \raisebox{5pt}{$\Biggl(\,\CutSetUnique_{n:\oneB}\Biggr)$}
  \bigg/\raisebox{-3pt}{$\cyclicgroup_n$}\,.
\end{equation}
We need to be more careful in this construction and mod
out by $\cyclicgroup_n\times \Rota$ in one step.
As an example for why this is not the case, assume we chose to 
include in $\CutSetUnique_{4:\oneB}$ the cut labels
\begin{equation}
    \begin{aligned}
      ((1),(2);()|(3),(4);()|(),();()|(),();())\,,\\
      ((1),(2);()|(),();()|(),();()|(3),(4);())\,,
    \end{aligned}
\end{equation}
These labels are not related by $\Rota$, and therefore belong 
to two different 
$\Rota$ equivalence classes.
Their associated cuts are related by shifting all external 
particles by two positions, 
so that only of the two should be included in 
$\GeneratingCuts_{4:\oneB}$.
However, this equivalence only appears at the level of cut labels 
for the combined 
action $\cyclicgroup_4\times\Rota$.
The set $\CutSetUnique_{4:\oneB}/\cyclicgroup_4$ would instead
include both labels 
as separate equivalence classes, and therefore lead to an 
overcounting of the cuts.

As in the case of the leading-color rational part,
we require symmetry factors for cuts that are invariant, 
in this case under $\Rota\times\cyclicgroup_n$.
In contrast to the leading-color case, there are two different types of 
such cuts of $\RT_{n:\oneB}$: those that are invariant under $\Rota$ (and therefore also
under $\Rota^2$ and $\Rota^3$), 
and those that are only invariant under $\Rota^2$.
These types of cuts require symmetry factors of $\tfrac{1}{4}$ and $\tfrac{1}{2}$ respectively,
as the associated cut appear four or two times in the cyclic sum of \eqn{eq:RT_n1B}.
As an example, the label,
\begin{equation}
    ((1),();()|(2),();()|(3),();()|(4),();())\,,
\end{equation}
belongs to the former class, as it is invariant under $\rho_1\circ\Rota$, 
$\rho_2\circ\Rota^2$ and $\rho_3\circ\Rota^3$.
Its symmetry factor is therefore $\tfrac{1}{4}$.
The label,
\begin{equation}
    ((1),(2);()|(),();()|(3),(4);()|(),();())\,,
\end{equation}
on the other hand belongs to the second class, as it is invariant under $\rho_2\circ\Rota^2$, 
making its symmetry factor $\tfrac{1}{2}$.
In summary, we have the following rule for the symmetry factors of $\TildeRT_{n:\oneB}$,
\begin{equation}
    S_c=\left\lbrace
    \begin{alignedat}{2}
      &\tfrac{1}{4},\quad &&\Rota[c]\overset{\cyclicgroup_n}{\sim} c\\
      &\tfrac{1}{2},\quad &&\Rota^2[c]\overset{\cyclicgroup_n}{\sim} c\\
      &1,\quad && \mathrm{otherwise}.
    \end{alignedat}
    \right.\,,
\end{equation}
where $\overset{\cyclicgroup_n}{\sim}$ represents equivalence under a cyclic permutation.

Finally, we again have to note that just as discussed in the previous
two section, any bubble cuts in $\GeneratingCuts_{n:\oneB}$ have to 
be supplemented by the all possible parent triangle cuts to recover the full bubble coefficients. 
\newcommand{\CoeffOneL}{\text{C}^{(1)}_{\lloop}}
\newcommand{\CoeffOneR}{\text{C}^{(1)}_{\rloop}}
\newcommand{\CoeffOneOne}{\text{C}^{(1)}_{1}}
\newcommand{\CoeffOneTwo}{\text{C}^{(1)}_{2}}
\newcommand{\CoeffTwo}{\text{C}^{(2)}}
\newcommand{\AmpsR}{\text{A}^{(0)}_{\rloop}}
\newcommand{\loopone}{1}
\newcommand{\looptwo}{2}

\subsection{The Rational Contribution}
\label{sec:rational_contributions}
We now define the function $\RT_c$, which we used in the previous
section to represent the rational contribution of a cut $c$.
As the labeling of cuts is dependent on the cut topology, we
discuss separately the leading-color single-, double-, and 
triple-trace cuts on the one hand,
and the subleading-color single-trace cuts on the other.

\subsubsection{Leading-Color Single-, Double-, Triple-Trace Cuts}

\label{sec:cut_computation_triple_trace}
We begin with the doubly punctured disk topology, for 
which a generic cut label $c$ has the form
\begin{equation}
   c = (\ESet_\lloop ; \ESet_\ctop ; \ESet_\rloop; \ESet_\cbottom
  |_\lloop  \ESet_{\central,\lloop};\ESet_{\inner,\lloop}
  |_\rloop \ESet_{\central,\rloop}; \ESet_{\inner,\rloop})\,.
  \label{eq:cut_label_triple_trace_compute}
\end{equation}
For empty traces in the double or leading-color single trace
case, either one or both of the $\ESet_{C,X}$, $\ESet_{\inner,X}$ are
made up of empty sequences.
The rational contribution from
the cut described by such a label is the
product of two one-loop $\mu^2$-integrals, $I^{(1)}_{c,\lloop}$ and
$I^{(1)}_{c,\rloop}$, together
with the associated integral coefficient.
In the separable approach, we determine this coefficient by performing
two consecutive one-loop integral coefficient computations, one for
each loop.
Given a cut label in the form of
\eqn{eq:cut_label_triple_trace_compute} we make the choice of always
computing the coefficient of the right loop first.
The result then enters in the computation of the coefficient belonging
to the left loop.

Let us be more explicit, focusing on the generic triple-trace cut
shown in \eqn{eq:triple-trace_generic_label}.
The cases of two-trace and leading-color single-trace cuts follow
similarly, by taking one or 
both $\ESet_{\central,X}$, $\ESet_{\inner,X}$
to be sets of empty sequences.
In our example, both loops are of the box type, so that
$I^{(1)}_{c,\lloop/\rloop}=I^{(1),D}_{\text{Box}}[\mu^2]$.
We use the parametrizations drawn from 
refs.~\cite{Forde:2007mi,Badger:2008cm}
and recorded in \app{GeneralizedUnitarityAppendix}.
Here, we impose the on-shell conditions in the right loop 
by defining $\ell_{\rloop}$ using the loop-momentum
parametrization of \eqn{eq:BoxLoopMomentumAnsatz} with
\begin{equation}
  \begin{aligned}
    K_{\rloop,1}&=k_{i_5}+\ldots+k_{j_5}+k_{\ibar_5}+\ldots+k_{\jbar_{5}}\,,\\
    K_{\rloop,2}&=k_{i_6}+\ldots+k_{j_6}+k_{\ibar_6}+\ldots+k_{\jbar_{6}}\,,\\
    K_{\rloop,3}&=k_{i_7}+\ldots+k_{j_7}+k_{\ibar_7}+\ldots+k_{\jbar_{7}}\,,\\
    K_{\rloop,4}&=-K_{\rloop,1}-K_{\rloop,2}-K_{\rloop,3}\,,\\
  \end{aligned}  
\end{equation}
We further set,
\begin{equation}
  \ell_{\rloop,1}=\ell_{\rloop},\quad \ell_{\rloop,2}=\ell_{\rloop}-K_{\rloop,1}
  ,\quad \ell_{\rloop,3}=\ell_{\rloop}-K_{\rloop,1}-K_{\rloop,2},\quad \ell_{\rloop,4}=\ell_{\rloop}+K_{\rloop,4}\,.
\end{equation}
We proceed similarly for the left loop, 
defining $\ell_\lloop$ according
to \eqn{eq:BoxLoopMomentumAnsatz}, with
\begin{equation}
  \begin{aligned}
    K_{\lloop,1}&=k_{i_1}+\ldots+k_{j_1}+k_{\ibar_1}+\ldots+k_{\jbar_{1}}\,,\\
    K_{\lloop,2}&=k_{i_2}+\ldots+k_{j_2}+k_{\ibar_2}+\ldots+k_{\jbar_{2}}\,,\\
    K_{\lloop,3}&=k_{i_3}+\ldots+k_{j_3}+k_{\ibar_3}+\ldots+k_{\jbar_{3}}\,,\\
    K_{\lloop,4}&=-K_{\lloop,1}-K_{\lloop,2}-K_{\lloop,3}\,,\\
  \end{aligned}  
\end{equation}
and
\begin{equation}
  \ell_{\lloop,1}=\ell_{\lloop},\quad \ell_{\lloop,2}=\ell_{\lloop}-K_{\lloop,1}
  ,\quad \ell_{\lloop,3}=\ell_{\lloop}-K_{\lloop,1}-K_{\lloop,2},\quad \ell_{\lloop,4}=\ell_{\lloop}+K_{\lloop,4}\,.
\end{equation}
We define the operator
\begin{equation}
  \CoeffOneR(c)\equiv\CoeffOneR(\{\ESet_{\rloop};\ESet_{\inner,\rloop}\},\{\ESet_{\central,\rloop};\ESet_{\cbottom};\ESet_{\central,\lloop};\ESet_{\ctop}\})\,,
  \label{eq:CR_def}
\end{equation}
which determines the integral coefficient of the right loop, consisting of the
tree amplitudes,
\begin{equation}
  \begin{aligned}
    &\hphantom{{}\times{}}\ATree((-\ell_{\rloop,1})_{\scalarP},i_5^+,\ldots,j_5^+,(\ell_{\rloop,2})_{\scalarP},\ibar_5^+,\ldots,\jbar_{5}^+)\\
    &\times\ATree((-\ell_{\rloop,2})_{\scalarP},i_6,\ldots,j_6,(\ell_{\rloop,3})_{\scalarP},\ibar_6,\ldots,\jbar_{6})\\
    &\times\ATree((-\ell_{\rloop,3})_{\scalarP},i_7,\ldots,j_7,(\ell_{\rloop,4})_{\scalarP},\ibar_7,\ldots,\jbar_{7})\\
    &\times\ATree(
    (-\ell_{\rloop,4})_{\scalarP},i_8^+,\ldots,j_8^+,
    (\ell_{\lloop,1})_{\scalar},\ibar_4^+,\ldots,\jbar_4^+,
    (-\ell_{\lloop,4})_{\scalar},i_4^+,\ldots,j_4^+,
    (\ell_{\rloop,1})_{\scalarP},\ibar_8^+,\ldots,\jbar_8^+)
  \end{aligned}
\end{equation}
The result of $\CoeffOneR(c)$ is still dependent on the parameters of
the left loop momentum.

From $\CoeffOneR(c)$, we determine the integral coefficient of the
left loop, which is the complete two-loop coefficient,
\begin{equation}
  \CoeffTwo(c)\equiv \CoeffOneL(c)\equiv \CoeffOneL\Big(\{\ESet_{\lloop};\ESet_{\inner,\lloop}\},\CoeffOneR(c)\Big).
\end{equation}
Here, $\CoeffOneL(c)$ is the operator which in the example of a
$\CutClass{\BoxBox}$ contribution determines the
coefficient from the product,
\begin{equation}
  \begin{aligned}
    &\hphantom{{}\times{}}\ATree((-\ell_{\lloop,1})_{\scalarP},i_1^+,\ldots,j_1^+,(\ell_{\lloop,1})_{\scalarP},\ibar_5^+,\ldots,\jbar_{1}^+)\\
    &\times\ATree((-\ell_{\lloop,2})_{\scalarP},i_2,\ldots,j_2,(\ell_{\lloop,3})_{\scalarP},\ibar_2,\ldots,\jbar_{2})\\
    &\times\ATree((-\ell_{\lloop,3})_{\scalarP},i_3,\ldots,j_3,(\ell_{\lloop,4})_{\scalarP},\ibar_3,\ldots,\jbar_{3})\\
    &\times \CoeffOneR(c)\,.
  \end{aligned}
\end{equation}

The rational contribution $\RT_{c}$ of a triple-trace cut $c$ is then made up
of the two integrals $I_{c,\lloop}^{(1)}$, $I_{c,\rloop}^{(1)}$,
the coefficient $\CoeffTwo(c)$,
and the factor $4$ relating $\RT_{\twoscalar}$ and
$\RT_{\withcontact}$ to the all-plus rational part,
\begin{equation}
\begin{aligned}
  \RT_{c}&=4\,I^{(1)}_{c,\lloop}I^{(1)}_{c,\rloop}\,\CoeffTwo(c)
  \\&=4\,I^{(1)}_{c,\lloop}I^{(1)}_{c,\rloop}\,\CoeffOneL\Big(\{\ESet_{\lloop};\ESet_{\inner,\lloop}\},\CoeffOneR(\{\ESet_{\rloop};\ESet_{\inner,\rloop}\},\{\ESet_{\central,\rloop};\ESet_{\cbottom};\ESet_{\central,\lloop};\ESet_{\ctop}\})\Big).
\end{aligned}
\end{equation}

\subsubsection{Subleading Single-Trace Cuts}
\label{sec:cut_computation_1B}
As described in section~\ref{sec:sub_single_traces_generating_sets},
the labels of subleading single-trace cuts have the generic form,
\begin{equation}
  (\ESet_{1};\ESet_{\central,1}|\ESet_{2};\ESet_{\central,2}|
  \ESet_{3};\ESet_{\central,3}|\ESet_{4};\ESet_{\central,4})\,.
\end{equation}
We compute the rational contribution of such a
cut in the same way as the triple-trace cuts.
To define the notation, we use the most generic cut,
shown in \fig{fig:1B_generic_cut}.  It is of the $\{\BoxBox\}$ type.
The treatment of cuts belonging to other classes follows similarly.

We can again treat the two loops one after the other.
We call the loop defined by $\ESet_{1}$, $\ESet_{3}$ loop $\loopone$,
while the one defined by $\ESet_{2}$, $\ESet_4$ we call loop $\looptwo$.
We further make the choice of always computing the coefficient of
loop $\looptwo$ first, which then enters in the computation of the
coefficient of loop $\loopone$.

Define the loop momentum $l_{\looptwo}$ via the box parametrization of
\eqn{eq:BoxLoopMomentumAnsatz} with
\begin{equation}
  \begin{aligned}
    K_{\looptwo,1}&=k_{i_{13}}+\ldots+k_{j_{13}}+k_{i_7}+\ldots+k_{j_7}\,,\\
    K_{\looptwo,2}&=k_{i_{14}}+\ldots+k_{j_{14}}+k_{i_6}+\ldots+k_{j_6}\,,\\
    K_{\looptwo,3}&=k_{i_{15}}+\ldots+k_{j_{15}}+k_{i_5}+\ldots+k_{j_5}\,,\\
    K_{\looptwo,4}&=-K_{\looptwo,1}-K_{\looptwo,2}-K_{\looptwo,3}\,,
  \end{aligned}
\end{equation}
and 
\begin{equation}
  \ell_{\looptwo,1}=\ell_{\looptwo},\quad \ell_{\looptwo,2}=\ell_{\looptwo}-K_{\looptwo,1}
  ,\quad \ell_{\looptwo,3}=\ell_{\looptwo}-K_{\looptwo,1}-K_{\looptwo,2},\quad \ell_{\looptwo,4}=\ell_{\looptwo}+K_{\looptwo,4}\,.
\end{equation}
Similarly, define $\ell_{\loopone}$ via \eqn{eq:BoxLoopMomentumAnsatz} using,
\begin{equation}
  \begin{aligned}
    K_{\loopone,\loopone}&=k_{i_{1}}+\ldots+k_{j_{1}}+k_{i_{11}}+\ldots+k_{j_{11}}\,,\\
    K_{\loopone,2}&=k_{i_{2}}+\ldots+k_{j_{2}}+k_{i_{10}}+\ldots+k_{j_{10}}\,,\\
    K_{\loopone,3}&=k_{i_{3}}+\ldots+k_{j_{3}}+k_{i_9}+\ldots+k_{j_9}\,,\\
    K_{\loopone,4}&=-K_{\loopone,1}-K_{\loopone,2}-K_{\loopone,3}\,,
  \end{aligned}
\end{equation}
and
\begin{equation}
  \ell_{\loopone,1}=\ell_{\loopone},\quad \ell_{\loopone,2}=\ell_{\loopone}-K_{\loopone,1}
  ,\quad \ell_{\loopone,3}=\ell_{\loopone}-K_{\loopone,1}-K_{\loopone,2},\quad \ell_{\loopone,4}=\ell_{\loopone}+K_{\loopone,4}\,.
\end{equation}
The coefficient of loop $2$ is then defined by the product of amplitudes,
\begin{equation}
  \begin{aligned}
    &\hphantom{{}\times{}}\ATree((-\ell_{\looptwo,1})_{\scalarP},i_{13}^+,\ldots,j_{13}^+,(\ell_{\looptwo,2})_{\scalarP},i_7^+,\ldots,j_{7}^+)\\
    &\times\ATree((-\ell_{\looptwo,2})_{\scalarP},i_{14},\ldots,j_{14},(\ell_{\looptwo,3})_{\scalarP},i_6,\ldots,j_{6})\\
    &\times\ATree((-\ell_{\looptwo,3})_{\scalarP},i_{15},\ldots,j_{15},(\ell_{\looptwo,4})_{\scalarP},i_5,\ldots,j_5)\\
    &\times\ATree(
    (-\ell_{\looptwo,4})_{\scalarP},i_{16}^+,\ldots,j_{16}^+,
    (\ell_{\loopone,1})_{\scalar},i_{12}^+,\ldots,j_{12}^+,
    \\&\hspace*{14mm}
      (\ell_{\looptwo,1})_{\scalarP},i_8^+,\ldots,j_8^+,
    (-\ell_{\loopone,4})_{\scalar},i_4^+,\ldots,j_4^+)\,.
  \end{aligned}
\end{equation}
The scalar lines are attached differently to the amplitude
in the last line than in the triple-trace case.
We call the operator computing this coefficient for a generic label $c$,
\begin{equation}
  \CoeffOneTwo(c)\equiv\CoeffOneTwo(\{\ESet_{2};\ESet_4\},
  \{\ESet_{\central,1},\ESet_{\central,2},\ESet_{\central,3},
  \ESet_{\central,4}\})\,.
\end{equation}

The coefficient of the two-loop cut is then given by the one-loop
coefficient of loop $\loopone$, defined by the product,
\begin{equation}
  \begin{aligned}
    &\hphantom{{}\times{}}\ATree((-\ell_{\loopone,1})_{\scalar},i_1^+,\ldots,j_1^+,(\ell_{\loopone,1})_{\scalar},i_{11}^+,\ldots,j_{11}^+)\\
    &\times\ATree((-\ell_{\loopone,2})_{\scalar},i_2,\ldots,j_2,\ell_{\loopone,3},i_{10},\ldots,j_{10})\\
    &\times\ATree((-\ell_{\loopone,3})_{\scalar},i_3,\ldots,j_3,\ell_{\loopone,,4},i_9,\ldots,j_{9})\\
    &\times \CoeffOneTwo(c)\,.
  \end{aligned}
\end{equation}
We again define an operator,
\begin{equation}
  \CoeffTwo(c)\equiv\CoeffOneOne(c)\equiv\CoeffOneOne(\{\ESet_{1};\ESet_{3}\},\CoeffOneTwo(c))\,,
\end{equation}
which computes this coefficient.
The rational contribution associated to a subleading 
single-trace cut labeled by $c$ is then
\begin{equation}
\begin{aligned}
  \RT_{c}&=4\,I^{(1)}_{c,\loopone}I^{(1)}_{c,\looptwo}\,\CoeffTwo(c)
  \\&=4\,I^{(1)}_{c,\loopone}I^{(1)}_{c,\looptwo}\,
  \CoeffOneOne\Big(\{\ESet_{1};\ESet_{3}\},
  \CoeffOneTwo(\{\ESet_{2};\ESet_{4}\},
  \{\ESet_{\central,1};\ESet_{\central,2};
  \ESet_{\central,3};\ESet_{\central,4}\})\Big)\,,
\end{aligned}
\end{equation}
where $I^{(1)}_{c,\loopone}$ and $I^{(1)}_{c,\looptwo}$ are 
one-loop
integrals for loops $\loopone$ and $\looptwo$ with appropriate
numerator powers of $\mu^2$.

\section{Tree-Level Amplitudes}
\label{TreeLevelAmplitudesSection}
\label{IngredientAmplitudesSection}

In order to compute the two-loop partial amplitudes using the cuts
described in the previous section, we need a
variety of tree amplitudes with four-dimensional gluons for the
external legs, and six-dimensional scalars for the internal (cut)
lines.
We need two different kinds of such amplitudes: those with a single
scalar line, which appear in either loop;
and those with two scalar lines of different flavors, appearing as the
`central' amplitude joining the two loops.
We will call the former two-scalar amplitudes, and the latter
four-scalar ones.
For the latter, we can further distinguish between contributions where
the two scalar lines are connected by gluon exchange or by a
four-scalar vertex. 
Ultimately, these two contributions appear together with the same
coupling ($g^2$), but it will be convenient to distinguish them.
Notably the subleading rational parts $\RT_{n:\oneB}$, receive
contributions only from the four-scalar vertex terms in the
four-scalar amplitudes.

One could imagine computing the tree amplitudes using the
six-dimensional techniques explained in
refs.~\cite{Cheung:2009dc,Bern:2010qa}.
However, we will see that it is sufficient to use the equivalent
four-dimensional amplitudes, replacing the six-dimensional massless
scalars with four-dimensional massive ones.
Some expressions for the required amplitudes are already available in
the literature.
We must just pay a bit of attention to the translation from
six-dimensional kinematic variables, and to the contact terms (usually not
considered in a scalar QCD context).

\subsection[]{Tree amplitudes with One Scalar Pair}
\label{subsec:two-scalar-trees}

We consider first six-dimensional tree amplitudes with a single
massless scalar
pair,
\begin{equation}
  A^{(0)}(1_\scalar, 2^+, \mathellipsis ,(n-1)^+,n_\scalar).
\end{equation}
As we only have one scalar line, the quartic scalar interactions play
no role in these amplitudes.
As the gluons carry only four-dimensional momenta, the
$(D-4)$-dimensional momentum components are conserved within the
scalar pair.
These components can appear only as squares of the extra-dimensional
components of the corresponding momenta,
\begin{equation}
  \mu^2=(p^{5}_1)^2+(p^{6}_1)^2=(p^{5}_n)^2+(p^{6}_n)^2\,.
\end{equation}
The six-dimensional scalars are massless, so we can interpret
$\mu^2$ as the mass of corresponding four-dimensional scalars,
$\mu^2=\overline{p}_1^2=\overline{p}_n^2$.
The required amplitudes are then exactly those of
four-dimensional massive scalar QCD, with squared scalar mass $\mu^2$.
The only kinematic variables that require translation are the
six-dimensional Mandelstams $\ssix$.
Their four-dimensional form depends on the number
of scalars $N_{\phi}$ appearing,
\begin{equation}
  \ssix_{1\ldots j}\rightarrow\left\lbrace
    \begin{alignedat}{2}
      &s_{1\ldots j},\quad && N_\phi=0\mod 2\\
      &\LProd_{1\ldots j},\quad && \text{otherwise}\,.
    \end{alignedat}
  \right.
\end{equation}
where,
\begin{equation}
  \LProd_{1\ldots j}=s_{1\ldots j}-\sum_{i=1}^j k_i^2\,.
  \label{eq:LProd_def}
\end{equation}
The $s_{1\ldots j}$ are four-dimensional Mandelstam invariants
defined by the (in the case of scalars massive) four-dimensional 
momenta $k_i$.

When determining two-scalar amplitudes from complex on-shell recursion,
gluonic factorization channels appear, in which we have to sum over
all polarization states.
For six-dimensional amplitudes, this sum would in principle include
two six-dimensional states, in addition to the two four-dimensional
positive and negative helcities.
However, when the external gluon momenta are four-dimensional as in
our case, the contributions of these additional states have to vanish.
This can be seen from the equivalence to amplitudes of massive
scalar QCD.

Results for four-dimensional two-scalar amplitudes may be found in
refs.~\cite{Badger:2005zh,Forde:2005ue,Rodrigo:2005eu,Schwinn:2006ca,
  Ferrario:2006np}; we use the forms of
ref.~\cite{Badger:2005zh}.
We list the relevant expressions in
\app{ScalarAmplitudesAppendix}.

\begin{figure}[htb]
  \captionsetup[subfigure]{justification=centering}
  \hspace*{1mm}
  \begin{subfigure}[]{0.3\textwidth}
    \includegraphics[height=2cm]{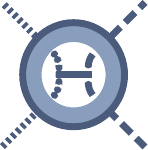}
    \caption{}
  \end{subfigure}
  \begin{subfigure}[]{0.3\textwidth}
    \includegraphics[height=2cm]{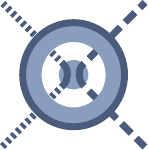}
    \caption{}
  \end{subfigure}
  \begin{subfigure}[]{0.3\textwidth}
    \includegraphics[height=2cm]{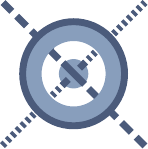}
    \caption{}
  \end{subfigure}
  \hspace*{10mm}
  \caption{The three distinct four-point four-scalar amplitudes:
    (a) $A^{\tree}_{\gluon}(1_\scalar 2_\scalar 3_{\scalarP} 4_{\scalarP})$
    (b) planar $A^{\tree}_{\contact}(1_\scalar 2_\scalar 3_{\scalarP} 4_{\scalarP})$
    (c) nonplanar $A^{\tree}_{\contact}(1_\scalar 2_{\scalarP} 3_{\scalar} 4_{\scalarP})$.
    The open circle inside the shaded annulus does \emph{not} represent a loop, but is rather
    a `window' depicting the type of four-point coupling between the two scalar lines.
  }
  \label{fig:four-scalar_tree_amplitudes}
\end{figure}

\subsection[]{Tree Amplitudes with Two Scalar Pairs}
\label{subsec:four-scalar-trees}
Next we consider four-scalar tree amplitudes, 
with distinct flavors
of scalars.  Here, we have two classes of amplitudes, 
distinguished by the manner by which the two scalar lines
are connected.
Amplitudes where the scalar
lines are connected by an exchanged gluon we will call
$\ATree_\gluon$, while those involving a four-scalar contact term we
will call $\ATree_\contact$.
While we could treat both cases together by considering the sum of
gluon and contact term contributions, we make this distinction to
better show their individual structure.
In computing
$\RT_{\twoscalarcontact}=\RT_{\twoscalar}+\RT_{\withcontact}$, however,
we always use the sum,
\begin{equation}
  \ATree=\ATree_{\gluon}+\ATree_{\contact}\,.
\end{equation}

The simplest amplitudes with two scalar pairs are four-point ones
with no external gluon legs.
There are three different kinds: the gluon-exchange one
$A^{\tree}_{\gluon}(1_\scalar 2_\scalar 3_{\scalarP} 4_{\scalarP})$,
a planar contact one
$A^{\tree}_{\contact}(1_\scalar 2_\scalar 3_{\scalarP} 4_{\scalarP})$,
and a nonplanar contact one
$A^{\tree}_{\contact}(1_\scalar 2_{\scalarP} 3_{\scalar} 4_{\scalarP})$.
They are shown pictorially in \fig{fig:four-scalar_tree_amplitudes}%
\footnote{We thank Ingrid Vazquez-Holm 
for the graphic-design idea.}.

The first of the three is required for $\RT_{\twoscalar}$, while the
latter two are needed for $\RT_{\withcontact}$.
Amplitudes of the type
$A^{\tree}_{\contact}(1_\scalar 2_{\scalarP} 3_{\scalar} 4_{\scalarP})$
are required only for the subleading single-trace rational parts
$\RT_{n:1\textrm{B}}$, as
hinted at in section~\ref{ColorStructureSection}.
If the scalar lines cross as shown in the third amplitude, the Feynman
rules of \eqn{eq:Feyn_rules_scalars} do not allow for a gluon-exchange
contribution.
Accordingly, partial
amplitudes of the type
$A^{\tree}_{\gluon}(1_\scalar 2_{\scalarP} 3_{\scalar} 4_{\scalarP})$ do not
exist.

In amplitudes of all three classes,
the $(D-4)$-dimensional momentum components are separately
conserved along each scalar line.
In the gluon-exchange amplitude
$A^{\tree}_{\gluon}$, the gluon propagator
contracts the momenta of the scalar lines, and in particular their
$(D-4)$-dimensional components.
These contractions survive only in fully six-dimensional 
amplitudes, and we cannot obtain such contributions using purely
four-dimensional techniques.  
In the language of
$D$-dimensional unitarity the missing terms will be proportional to
$\mu_{12}$.  For one-loop squared topologies, $\mu_{12}$ integral
coefficients are guaranteed to vanish by Lorentz 
invariance\footnote{Higher powers cannot arise in Yang--Mills
theory.}.
For our purposes, however, the
terms arising from four-dimensional contractions suffice.
We can thus use amplitudes computed in massive, four-dimensional,
scalar QCD.

If we were to start from six-dimensional amplitudes, we would again
have to correctly translate the Mandelstams $\ssix$ to
four-dimensional quantities.
Similar to the case of two-scalar amplitudes, this translation depends on
the number of scalars of each pair appearing.
Denoting these numbers as $N_\phi$ and $N_{\phi'}$, we find the translation,
\begin{equation}
  \ssix_{1\ldots j}\rightarrow\left\lbrace
    \begin{alignedat}{2}
      &s_{1\ldots j},\quad && (N_\phi,N_{\phi'})=(0,0)\mod 2,\\
      &\LProd_{1\ldots j},\quad && \text{otherwise}\,.
    \end{alignedat}
  \right.
  \label{eq:ssix_to_four_four_scalars}
\end{equation}

The amplitudes in \fig{fig:four-scalar_tree_amplitudes} are each
computed from a single Feynman diagram,
with results,
\begin{align}
  \label{eq:SSss_gluon_amplitude}
  A^{(0)}_{\gluon}(1_{\scalar}2_{\scalar}3_{\scalarP}4_{\scalarP})&=
                                                                    -\left(\frac{1}{2}+\frac{\LProd_{23}}{s_{12}}\right)\,,\\
  \label{eq:SSss_contact_amplitude}
  A^{(0)}_{\contact}(1_{\scalar}2_{\scalar}3_{\scalarP}4_{\scalarP})&=
                                                                      -\frac{1}{2}\,,\\
  \label{eq:SsSs_contact_amplitude}
  A^{(0)}_{\contact}(1_{\scalar}2_{\scalarP}3_{\scalar}4_{\scalarP})&=
                                                                      1\,.
\end{align}
The expression for
$A^{(0)}_{\gluon}(1_{\scalar}2_{\scalar}3_{\scalarP}4_{\scalarP})$ is
consistent with the one used in ref.~\cite{Bern:2000dn}, as well as the result
of ref.~\cite{Carrasco:2020ywq}%
\footnote{The results in ref.~\cite{Carrasco:2020ywq} are defined 
up to prefactors, which in this case needs to be $2$ for 
agreement with our expression.}.

\begin{figure}
  \centering
  \begin{minipage}[b]{0.3\linewidth}
    \centering
    \scalebox{1.2}{
    \includegraphics[valign=c]
    {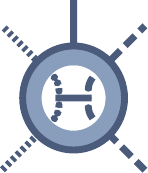}}\\
    $A^{(0)}_{\gluon}(1_{\scalar} 2_{\scalar} 3^+ 4_{\scalarP} 5_{\scalarP})$
  \end{minipage}
  \begin{minipage}[b]{0.3\linewidth}
    \centering
    \scalebox{1.2}{
    \includegraphics[valign=c]
    {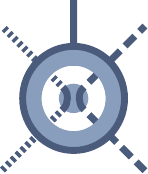}}\\
    $A^{(0)}_{\contact}(1_{\scalar} 2_{\scalar} 3^+ 4_{\scalarP} 5_{\scalarP})$
  \end{minipage}
  \begin{minipage}[b]{0.3\linewidth}
    \centering
    \scalebox{1.2}{
    \includegraphics[valign=c]
    {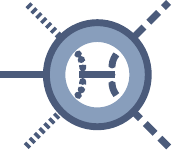}}\\
     $A^{(0)}_{\gluon}(1_{\scalar} 2^+ 3_{\scalar} 4_{\scalarP} 5_{\scalarP})$
  \end{minipage}\\[1em]
  \begin{minipage}[b]{0.3\linewidth}
    \centering
    \scalebox{1.2}{
    \includegraphics[valign=c]
    {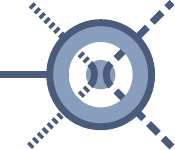}}\\
    $A^{(0)}_{\contact}(1_{\scalar} 2^+ 3_{\scalar} 4_{\scalarP} 5_{\scalarP})$
  \end{minipage}
  \begin{minipage}[b]{0.3\linewidth}
    \centering
    \scalebox{1.2}{
    \includegraphics[valign=c]
    {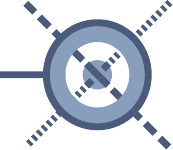}}\\
    $A^{(0)}_{\contact}(1_{\scalar} 2^+ 3_{\scalarP} 4_{\scalar} 5_{\scalarP})$
  \end{minipage}
  \caption{All independent four-scalar amplitudes with an additional gluon attached.}
  \label{fig:FourScalarFivePointAmplitudes}
\end{figure}

Next we consider five-point amplitudes with a single gluon in addition to the
two scalar pairs.
There are five unique amplitudes, shown in
\Fig{fig:FourScalarFivePointAmplitudes}.
For the gluon-exchange amplitudes we again use Feynman diagrams to obtain the
compact expressions,
\begin{align}
&\begin{aligned}
    \label{eq:SS+ss_gluon_Feynman}
    A^{(0)}_{\gluon}&(1_{\scalar}2_{\scalar}3^+4_{\scalarP}5_{\scalarP})= \\
    &
    -\frac{\LProd_{2 4} (\LProd_{2 3} \spbb{3 | 4 5 | 3}+\LProd_{3 4}
      \spbb{3 | 1 2 | 3})}{2 \sfour_{1 2} \LProd_{2 3} \LProd_{3 4}
      \sfour_{4 5}}\\
    &-\frac{\spbb{3 | 2 4 | 3} (\sfour_{1 2} (\LProd_{3 4}-\LProd_{1 4})+
      \sfour_{4 5} (\LProd_{2 3}-\LProd_{2 5})+ 2 \LProd_{2 3} \LProd_{3 4}-\sfour_{1 2} \sfour_{4 5})}{2 \sfour_{1 2} \LProd_{2 3} \LProd_{3 4}
      \sfour_{4 5}}\,,
  \end{aligned}\\
    &\begin{aligned}
        A^{(0)}_{\gluon}(1_{\scalar}2^+3_{\scalar}4_{\scalarP}5_{\scalarP}) & =
       -\frac{\spbb{2|13|2}(\LProd_{34}-\LProd_{35})
        +\LProd_{12}\spbb{2|(4-5)3|2}}{2\LProd_{12}\sfour_{45}
       \LProd_{23}}\,.\hspace*{5mm}
    \end{aligned}
\end{align}
We verified the expression in eq.~\eqref{eq:SS+ss_gluon_Feynman} 
numerically using Berends--Giele recursion.
It is also in agreement with the result of
ref.~\cite{Carrasco:2020ywq}%
\footnote{ As in the case of eq.~\eqref{eq:SSss_gluon_amplitude}, the result of
  ref.~\cite{Carrasco:2020ywq} requires a factor of $2$ for agreement, due
  to being defined up to prefactors.  }.
We obtain the contact-term
amplitudes $\ATree_{\contact}$ from off-shell scalar currents.
In \app{OffShellCurrentAppendix} we use this
construction to compute the contact-term amplitudes with one and two
gluons.
As the currents are agnostic to the scalar flavor, we can express
these amplitudes as a product of the contact term with a common
kinematic factor,
\begingroup
\allowdisplaybreaks
\begin{align}
&\hphantom{6_{\scalar_4}\!}\begin{aligned}
  \label{eq:SS+ss_contact_from_currents}
  \ATree_{\contact}&(1_{\scalar_1} 2_{\scalar_2} 3^+ 4_{\scalar_3} 5_{\scalar_4})
  =
  \\&\hspace*{-3.5mm} V(\scalar_1 \scalar_2 \scalar_3 \scalar_4)\times\left[
    \frac{\spbb{3|24|3}}{\ssix_{23}\ssix_{34}}\right]\,,
\end{aligned}\\
&\begin{aligned}
  \ATree_{\contact}&(1_{\scalar_1} 2_{\scalar_2} 3^+ 4^+ 5_{\scalar_3} 6_{\scalar_4})
  =
  \\&V(\scalar_1 \scalar_2 \scalar_3 \scalar_4)\times\left[
    \frac{\spbb{3|25|4}}{\ssix_{2 3} \ssix_{4 5}\spaa{3 4}}-
    \frac{\spbb{3 4}}{\spaa{3 4}} \left(\frac{m_2^2}{\ssix_{2 3} \ssix_{2 3 4}}+
    \frac{m_5^2}{\ssix_{3 4 5} \ssix_{4 5}}\right)\right]\,,\\
\end{aligned}\\
&\begin{aligned}
  \ATree_{\contact}&(1_{\scalar_1} 2_{\scalar_2} 3^+ 4_{\scalar_3} 5_{\scalar_4} 6^+)
  =
  \\&V(\scalar_1 \scalar_2 \scalar_3 \scalar_4)\times\left[
    \frac{\spbb{3|24|3}}{\ssix_{23}\ssix_{34}}\frac{\spbb{6|51|6}}{\ssix_{56}\ssix_{61}}\right],\\
\end{aligned}\\
&\begin{aligned}
  \ATree_{\contact}&(1_{\scalar_1} 2^+ 3_{\scalar_2}  4^+ 5_{\scalar_3} 6_{\scalar_4})
  =
  \\&V(\scalar_1 \scalar_2 \scalar_3 \scalar_4)\times
    \left[\frac{m_{3}^2 \spbb{2 4}^2}{\ssix_{2 3} \ssix_{2 3 4} \ssix_{3 4}}
    +\frac{\spbb{2| 1 3 | 2} \spbb{4 | 3 5 | 4}}{\ssix_{1 2} \ssix_{2 3} \ssix_{3 4} \ssix_{4 5}}\right].
\end{aligned}
\end{align}
\endgroup
In these expressions, the six-dimensional Mandelstams $\ssix$ need to be replaced 
with four-dimensional kinematic variables according to \eq{eq:ssix_to_four_four_scalars},
once the scalar flavors $\scalar_i$ are fixed.
We also derived a compact expression for the seven-point contact term amplitude, where 
all gluons are adjacent,
\begin{equation}
    \begin{aligned}
        \ATree_{\contact}&(1_{\scalar_1} 2_{\scalar_2} 3^+ 4^+ 5^+ 6_{\scalar_3} 7_{\scalar_4})=\\
        &V(\scalar_1 \scalar_2 \scalar_3 \scalar_4)\times\biggl[
        \frac{m^2_{1} \spbb{5 | (3+4) 2 | 3}}{\ssix_{2 3} \ssix_{2 3 4} \ssix_{2 3 4 5} \spaa{3 4} \spaa{4 5}}+
        \frac{m^2_{1} \spbb{5| 6 4 | 3}}{\ssix_{2 3} \ssix_{2 3 4} \ssix_{5 6} \spaa{3 4} \spaa{45}}
        +\frac{m_{7}^2 \spbb{5 | 4 2 | 3} }{\ssix_{2 3} \ssix_{4 5 6} \ssix_{5 6} \spaa{3 4} \spaa{4 5}}\\
        &\hspace{7em}+\frac{m^2_{7} \spbb{5 | 6 (4+5) | 3}}{\ssix_{3 4 5 6} \ssix_{4 5 6} \ssix_{5 6} \spaa{3 4} \spaa{4 5}}
        -\frac{\spbb{5 | 6 2 | 3}}{\ssix_{2 3} \ssix_{5 6} \spaa{3 4} \spaa{4 5}}
        \biggr]
    \end{aligned}
\end{equation}

For gluon-exchange amplitudes with more than one gluon, as well as
contact-term amplitudes with more than two gluons we make use of BCFW on-shell recursion~\cite{Britto:2005fq}.
If such an amplitude has two adjacent gluons,
we can use a standard BCFW gluon-gluon shift.
We have verified using
Berends--Giele recursion that the relevant amplitudes with up to three
gluons scale as $\frac{1}{z}$ or better under such shifts.
If an amplitude does not possess a pair of adjacent gluons, we can
still use recursion, now shifting the momenta of an adjacent 
gluon and
massive scalar pair.
Such shifts have been used for example in
refs.~\cite{Badger:2005zh,Schwinn:2007ee} to compute tree amplitudes
with a single massive scalar line.
In the case of four-scalar amplitudes, such shifts can also be used,
provided that only the gluon's angle spinor is shifted, and the
shifted gluon and scalar are adjacent.
We will demonstrate the method by recomputing
$A^{(0)}_{\gluon}(1_{\scalar}2_{\scalar}3^+4_{\scalarP}5_{\scalarP})$ and
$A^{(0)}_{\contact}(1_{\scalar}2_{\scalar}3^+4_{\scalarP}5_{\scalarP})$.

As in ref.~\cite{Schwinn:2007ee}, we choose to construct the shifted
momenta using the massless projection of the massive momentum with
respect to the massless one.
For the amplitude in question we use a $\spab{3,2^\flat}$-shift of the
form
\begin{equation}
  \label{eq:BCFW_SG_flat_shift}
  \begin{gathered}
    k_2\to\hat{k}_2=k_2-\frac{z}{2}\spba{3|\gamma|2^\flat},\quad
    k_3\to\hat{k}_3=k_3+\frac{z}{2}\spba{3|\gamma|2^\flat},\\
    \lambda_3\to\lambda_3=\lambda_3+z\lambda_2^\flat,\quad
    \tilde{\lambda}_3\to\hat{\tilde{\lambda}}_3=\tilde{\lambda}_3.
  \end{gathered}
\end{equation}
The momentum $k_2^\flat$ is the massless projection of $k_2$ on
$k_3$~\cite{Kosower:2004yz},
\begin{equation}
  k_2^{\flat}= k_2-\frac{\mu^2_2}{2 (k_2\cdot k_3)}k_3.
\end{equation}
From this definition we see that the shifted momenta $\hat{k}_2$,
$\hat{k}_3$ satisfy all the requirements for a BCFW shift
\begin{equation}
  \begin{gathered}
    \hat{k}_2^2=\hat{k}_2\cdot k_2 = k_2^2,\qquad
    \hat{k}_3^2=\hat{k}_3\cdot k_3 = k_3^2,\\
    \hat{k}_2\cdot \hat{k}_3 =\hat{k}_2\cdot k_3 =
    \hat{k}_3\cdot k_2 = k_{2}\cdot k_3.
  \end{gathered}
\end{equation}
Using this shift, the gluon-exchange amplitude can be computed via
\begin{equation}
  \label{eq:A5SS+ss_gluon_BCFW}
  \begin{aligned}
    A^{(0)}_{\gluon}(1_{\scalar} 2_{\scalar}  3^+ 4_{\scalarP} 5_{\scalarP})=&
    -\sum_{h=\pm}A^{(0)}(\hat{2}_{\scalar},\hat{K}^h,1_{\scalar})
    \frac{1}{\sfour_{12}}
    A^{(0)}(5_{\scalarP},(-\hat{K})^{-h},\hat{3}^+,4_{\scalarP})\\
    &-A^{(0)}_{\gluon}(1_{\scalar},\hat{2}_{\scalar},\hat{K}_{\scalarP},5_{\scalarP})
    \frac{1}{\LProd_{34}}A^{(0)}_3((-\hat{K})_{\scalarP},\hat{3}^+,4_{\scalarP}).
  \end{aligned}
\end{equation}
Note that we are only summing over the four-dimensional polarization
states.
Were we computing the full six-dimensional amplitude, we would at this
point also have to include the six-dimensional states%
\footnote{When summing over these states defined as in
  refs.~\cite{Cheung:2009dc,Bern:2010qa}, they need to be accompanied by a
  relative sign compared to the four-dimensional states, as the polarization
  vectors in the completeness relation in eq.~(33) of ref.~\cite{Cheung:2009dc}
  are contracted by anti-symmetric tensors.}.
However, for simplicity we
discard these terms, as they would lead to cross-terms $\mu_{14}$,
$\mu_{15}$, $\mu_{15}$, $\mu_{25}$, which will be irrelevant for our
computations.  Using the result for the four-scalar amplitude in
eq.~\eqref{eq:SSss_gluon_amplitude} as well as the two-scalar amplitudes in
Appendix~\ref{ScalarAmplitudesAppendix}, we obtain
\begin{multline}
  \label{eq:SS+ss_amplitude_BCFW}
  A^{(0)}_{\gluon}(1_{\scalar}2_{\scalar}3^+4_{\scalarP}5_{\scalarP})
  =\frac{1}{\Delta}\left(
    \frac{\spbb{3|45|3}^2m_2^2+\spbb{3|21|3}^2m_4^2}{\LProd_{12}\LProd_{45}}-
    \frac{\spbb{3|42|3}^2\LProd_{15}}{\LProd_{23}\LProd_{34}}\right)-
  \frac{\spbb{3|42|3}}{2\LProd_{23}\LProd_{34}},
\end{multline}
where $\Delta=(\LProd_{34}\spbb{3|12|3}+\LProd_{12}\spbb{3|42|3})$ is a spurious pole.
This expression matches the Feynman diagram result of
eq.~\eqref{eq:SS+ss_gluon_Feynman} numerically.

Using the same approach, we can also obtain the contact-term
amplitudes.
For the amplitude in eq.~\eqref{eq:A5SS+ss_gluon_BCFW} we find that,
\begin{equation}
  \label{eq:A5SS+ss_contact_BCFW}
  A^{(0)}_{\contact}
  (1_{\scalar} 2_{\scalar} 3^+ 4_{\scalarP} 5_{\scalarP})
  =-A^{(0)}_{\contact}
  (1_{\scalar}\hat{2}_{\scalar} \hat{K}_{\scalarP} 5_{\scalarP})
  \frac{1}{\LProd_{34}}A^{(0)}_3((-\hat{K})_{\scalarP},\hat{3}^+ 4_{\scalarP}).
\end{equation}
The result is automatically free of spurious poles, and agrees with
the expression of \eqn{eq:SS+ss_contact_from_currents}

While recursion allows us to obtain analytic expressions 
for tree amplitudes with
an arbitrary number of gluons, the resulting expressions almost 
always contain spurious poles
which cancel non-trivially.  In addition, as four-scalar 
amplitudes from
previous steps appear in the recursion, these spurious 
poles accumulate and
lead to large expressions in denominators.
As a consequence, we find that particularly for four-scalar 
amplitudes $\ATree_{\gluon}$ with more than two external gluons, 
expressions obtained from Berends--Giele recursion leads to 
better performance in automated computations.

\section{Analytic Computation of the Four-Gluon Amplitude}
\label{FourGluonSection}

In this section, we give an explicit example of the 
separable approach.  We go through the calculation
of the rational parts of the two-loop four-gluon
all-plus amplitude.
Its color decomposition for gauge group $\mathrm{SU}(N_c)$ takes the form,
\begin{equation}
  \begin{aligned}
    \mathcal{A}^{(2)}(1^+2^+3^+4^+)&=
    N_c^2\sum_{\sigma\in\modout{S_4}{\cyclicgroup_4}}\TTr(\sigma(1,2,3,4))\ATwo_{4:1}(\sigma(1,2,3,4))\\    &\hphantom{{}={}}+N_c\sum_{\sigma\in\modout{S_4}{P_{4:3}}}\TTr(\sigma(1,2))\TTr(\sigma(3,4))\ATwo_{4:3}(\sigma(1,2);\sigma(3,4))\\
    &\hphantom{{}={}}+\sum_{\sigma\in\modout{S_4}{\cyclicgroup_4}}\TTr(\sigma(1,2,3,4))\ATwo_{4:\oneB}(\sigma(1,2,3,4))
  \end{aligned}
\end{equation}
so that we have to determine three rational parts:
$\RT_{4:1}$, $\RT_{4:3}$ and $\RT_{4:\oneB}$.
\subsection{Leading-Color \texorpdfstring{$\RTwo_{4:1}$}{R(2)4:1}}
\label{subsubsec:4pt_example_leading-color}
\begin{figure}[thb]
  \centering
  \includegraphics[valign=c]{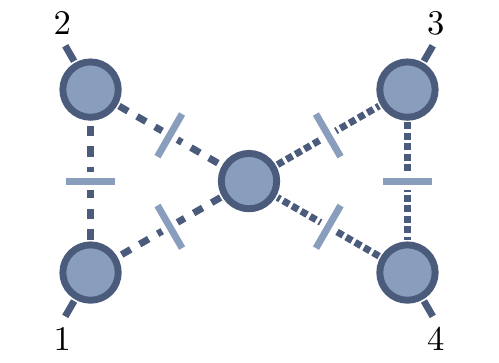}
  \includegraphics[valign=c]{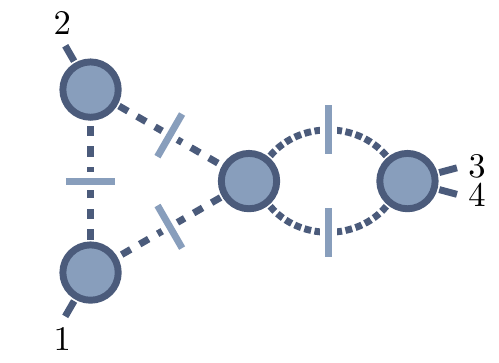}
  \includegraphics[valign=c]{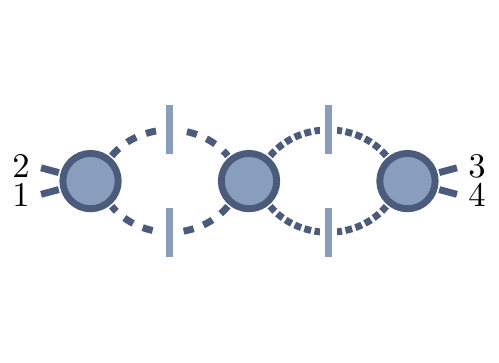}
    \caption{Cuts associated to the labels $c_{4:1;1}$, $c_{4:1;2}$, $c_{4:1;3}$ of the generating set $\GeneratingCuts_{4:1}$.
      The cuts of $c_{4:1;1}$ and $c_{4:1;3}$ are invariant under $\rho_2\circ\Exch$, and
    require symmetry factors of $\tfrac{1}{2}$.}
  \label{fig:R_41_cuts}
\end{figure}

We first discuss the computation of the leading-color rational part
$\RTwo_{4:1}$.
Its generating set is made up of three cut labels,
\begin{equation}
  \GeneratingCuts_{4:1}=\{c_{4:1;1},c_{4:1;2},c_{4:1;3}\}\,,
\end{equation}
which are
\begin{equation}
  \begin{aligned}
    &c_{4:1;1}=((1),(2);();(3),(4);()),\\
    &c_{4:1;2}=((1),(2);();(3,4);()),\\
    &c_{4:1;3}=((1,2);();(3,4);()).
  \end{aligned}
\end{equation}
These cuts are shown in \fig{fig:R_41_cuts}.
As we are dealing with a leading-color rational part of an amplitude
with an even number of particles, some cuts require symmetry factors, as
explained in section~\ref{subsec:leading_color_cuts}.
Both $c_{4:1;1}$ and $c_{4:1;3}$ require such factors, as
\begin{equation}
  \begin{aligned}
    &\rho_{2}\circ\Exch[((1),(2);();(3),(4);())]=((1),(2);();(3),(4);())\,,\\
    &\rho_{2}\circ\Exch[((1,2);();(3,4);())]=((1,2);();(3,4);())\,.\\
  \end{aligned}
\end{equation}
Accordingly,
\begin{equation}
  S_{c_{4:1;1}}=S_{c_{4:1;3}}=\frac{1}{2},\quad S_{c_{4:1;2}}=1\,.
\end{equation}

\begin{figure}[h]
  \textrm{(I)}\quad
  \includegraphics[width=0.25\textwidth,valign=c]
  {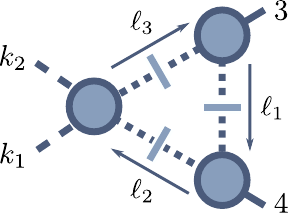},\hspace{3em}
  \textrm{(II)}\quad
  \includegraphics[width=0.28\textwidth,valign=c]
  {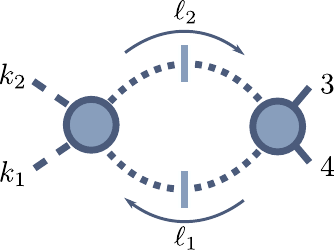}.
  \caption{The two unique cuts for the right-hand loop of $\RT_{4:1}$.}
  \label{fig:4-pt_example_rhs_cuts}
\end{figure}

Given this generating set, we can determine $\RT_{4:1}(1^+,2^+,3^+,4^+)$ via
\begin{equation}
  \RTwo_{4:1}(1^+2^+3^+4^+)=\sum_{\sigma\in \cyclicgroup_4}\TildeRT_{4:1}(\sigma[1,2,3,4]),
\end{equation}
with $\TildeRT_{4:1}(1,2,3,4)$ given by,
\begin{equation}
 \TildeRT_{4:1}(1,2,3,4)=\frac{1}{2}R_{c_{4:1;1}}(1,2,3,4)+R_{c_{4:1;2}}(1,2,3,4)+\frac{1}{2}R_{c_{4:1;3}}(1,2,3,4)\,.
  \label{eq:4pt_example_leading-color_RTilde}
\end{equation}
Thanks to the separability of the two loops we can first focus our
attention on the right-hand loop in each cut.
We need to consider only two unique contributions,
\begin{equation}
  \begin{aligned}
    \text{(I):}\ &\CoeffOneR(\{(3),(4);(),()\},\{();();();()\})\,,\\
    \text{(II):}\ &\CoeffOneR(\{(3,4);(),()\},\{();();();()\})
  \end{aligned}
\end{equation}
as shown in \fig{fig:4-pt_example_rhs_cuts}.
For the triangle coefficient (I), we use the
loop-momentum parametrization of
\eqns{eq:D-dim_triangle_loop_momentum_parametrization}{eq:D-dim_triangle_loop_momentum_parametrization_cc},
which for the present case of $K_1=k_3$, 
$K_2=k_4$ simplifies to,
\begin{equation}\label{eq:4-pt_example_rhs_triangle_l}
  \begin{aligned}
    \ell_1^\mu=\frac{1}{2}\left(t\spab{4|\gamma^\mu|3}-
      \frac{\mu^2}{s_{34}t}\spab{3|\gamma^\mu|4}\right),\\
    \ell_1^{\ast\mu}=\frac{1}{2}\left(t\spab{3|\gamma^\mu|4}-
      \frac{\mu^2}{s_{34}t}\spab{4|\gamma^\mu|3}\right).
  \end{aligned}
\end{equation}
The routing of the loop momenta used here is shown in \fig{fig:4-pt_example_rhs_cuts}.
Using the expressions for the scalar tree amplitudes from
Appendix~\ref{ScalarAmplitudesAppendix}, a brief computation leads to,
\begin{equation}\label{eq:4-pt_example_rhs_triangle}
  \begin{aligned}
    &\hphantom{{}={}}\CoeffOneR(\{(3),(4);(),()\},\{();();();()\})\\
    &=\frac{1}{2}\sum_{\ell,\ell^\ast}\Inf_{\mu^2,t}\Big[
    A^{(0)}(k_{1\scalar} k_{2\scalar}\ell_{3\scalarP}
    (-\ell_2)_{\scalarP}) A^{(0)}(3^+\ell_{1\scalarP}(-\ell_3)_{\scalarP})
    A^{(0)}(4^+\ell_{2\scalarP}(-\ell_1)_{\scalarP})\Big]
    \Big\vert_{\mu^2,t^0}\\
    &=\frac{1}{2}\Inf_{\mu^2,t}\Big[\sum_{\ell,\ell^\ast}
    \frac{\spba{3|\ell_1|4}}{\spaa{34}}
    \frac{\spba{4|\ell_1|3}}{\spaa{43}}\frac{\LProd_{k_1\ell_3}}{s_{34}}\Big]
    \Big\vert_{\mu^2,t^0}\\
    &=\frac{1}{2}\Inf_{\mu^2,t}\Big[\frac{\mu^2}{\spaa{34}^2}
    \Big(2\LProd_{k_1 3}+\Big(t\spab{4|k_1|3}-\frac{\mu^2}{s_{34}t}
    \spab{3|k_1|4}\Big)+\Big(t\spab{3|k_1|4}-\frac{\mu^2}{s_{34}t}\spab{4|k_1|3}
    \Big)\Big)\Big]\Big\vert_{\mu^2,t^0}\\
    &= \frac{\LProd_{k_1 3}}{\spaa{34}^2}.
  \end{aligned}
\end{equation}

For the bubble cut (II) we use the parametrization of%
~\eqn{eq:dim-D_bubble_loop_parametrization} with $K=k_3+k_4$.
In addition to the bubble cut, the bubble coefficient (II) 
in principle
requires the contribution from a parent triangle with $K'=-k_3$.
However, choosing the bubble reference momentum $\chi$ in
the loop-momentum parametrization of \eqn{eq:dim-D_bubble_loop_parametrization}
to be $k_3$, the integrals $T^i$ of~\eqn{eq:bubble_triangle_integrals}
vanish.
The triangle therefore does not contribute, and only the bubble cut
itself is required.

For this choice of $\chi$, the generic bubble loop-momentum
parametrization of~\eqn{eq:dim-D_bubble_loop_parametrization}
simplifies to,
\begin{equation}
  \ell^{\mu}_1=(y-1) k_4^{\mu}-yk_3^{\mu}+\frac{1}{2}\left(t\spab{4|\gamma^\mu|3}+
    \left(\frac{y(1-y)}{t}-\frac{\mu^2}{ts_{34}}\right)\spab{3|\gamma^\mu|4}\right)\,.
  \label{eq:4-pt_example_rhs_bubble_l}
\end{equation}
The bubble coefficient is then, 
\begin{equation}\label{eq:4-pt_example_rhs_bubble}
  \begin{aligned}
    &\hphantom{{}={}}\CoeffOneR(\{(3,4);(),()\},\{();();();()\})\\
    &=\Inf_{\mu^2,y,t}\Big[
    A^{(0)}(k_{1\scalar} k_{2\scalar} l_{2\scalarP} (-l_1)_{\scalarP})
    A^{(0)}(3^+ 4^+(\ell_1)_{\scalarP}(-\ell_2)_{\scalarP})
    \Big]\Big\vert_{t^0,y^i\to Y_i,\mu^2}\\
    &=-\Inf_{\mu^2,y,t}\Big[
    \frac{\LProd_{k_1(-\ell_1)}}{s_{34}}\frac{\mu^2}{\LProd_{4\ell_1}}\frac{\spbb{34}}{\spaa{34}}
    \Big]\Big\vert_{t^0,y^i\to Y_i,\mu^2}\\
    &=\Inf_{\mu^2,y,t} \Big[
    \frac{\mu^2}{y \spaa{34}^2 s_{34}}\\
    &\hspace{2em} \times \Big[ (y-1) \LProd_{k_14} - y  \LProd_{k_13} +t\spba{3|k_1|4}+\frac{y(1-y)
      s_{34}-\mu^2}{s_{34}t}\spba{4|k_1|3}\Big]\Big]\Big\vert_{t^0,y^i\to Y_i,\mu^2}\\
    &=\frac{\LProd_{k_14}-\LProd_{k_13}}{\spaa{34}^2s_{34}}.
  \end{aligned}
\end{equation}
In this case the $\Inf_{y}$ operation only yields 
a $y^0$ term, whose
parameter integral $Y_0$ is $1$.

We can now use these results to compute the three two-loop
coefficients,
\begin{equation}
  \begin{aligned}
    \CoeffTwo(c_{4:1;1})&=
    \CoeffOneL\Big(\{(1),(2);(),()\},\CoeffOneR(\{(3),(4);(),()\},\{();();();()\})\Big)\,,\\
    \CoeffTwo(c_{4:1;2})&=
    \CoeffOneL\Big(\{(1),(2);(),()\},\CoeffOneR(\{(3,4);(),()\},\{();();();()\})\Big)\,,\\
    \CoeffTwo(c_{4:1;1})&=
    \CoeffOneL\Big(\{(1,2);(),()\},\CoeffOneR(\{(3,4);(),()\},\{();();();()\})\Big)\,.
  \end{aligned}
\end{equation}
\begin{figure}
  $\mathrm{(I)}\quad
  \includegraphics[width=0.25\textwidth,valign=c]
  {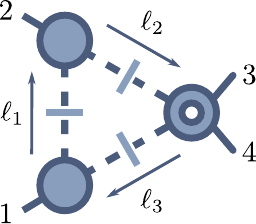},
  \hspace{3em}
  \mathrm{(II)}\quad
  \includegraphics[width=0.28\textwidth,valign=c]
  {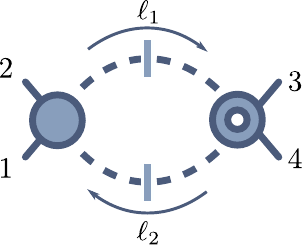}$
  \caption{The two distinct unitarity cuts of the right-hand loop that are
  required for the cut labels of $\GeneratingCuts_{4:1}$.}
  \label{fig:4-pt_example_lhs_cuts}
\end{figure}
The rational parts of  $c_{4:1;1}$ and $c_{4:1;2}$ require triangle cuts
of the form shown in (I) of \fig{fig:4-pt_example_lhs_cuts}.
We use the loop-momentum parametrization
\begin{equation}\label{eq:4-pt_example_lhs_triangle_l}
  \begin{aligned}
    \ell_1^\mu&=\frac{1}{2}\left(t\spab{2|\gamma^\mu|1}-\frac{\mu^2}{s_{12}t}
      \spab{1|\gamma^\mu|2}\right)\,,\\
    \ell_1^{\ast\mu}&=\frac{1}{2}\left(t\spab{1|\gamma^\mu|2}-
      \frac{\mu^2}{s_{12}t}\spab{2|\gamma^\mu|1}\right)\,,
  \end{aligned}
\end{equation}
with the momentum routing shown in \fig{fig:4-pt_example_lhs_cuts} (I).
The coefficients of $c_{4:1;1}$ and $c_{4:1;2}$ are then,
\begin{align}
  \begin{split}\label{eq:R_41_cut_1}
    \CoeffTwo(c_{4:1;1})
    &=\frac{1}{2}\sum_{\ell_1,\ell_1^\ast}\Inf_{\mu^2,t}\bigg[
    A^{(0)}((-\ell_3)_{\scalar} 1^+ (\ell_1)_{\scalar})
    A^{(0)}((-\ell_1)_{\scalar} 2^+ (\ell_2)_{\scalar})\\
    &\hspace{10em}\times \CoeffOneR(\{(3),(4);(),()\},\{();();();()\})\bigg]\bigg]\bigg\vert_{t^0,\mu^2}\\
    &=\frac{1}{2}\frac{s_{12}}{\spaa{12}^2\spaa{34}^2}
    \sum_{\ell_1,\ell_1^\ast}\Inf_{\mu^2,t}
    \Big[\LProd_{\ell_3 3}
    \Big]\Big\vert_{t^0,\mu^2}\\
    &=\frac{1}{2}\frac{s_{12}}{\spaa{12}^2\spaa{34}^2}\Inf_{\mu^2,t}\Big[\mu^2
    \Big(2s_{13}+\Big(t\spab{2|3|1}-\frac{\mu^2}{s_{12}t}\spab{1|3|2}\Big)\\
    &\hspace{15em}+\Big(t\spab{1|3|2}-\frac{\mu^2}{s_{12}t}\spab{2|3|1}\Big)\Big)
    \Big]\Big\vert_{\mu^2,t^0}\\
    &=\frac{s_{12}s_{13}}{\spaa{12}^2\spaa{34}^2}\,,
  \end{split}\\[2em]
  \begin{split}\label{eq:R_41_cut_2}
    \CoeffTwo(c_{4:1;2})
    &=\frac{1}{2}\sum_{\ell_1,\ell_1^\ast}\Inf_{\mu^2,t}\bigg[
    A^{(0)}((-\ell_3)_{\scalar} 1^+ (\ell_1)_{\scalar})
    A^{(0)}((-\ell_1)_{\scalar} 2^+ (\ell_2)_{\scalar})\\
    &\hspace{10em}\times \CoeffOneR(\{(3,4);(),()\},\{();();();()\})\bigg]\bigg]\bigg\vert_{t^0,\mu^2}\\
    &=\frac{1}{2}\frac{s_{12}}{\spaa{12}^2\spaa{34}^2}
    \sum_{\ell_1,\ell_1^\ast}\Inf_{\mu^2,t}
    \Big[\LProd_{\ell_3 4}-\LProd_{\ell_3 3}\Big]\Big\vert_{t^0,\mu^2}\\
    &=\frac{s_{14}-s_{13}}{\spaa{12}^2\spaa{34}^2}\,.
  \end{split}
\end{align}
The rational contributions of these cuts are then,
\newcommand{\RcFOOne}[4]{\frac{s_{#1 #2}s_{#1 #3}}{\spaa{#1 #2}^2\spaa{#3 #4}^2}}
\newcommand{\RcFOTwo}[4]{\frac{1}{6}\frac{s_{#1 #2}(s_{#1 #4}-s_{#1 #3})}{\spaa{#1 #2}^2\spaa{#3 #4}^2}}
\newcommand{\RcFOThree}[4]{\frac{2}{9}\frac{s_{#1 #2}(s_{#1 #3}-s_{#1 #4})}{\spaa{#1 #2}^2\spaa{#3 #4}^2}}
\begin{equation}
  \begin{aligned}
    \RT_{c_{4:1;1}}(1,2,3,4)&=4 I^{(2),D}_{\TriTri}(c_{4:1;1})\, \CoeffTwo(c_{4:1;1})=\CoeffTwo(c_{4:1;1})=
    \RcFOOne{1}{2}{3}{4}\,,\\
    \RT_{c_{4:1;2}}(1,2,3,4)&=4 I^{(2),D}_{\TriBub}(c_{4:1;2})\, \CoeffTwo(c_{4:1;2})=\frac{s_{12}}{6}\,\CoeffTwo(c_{4:1;2})=
    \RcFOTwo{1}{2}{3}{4}\,.\\
  \end{aligned}
  \label{eq:R2_4:1;12}
\end{equation}
For the parameterization of the left loop momentum of $c_{4:1;3}$,
we choose $\chi=k_1$ as the reference momentum, so that 
the bubble coefficient again does
not require parent-triangle cut contributions.
With this parametrization,
\begin{equation}\label{eq:4-pt_example_lhs_bubble_l}
  \ell^{\mu}_1=(y-1) k_2^{\mu}-yk_1^{\mu}+\frac{1}{2}\
  \left(t\spab{2|\gamma^\mu|1}+\left(\frac{y(1-y)}{t}-
      \frac{\mu^2}{ts_{12}}\right)\spab{1|\gamma^\mu|2}\right).
\end{equation}
The two-loop coefficient then evaluates to
\begin{equation}\label{eq:R_41_cut_3}
  \begin{aligned}
    \CoeffTwo(c_{4:1;3})
    &=\Inf_{\mu^2,t,y}\bigg[
    A^{(0)}(1^+ 2^+ (\ell_1)_{\scalar}(-\ell_2)_{\scalar} )\\
    &\hspace{10em}\times \CoeffOneR(\{(3,4);(),()\},\{();();();()\})\bigg]
    \bigg]\bigg\vert_{y^i\to Y_i, t^0, \mu^2}\\
    &=\Inf_{\mu^2,t,y}
    \Big[
    \frac{\spbb{12}}{\spaa{12}}\frac{\mu^2}{s_{2\ell_1}}
    \frac{\LProd_{4\ell_2}-\LProd_{3\ell_2}}{\spaa{34}^2s_{34}}
    \Big]\Big\vert_{y^i\to Y^i,t^0,\mu^2}\\
    &=2\frac{s_{13}-s_{14}}{\spaa{12}^2\spaa{34}^2s_{12}}\,.
  \end{aligned}
\end{equation}
The corresponding rational part is
\begin{equation}
  \RT_{c_{4:1;3}}(1,2,3,4)=4 I^{(2),D}_{\BubBub}(c_{4:1;3})\, \CoeffTwo(c_{4:1;3})=\frac{s_{12}^2}{9}\,\CoeffTwo(c_{4:1;3})=
  \RcFOThree{1}{2}{3}{4}\,.
  \label{eq:R2_4:1;3}
\end{equation}
Summing the results of~\eqns{eq:R2_4:1;12}{eq:R2_4:1;3} with their
associated symmetry factors we find,
\begin{equation}
  \begin{aligned}
    \TildeRT_{4:1}(1^+2^+,3^+4^+)=
    \frac{5s_{12}s_{23}+s_{23}^2}{18\spaa{12}\spaa{23}\spaa{34}\spaa{41}},
  \end{aligned}
\end{equation}
After summing over cyclic permutations of the external kinematics we obtain
\begin{equation}\label{eq:4pt_example_R241}
  \begin{aligned}
    \RTwo_{4:1}(1^+ 2^+ 3^+ 4^+)=\sum_{\sigma\in \cyclicgroup_4} \TildeRT_{4:1}(\sigma[1,2,3,4])=\frac{s_{13}^2+8s_{12}s_{23}}{9\spaa{12}\spaa{23}\spaa{34}\spaa{41}},
  \end{aligned}
\end{equation}
which agrees with the result of refs.~\cite{Bern:2000dn,Bern:2002tk}.

\subsubsection{Subleading-color \texorpdfstring{$\RTwo_{4:3}$}{R(2)4:3}}
\label{subsubsec:4pt_example_subleading-color_3}
\begin{figure}[h]
  \centering
  \begin{subfigure}{\linewidth}
    \includegraphics[width=0.24\linewidth,valign=c]{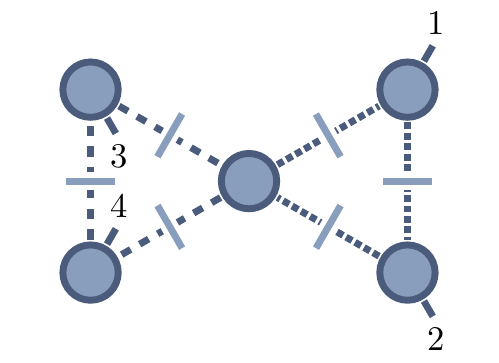}
    \includegraphics[width=0.24\linewidth,valign=c]{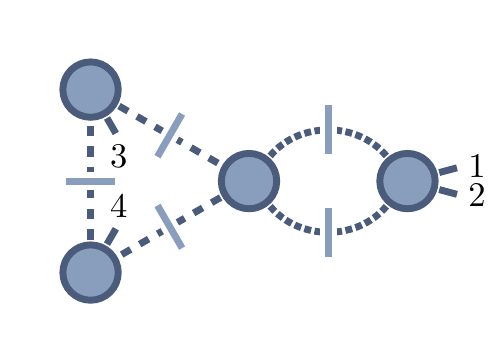}
    \includegraphics[width=0.24\linewidth,valign=c]{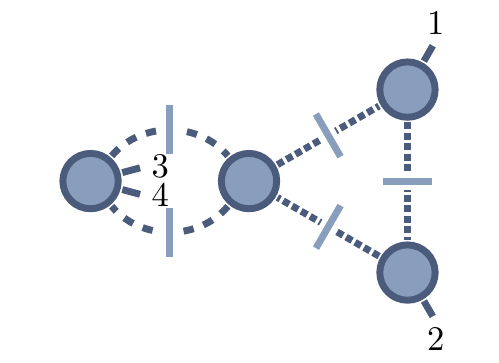}
    \includegraphics[width=0.24\linewidth,valign=c]{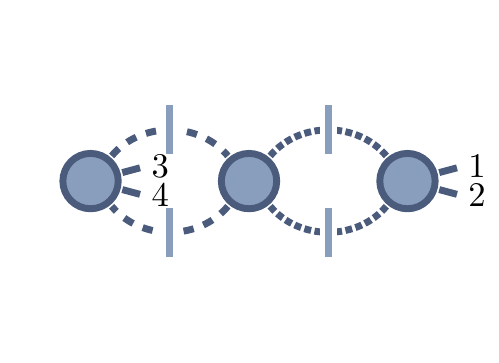}
    \caption{The cut labels $c_{4:3;1}$, $c_{4:3;2}$, $c_{4:3;3}$ and
      $c_{4:3;4}$ (in order) make up
      $\CutSet^{\{\TTr{1,2},\TTr{3,4},N_c\}}$.}
    \label{fig:all_labels_R43_1}
  \end{subfigure}
  \begin{subfigure}{\linewidth}
    \includegraphics[width=0.24\linewidth,valign=c]{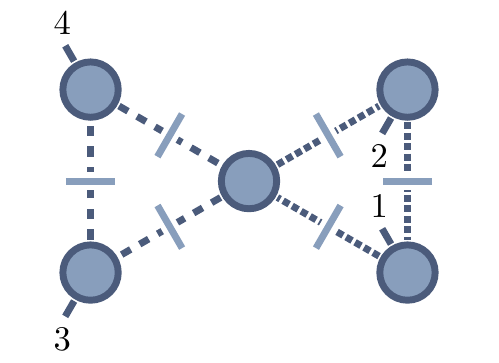}
    \includegraphics[width=0.24\linewidth,valign=c]{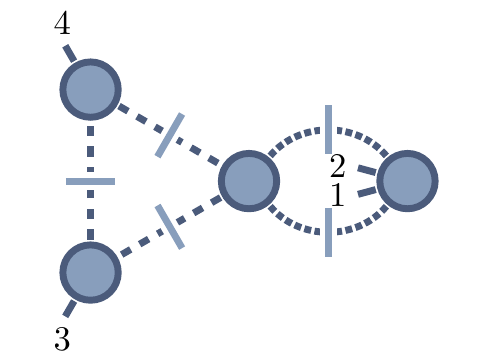}
    \includegraphics[width=0.24\linewidth,valign=c]{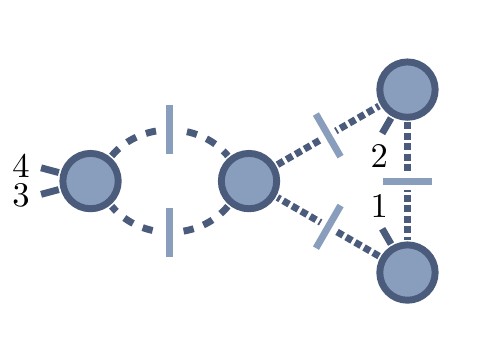}
    \includegraphics[width=0.24\linewidth,valign=c]{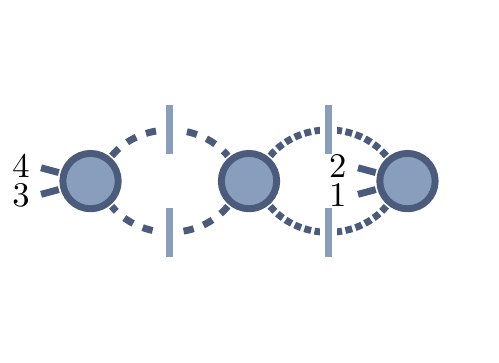}
    \caption{The cut labels $c_{4:3;5}$, $c_{4:3;6}$, $c_{4:3;7}$ and
      $c_{4:3;8}$ (in order) make up
      $\CutSet^{\{\TTr{3,4},N_c,\TTr{1,2}\}}$.}
    \label{fig:all_labels_R43_2}
  \end{subfigure}
  \begin{subfigure}{\linewidth}
    \includegraphics[width=0.24\linewidth,valign=c]{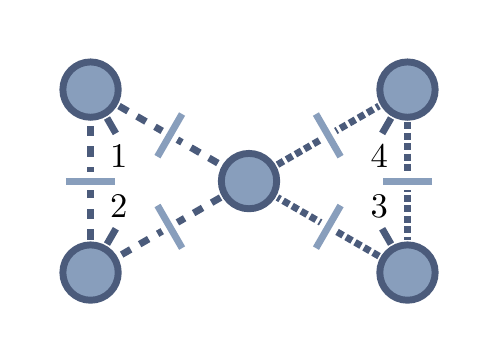}
    \includegraphics[width=0.24\linewidth,valign=c]{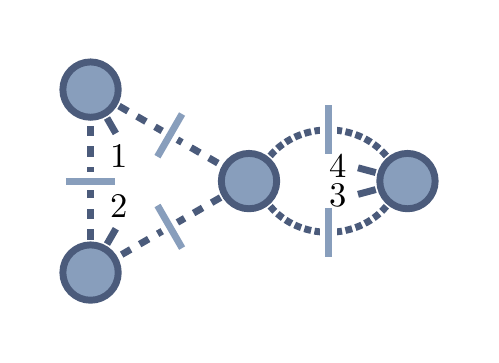}
    \includegraphics[width=0.24\linewidth,valign=c]{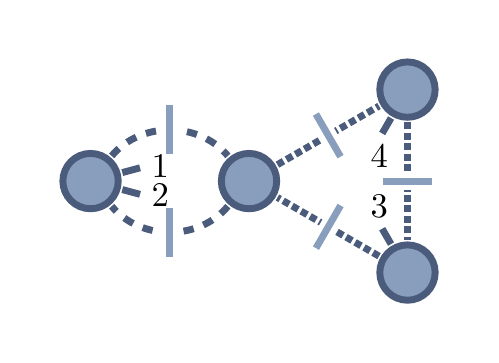}
    \includegraphics[width=0.24\linewidth,valign=c]{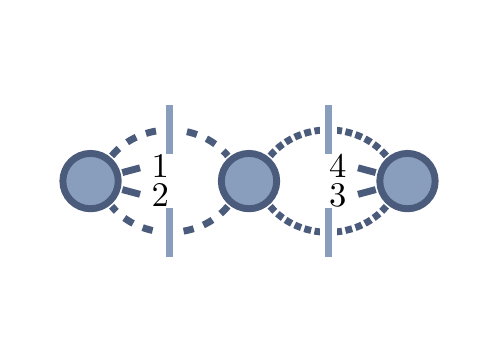}
    \caption{The cut labels $c_{4:3;9}$, $c_{4:3;10}$, $c_{4:3;11}$ and
      $c_{4:3;12}$ (in order) make up
      $\CutSet^{\{N_c,\TTr{1,2},\TTr{3,4}\}}$.}
    \label{fig:all_labels_R43_3}
  \end{subfigure}
  \caption{Sets of cut labels that make up the generating set
    $\GeneratingCuts_{4:3}$.}
  \label{fig:all_labels_R43}
\end{figure}

To compute the subleading rational part $\RT_{4:3}(1,2;3,4)$, we
construct a generating set following~\eqn{eq:cutsetunique_nkl_simplified},
\begin{equation}
  \GeneratingCuts_{4:3}=
  \CutSet^{\{\TTr(1,2),\TTr(3,4),N_c\}}\cup
  \CutSet^{\{\TTr(3,4),N_c,\TTr(1,2)\}}\cup
  \CutSet^{\{N_c,\TTr(1,2),\TTr(3,4)\}}\,.
\end{equation}
Each of the sets contains four labels,
\begin{equation}
  \begin{aligned}
    \CutSet^{\{\TTr(1,2),\TTr(3,4),N_c\}}&=\{c_{4:3;1},c_{4:3;2},c_{4:3;3},c_{4:3;4}\}\\
    \CutSet^{\{\TTr(3,4),N_c,\TTr(1,2)\}}&=\{c_{4:3;5},c_{4:3;6},c_{4:3;7},c_{4:3;8}\}\\
    \CutSet^{\{N_c,\TTr(1,2),\TTr(3,4)\}}&=\{c_{4:3;9},c_{4:3;10},c_{4:3;11},c_{4:3;12}\}\\
  \end{aligned}
\end{equation}
which take the form,
\begin{equation}
  \resizebox{\hsize}{!}{$
    \begin{alignedat}{4}
      &c_{4:3;1}&&=((), ();();(1), (2);()|_L(); (4),(3)|_R();(), ())\,,\quad
      &&c_{4:3;2}&&=((), ();();(1, 2);()|_L(); (4),(3),|_R();())\,,\\
      &c_{4:3;3}&&=(();();(1), (2);()|_L();(4,3)|_R();(), ())\,,\quad
      &&c_{4:3;4}&&=(();();(1, 2);()|_L();(4,3)|_R();())\,,\\
      &c_{4:3;5}&&=((3), (4);();(), ();()|_L();(), ()|_R(); (2),(1))\,,\quad
      &&c_{4:3;6}&&=((3), (4);();();()|_L();(), ()|_R();(2,1))\,,\\
      &c_{4:3;7}&&=((3, 4);();(), ();()|_L();()|_R(); (2),(1))\,,\quad
      &&c_{4:3;8}&&=((3, 4);();();()|_L();()|_R();(2,1))\,,\\
      &c_{4:3;9}&&=((), ();();(), ();()|_L(); (2),(1)|_R(); (4),(3))\,,\quad
      &&c_{4:3;10}&&=((), ();();();()|_L();(2),(1)|_R();(4,3))\,,\\
      &c_{4:3;11}&&=(();();(), ();()|_L();(2,1)|_R(); (4),(3))\,,\quad
      &&c_{4:3;12}&&=(();();();()|_L();(2,1)|_R();(4,3))\,.
    \end{alignedat}
    $}
\end{equation}
The corresponding cuts are shown in \fig{fig:all_labels_R43}.
Summing their rational contributions determines
$\TildeRT_{4:3}(1^+ 2^+;3^+ 4^+)$,
\begin{equation}
  \TildeRT_{4:3}(1^+ 2^+;3^+ 4^+)=\sum_{c\in\GeneratingCuts_{4:3}}\RT_{c}(1,2;3,4)\,,
\end{equation}
which in turn allows us to compute $\RT_{4:3}(1^+,2^+;3^+,4^+)$ via,
\begin{equation}
  \RTwo_{4:3}(1^+2^+;3^+4^+)=\sum_{
      \sigma_1,\sigma_2\in \cyclicgroup_2
  }
  \TildeRT_{4:3}
  (\sigma_1[1, 2] ; \sigma_2[3,4]).
  \label{eq:R43FromTildeR43}
\end{equation}
Fortunately we do not need to compute these cuts explicitly.
Instead we can relate them to the leading-color cuts of \eqns{eq:R2_4:1;12}{eq:R2_4:1;3}
using the tree-amplitude identities,
\begin{equation}\label{eq:4pt_example_amplitude_reversal}
  A^{(0)}(1_{\scalar}2^+3_{\scalar})=-A^{(0)}(3_{\scalar}2^+1_{\scalar}),\quad
  A^{(0)}(1_{\scalar}2^+3^+4_{\scalar})=A^{(0)}(4_{\scalar}3^+2^+1_{\scalar})\,.
\end{equation}
We then find,
\begin{equation}
  \resizebox{\hsize}{!}{$
    \begin{alignedat}{4}
      &\RT_{c_{4:3;1}}(1,2;3,4) &&= \RT_{c_{4:1;1}}(1,2,4,3)=\RcFOOne{1}{2}{4}{3}\,,\quad
      &&\RT_{c_{4:3;2}}(1,2;3,4) &&= \RT_{c_{4:1;2}}(4,3,1,2)=\RcFOTwo{4}{3}{1}{2}\,,\\
      &\RT_{c_{4:3;3}}(1,2;3,4) &&= \RT_{c_{4:1;2}}(1,2,4,3)=\RcFOTwo{1}{2}{4}{3}\,,\quad
      &&\RT_{c_{4:3;4}}(1,2;3,4) &&= \RT_{c_{4:1;3}}(1,2,4,3)=\RcFOThree{1}{2}{4}{3}\,,\\
      &\RT_{c_{4:3;5}}(1,2;3,4) &&= \RT_{c_{4:1;1}}(3,4,2,1)=\RcFOOne{3}{4}{2}{1}\,,\quad
      &&\RT_{c_{4:3;6}}(1,2;3,4) &&= \RT_{c_{4:1;2}}(3,4,2,1)=\RcFOTwo{3}{4}{2}{1}\,,\\
      &\RT_{c_{4:3;7}}(1,2;3,4) &&= \RT_{c_{4:1;2}}(2,1,3,4)=\RcFOTwo{2}{1}{3}{4}\,,\quad
      &&\RT_{c_{4:3;8}}(1,2;3,4) &&= \RT_{c_{4:1;3}}(3,4,2,1)=\RcFOThree{3}{4}{2}{1}\,,\\
      &\RT_{c_{4:3;9}}(1,2;3,4) &&= \RT_{c_{4:1;1}}(2,1,4,3)=\RcFOOne{2}{1}{4}{3}\,,\quad
      &&\RT_{c_{4:3;10}}(1,2;3,4) &&= \RT_{c_{4:1;2}}(2,1,4,3)=\RcFOTwo{2}{1}{4}{3}\,,\\
      &\RT_{c_{4:3;11}}(1,2;3,4) &&= \RT_{c_{4:1;2}}(4,3,2,1)=\RcFOTwo{4}{3}{2}{1}\,,\quad
      &&\RT_{c_{4:3;12}}(1,2;3,4) &&= \RT_{c_{4:1;3}}(2,1,4,3)=\RcFOThree{2}{1}{4}{3}\,.\\
    \end{alignedat}
    $}
\end{equation}
Summing these expressions gives,
\begin{equation}
  \TildeRT_{4:3}(1, 2;3, 4)=-\frac{1}{9}
  \frac{s_{12}}{\spaa{12}^2\spaa{34}^2}(14s_{12}+s_{13})\,.
\end{equation}
Carrying out the two cyclic sums of \eqn{eq:R43FromTildeR43}, we
obtain after some algebra,
\begin{equation}\label{eq:4pt_example_R243}
  \begin{aligned}
    \RTwo_{4:3}(1^+ 2^+ ; 3^+ 4^+)&
    =-\frac{1}{6}\frac{\spbb{12}^2}{\spaa{34}^2}
  \end{aligned}
\end{equation}
This expression is in agreement with ref.~\cite{Bern:2002tk}.

\subsubsection{Subleading-color \texorpdfstring{$\RTwo_{4:1\mathrm{B}}$}{R(2)41B}}
\label{subsubsec:4pt_example_subleading-color_1B}
\begin{figure}[h]
  \begin{subfigure}{\linewidth}
    \includegraphics[valign=c,width=0.19\linewidth]{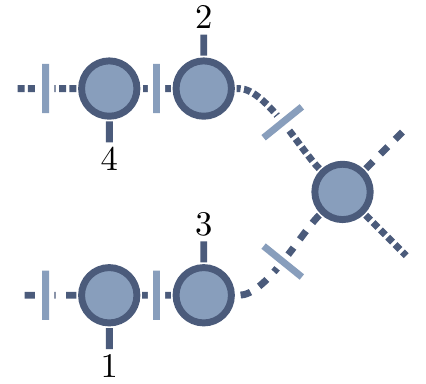}
    \includegraphics[valign=c,width=0.19\linewidth]{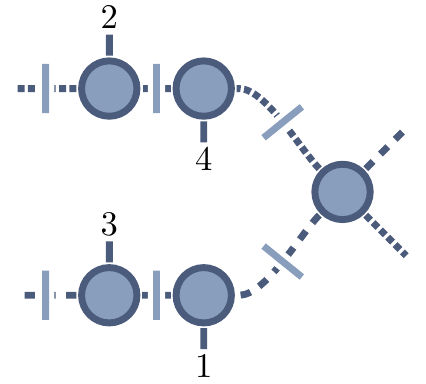}
    \includegraphics[valign=c,width=0.19\linewidth]{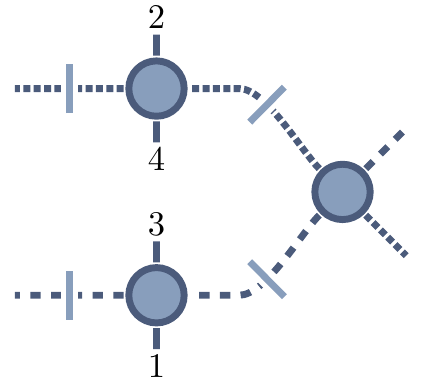}
    \caption{Cuts associated to the labels  $c_{4:\oneB;1}$, $c_{4:\oneB;2}$, $c_{4:\oneB;3}$ in
      $\GeneratingCuts_{4:\oneB}$.
      These labels are invariant under $\rho_{-1}\circ\Rota$, and therefore require a symmetry factor of $\tfrac{1}{4}$.}
    \label{fig:4-pt_example_R_41B_cuts_1o4}
  \end{subfigure}
  \begin{subfigure}{\linewidth}
    \includegraphics[valign=c,width=0.19\linewidth]{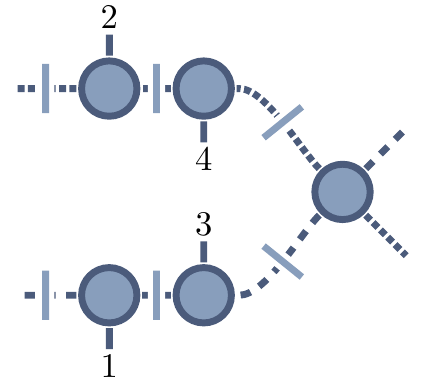}
    \includegraphics[valign=c,width=0.19\linewidth]{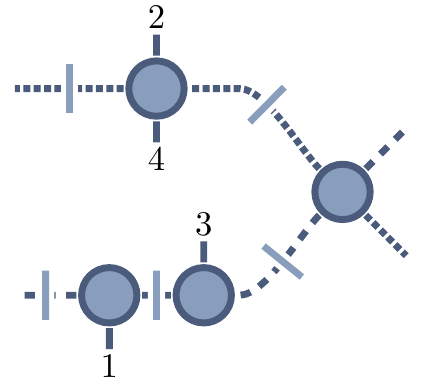}
    \includegraphics[valign=c,width=0.19\linewidth]{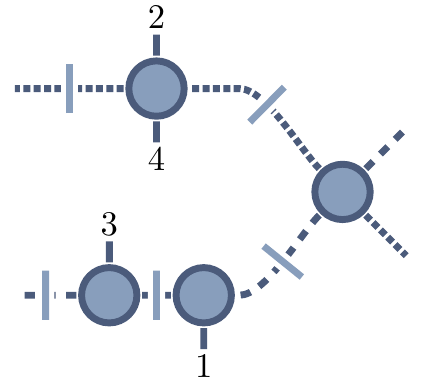}
    \caption{Cuts associated to the labels  $c_{4:\oneB;4}$, $c_{4:\oneB;5}$, $c_{4:\oneB;6}$ in
      $\GeneratingCuts_{4:\oneB}$.
      These labels are invariant under $\rho_{-2}\circ\Rota^2$, and therefore require a symmetry factor of $\tfrac{1}{2}$.}
    \label{fig:4-pt_example_R_41B_cuts_1o2}
  \end{subfigure}
  \begin{subfigure}{\linewidth}
    \includegraphics[valign=c,width=0.19\linewidth]{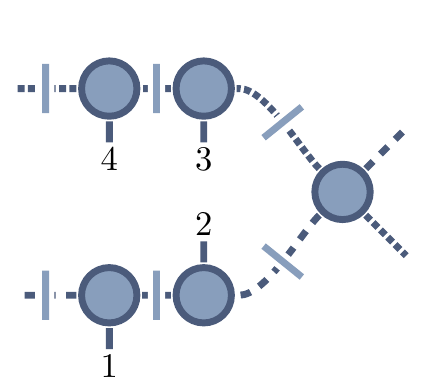}
    \includegraphics[valign=c,width=0.19\linewidth]{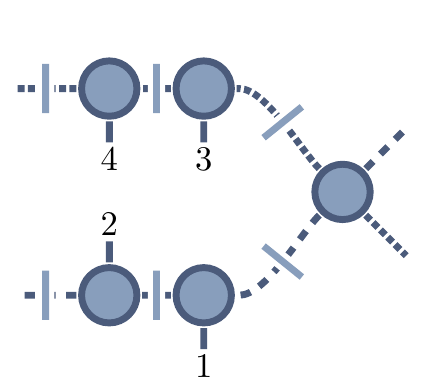}
    \includegraphics[valign=c,width=0.19\linewidth]{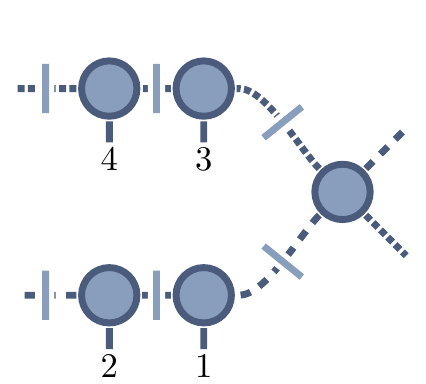}
    \includegraphics[valign=c,width=0.19\linewidth]{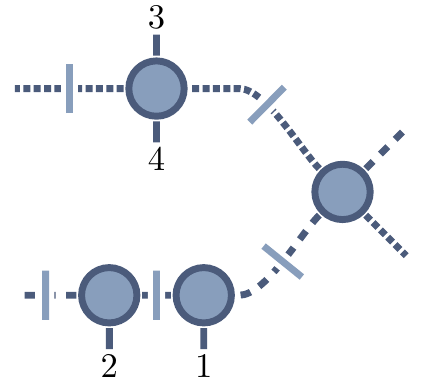}
    \includegraphics[valign=c,width=0.19\linewidth]{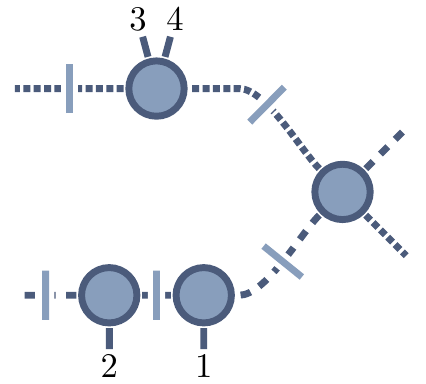}
    \includegraphics[valign=c,width=0.19\linewidth]{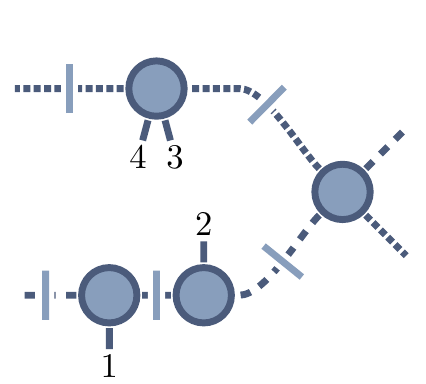}
    \includegraphics[valign=c,width=0.19\linewidth]{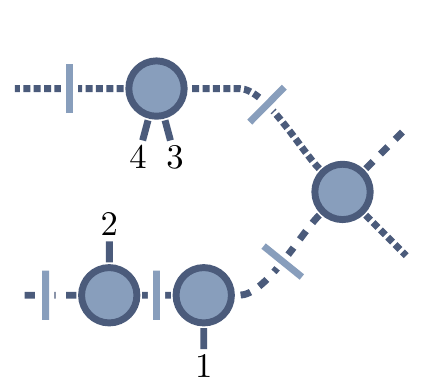}
    \includegraphics[valign=c,width=0.19\linewidth]{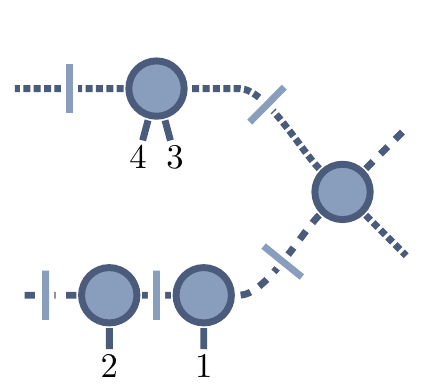}
    \includegraphics[valign=c,width=0.19\linewidth]{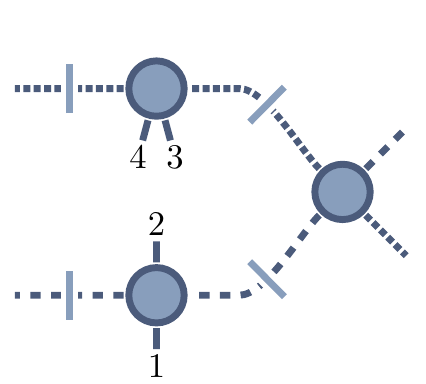}
    \includegraphics[valign=c,width=0.19\linewidth]{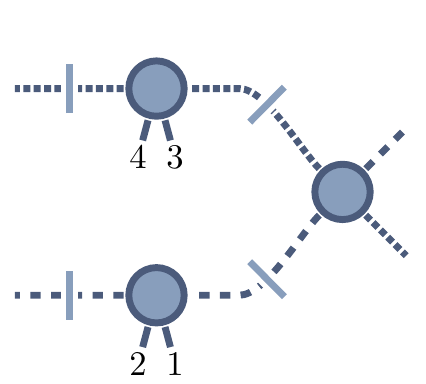}
    \caption{Cuts associated to the labels  $c_{4:\oneB;7},\ldots, c_{4:\oneB;16}$ in
      $\GeneratingCuts_{4:\oneB}$.
      These labels are not invariant under $\cyclicgroup_4\times\Rota$, and therefore do not require a symmetry factor.}
    \label{fig:4-pt_example_R_41B_cuts_1}
  \end{subfigure}
  \caption{cut labels in the generating set $\GeneratingCuts_{4:\oneB}$,
  classified by the symmetry factors they require.}
  \label{fig:4-pt_example_R_41B_cuts}
\end{figure}

Finally, we determine the subleading-color
single-trace rational part
$\RTwo_{4:1\mathrm{B}}$ using the separable approach.
The generating set $\GeneratingCuts_{4:\oneB}$ consists of the labels,
\begin{equation}
  \resizebox{\hsize}{!}{$
  \begin{alignedat}{4}
    &c_{4:\oneB;1}&&=((),(1);()|(),(2);()|(),(3);()|(),(4);()),\qquad
    &&c_{4:\oneB;2}&&=((1),();()|(2),();()|(3),();()|(4),();())\,,\\
    &c_{4:\oneB;3}&&=((1);()|(2);()|(3);()|(4);())\,,
    &&c_{4:\oneB;4}&&=((),(1);()|(2),();()|(),(3);()|(4),();())\,,\\
    &c_{4:\oneB;5}&&=((),(1);()|(2);()|(),(3);()|(4);())\,,
    &&c_{4:\oneB;6}&&=((1),();()|(2);()|(3),();()|(4);())\,,\\
    &c_{4:\oneB;7}&&=((),(1);()|(),();()|(),(2);()|(3),(4);())\,,
    &&c_{4:\oneB;8}&&=((1),();()|(),();()|(2),();()|(3),(4);())\,,\\
    &c_{4:\oneB;9}&&=((1),(2);()|(),();()|(),();()|(3),(4);())\,,
    &&c_{4:\oneB;10}&&=((1),(2);()|(3);()|(),();()|(4);())\,,\\
    &c_{4:\oneB;11}&&=((1),(2);()|(3,4);()|(),();()|();())\,,
    &&c_{4:\oneB;12}&&=((),(1);()|();()|(),(2);()|(3,4);())\,,\\
    &c_{4:\oneB;13}&&=((1),();()|();()|(2),();()|(3,4);())\,,
    &&c_{4:\oneB;14}&&=((1),(2);()|();()|(),();()|(3,4);())\,,\\
    &c_{4:\oneB;15}&&=((1);()|();()|(2);()|(3,4);())\,,
    &&c_{4:\oneB;16}&&=((1,2);()|();()|();()|(3,4);())\,.\\
  \end{alignedat}
  $}
\end{equation}
which are shown in \fig{fig:4-pt_example_R_41B_cuts}.
Of these labels, the first six require symmetry factors.
The labels $c_{4:\oneB;1}$, $c_{4:\oneB;2}$, and $c_{4:\oneB;3}$ each come with a factor of $\frac{1}{4}$, because
\begin{equation}
  \rho_{-1}\circ\Rota[c_{4:\oneB;i}]=c_{4:\oneB;i},\quad i=1,2,3\,.
\end{equation}
The labels $c_{4:\oneB;4}$, $c_{4:\oneB;5}$, and $c_{4:\oneB;6}$ come with a factor of
$\frac{1}{2}$, because
\begin{equation}
  \rho_{-2}\circ\Rota^2[c_{4:\oneB;i}]=c_{4:\oneB;i},\quad i=4,5,6\,.
\end{equation}
The remaining cuts are not invariant under $\cyclicgroup_4\times \Rota$,
and therefore do not require such factors.
In summary, we have,
\begin{equation}
  S_{c_{4:\oneB;i}}=\left\{
      \begin{alignedat}{2}
        &\tfrac{1}{4},\quad && i=1,2,3,\\
        &\tfrac{1}{2},\quad && i=4,5,6,\\
        &1,\quad && \text{otherwise}.
      \end{alignedat}
    \right.
    \label{eq:R41B_symmetry_factors}
\end{equation}
We then have
\begin{equation}
  \RTwo_{4:1\mathrm{B}}(1^+2^+3^+4^+)=\sum_{\sigma\in \cyclicgroup_4}
  \TildeRT_{4:1\mathrm{B}}(\sigma[1,2,3,4])\,,
\end{equation}
with
\begin{equation}
   \TildeRT_{4:1\mathrm{B}}(1,2,3,4)=\sum_{c\in\GeneratingCuts_{4:\oneB}}S_c\RT_c(1,2,3,4)\,.
\end{equation}

\begin{figure}
  (I)\quad
  \includegraphics[width=0.2\textwidth,valign=c]
  {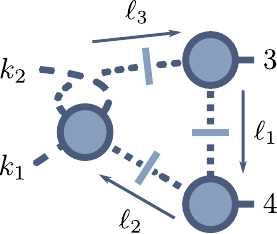}\,,\hspace{2em}
  (II)\quad
  \includegraphics[width=0.2\textwidth,valign=c]
  {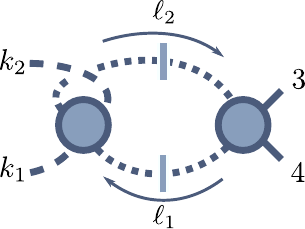}\,,\hspace{2em}
  (III)\quad 
  \includegraphics[width=0.2\textwidth,valign=c]
  {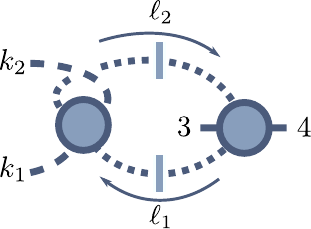}\,.
  \caption{The three distinct unitarity cuts (I, II, III) 
  of loop $\looptwo$ needed for the cut labels in $\GeneratingCuts_{4:\oneB}$.}
  \label{fig:4-pt_example_rhs_cuts_R41B}
\end{figure}

To compute the rational contributions $\RT_c$ of the cuts $c\in\GeneratingCuts_{4:\oneB}$,
we follow the procedure of section~\ref{sec:cut_computation_1B}.
We start by computing the one-loop
coefficient of loop $\looptwo$.
Thanks to the amplitude relations of
\eqn{eq:4pt_example_amplitude_reversal}, we do not need to compute
the coefficient of every cut individually.
Instead, we only need to consider three distinct types of cuts.
By connecting the scalar line of loop $\looptwo$, we can represent them in
the familiar form as shown in \fig{fig:4-pt_example_rhs_cuts_R41B}.
In the label notation, these correspond to the sets
\begin{equation}
  \begin{alignedat}{2}
    \mathrm{(I):}&\quad \ESet_2=(),(),&&\qquad \ESet_4=(3),(4)\,,\\
    \mathrm{(II):}&\quad \ESet_2=(),&&\qquad \ESet_4=(3,4)\,,\\
    \mathrm{(III):}&\quad \ESet_2=(3),&&\qquad \ESet_4=(4)\,,
  \end{alignedat}
\end{equation}
where for all three, $\ESet_{\central,i}=()$.
Using the same loop-momentum parametrization as
in~eqs.~\eqref{eq:4-pt_example_rhs_triangle_l}
and~\eqref{eq:4-pt_example_rhs_bubble_l}, we obtain for these cuts,

\begin{align}
  \begin{split}\label{eq:4-pt_example_rhs_triangle_R41B}
    &\hphantom{={}}\CoeffOneTwo(\{(),();(3),(4)\},\{(),(),(),()\})\\
    &=\frac{1}{2}\sum_{\ell,\ell^\ast}\Inf_{\mu^2,t}\Big[
    A^{(0)}(k_{1\scalar} (\ell_3)_{\scalarP}k_{2\scalar}
    (-\ell_2)_{\scalarP})
    A^{(0)}(3^+(\ell_1)_{\scalarP}(-\ell_3)_{\scalarP})
    \\&\hspace{30mm}\times A^{(0)}(4^+(\ell_2)_{\scalarP}(-\ell_1)_{\scalarP})\Big]
    \Big\vert_{\mu^2,t^0}\\
    &=-\frac{1}{2}\Inf_{\mu^2,t}\Big[\sum_{\ell,\ell^\ast}
    \frac{\spba{3|\ell_1|4}}{\spaa{34}}
    \frac{\spba{4|\ell_1|3}}{\spaa{43}}\Big]
    \Big\vert_{\mu^2,t^0}\\
    &=-\Inf_{\mu^2,t}\Big[\frac{\mu^2 s_{34}}{\spaa{34}^2}
    \Big]\Big\vert_{\mu^2,t^0}\\&=\frac{\spbb{34}}{\spaa{34}},
  \end{split}\\[1em]
  \begin{split}\label{eq:4-pt_example_rhs_bubble_1_R41B}
    &\hphantom{={}}\CoeffOneTwo(\{();(3,4)\},\{(),(),(),()\})\\
    &=\Inf_{\mu^2,y,t}\Big[
    A^{(0)}(k_{1\scalar}  (l_2)_{\scalarP}k_{2\scalar} (-l_1)_{\scalarP})
    A^{(0)}(3^+ 4^+(\ell_1)_{\scalarP}(-\ell_2)_{\scalarP})
    \Big]\Big\vert_{t^0,y^i\to Y_i,\mu^2}\\
    &=\Inf_{\mu^2,y,t}\Big[
    \frac{\mu^2}{\LProd_{4\ell_1}}\frac{\spbb{34}}{\spaa{34}}
    \Big]\Big\vert_{t^0,y^i\to Y_i,\mu^2}
    =\Inf_{\mu^2,y,t} \Big[
    \frac{\mu^2}{y \spaa{34}^2}\Big]\Big\vert_{t^0,y^i\to Y_i,\mu^2}=0,
  \end{split}\\[1em]
  \begin{split}\label{eq:4-pt_example_rhs_bubble_2_R41B}
    &\hphantom{={}}\CoeffOneTwo(\{(3);(4)\},\{(),(),(),()\})\\
    &=\Inf_{\mu^2,y,t}\Big[
    A^{(0)}(k_{1\scalar}  (l_2)_{\scalarP}k_{2\scalar} (-l_1)_{\scalarP})
    A^{(0)}(3^+(-\ell_2)_{\scalarP}4^+(\ell_1)_{\scalarP})
    \Big]\Big\vert_{t^0,y^i\to Y_i,\mu^2}\\
    &=\Inf_{\mu^2,y,t}\Big[
    \frac{\mu^2\spbb{34}^2}{\LProd_{3\ell_1}\LProd_{3\ell_2}}
    \Big]\Big\vert_{t^0,y^i\to Y_i,\mu^2}
    =-\Inf_{\mu^2,y,t} \Big[
    \frac{\mu^2}{y(1-y)\spaa{34}^2}\Big]\Big\vert_{t^0,y^i\to Y_i,\mu^2}\\
    &=0.
  \end{split}
\end{align}
The bubble coefficients (II) and (III) would in
principle also require the computation of one or two triangle
cuts respectively.
However, the choice $\chi=k_3$ makes such triangle cuts vanish 
in either case, and the bubble cut again suffices.

As the bubble coefficients (II) and (III) vanish, 
the rational contributions corresponding to
most cut labels in $\GeneratingCuts_{4:\oneB}$ vanish as well.
In fact, as the triangle coefficient (I) is independent of $k_1$ and $k_2$,
we can immediately see that any bubble coefficient of loop $\loopone$
has to vanish as well.
Thus the only non-vanishing cuts of $\TildeRT_{4:1\mathrm{B}}$ are those of
the $\CutClass{\TriTri}$ class, specifically
$c_{4:\oneB;1}$, $c_{4:\oneB;2}$, $c_{4:\oneB;4}$, $c_{4:\oneB;7}$,
$c_{4:\oneB;8}$, and $c_{4:\oneB;9}$.
We will see later in \sect{SubleadingSingleTraceSeparable}
that the vanishing of all one-loop squared coefficients
except for $\CutClass{\TriTri}$ ones is a generic feature of $\RT_{n:\oneB}$
rational parts, which holds for any number of external gluons.

Using tree-amplitude relations, we find the following relations
between the rational contributions of the labels in $\GeneratingCuts_{4:\oneB}$,
\begin{equation}
  \RT_{c_{4:\oneB;1}}=\RT_{c_{4:\oneB;2}}=\RT_{c_{4:\oneB;9}}=-\RT_{c_{4:\oneB;4}}=-\RT_{c_{4:\oneB;7}}=-\RT_{c_{4:\oneB;8}}\,.
  \label{eq:R41B_TriTri_Rational_Relations}
\end{equation}
We thus only need to determine a single one-loop squared coefficient.
We pick $c_{4:\oneB;9}$.
Using the momentum routing already seen in
\fig{fig:4-pt_example_lhs_cuts} and parametrization of
\eqn{eq:4-pt_example_lhs_triangle_l},
we can determine the triangle coefficient of loop $\loopone$---and therefore the
two-loop coefficient of $c_{4:\oneB;9}$---to be,
\begin{equation}\label{eq:R_41B_cut_6}
  \begin{aligned}
    \CoeffTwo(c_{4:\oneB;9})&=\CoeffOneOne(\{(1),(2);(),()\},\CoeffOneTwo(\{(),();(3),(4)\},\{(),(),(),()\}))\\
    &=\frac{1}{2}\sum_{\ell_1,\ell_1^\ast}\Inf_{\mu^2,t}\bigg[
    A^{(0)}((-\ell_3)_{\scalar} 1^+ (\ell_1)_{\scalar})
    A^{(0)}((-\ell_1)_{\scalar} 2^+ (\ell_2)_{\scalar})\\
    &\hspace{12em}\times \CoeffOneTwo(\{(),();(3),(4)\},\{(),(),(),()\})
    \bigg]\bigg\vert_{t^0,\mu^2}\\
    &=\frac{1}{2}\sum_{\ell_1,\ell_1^\ast}\Inf_{\mu^2,t}
    \Big[
    \frac{\spba{1|\ell_1|2}}{\spaa{12}}\frac{\spba{2|\ell_1|1}}{\spaa{21}}
    \frac{\spbb{34}}{\spaa{34}}
    \Big]\Big\vert_{t^0,\mu^2}\\
    &=\frac{\spbb{12}\spbb{34}}{\spaa{12}\spaa{34}}\,.
  \end{aligned}
\end{equation}
The rational contribution of $c_{4:\oneB;9}$ is thus,
\begin{equation}
  \RT_{c_{4:\oneB;9}}=4 I^{(2),D}_{\TriTri}\CoeffTwo(c_{4:\oneB;9})=\frac{\spbb{12}\spbb{34}}{\spaa{12}\spaa{34}}\,.
\end{equation}
Combining this result with the relations of \eqn{eq:R41B_TriTri_Rational_Relations}
and the symmetry factors of \eqn{eq:R41B_symmetry_factors}, we find
\begin{equation}\label{eq:4pt_example_sub-leading-color_1B_RTilde}
  \begin{aligned}
    \TildeRT_{4:1\mathrm{B}}(1^+2^+3^+4^+)&=\frac{1}{4}\left[\RT_{c_{4:\oneB;1}}+\RT_{c_{4:\oneB;2}}\right]+\frac{1}{2}\RT_{c_{4:\oneB;4}}+\RT_{c_{4:\oneB;7}}+\RT_{c_{4:\oneB;8}}+\RT_{c_{4:\oneB;9}}\\
    &=\frac{\spbb{12}\spbb{34}}{\spaa{12}\spaa{34}}
    \left[\frac{2}{4}+\frac{1}{2}-1-1+1\right]=0\,,
  \end{aligned}
\end{equation}
and as a consequence,
\begin{equation}
  \RTwo_{4:1\mathrm{B}}(1^+2^+3^+4^+)=\sum_{\sigma\in \cyclicgroup_4}
  \TildeRT_{4:1\mathrm{B}}(\sigma(1,2,3,4)) = 0\,.
\end{equation}
This is again in agreement with the result of
refs.~\cite{Glover:2001af,Bern:2002tk}.
\FloatBarrier

\section{Mathematica Implementation}
\label{MathematicaImplementationSection}

To verify the one-loop squared approach against literature results beyond four points, we automated the generation of cuts and 
symmetry factors required for the
generating sets $\GeneratingCuts_{n:1}$, $\GeneratingCuts_{n:i}$,
$\GeneratingCuts_{n:r,k}$ and $\GeneratingCuts_{n:\oneB}$ in \mm.  
The code is general, and able to produce generating sets for an arbitrary number of gluons.

We wrote a set of \mm\ packages to automate the evaluation of
unitarity cuts on specified (rational) kinematic points%
\footnote{The entire codebase is included in the auxiliary files. 
The most current version is available at
\begin{center}
    \url{https://github.com/spoegel/SpinorHelicityPackages}
\end{center}}.  
These packages are extensions of the
\texttt{SpinorHelicity6D} package of ref.~\cite{AccettulliHuber:2019abj},
which provides an implementation of the four- and six-dimensional spinors.

To determine the integral coefficients as reviewed in \app{GeneralizedUnitarityAppendix}, we require an implementation of the $\Inf$ operation.
As a reminder, given a function $f(x)$ that
scales like $x^p$ for $x \to \infty$, we obtain $\Inf_x[f(x)])$ from
its series expansion around large values of $x$,
\begin{equation}
  f(x) = \sum_{i=0}^{p}c_i x^i + \mathcal{O}\left( \frac{1}{x} \right)
  \equiv\Inf_x[f(x)]+ \mathcal{O}\left( \frac{1}{x} \right)\,.
\end{equation}
We use \mm\ to perform these series expansions symbolically.  While this
approach is not optimal numerically, it retains full
flexibility in choosing the kinematics.  We can use rational
kinematics to make exact comparisons with literature results.  
We can also make use of the symbolic capabilities of \mm\ on 
partially or fully parametrized kinematics.  
Examples of the latter include the verifying collinear behavior
exactly, obtaining analytic results from completely
parametrized kinematics, and analyzing the large-$z$ behavior 
under a BCFW shift.

The products of trees in $D$-dimensional unitarity cuts are always
rational functions in the loop-momentum parameters $\mu^2$, $t$ and $y$,
with coefficients generically algebraic in the external kinematics.  A potential problem that arises when carrying out the series
expansions symbolically is the appearance of
deeply nested intermediate expressions.  These can appear
in the
evaluation of cuts even for purely numerical kinematic points.  
We therefore
avoid \mm's built-in \texttt{Series} and \texttt{SeriesCoefficient}, using instead our own implementation of these functions \texttt{RationalSeries}
and \texttt{RationalSeriesCoefficient}.  They are optimized for expansions of
rational functions.  Our functions perform the series expansions recursively,
caching intermediate results for improved performance.

In addition, we store any coefficients depending only on 
the external
kinematics in symbolic objects called \symbolObj.  These can be 
defined
recursively, and depend on \symbolObj\ objects of previous steps in 
the computation.  The \symbolObj\ objects can therefore be arranged 
in a directed
acyclic graph.  Each \symbolObj\ represents an
algebraic number for purely numerical kinematics, and
an algebraic function of the parameters in parametrized kinematics.
We assign the \textit{same\/} \symbolObj\ for repeated
appearances of a specific subexpression, so that every
\symbolObj\ represents a unique kinematic structure.  
We also avoid unnecessary
computations by performing a zero test on each newly defined
\symbolObj\ object; the
\mm\ function \texttt{PossibleZeroQ} does this
efficiently and exactly, given the option
\texttt{Method->ExactAlgebraics}.  
For this zero test we set all kinematic parameters
to random integer values of sufficient size to avoid accidental
cancellations.

For cut coefficients, only the final result is known to be 
rational in the
kinematics: on purely numeric rational kinematic points it 
will be a rational number, while for rational kinematics 
involving analytic parameters, it is a
rational function in these parameters.  Intermediate expressions 
generically involve square roots, which are only guaranteed to
cancel in the
final result.  Performing simplifications in intermediate steps 
is therefore
costly.  Furthermore, the final result will not necessarily 
depend on all of
the \symbolObj\ objects, as they may cancel or contribute 
only to powers in
the loop-momentum parameters not required for the 
desired coefficient.  We
therefore perform the series expansion without any
simplifications along the way, storing the kinematic 
dependence in the
\symbolObj\ objects mentioned before.  The end result of the 
expansion
will be a single \symbolObj\ object.  We then simplify the result by
traversing the dependency graph of this final \symbolObj\ object, 
performing
simplifications from the bottom up, or depth-first.  This ensures 
that we
perform only simplifications required for the final result, 
and that every
subexpression is processed only once.

\section{Higher-Point Amplitudes}
\label{HigherPointSection}

Following our earlier presentation of the separable approach 
in the example of the four-gluon partial
amplitudes, we turn to a discussion of
higher-multiplicity results and their comparison to
expressions available in the literature.
Explicit forms for the generating sets of cut label for up 
to nine gluons can be found in the auxiliary files provide with
this article.

\subsection{The Five-Gluon Amplitude Analytically}
\label{subsec:five-gluon-amplitude-analytically}

\begin{figure}
  \foreach \x in {1,...,12}{%
    \includegraphics[width=0.24\textwidth,valign=c]
    {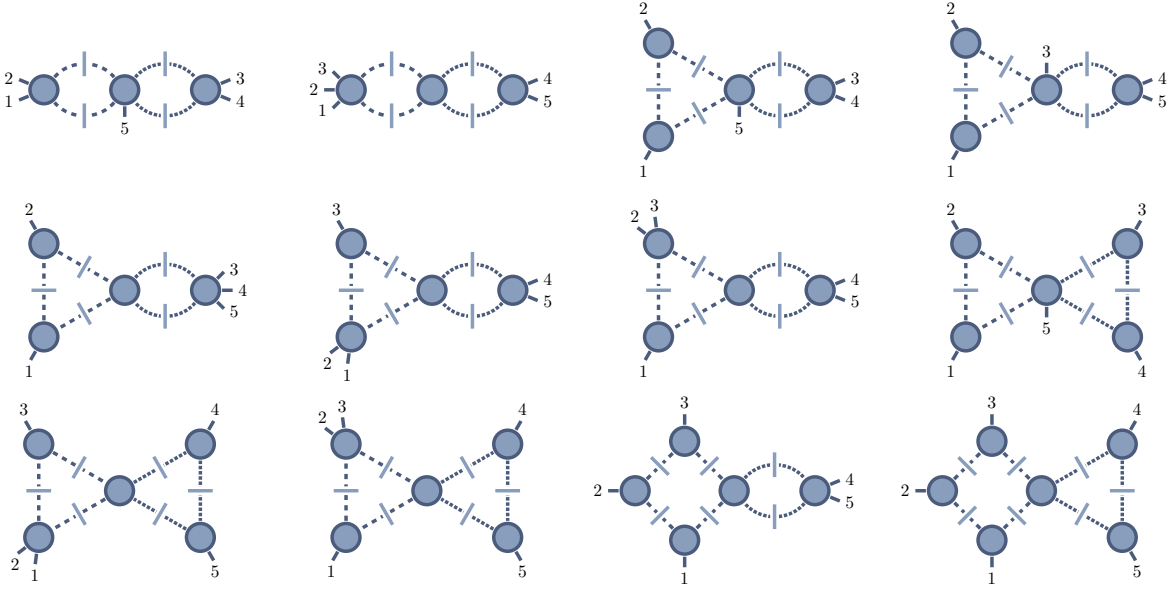}
  }
  \caption{Unitarity cuts in the generating set $\GeneratingCuts_{5:1}$ of the
    leading-color rational part $\RT_{5:1}$.}
  \label{fig:R251_cuts}
\end{figure}

\begin{figure}
  \foreach \x in {32,...,62}{%
    \includegraphics[width=0.24\textwidth,valign=c]
    {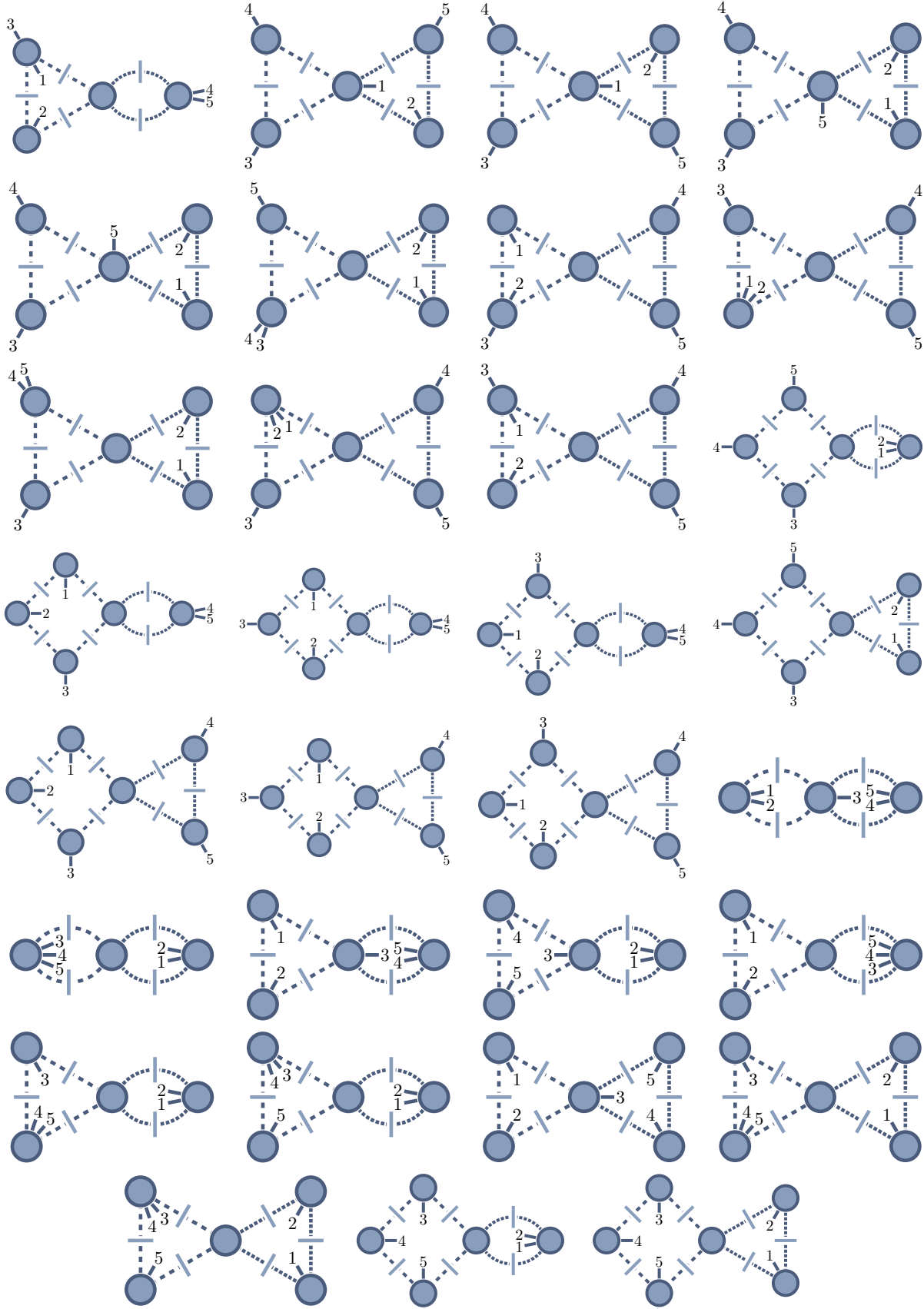}\vspace*{0.4mm}
  }
  \caption{The remaining $31$ unitarity cuts in the generating set $\GeneratingCuts_{5:3}$ of $\RT_{5:3}$.}
  \label{fig:R253_cuts_b}
\end{figure}

The color decomposition of the five-gluon amplitude has the form
\begin{equation}
  \begin{aligned}
    \mathcal{A}^{(2)}(1^+2^+3^+4^+5^+)&=
    N_c^2\sum_{\sigma\in\modout{S_5}{\cyclicgroup_5}}\TTr(\sigma(1,2,3,4,5))\ATwo_{5:1}(\sigma(1,2,3,4,5))\\    &\hphantom{{}={}}+N_c\sum_{\sigma\in\modout{S_5}{P_{5:3}}}\TTr(\sigma(1,2))\TTr(\sigma(3,4,5))\ATwo_{5:3}(\sigma(1,2);\sigma(3,4,5))\\
    &\hphantom{{}={}}+\sum_{\sigma\in\modout{S_5}{\cyclicgroup_5}}\TTr(\sigma(1,2,3,4,5))\ATwo_{5:\oneB}(\sigma(1,2,3,4,5))\,.
  \end{aligned}
\end{equation}
Just as in the four-gluon case, we require the rational 
parts of three
partial amplitudes: $\RTwo_{5:1}$,
$\RTwo_{5:3}$, and $\RTwo_{5:1\mathrm{B}}$.
Following the discussion of
section~\ref{ColorStructureSection}, we build generating sets
$\GeneratingCuts_{5:1}$, $\GeneratingCuts_{5:3}$, $\GeneratingCuts_{5:1\oneB}$,
which contain $12$, $62$, and $133$ unitarity cuts respectively.
The
cuts of $\GeneratingCuts_{5:1}$ and
$\GeneratingCuts_{5:3}$ are shown in
Figures~\ref{fig:R251_cuts},~\ref{fig:R253_cuts_a},~and~\ref{fig:R253_cuts_b}, while those of
$\TildeRT_{5:1\mathrm{B}}$ are not shown on account of
their large number.
We include the labels of the cuts in the three generating sets 
in auxiliary files.
For an odd
number of momenta, none of the cuts require symmetry factors.

Beginning with five-gluon amplitudes, we rely on our 
implementation in
\textit{Mathematica} to compute the unitarity cuts due to their increased
number and complexity.
As our code is able to perform computations symbolically, we 
evaluated
the cuts using parametrized kinematics obtained
from momentum twistors, giving us analytic results.
In Appendix~\ref{MomentumTwistorAppendix} we provide the details of 
this
parameterization.
We provide the computed expressions for $\TildeRT_{5:1}$, $\TildeRT_{5:3}$ and $\TildeRT_{5:\oneB}$ in the auxiliary files. 

We verify our approach using the known results for the three partial
amplitudes of 
refs.~\cite{Dunbar:2016aux,Dunbar:2019fcq,Dunbar:2020wdh}.
Our expressions are in numerical agreement with these results%
\footnote{We compared with the expressions in the 
published versions.}
and satisfy the color relations laid out in ref.~\cite{Edison:2011ta}%
\footnote{The equivalent expression given in ref.~\cite{Dunbar:2019fcq} is
  missing an overall sign.}.
\begin{figure}
  \foreach \x in {1,...,31}{%
    \includegraphics[width=0.24\textwidth,valign=c]
    {generating_sets/R53/graph_\x}\vspace*{0.4mm}
  }
  \caption{The first $31$ unitarity cuts in the generating set $\GeneratingCuts_{5:3}$ of $\RT_{5:3}$.}
  \label{fig:R253_cuts_a}
\end{figure}

\subsection{The Six- and Seven-Gluon Amplitudes: Numeric Results}
\label{subsec:67pt_example_numeric-results}
For $n=6,7$, the two-loop color decomposition has the form,
\begin{equation}
  \begin{aligned}
    &\hphantom{={}}\mathcal{A}^{(2)}(1^+2^+3^+4^+5^+\ldots n^+)\\
    &=N_c^2\sum_{\sigma\in\modout{S_n}{\cyclicgroup_n}}\TTr(\sigma(1,2,3,4,5,\ldots,n))\ATwo_{n:1}(\sigma(1,2,3,4,5,\ldots,n))\\    &\hphantom{{}={}}+N_c\sum_{\sigma\in\modout{S_n}{P_{n:3}}}\TTr(\sigma(1,2))\TTr(\sigma(3,4,5,\ldots))\ATwo_{n:3}(\sigma(1,2);\sigma(3,4,5,\ldots,n))\\
    &\hphantom{={}}+N_c\sum_{\sigma\in\modout{S_n}{P_{n:4}}}\TTr(\sigma(1,2,3))\TTr(\sigma(4,5,\ldots))\ATwo_{n:4}(\sigma(1,2,3);\sigma(4,5,\ldots,n))\\
    &\hphantom{={}}+\sum_{\sigma\in\modout{S_n}{P_{n:2,2}}}\TTr(\sigma(1,2))\TTr(\sigma(3,4))\TTr(\sigma(5,\ldots,n))\ATwo_{n:2,2}(\sigma(1,2);\sigma(3,4);\sigma(5,\ldots,n))\\
    &\hphantom{={}}+\sum_{\sigma\in\modout{S_n}{\cyclicgroup_n}}\TTr(\sigma(1,2,3,4,5,\ldots,n))\ATwo_{n:\oneB}(\sigma(1,2,3,4,5,\ldots,n))\,.
  \end{aligned}
\end{equation}
For six and seven gluons, we therefore have to determine five types of rational parts: $\RT_{n:1}$, $\RT_{n:3}$,
$\RT_{n:4}$, $\RT_{n:2,2}$ and $\RT_{n:\oneB}$.
While closed form expressions of these are known in the six-gluon case~\cite{Dunbar:2019fcq,Dalgleish:2020mof}, for seven gluons only $\RT_{7:1}$ has 
been determined through direct computation~\cite{Dunbar:2017nfy,Dunbar:2017azf}.

We determine the generating sets of for these rational parts using our automated code. 
The sizes of these sets are given in \tab{tab:n_generating_sets}.
The cuts of the generating set $\GeneratingCuts_{6:1}$ are shown in \figs{fig:R261_cuts_a}{fig:R261_cuts_b}.
It also lists the number of cuts in all generating sets up for 
up to nine gluons.
The reader may find all generating sets in machine-readable 
form in the auxiliary files.
\begin{figure}
  \foreach \x in {1,...,24}{%
    \includegraphics[width=0.24\textwidth,valign=c]
    {generating_sets/R61/graph_\x}\vspace*{0.5mm}
  }
  \caption{The first 24 unitarity cuts in the generating set $\GeneratingCuts_{6:1}$ of $\RT_{6:1}$.}
  \label{fig:R261_cuts_a}
\end{figure}
\begin{figure}
  \foreach \x in {25,...,47}{%
    \includegraphics[width=0.24\textwidth,valign=c]
    {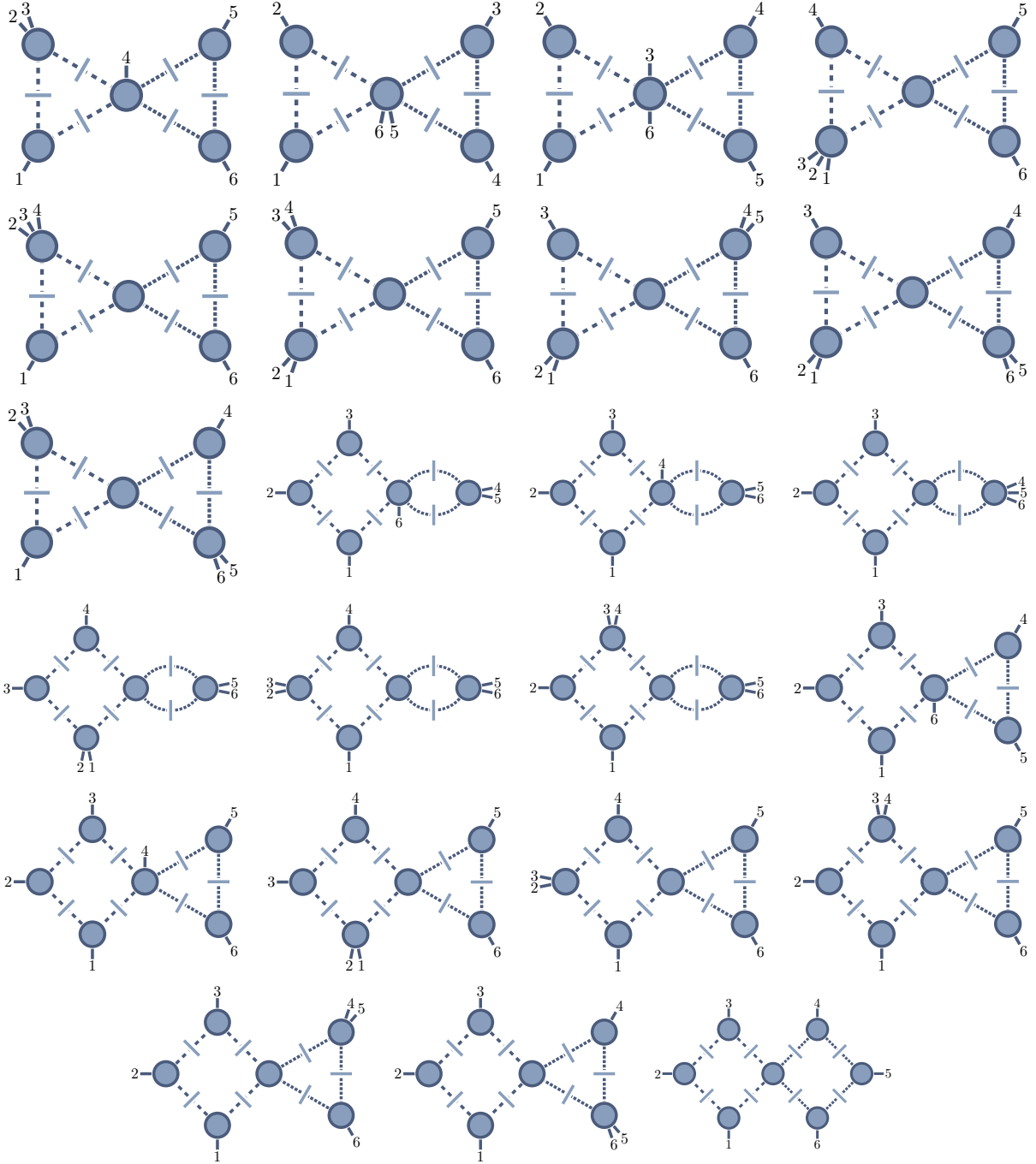}\vspace*{0.5mm}
  }
  \caption{The remaining 23 unitarity cuts in the generating set $\GeneratingCuts_{6:1}$ of $\RT_{6:1}$.}
  \label{fig:R261_cuts_b}
\end{figure}

Using our \mm\ implementation, we compute $\RT_{6:1}$, 
$\RT_{6:3}$, $\RT_{6:2,2}$, $\RT_{6:\oneB}$ and $\RT_{7:\oneB}$ 
numerically on 
rational kinematic points using the separable approach.
In all cases, we find exact agreement with the analytic expressions
reported in the literature.
\begin{table}
  \centering
  \begin{tabular}[t]{c|c|c|c|c|c|c}
    $n$ & $4$ & $5$ & $6$ & $7$& $8$& $9$\\ \hline
    $\TildeRT_{n:1}$ & $3(7)$ & $12(33)$ & $47(149)$ & $126 (413)$ & $315 (1083)$& $ 676 (2314)$\\
    $\TildeRT_{n:3}$ & $12(27)$ & $62(171)$ & $242(729)$ & $773 (2440)$ & $2111 (6801)$& $5101 (16479)$\\
    $\TildeRT_{n:4}$ & --- & --- & $240(735)$ & $768 (2422)$& $2163 (6879)$& $5459 (17315)$\\
    $\TildeRT_{n:2,2}$ & --- & --- & $1023(3168)$ & $3300 (10760)$& $8911 (29120)$& $21694 (69696)$\\
    $\TildeRT_{n:5}$ & --- & --- & --- & ---& $2099 (6621)$& $5259 (16414)$\\
    $\TildeRT_{n:2,3}$ & --- & --- & --- & --- & $9893 (32860)$& $25044 (81730)$\\
    $\TildeRT_{n:3,3}$ & --- & --- & --- & --- & --- & $28242 (93576)$\\
    $\TildeRT_{n:1\mathrm{B}}$ & $16(42)$ & $133(385)$ & $847(2678)$ & $3909 (12751)$ & $15199 (50078)$& $51030 (165603)$
  \end{tabular}
  \caption{ The number of cuts in the generating sets 
  defined in eqs.~\eqref{eq:G_n1},~\eqref{eq:G_nkl}~and~\eqref{eq:G_n1B} for
    all rational parts of partial amplitudes with up to nine gluons.
    Bubble coefficients require additional triangle cuts, 
    and we indicate the total number of cuts including these
    triangles in parenthesis.}
  \label{tab:n_generating_sets}
\end{table}

\FloatBarrier
\subsection{%
Subleading-Color Single-Trace
Rational Parts in the Separable Approach}
\label{SubleadingSingleTraceSeparable}
DPS conjectured an all-multiplicity form~\cite{Dunbar:2020wdh} for 
the subleading single-trace amplitudes
$\RT_{n:\oneB}$, making them the first two-loop
Yang--Mills amplitudes (conjecturally) known
for an arbitrary number of gluons.
This conjecture is possible due to special properties
of these amplitudes.
These properties lead to a simpler form, easier to determine
than that  of the remaining color-ordered amplitudes.

A significant simplification is the absence of multi-particle poles
in these rational parts.
This property was first observed in ref.~\cite{Dunbar:2020wdh}.
Here we provide an
argument from the point of view of the separable approach.
In addition, as we shall see below,
in the separable approach, all but triangle cuts
of $\RT_{n:\oneB}$ vanish.
We already observed this fact in the derivation of 
$\RT_{4:\oneB}$ in \sect{FourGluonSection}.
We provide a general argument based on power counting, specializing
the analysis given in Appendix A of ref.~\cite{Badger:2008cm}.
Finally we discuss high-multiplicity agreement of the
conjecture stated in ref.~\cite{Dunbar:2020wdh} and our computations
using the separable approach.

\subsubsection{Absence of Multi-Particles Poles}
A striking feature of the $\oneB$ amplitudes 
is the absence of multi-particle poles, as well as the 
absence of double poles
on complex momentum configurations.
This feature was exploited in the DPS conjecture by requiring the
correct two-particle factorization behavior.

In the separable approach, multi-particle poles can only 
originate from
two-scalar amplitudes of the type $\ATwo_{\twoscalar}$.
For the all-plus helicity configuration, only a factorization of the
form
$\AOne_{\onescalar} \times \AOne_{\onescalar}$ is allowed, as
for the factorizations 
$\ATwo_{\twoscalar} \times \ATree$ and 
$\ATree \times \ATwo_{\twoscalar}$
the tree amplitude vanishes.
For $K=k_i+\ldots+k_{i+m}$ going on shell, $\ATwo_{\twoscalar}$
factorizes along
the gluon propagator connecting the two loops,
\begin{multline}
  \ATwo_{\twoscalar}(1^+2^+\ldots i^+\ldots (i+m)^+ \ldots n^+)\
  \overset{s_{i\mathellipsis i+m}=K^2\to 0}{\longrightarrow}\\
  \sum_{h=\pm}\AOne_{\onescalar}(1^+2^+\mathellipsis K^h \mathellipsis n^+)
  \frac{1}{K^2}\AOne_{\onescalar}(i^+\mathellipsis (i+m)^+(-K)^{\overline{h}} )
\end{multline}
As $\ATwo_{\withcontact}$ does not possess such a propagator, it is
finite in this limit.
However, we saw that the rational parts $\RT_{n:\oneB}$ only receive
contributions from $\ATwo_{\withcontact}$, as the color structure is
incompatible with $\ATwo_{\twoscalar}$.
We can therefore conclude, that $\RT_{n:\oneB}$ does not
contain any multi-particle poles, in agreement with the findings of
ref.~\cite{Dunbar:2020wdh}.

The absence of multi-particle poles in $\ATwo_{\withcontact}$ further
implies, that the complexity seen in rational parts of the remaining
partial amplitudes stems from $\ATwo_{\twoscalar}$.
The fact that $\ATwo_{\twoscalar}$ is a quantity derivable
from scalar QCD may be of use in studying its all-multiplicity
behavior.

\subsubsection{Vanishing Cuts}
\label{app:RTOneBBowTies}

In the one-loop squared construction of the subleading single-trace
rational parts $\RT_{n:\oneB}$ we find numerically that all cuts except
those of the double-triangle class vanish.
We can view this as a consequence of a more general principle:
the only cut coefficients that contribute to the rational parts
$\RT_{\withcontact}$ are of the $\CutClass{\TriTri}$ type. 
The coefficients of one-loop squared integrals involving boxes or
bubbles vanish.
As $\RT_{n:\oneB}$ is made up entirely of $\RT_{\withcontact}$, this
rational part can be reduced to products of
triangle integrals with appropriate coefficients.

The vanishing of box and bubble coefficients can be understood from
power counting arguments.
While a gluon coupling to a scalar line contributes one power of the
loop momentum to the integrands numerator, 
the contact term does not.
To  see explicitly that this leads to the vanishing of
box and bubble cuts, we follow the analysis found in
Appendix A of Ref.~\cite{Badger:2008cm}, with the requirement that one
of the vertices is a contact term.

\begin{figure}
  \centering
  \captionsetup[subfigure]{justification=centering}
  \begin{subfigure}{0.49\linewidth}
    \includegraphics[height=9em]{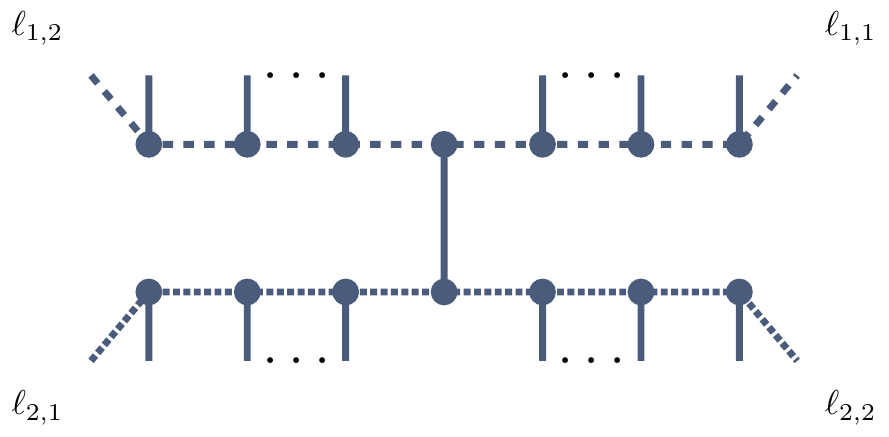}
    \caption{A diagram in the four-scalar amplitude
      $\ATree_{\gluon}$.}
    \label{fig:leadingScalarDiagramFourScalarGluon}
  \end{subfigure}
  \begin{subfigure}{0.49\linewidth}
    \includegraphics[height=9em]{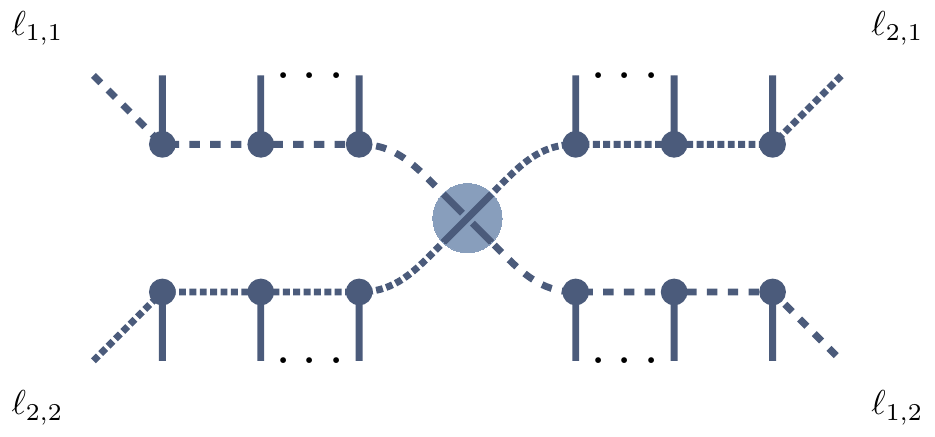}
    \caption{A diagram in the four-scalar amplitude
      $\ATree_{\contact}$, as it appears in $\RT_{n:\oneB}$.}
    \label{fig:leadingScalarDiagramFourScalarContact}
  \end{subfigure}
  \caption{Examples of Feynman diagrams contributing to
  four-scalar tree amplitudes,
    that are leading in the large-loop-momentum limit for both loops.}
  \label{fig:leadingScalarDiagrams}
\end{figure}

Let us first consider a box cut of one of the loops.
We need to determine
\begin{equation}
  \Inf_{\mu^2}\left[\ATree_{1}\ATree_2\ATree_3\ATree_4\right]_{\mu^4}.
\end{equation}
Expanding the loop momentum of \eqn{eq:BoxLoopMomentumAnsatz} in the
large-$\mu^2$ limit, we find that it can generally be expressed as,
\begin{equation}
  \ell^{\nu}_{\mathrm{Box}}=\alpha_1^{\nu}+\alpha_2^{\nu}\sqrt{\mu^2}+\frac{1}{\sqrt{\mu^2}}\alpha_3^{\nu}+\mathcal{O}\bigg(\frac{1}{\sqrt{\mu^2}^3}\bigg)\,,
\end{equation}
where the $\alpha^{\nu}_i$ depend only on the external kinematics.
Of the the four amplitudes $\ATree_i$, three are of the two-scalar
type, while one is the four-scalar central amplitude connecting the
loops.
In the large-loop-momentum limit, each cubic gluon vertex on the
scalar line provides a power of $\ell_{\mathrm{Box}}$.
Due to the on-shell condition $\ell_{\mathrm{Box}}^2=0$, each scalar
propagator carrying the loop momentum contributes $1/\ell_{\mathrm{Box}}$.
The dominant contribution (with no special choice of helicity
reference momentum) arises from those Feynman diagrams in the 
tree amplitudes where the ratio of
number of gluon-scalar-scalar vertices to scalar propagators 
is maximal.
Examples of such Feynman diagrams for $\ATree_{\gluon}$ and $\ATree_{\contact}$ are shown in
\fig{fig:leadingScalarDiagrams}.

A product $P^m_{\mathrm{Box}}$ of $m$ powers of $\ell_{\mathrm{Box}}$ has
the generic form
\begin{equation}
  P^m_{\mathrm{Box}}(x)=\sum_{i=-m}^{m}x_i\left(\sqrt{\mu^2}\right)^i
\end{equation}
In a box cut involving a four-scalar gluon-exchange amplitude
 $\ATree_{\gluon}$ we obtain the same power counting as 
in the analysis of
ref.~\cite{Badger:2008cm}.
The dominant terms in such terms involve diagrams of the form
shown in \fig{fig:leadingScalarDiagramFourScalarGluon}.
Determining the box coefficient, up to prefactors that depend
on the external kinematics we obtain,
\begin{equation}
  \Inf_{\mu^2}\left[\frac{P^m_{\mathrm{Box}}(x)}{P^{m-4}_{\mathrm{Box}}(y)}\right]_{\mu^4}=\frac{x_m}{y_{m-4}}\,.
\end{equation}
If however a contact-term amplitude $\ATree_{\contact}$
connects the loops, the contact term takes the place of one
gluon-scalar-scalar vertex on each scalar line.
This shifts the ratio of powers of loop momenta 
in the numerator and denominator.
The leading terms have one fewer power of $\ell_{\mathrm{Box}}$ in
the numerator compared to gluon-exchange case, so that
\begin{equation}
  \Inf_{\mu^2}\left[\frac{P^{m-1}_{\mathrm{Box}}(x)}{P^{m-4}_{\mathrm{Box}}(y)}\right]_{\mu^4}=0\,.
\end{equation}
Thus, any box cut involving the four-scalar
contact term has to vanish.

We can perform the same analysis
for bubble cuts to see that they too
vanish in the presence of a contact term.
The parametrized bubble loop momentum of
\eqn{eq:dim-D_bubble_loop_parametrization} can be written as
\begin{equation}
  \ell_{\mathrm{Bubble}}^{\nu}=\alpha_{1}^{\nu}+y\alpha_2^{\nu}
  +t\alpha_3^{\nu}+\frac{y}{t}\alpha_4^{\nu}+\frac{y^2}{t}\alpha_5^{\nu}
  +\frac{\mu^2}{t}\alpha_6^{\nu}\,,
\end{equation}
while an $m$-fold product $P^m_{\mathrm{Bubble}}(x)$ has the 
generic form
\begin{equation}
  P^m_{\mathrm{Bubble}}(x)=\sum_{i=0}^{m}\sum_{j=-m}^{m-2i}\sum_{k=\max(0,-j-i)}^{m-j-2i}x_{i,j,k}\,\mu^{2i}t^jy^k\,.
\end{equation}
If the four-scalar amplitude in the bubble cut is of the
gluon-exchange type, the leading part of the cut is made up of terms
of the form,
\begin{equation}
  \label{eq:bubbleCutPolynomialGluon}
  \Inf_{y,t,\mu^2}\left[\frac{P^m_{\mathrm{Bubble}}(x)}{P^{m-2}_{\mathrm{Bubble}}(y)}\right]_{y^i\to Y_i,t^0,\mu^2}.
\end{equation}
The expanded, generally non-vanishing, form of
\eqn{eq:bubbleCutPolynomialGluon} can be found in
ref.~\cite{Badger:2008cm}.
If a four-scalar amplitude $\ATree_{\contact}$ is present, 
the power of
loop momentum in the numerator is again reduced by one.
The bubble cut then behaves as
\begin{equation}
  \Inf_{y,t,\mu^2}\left[\frac{P^{m-1}_{\mathrm{Bubble}}(x)}{P^{m-2}_{\mathrm{Bubble}}(y)}\right]_{y^i\to Y_i,t^0,\mu^2}=0\,,
\end{equation}
meaning that such cuts also vanish for a contact-term.

That leaves us with only triangle cuts.
There are two types of such cuts: those that contribute to the
coefficients of triangle basis integrals, and those required for
coefficients of bubble integrals.
In either case, the parametrization of the loop momentum
of \eqns{eq:D-dim_triangle_loop_momentum_parametrization}%
{eq:D-dim_triangle_loop_momentum_parametrization_cc} can be written as 
\begin{equation}
  \ell_{\mathrm{Triangle}}^{\nu}=\alpha_{1}^{\nu} +t\alpha_2^{\nu}
  +\frac{1}{t}\alpha_3^{\nu}
  +\frac{\mu^2}{t}\alpha_4^{\nu}\,,
\end{equation}
with a generic $m$-fold product being
\begin{equation}
  P^m_{\mathrm{Triangle}}(x)=
  \sum_{i=0}^{m}\sum_{j=-m}^{m-2i}x_{i,j}\,\mu^{2i}t^j\,.
\end{equation}
For a four-scalar amplitude with a gluon exchange, we find
that triangle coefficients receive contributions of the form
\begin{multline}
  \Inf_{t,\mu^2}\left[\frac{P^m_{\mathrm{Triangle}}(x)}{P^{m-3}_{\mathrm{Triangle}}(y)}\right]_{t^0,\mu^2}\\
    =\frac{x_{1,m-3}}{y_{0,m-3}}-\frac{x_{0,m} y_{1,m-6}}{y_{0,m-3}^2}-\frac{x_{0,m-1} y_{1,m-5}}{y_{0,m-3}^2}+\frac{2 x_{0,m} y_{0,m-4}
      y_{1,m-5}}{y_{0,m-3}^3}-\frac{x_{1,m-2} y_{0,m-4}}{y_{0,m-3}^2}
\end{multline}
while those of bubble coefficients are
\begin{equation}
  \Inf_{t,\mu^2}\left[\frac{P^m_{\mathrm{Triangle}}(x)}
    {P^{m-3}_{\mathrm{Triangle}}(y)}\right]_{t^i\to T_i,\mu^2}=c_1\left(\frac{x_{1,m-2}}{y_{0,m-3}}-\frac{x_{0,m}
   y_{1,m-5}}{y_{0,m-3}^2}\right)+c_3\left(\frac{x_{0,m}}{y_{0,m-3}}\right)
\end{equation}
The $c_i$ are kinematic factors originating from the
$T_i$ parameter integrals.
On the other hand, for a contact term tree amplitude we find
respectively
\begin{equation}
  \Inf_{t,\mu^2}\left[\frac{P^{m-1}_{\mathrm{Triangle}}(x)}{P^{m-3}_{\mathrm{Triangle}}(y)}\right]_{t^0,\mu^2}=
  \frac{x_{1,m-3}}{y_{0,m-3}}-\frac{x_{0,m-1} y_{1,m-5}}{y_{0,m-3}^2}
\end{equation}
and
\begin{equation}
  \Inf_{t,\mu^2}\left[\frac{P^m_{\mathrm{Triangle}}(x)}
    {P^{m-3}_{\mathrm{Triangle}}(y)}\right]_{t^i\to T_i,\mu^2}=0.
\end{equation}
We can therefore see that for any one-loop amplitude 
requiring a contact term, only the coefficients of triangle
integrals are non-vanishing.  In our setting, this implies that
only coefficients of the bowtie integral are nonvanishing
in the rational terms of 
the subleading-color single-trace two-loop amplitude.

\subsubsection{Comparison with Conjecture for \texorpdfstring{$\RTwo_{7:1\mathrm{B}}$}{R(2)7:1B} \texorpdfstring{$\RTwo_{8:1\mathrm{B}}$}{R(2)8:1B} and \texorpdfstring{$\RTwo_{9:1\mathrm{B}}$}{R(2)9:1B}}
\label{subsec:7pt_R_71B_conjecture}

The five- and six-gluon
rational terms $\RTwo_{5:\oneB}$, $\RTwo_{6:\oneB}$ to which
we have compared our results
are known from explicit computations in 
refs.~\cite{DalgleishDunbar:2020}.
Starting from the seven-gluon case, $\RTwo_{n:\oneB}$ is only known
in the form of the all-$n$ conjecture of ref.~\cite{Dunbar:2020wdh}.

We can use the separable approach to provide an independent cross-check of this result.
Finding
agreement would provide evidence for both the correctness of the the
conjecture of ref.~\cite{Dunbar:2020wdh}, while at the same time validating
the use of the separable approach in the computation of the rational
contributions $\RTwo_{n:1\mathrm{B}}$.

We first consider the seven gluon rational terms $\RT_{7:\oneB}$.
By specializing the all-$n$ form of ref.~\cite{Dunbar:2020wdh} to
$n=7$ we obtain
\begin{equation}
  \RTwo_{7:1\mathrm{B}} = \RTwo_{7:1\mathrm{B},1}+\RTwo_{7:1\mathrm{B},2}\,,
\end{equation}
where,
\begin{equation}
  \label{eq:R271B2_conjecture}
  \begin{aligned}
    &\RTwo_{7:1\mathrm{B},1}=2
    \frac{\sum_{1\le i< j< k <l \le 7}\trF(i j k l)}
    {\spaa{1 2}\spaa{2 3}\spaa{3 4}\spaa{4 5}\spaa{5 6}\spaa{6 7}\spaa{71}}\\
    &\RTwo_{7:1\mathrm{B}2}=\\
    -4\Big[ &C_{3745} \trF [(1+2+3) 5 4 7]
    -C_{3746} \trF [(1+2+3) 6 4 7]+C_{3756} \trF [(1+2+3) 6 5 7]\\
    &+C_{2634} \trF [(1+2) 4 3 (6+7)]+C_{2734} \trF [(1+2) 4 37]
    -C_{2635} \trF [(1+2)  5 3 (6+7)]\\
    &-C_{2735} \trF [(1+2)  5 3 7]+C_{2645} \trF [(1+2) 5 4 (6+7)]
    +C_{2745} \trF [(1+2)  5 4 7]\\
    &+C_{2736} \trF [(1+2)  6 37]-C_{2746} \trF [(1+2) 6 4 7]
    +C_{2756} \trF [(1+2)  6 5 7]\\
    &+C_{1523} \trF [1 3 2 (5+6+7)]
    +C_{1623} \trF [1 3 2 (6+7)]+C_{1723} \trF [1 3 2 7]\\
    &-C_{1524} \trF [1 4 2 (5+6+7)]-C_{1624} \trF [1 4 2 (6+7)]
    -C_{1724} \trF [1 4 2 7]\\
    &+C_{1534} \trF [1 4 3 (5+6+7)]
    +C_{1634} \trF [1 4 3 (6+7)]+C_{1734} \trF [1 4 3 7]\\
    &+C_{1625} \trF [1 5 2 (6+7)]+C_{1725} \trF [1 5 2 7]
    -C_{1635} \trF [1 5 3 (6+7)]\\
    &-C_{1735} \trF [1 5 3 7]+C_{1645} \trF [1 5 4 (6+7)]
    +C_{1745} \trF [1 5 4 7)]-C_{1726} \trF [1 6 2 7]\\
    &+C_{1736} \trF [1 6 3 7]
    -C_{1746} \trF [1 6 4 7]+C_{1756} \trF [1 6 5 7]\Big].
  \end{aligned}
\end{equation}
Here, we have introduced,
\begin{equation}
    \tr_5(ijkl)=\trM(ijkl)-\trP(ijkl)\,;
\end{equation}
we also define,
\begin{equation}
    \begin{aligned}
      \trM(ijkl)&=\spab{i|jkl|i}\,,\\
      \trP(ijkl)&=\spba{i|jkl|i}\,.
    \end{aligned}
\end{equation}
The $C_{r s i j}$ are sums over terms with Parke--Taylor denominators,
\def\PTF#1{\langle\!\langle #1\rangle\!\rangle}
\begin{equation}
  C_{r s i j}=\sum_{\alpha\in S_{r s i j}}\CPT(1,\mathellipsis ,r,j,\{\alpha\},i,
  s\mathellipsis,7),
\end{equation}
where
\begin{equation}
  \CPT(a_1,\mathellipsis,a_n)=
  \frac{1}{\PTF{a_1 a_2\mathellipsis a_7}}\,.
\end{equation}
Here we introduce the notation,
\begin{equation}
    \PTF{a_1 a_2\mathellipsis a_n} =
    \spaa{a_1a_2}\spaa{a_2a_3}\mathellipsis
        \spaa{a_{n-1}a_n}\spaa{a_n a_1}\,.
\label{PTFdef}
\end{equation}
Define the sets $S_1$, $S_2$ and $S_3$ by splitting
$\{1,\mathellipsis,7\}$ as follows,
\begin{equation}
  \{1,\mathellipsis,7\} = \{1,\mathellipsis,r\} \cup S_1 \cup \{i\} \cup S_2
  \cup \{j\} \cup S_3 \cup \{s,\mathellipsis,7\},
\end{equation}
the sets $S_{r s i j}$ are obtained via the shuffle product,
\begin{equation}
  S_{r s i j} = S_1 \shuffle S_2^T \shuffle S_3,
\end{equation}
with the transposition denoting a reversal of the elements in $S_2$.  An
explicit form of $\RTwo_{7:1\mathrm{B},2}$ after evaluation of the $C_{r s
  i j}$ can be found in eq.~\eqref{eq:R2_71B2_explicit}.

The generating set $\GeneratingCuts_{7:1\mathrm{B}}$ is made up of $3909$
unitarity cuts.
When including the triangle contributions
to bubble coefficients, there are $12751$ cuts to evaluate per
permutation, or $89257$ cuts in total. 
For each permutation, only $416$ cuts are non-zero, all of which
again belong to the bowtie topology.
We evaluate the cuts semi-analytically on a one-dimensional kinematic
slice, parameterized by a variable $\delta$.
Thanks to the symbolic approach of our
code, we preserve the full analytic dependence of our result on $\delta$
during the computation.
Choosing all remaining kinematic degrees of freedom
to be random rational numbers, we can perform an exact comparison with
the conjectured form on the entire slice.
In addition we choose kinematics such that the limit
$\delta\to 0$ probes the collinear momentum configuration $6||7$.
The explicit form of this kinematic slice is given
Appendix~\ref{SevenGluonSubleadingAppendix}.

We find that the expression from the separable approach and the
conjectured form~\cite{Dunbar:2020wdh} 
as rational functions of $\delta$ are in full
analytic agreement.
We also find the correct collinear behavior
\begin{equation}
  \RTwo_{7:1\mathrm{B}}(1^+2^+3^+4^+5^+6^+7^+)\overset{6||7}{\longrightarrow}
  \RTwo_{6:1\mathrm{B}}(1^+2^+3^+4^+5^+K^+)\times\splittingFunctionTree_{-}(6^+7^+).
\end{equation}
This is expected, as the DPS conjecture 
is based on demanding the correct two-particle collinear limits.

Furthermore, we were able to compute $\RT_{8:\oneB}$ and
$\RT_{9:\oneB}$ directly
on rational kinematic points using the separable
approach.
In the eight-point case we evaluated all one-loop squared cuts.
We find again that again the coefficients of all but 
bowtie integrals
vanish, in agreement with the analysis of the previous section.
In the nine-gluon case, we limited the
computation to just the bowtie contributions due to the large number
of cuts in the full generating set.
 
For both $\RT_{8:\oneB}$ and $\RT_{9:\oneB}$ we find exact, rational numerical
agreement of with the expressions of the conjecture.

\FloatBarrier
\section{Conclusions}
\label{ConclusionSection}

The study of scattering amplitudes in the most symmetric
Yang--Mills theory, the $\mathcal{N}=4$ supersymmetric one,
has taught us a great deal about how to calculate more
efficiently in all Yang--Mills theories and in gravity
as well.  These amplitudes are in a sense the simplest
and also exhibit the most structure.
Their study has also led to insights on deeper structures
in these theories and on the connections between them.

In this article, we have explored rational terms in 
amplitudes, which at present are at the opposite extreme, 
exhibiting the least
symmetry or structure
of all contributions to Yang--Mills amplitudes.
We studied the simplest such terms at two loops,
in the all-plus gluon amplitudes.
We relied on the BMP separability conjecture and
used a straightforward extension of generalized unitarity
techniques for computing one-loop rational terms in order
to compute them.  We computed the rational
terms in the four- and five-point two-loop amplitudes
analytically, and those in the six- and seven-point
amplitudes numerically.  Our results agree with
those obtained by Dunbar, Dalgleish, Jehu, Perkins and Strong through a recursive
approach.  They also agree numerically
with the all-$n$ conjecture of
Dunbar, Perkins and Strong for the subleading-color single-trace 
amplitude at eight and nine points.  In addition to
providing evidence for the correctness of the results
in refs.~\cite{Dunbar:2017nfy,%
Dunbar:2017azf,Dalgleish:2020mof,Dunbar:2019fcq} and conjecture in 
ref.~\cite{Dunbar:2020wdh}, our calculations
also provide evidence for the correctness of the separability
conjecture~\cite{Badger:2016ozq} both for 
leading- and subleading-color
amplitudes.  The ideas developed here may also help simplify
the calculation of other rational terms at two loops,
in particular to the other simple helicity configuration,
with a lone negative-helicity gluon.

\acknowledgments

We thank 
Manuel Accettulli Huber, 
Simon Badger, 
Stefano De Angelis, 
Lance Dixon,
Maxence Grandadam, 
Ingrid Vazquez-Holm, 
Gregory Korchemsky, 
Gustav Mogull, 
Ben Page, 
Tiziano Peraro, 
and Gregory Soyez
for helpful discussions, 
and Lance Dixon for ongoing encouragement as well.  
The research described here
has received funding from the European Union's 
Horizon~2020 research and innovation program
under the Marie Sk\l{}odowska-Curie grant agreement
No.~764850 ``SAGEX''.
DAK's work was supported in part by the French 
\textit{Agence 
Nationale pour la
 Recherche\/}, under grant ANR--17--CE31--0001--01,
 and in part by the European Research Council, under
 grant ERC--AdG--885414. 
SP's work has been supported in part by the Mainz Institute for Theoretical Physics (MITP) of the Cluster of Excellence PRISMA+. The article is partly based on material from SP's doctoral thesis at the Institut de Physique Th\'eorique (IPhT).

\appendix

\section{Conventions}
\label{ConventionsAppendix}

We follow ref.~\cite{Henn:2014yza} for our spinor conventions.
  We use the mostly minus convention for the metric
$\eta^{\mu\nu}=\text{diag}(+-\mathellipsis-)$, and four-dimensional spinor
products satisfy the relation
\begin{equation}
  2(k_i\cdot k_j)=\spaa{ij}\spbb{ji}.
\end{equation}
In color structures, we use the standard normalization
for the generators of the gauge groups
$\mathrm{SU}(N_c)$ and $\mathrm{U}(N_c)$,
\begin{equation}
  \label{eq:TrTaTb}
  \Tr(T^a T^b)=\delta^{ab}.
\end{equation}
We use the following normalization for $L$-loop amplitudes in $D$ dimensions,
\begin{equation}
  \label{eq:amplitude_normalization}
  A^{(L)}=(-i)(4\pi)^{L D/2}\times\left( \text{sum of Feynman diagrams} \right).
\end{equation}
For one-loop Feynman integrals we use the normalization
\begin{equation}\label{eq:integral_def}
  I^D_n\big[\mathcal{N}[\ellb,\mu^2]\big] = (-i)(-1)^n (4\pi)^{D/2}
  \mathcal{I}_n[\mathcal{N}[\ellb,\mu^2]],
\end{equation}
where
\begin{equation}\label{eq:caligraphic_integral_def}
  \begin{aligned}
    \mathcal{I}_n\big[\mathcal{N}[\ellb,\mu^2]\big]&=\int\frac{\diff^D \ell}{(2\pi)^D}
    \frac{\mathcal{N}[\ellb,\mu^2]}
    {\ell^2(\ell-K_1)^2\mathellipsis)(\ell-K_1-\mathellipsis-K_{n-1})^2}\\
    &=\int\frac{\diff^{D/2-2} \mu^2}{(2\pi)^{D-4}}\int
    \frac{\diff^4 \ellb}{(2\pi)^4}\frac{\mathcal{N}[\ellb,\mu^2]}
    {\big(\ellb^2-\mu^2\big)\big((\ellb-K_1)^2-\mu^2\big)\mathellipsis
      \big((\ellb-K_{1\mathellipsis n-1})^2-\mu^2\big)}\\
  \end{aligned}
\end{equation}
Here $\ellb$ denotes the four-dimensional, and $\elle$ the
$(D-4)$-dimensional components of loop momentum.
In one-loop integrals with
four-dimensional external kinematics, $\elle$ can only appear as its
square, which we label as $\mu^2 \equiv -\elle^2$.
The definitions of
eqs.~\eqref{eq:integral_def}
and~\eqref{eq:caligraphic_integral_def} mirror the
notation found in ref.~\cite{Bern:1992em}, with the addition of a numerator
function $\mathcal{N}$, which can depend on $\ellb$ and $\mu^2$.

The integrals relevant for the computation of one-loop rational contributions
are~\cite{Bern:1998sv,Badger:2008cm} given in
eq.~\eqref{eq:D-dim_mu2_integrals}.

\section{Scalar Tree Amplitudes}
\label{ScalarAmplitudesAppendix}

As desctibed in section~\ref{TreeLevelAmplitudesSection}, we require
amplitudes with a single massive scalar pair.
These are well known, and below we collect the ones
relevant for the computations in this paper.  For adjacent 
scalars and up to four gluons, we use the expressions from
ref.~\cite{Badger:2005zh}.
Arbitrary-multiplicity expressions exist as well for the
all-plus%
~\cite{Forde:2005ue,Rodrigo:2005eu,Schwinn:2006ca,Ferrario:2006np} and
single-minus~\cite{Forde:2005ue} helicity configurations.
In the case of
tree amplitudes with non-adjacent scalars we use Kleiss--Kuijff
relations~\cite{Kleiss:1988ne} to obtain them from adjacent-scalar
ones.
We numerically verified all tree amplitudes using the
Berends--Giele recursion relations.
\begingroup
\allowdisplaybreaks
\begin{align}
  A^{(0)}(1_\scalar 2^+ 3^+ 4_\scalar)&=
                                            \frac{\scalarMSqu }{\LProd_{1 2}}\frac{\spbb{2 3}}{\spaa{2 3}}\,,\\
  A^{(0)}(1_\scalar 2^+ 3^- 4_\scalar)&=
                                            -\frac{\spab{3|1|2}\spab{3|4|2}}{\LProd_{1 2}\sfour_{2 3}}\,,\\
  A^{(0)}(1_\scalar 2^+ 3^+ 4^+ 5_\scalar)&=
                                                -\frac{\scalarMSqu \spbb{4 | (2+3) 1 | 2}}{\LProd_{1 2} \LProd_{4 5} \spaa{2 3} \spaa{3 4}}\,,\\
  A^{(0)}(1_\scalar 2^+ 3^+ 4^- 5_\scalar)&=
                                                -\frac{\scalarMSqu \spbb{2 3}^3}{\sfour_{1 5} \spbb{3 4}
                                                \spbb{4 | (2+3) 1 | 2}}+\frac{\spab{4 | 5 (2+3) 1 | 2}^2}{\LProd_{1 2} \LProd_{4 5}
                                                \spaa{2 3} \spaa{3 4} \spbb{4 | (2+3) 1 | 2}}\,,\\
  A^{(0)}(1_\scalar 2^+ 3^- 4^+ 5_\scalar)&=
                                                -\frac{\scalarMSqu \spbb{2 4}^4}{\sfour_{1 5} \spbb{2 3}\spbb{3 4}
                                                \spbb{4 | (2+3) 1 | 2}}+
                                                \frac{\spab{3 | 1 | 2}^2 \spab{3 | 5 | 4}^2}
                                                {\LProd_{1 2} \LProd_{4 5} \spaa{2 3} \spaa{3 4} \spbb{4 | (2+3) 1 | 2}}\,,\\
  A^{(0)}(1_\scalar 2^+ 3^+ 4^+ 5^+ 6_\scalar)&=
                                                    -\frac{\scalarMSqu\spbb{5 | 6 (4+5) (2+3) 1 | 2}}
                                                    {\LProd_{1 2} \LProd_{1 2 3} \LProd_{5 6} \spaa{2 3} \spaa{3 4} \spaa{4 5}}\,.
\end{align}
\endgroup

To match our own BCFW result for
$A^{(0)}(1_\scalar 2^+ 3^+ 4^+ 5^- 6_\scalar)$ with that of
ref.~\cite{Badger:2005zh}, we have to flip the sign
of the second term in eq.~(3.20) of that reference, so that
\begin{equation}
  \begin{aligned}
    A^{(0)}(1_\scalar 2^+ 3^+ 4^+ 5^- 6_\scalar)&=
    \frac{(\LProd_{1 2 3} \spab{5 | 6 (2+3+4) 1 | 2}-
      \scalarMSqu\spab{5 | 6 4 3 | 2})^2}{\LProd_{1 2} \LProd_{1 2 3} \LProd_{1 2 3 4}
      \spaa{2 3} \spaa{3 4}\spaa{4 5} \spbb{5 | 6 (4+5) (2+3) 1 | 2}}\\
    &\hphantom{={}}-\frac{ \scalarMSqu\spbb{4| (2+3) 1 | 2}^3  }{\LProd_{1 2}
      \spaa{2 3} \spab{3 | (4+5) (1+6) 1 | 2} \spbb{4 5}
      \spbb{5 | 6 (4+5) (2+3) 1 | 2}}\\
    &\hphantom{={}}-\frac{ \scalarMSqu\spab{5 | (3+4) | 2}^3}{\sfour_{2 3 4 5}
      \sfour_{3 4 5} \spaa{3 4} \spaa{4 5}\spab{3 | (4+5) (1+6) 1 | 2}}\,.
  \end{aligned}
\end{equation}
For $A^{(0)}(1_\scalar 2^+ 3^+ 4^- 5^+ 6_\scalar)$, we were not 
able to
align the expression in eq.~(3.21) of ref.~\cite{Badger:2005zh} 
with our numerical result.
While we can verify the last two lines of eq.~(3.21) 
to be the $\sfour_{234}$ and $\sfour_{16}$ channels in the
BCFW computation, the first two lines, from the $\sfour_{2 3}$
channel, do not match.
We therefore replace these lines with our less-compact result for
this channel,
\begin{equation}
  \begin{aligned}
    A^{(0)}&(1_\scalar 2^+ 3^+ 4^- 5^+ 6_\scalar)=\\
    &\frac{\spbb{2 3}}{\LProd_{1 2} \spaa{23} }\frac{\left[\frac{\scalarMSqu \spbb{5 | (2+3) 1 | 2}^4 }
    {\spbb{4 |(2+3) 1| 2}\spbb{4 5} (\sfour_{1 6} \spbb{2 | 3 1 | 2}+\sfour_{2 3} \spbb{2 | 6 1 | 2})}
    +\frac{\spab{4 | (1+2+3) (2+3) 1 | 2}^2 \spab{4 | 6 | 5}^2}{\LProd_{1 2 3} \LProd_{5 6}
      \spaa{4 5} \spab{4 | 3 | 2}}\right]}{\left(\frac{\sfour_{2 3}
          \spbb{2 | 1 (1+2+3)(2+3) 1 | 2}
          \spbb{2 5}}{\spbb{2 | 1 3 | 2}}+\spbb{5 | (2+3+4) (1+2+3) (2+3) 1 | 2}\right)}\\
    -&\frac{\scalarMSqu \spab{4 | (3+5) | 2}^4}{\sfour_{1 6} \sfour_{3 4 5} \spaa{3 4} \spaa{4 5}
      \spab{3 | (4+5) (1+6) 1 | 2}\spab{5 | (3+4) | 2}}\\
    -&\frac{\scalarMSqu \spbb{2 3}^3 \spbb{5 | (1+6) 1 | 2}}
    {\LProd_{5 6} \sfour_{2 3 4} \spab{5 | (3+4) | 2} \spbb{3 4}
      \spbb{4 | (2+3) 1 | 2}}\,.
  \end{aligned}
\end{equation}
For the seven-gluon rational contributions we also require 
scalar amplitudes
with five positive-helicity gluons.  We obtained the adjacent 
scalar amplitude
using the all-multiplicity form of ref.~\cite{Ferrario:2006np},
\begin{multline}
    A^{(0)}(1_\scalar 2^+ 3^+ 4^+ 5^+ 6^+ 7_\scalar)=\\
    \frac{\scalarMSqu\left(-\sfour_{2 3} \spbb{2 | 1(5+6+7) (5+6) 7 | 6}+
        \spbb{2 | 1 (2+3) 1 (2+3+4) (5+6) 7 | 6}\right)}{\six_{1 2}\six_{1 2 3} \six_{5 6 7} \six_{6 7}
      \spaa{2 3}\spaa{3 4}\spaa{4 5} \spaa{5 6}},
\end{multline}
while the non-adjacent scalar amplitudes are again obtained from Kleiss--Kuijf
relations.

\section{Contact-Term Amplitude from Off-Shell Scalar Currents}
\label{OffShellCurrentAppendix}

Here we discuss an alternative method of obtaining compact 
expressions for the four-scalar tree amplitudes $\ATree_{\contact}$,
where the two scalar lines are connected via a contact term.
As the contact term only amounts to a factor, we can construct these
tree amplitudes by connecting four off-shell scalar currents to
the contact-term.
The required currents contain a scalar line with a number
of positive helicity gluons attached.
In these currents, the mass and momentum squared of the off-shell
scalars are no longer related.
The current is also not gauge invariant, such that its
form generally depends on the choice of reference momentum for the
polarization vectors of the gluons.
Fortunately off-shell scalars are comparatively simple, so that
we are able to derive their form using Feynman diagrams.

The scalar current with only a single positive gluon is
\begin{equation}
  J_{\scalar}(2^+3_{\scalar})=-\frac{\spba{2|3|q}}{\spaa{2q}}\,.
\end{equation}
This matches exactly the on-shell amplitude, as it only consists of a
single Feynman diagram.
Note that we define the currents $J_{\scalar}$ to be the sum of
Feynman diagrams normalized by a factor of $(-i)$, just as in the case
of on-shell amplitudes.
We therefore sew the currents to the
contact term using propagators of the form
$\frac{-1}{\LProd_{ijk\mathellipsis}}$, with $\LProd_{i\ldots}$
as defined in \eq{eq:LProd_def}.
This current is sufficient to obtain the amplitude
$\ATree_{\contact}(1_{\scalar} 2_{\scalar} 3^+ 4_{\scalarP} 5_{\scalarP})$,
\begin{equation}\label{eq:A5SSPss}
  \begin{aligned}
    \ATree_{\contact}(1_{\scalar} 2_{\scalar} 3^+ 4_{\scalarP} 5_{\scalarP})&=
    V(\scalar \scalar \scalarP \scalarP)
    \frac{-1}{\LProd_{23}}J_\scalar(2_{\scalar} 3^+)+V(\scalar \scalar \scalarP \scalarP)
    \frac{-1}{\LProd_{34}}J_\scalarP(3^+ 4_{\scalarP})\\
    &=\frac{1}{2\LProd_{23}\LProd_{34}\spaa{3q}}\left
      (\spba{3|2|q}\spab{3|4|3}-\spba{3|4|q}\spab{3|2|3}\right)\\
    &=\frac{1}{2\LProd_{23}\LProd_{34}\spaa{3q}}\left
      (\spba{3|2|3}\spab{q|4|3}-\spbb{3|24|3}\spaa{3q}-\spba{3|4|q}\spab{3|2|3}\right)\\
    &=-\frac{\spbb{3|24|3}}{2\LProd_{23}\LProd_{34}}
  \end{aligned}
\end{equation}

In the six-point case, i.e.\ amplitudes with two gluons, we
also need the two-gluon currents
$J_{\scalar}(2^+ 3^+ 4_{\scalar})$ and $J_{\scalar}(2^+3_{\scalar} 4^+)$.
If we choose the reference momenta $q_2=k_3$,
$q_3=k_2$, we obtain for $J_{\scalar}(2^+ 3^+ 4_{\scalar})$,
\begin{equation}
  J_{\scalar}(2^+ 3^+ 4_{\scalar})=\frac{m_4^2 \spbb{2 3}}{\LProd_{3 4} \spaa{2 3}}+
  \frac{\LProd_{2 3 4}}{2 \spaa{2 3}^2}\,.
\end{equation}
This closely resembles the on-shell result,
\begin{equation}
  \ATree(1_{\scalar} 2^+ 3^+ 4_{\scalar})=\frac{m_4^2 \spbb{2 3}}{\LProd_{3 4} \spaa{2 3}}\,.
\end{equation}
Since we have made a specific choice for the reference momenta we have to
make the same choice for the remaining currents.
For $J_{\scalar}(2^+ 3_{\scalar} 4^+)$ we get,
\begin{equation}
  J_{\scalar}(2^+ 3_{\scalar} 4^+)=\frac{m_3^2 \spbb{2 4}}{\LProd_{3 4}\spaa{2 4} }
  -\frac{\LProd_{2 3 4} \spba{2 | 3 | 4} \spba{4 | 3 | 2}}{\LProd_{2 3} \LProd_{3 4} \spaa{2 4}^2}\,,
\end{equation}
where the first term again mirrors the on-shell amplitude
\begin{equation}
  \ATree(1_{\scalar} 2^+ 3_{\scalar} 4^+) = \frac{m_3^2 \spbb{2 4}}{\LProd_{3 4}\spaa{2 4} }\,.
\end{equation}

We can then construct the on-shell amplitudes.  For
$\ATree_{\contact}(1_{\scalar} 2_{\scalar} 3^+ 4^+ 5_{\scalarP} 6_{\scalarP})$ we
obtain
\begin{equation}
  \begin{aligned}
    \ATree_{\contact}(1_{\scalar} 2_{\scalar} 3^+ 4^+ 5_{\scalarP} 6_{\scalarP})&=
    V(\scalar \scalar \scalarP \scalarP)
    \frac{-1}{\LProd_{234}}J_\scalar(2_{\scalar} 3^+4^+)\\
    &\hphantom{=}+V(\scalar \scalar \scalarP \scalarP)
    \frac{-1}{\LProd_{345}}J_\scalarP(3^+ 4^+ 5_{\scalarP})\\
    &\hphantom{=}+V(\scalar \scalar \scalarP \scalarP)
    \left(\frac{-1}{\LProd_{23}}J_\scalar(2_{\scalar} 3^+)\right)\times
    \left(\frac{-1}{\LProd_{45}}J_\scalarP(4^+ 5_{\scalarP})\right)\\
    &=-\frac{1}{2\spaa{3 4}}\left[
      \frac{\spbb{3|25|4}}{\LProd_{2 3} \LProd_{4 5}}-
      \spbb{3 4} \left(\frac{m_2^2}{\LProd_{2 3} \LProd_{2 3 4}}+
        \frac{m_5^2}{\LProd_{3 4 5} \LProd_{4 5}}\right)\right]
  \end{aligned}
\end{equation}
Similarly,
$\ATree_{\contact}(1_{\scalar} 2_{\scalar} 3^+ 4_{\scalarP} 5_{\scalarP} 6^+)$ and
$\ATree_{\contact}(1_{\scalar} 2^+ 3_{\scalar} 4_{\scalarP} 5^+ 6_{\scalarP})$
evaluate to
\begin{equation}
  \begin{aligned}
    \ATree_{\contact}(1_{\scalar} 2_{\scalar} 3^+ 4_{\scalarP} 5_{\scalarP} 6^+)&=
    V(\scalar \scalar \scalarP \scalarP)\\
    &\hphantom{=}\times\left(
      \frac{-1}{\LProd_{23}}J_\scalar(2_{\scalar} 3^+)+
      \frac{-1}{\LProd_{34}}J_\scalarP(3^+ 4_{\scalarP})
    \right)\\
    &\hphantom{=}\times\left(
      \frac{-1}{\LProd_{56}}J_\scalarP(5_{\scalarP} 6^+)+
      \frac{-1}{\LProd_{61}}J_\scalar(6^+ 1_{\scalar})
    \right)\\
    &=-\frac{1}{2}\frac{\spbb{3|24|3}}{\LProd_{23}\LProd_{34}}\frac{\spbb{6|51|6}}{\LProd_{56}\LProd_{61}},
  \end{aligned}
\end{equation}
\begin{equation}
  \begin{aligned}
    \ATree_{\contact}(1_{\scalar} 2^+ 3_{\scalar} 4_{\scalarP} 5^+ 6_{\scalarP})&=
    V(\scalar \scalar \scalarP \scalarP)\\
    &\hphantom{=}\times\left(
      \frac{-1}{\LProd_{21}}J_\scalar(1_{\scalar} 2^+)+
      \frac{-1}{\LProd_{23}}J_\scalar(2^+ 3_{\scalar})
    \right)\\
    &\hphantom{=}\times\left(
      \frac{-1}{\LProd_{45}}J_\scalarP(4_{\scalarP} 5^+)+
      \frac{-1}{\LProd_{56}}J_\scalarP(5^+ 6_{\scalarP})
    \right)\\
    &=-\frac{1}{2}\frac{\spbb{2|13|2}}{\LProd_{21}\LProd_{23}}\frac{\spbb{5|46|5}}{\LProd_{45}\LProd_{56}},
  \end{aligned}
\end{equation}
where we used the same steps as in eq.~\eqref{eq:A5SSPss}.
For these two amplitudes, the sum over attachments factorizes, as
the gluons can not appear in the same current.
Note that in $\ATree_{\contact}(1_\scalar 2^+ 3_\scalar 4_\scalarP 5^+ 6_\scalarP)$ the currents in each factor belong to the same scalar line, while in $\ATree_{\contact}(1_\scalar 2_\scalar 3^+ 4_\scalarP 5_\scalarP 6^+)$ they do not.

Lastly we determine $\ATree_{\contact}(1_{\scalar} 2^+ 3_{\scalar} 4^+ 5_{\scalarP}
6_{\scalarP})$,
\begin{equation}
  \begin{aligned}
    \ATree_{\contact}(1_{\scalar} 2^+ 3_{\scalar} 4^+ 5_{\scalarP} 6_{\scalarP})&=
    V(\scalar \scalar \scalarP \scalarP)\times\left(
      \frac{-1}{\LProd_{12}}J_\scalar(1_{\scalar} 2^+)\right)
    \times\left(\frac{-1}{\LProd_{34}}J_\scalar(3_{\scalar} 4^+)\right)\\
    &\hphantom{=}+V(\scalar \scalar \scalarP \scalarP)\times\left(
      \frac{-1}{\LProd_{12}}J_\scalar(1_{\scalar} 2^+)\right)
    \times\left(\frac{-1}{\LProd_{45}}J_\scalarP( 4^+ 5_{\scalarP})\right)\\
    &\hphantom{=}+V(\scalar \scalar \scalarP \scalarP)\times\left(
      \frac{-1}{\LProd_{23}}J_\scalar(2^+ 3_{\scalar})\right)
    \times\left(\frac{-1}{\LProd_{45}}J_\scalarP( 4^+ 5_{\scalarP})\right)\\
    &\hphantom{=}+V(\scalar \scalar \scalarP \scalarP)\times\left(
      \frac{-1}{\LProd_{234}}J_\scalar(2^+ 3_{\scalar} 4^+)\right)\\
    &=-\frac{1}{2}\Bigg[\frac{m_3^2 \spbb{2 4}^2}{\LProd_{2 3} \LProd_{2 3 4} \LProd_{3 4}}
    -\frac{\spab{2 | 3 | 4} \spab{4| 1| 2}}{\LProd_{1 2} \LProd_{3 4} \spaa{2 4}^2}
    +\frac{\spab{2 | 5 | 4} \spab{4 | 1 | 2}}{\LProd_{1 2} \LProd_{4 5} \spaa{2 4}^2}\\
    &\hphantom{=-\frac{1}{2}\Bigg[}
    +\frac{\spab{2 | 3 | 4} \spab{4 | 3 | 2}}{\LProd_{2 3} \LProd_{3 4} \spaa{2 4}^2}
    -\frac{\spab{2 | 5 | 4} \spab{4 | 3 | 2}}{\LProd_{2 3} \LProd_{4 5} \spaa{2 4}^2}\Bigg]\\
    &=-\frac{1}{2}\left[\frac{m_{3}^2 \spbb{2 4}^2}{\LProd_{2 3} \LProd_{2 3 4} \LProd_{3 4}}
      +\frac{\spbb{2| 1 3 | 2} \spbb{4 | 3 5 | 4}}{\LProd_{1 2} \LProd_{2 3} \LProd_{3 4} \LProd_{4 5}}\right]
  \end{aligned}
\end{equation}
As the currents are agnostic to the scalar flavor, 
we can summarize the results
as
\begingroup
\allowdisplaybreaks
\begin{align}
&\hphantom{6_\scalar}\begin{aligned}
  \ATree_{\contact}&(1_{\scalar_1} 2_{\scalar_2} 3^+ 4_{\scalar_3} 5_{\scalar_4})
  =\\&V(\scalar_1 \scalar_2 \scalar_3 \scalar_4)\times\left[
    \frac{\spbb{3|24|3}}{\ssix_{23}\ssix_{34}}\right]\,,\\
\end{aligned}\\
&\begin{aligned}
  \ATree_{\contact}&(1_{\scalar_1} 2_{\scalar_2} 3^+ 4^+ 5_{\scalar_3} 6_{\scalar_4})
  =\\&V(\scalar_1 \scalar_2 \scalar_3 \scalar_4)\times\left[
    \frac{\spbb{3|25|4}}{\ssix_{2 3} \ssix_{4 5}\spaa{3 4}}-
    \frac{\spbb{3 4}}{\spaa{3 4}} \left(\frac{m_2^2}{\ssix_{2 3} \ssix_{2 3 4}}+
    \frac{m_5^2}{\ssix_{3 4 5} \ssix_{4 5}}\right)\right]\,,\\
\end{aligned}\\
&\begin{aligned}
  \ATree_{\contact}&(1_{\scalar_1} 2_{\scalar_2} 3^+ 4_{\scalar_3} 5_{\scalar_4} 6^+)
  =\\&V(\scalar_1 \scalar_2 \scalar_3 \scalar_4)\times\left[
    \frac{\spbb{3|24|3}}{\ssix_{23}\ssix_{34}}\frac{\spbb{6|51|6}}{\ssix_{56}\ssix_{61}}\right]\,,\\
\end{aligned}\\
&\begin{aligned}
  \ATree_{\contact}&(1_{\scalar_1} 2^+ 3_{\scalar_2}  4^+ 5_{\scalar_3} 6_{\scalar_4})
  =\\&V(\scalar_1 \scalar_2 \scalar_3 \scalar_4)\times
    \left[\frac{m_{3}^2 \spbb{2 4}^2}{\ssix_{2 3} \ssix_{2 3 4} \ssix_{3 4}}
    +\frac{\spbb{2| 1 3 | 2} \spbb{4 | 3 5 | 4}}{\ssix_{1 2} \ssix_{2 3} \ssix_{3 4} \ssix_{4 5}}\right]\,.
\end{aligned}
\end{align}
\endgroup
As we have not specified the scalar flavors, we used the six-dimensional Mandelstam variables.
Once the $\phi_i$ are fixed, we can used the relation of \eq{eq:ssix_to_four_four_scalars} to find the corresponding 
four-dimensional replacement.

\section{Seven-Gluon Subleading Single-Trace Amplitude}
\label{SevenGluonSubleadingAppendix}
The expression for $\RTwo_{7:1\mathrm{B},2}$ in
eq.~\eqref{eq:R271B2_conjecture} is given in the style of the all-$n$
conjecture of ref.\cite{Dunbar:2020wdh}.  As this slightly obscures the final
form in terms of spinor products, we provide here also an explicit expression
after evaluating the $C_{rsij}$ and applications of the Schouten identity,
\begin{equation}
  \label{eq:R2_71B2_explicit}
  \begin{aligned}
    (-i)&\RTwo_{7:1\mathrm{B}2}=\frac{4}{\spaa{12}\spaa{23}\spaa{34}\spaa{45}\spaa{56}\spaa{67}\spaa{71}}\times\\
    &\Big[
    \frac{\spaa{1 2} \spaa{2 3} \spaa{4 5} }{\spaa{1 3} \spaa{2 4} \spaa{2 5}}\trF (1 3 2 5)+
    \frac{\spaa{1 2} \spaa{2 3} \spaa{4 6} }{\spaa{1 3} \spaa{2 4} \spaa{2 6}}\trF (1 3 2 6)+
    \frac{\spaa{1 2} \spaa{2 3} \spaa{4 7} }{\spaa{1 3} \spaa{2 4} \spaa{2
        7}}\trF (1 3 2 7)\\
    &+\frac{\spaa{1 2} \spaa{4 5} }{\spaa{1 4} \spaa{2 5}}\trF (1 4 2 5)+
    \frac{\spaa{1 2} \spaa{4 6} }{\spaa{1 4} \spaa{2 6}}\trF (1 4 2 6)+
    \frac{\spaa{1 2} \spaa{4 7} }{\spaa{1 4} \spaa{2 7}}\trF (1 4 2 7)\\
    &+\frac{\spaa{1 2} \spaa{3 4} \spaa{4 5} }{\spaa{1 4} \spaa{2 4} \spaa{3 5}}\trF (1 4 3 5)+
    \frac{\spaa{3 4} (\spaa{1 2} \spaa{3 6} \spaa{4 5}+\spaa{1 3} \spaa{2 4} \spaa{5 6}) }
    {\spaa{1 4} \spaa{2 4} \spaa{3 5} \spaa{3 6}}\trF(1 4 3 6)\\
    &+\frac{\spaa{3 4} (\spaa{1 2} \spaa{3 7} \spaa{4 5}+\spaa{1 3} \spaa{2 4} \spaa{5 7})}
    {\spaa{1 4} \spaa{2 4} \spaa{3 5} \spaa{3 7}}\trF (1 4 3 7)+
    \frac{\spaa{1 2} \spaa{5 6} }{\spaa{1 5} \spaa{2 6}}\trF (1 5 2 6)\\
    &+\frac{\spaa{1 2} \spaa{5 7} }{\spaa{1 5}\spaa{2 7}}\trF (1 5 2 7)+
    \frac{\spaa{1 3} \spaa{5 6} }{\spaa{1 5} \spaa{3 6}}\trF (1 5 3 6)+
    \frac{\spaa{1 3} \spaa{5 7} }{\spaa{1 5} \spaa{3 7}}\trF (1 5 3 7)\\
    &+\frac{\spaa{1 3} \spaa{4 5} \spaa{5 6} }{\spaa{1 5} \spaa{3 5} \spaa{4 6}}\trF (1 5 4 6)+
    \frac{\spaa{4 5} (\spaa{1 3} \spaa{4 7} \spaa{5 6}+\spaa{1 4} \spaa{3 5} \spaa{6 7})}
    {\spaa{1 5} \spaa{3 5} \spaa{4 6} \spaa{4 7}}\trF(1 5 4 7)\\
    &+\frac{\spaa{1 2} \spaa{6 7} }{\spaa{1 6} \spaa{2 7}}\trF (1 6 2 7)+
    \frac{\spaa{1 3} \spaa{6 7} }{\spaa{1 6} \spaa{3 7}}\trF (1 6 3 7)+
    \frac{\spaa{1 4} \spaa{6 7} }{\spaa{1 6}\spaa{4 7}}\trF (1 6 4 7)\\
    &+\frac{\spaa{1 4} \spaa{5 6} \spaa{6 7} }{\spaa{1 6} \spaa{4 6} \spaa{5 7}}\trF (1 6 5 7)+
    \frac{\spaa{2 3} \spaa{3 4} \spaa{5 6} }{\spaa{2 4} \spaa{3 5} \spaa{3 6}}\trF (2 4 3 6)+
    \frac{\spaa{2 3} \spaa{3 4} \spaa{5 7} }{\spaa{2 4} \spaa{3 5} \spaa{3 7}}\trF (2 4 3 7)\\
    &+\frac{\spaa{2 3} \spaa{5 6} }{\spaa{2 5} \spaa{3 6}}\trF (2 5 3 6)+
    \frac{\spaa{2 3} \spaa{5 7} }{\spaa{2 5} \spaa{3 7}}\trF(2 5 3 7)+
    \frac{\spaa{2 3} \spaa{4 5} \spaa{5 6} }{\spaa{2 5} \spaa{3 5} \spaa{4 6}}\trF (2 5 4 6)\\
    &+\frac{\spaa{4 5} (\spaa{2 3} \spaa{4 7} \spaa{5 6}+\spaa{2 4} \spaa{3 5} \spaa{6 7}) }
    {\spaa{2 5} \spaa{3 5} \spaa{4 6} \spaa{4 7}}\trF (2 5 4 7)+
    \frac{\spaa{2 3} \spaa{6 7} }{\spaa{2 6} \spaa{3 7}}\trF (2 6 3 7)\\
    &+\frac{\spaa{2 4} \spaa{6 7} }{\spaa{2 6} \spaa{4 7}}\trF (2 6 4 7)+
    \frac{\spaa{2 4} \spaa{5 6} \spaa{6 7} }{\spaa{2 6} \spaa{4 6} \spaa{5 7}}\trF (2 6 5 7)+
    \frac{\spaa{3 4} \spaa{4 5} \spaa{6 7} }{\spaa{3 5} \spaa{4 6} \spaa{4 7}}\trF (3 5 4 7)\\
    &+\frac{\spaa{3 4} \spaa{6 7} }{\spaa{3 6} \spaa{4 7}}\trF (3 6 4 7)+
    \frac{\spaa{3 4}\spaa{5 6} \spaa{6 7} }{\spaa{3 6} \spaa{4 6} \spaa{5 7}}\trF (3 6 5 7)\Big].
  \end{aligned}
\end{equation}

As discussed in section~\ref{subsec:7pt_R_71B_conjecture}, we used a
one-dimensional rational kinematic slice to verify the result of our
computation against the conjectured form of $\RTwo_{7:1\mathrm{B}}$ with
high accuracy.  The explicit spinors on this slice are,
\begin{equation}
  \begin{gathered}
    \left(
      \begin{array}{c}
        \spa{1}\\
        \spb{1}
      \end{array}
    \right)=\left(
      \begin{array}{c}
        \frac{439436}{7631} \\
        -38 \\
        -\frac{31698 \left(285029 \delta ^2-7419\right)}{1865571745 \delta ^2-8709906} \\
        \frac{44612 \left(285029 \delta ^2-7419\right)}{1865571745 \delta ^2-8709906} \\
      \end{array}
    \right),\quad
    \left(
      \begin{array}{c}
        \spa{2}\\
        \spb{2}
      \end{array}
    \right)=\left(
      \begin{array}{c}
        \frac{40}{13} \left(\frac{22991}{8689 \delta ^2+2348}+15\right) \\
        -50 \\
        \frac{\left(8689 \delta ^2+2348\right) \left(216723227 \delta ^2-704805\right)}{43549530-9327858725 \delta ^2}
        \\
        \frac{\left(8689 \delta ^2+2348\right) \left(264559501 \delta ^2-370950\right)}{18655717450 \delta ^2-87099060}
        \\
      \end{array}
    \right),\\
    \left(
      \begin{array}{c}
        \spa{3}\\
        \spb{3}
      \end{array}
    \right)=\left(
      \begin{array}{c}
        36 \\
        -39 \\
        \frac{10094}{39} \\
        -\frac{6161}{39} \\
      \end{array}
    \right),\quad
    \left(
      \begin{array}{c}
        \spa{4}\\
        \spb{4}
      \end{array}
    \right)=\left(
      \begin{array}{c}
        66 \\
        -97 \\
        -72 \\
        19 \\
      \end{array}
    \right),\quad
    \left(
      \begin{array}{c}
        \spa{5}\\
        \spb{5}
      \end{array}
    \right)=\left(
      \begin{array}{c}
        36 \\
        -52 \\
        -1 \\
        19 \\
      \end{array}
    \right),\\
    \left(
      \begin{array}{c}
        \spa{6}\\
        \spb{6}
      \end{array}
    \right)=\left(
      \begin{array}{c}
        -\frac{4}{5} (27 \delta -38) \\
        \frac{3}{5} (39 \delta -88) \\
        -\frac{2}{65} (5047 \delta +52) \\
        \frac{6161 \delta }{65}+\frac{84}{5} \\
      \end{array}
    \right),\quad
    \left(
      \begin{array}{c}
        \spa{7}\\
        \spb{7}
      \end{array}
    \right)=\left(
      \begin{array}{c}
        \frac{6}{5} (24 \delta +19) \\
        -\frac{6}{5} (26 \delta +33) \\
        \frac{40376 \delta }{195}-\frac{6}{5} \\
        \frac{63}{5}-\frac{24644 \delta }{195} \\
      \end{array}
    \right).
  \end{gathered}
\end{equation}

\section{Momentum Twistor Parametrization}
\label{MomentumTwistorAppendix}
In determining analytic expressions for $\RTwo_{5:1}$ and $\RTwo_{5:3}$ we
used parametrized kinematics based on its representation in momentum twistor
space.
Below we give a short summary of momentum twistors and one possible
choice of an $n$-momentum parametrization, which is the one used for the
results of section~\ref{subsec:five-gluon-amplitude-analytically}.

Starting with an ordered set of massless four-dimensional momenta
$(k_1,\mathellipsis,k_n)$ satisfying \(\sum k_i=0\), one can associate to the
\(k_i\) a set of
elements \(y_i\) in dual momentum space, satisfying the relation
\(y_i-y_{i-1}=k_{i}\).
The \(y_i\) are constructed explicitly as \(y_i=\sum_{j=0}^{i}k_j\), where \(k_0\)
can be chosen arbitrarily.
Since the momenta are massless we further have the property that
\((y_i-y_{i-1})^2=0\).
Ref.~\cite{Hodges:2009hk} details the construction of the twistor version of
this dual momentum space, called projective momentum-twistor space.
Momentum twistors are elements of \(\mathbb{CP}^{3}\), and for every momentum
\(k_i\) in our set we can find an associated momentum twistor with
homogeneous coordinates $Z^I_j=(\lambda_j^\alpha,\mu_j^{\dot{\alpha}})$.
Here $\lambda_i$ is the angle spinor of \(k_i\) defined as usual, while $\mu_i$ is a
spinor transforming in the conjugate $\text{SU}(2)$ representation.
The \(\lambda_i\) and \(\mu_i\) satisfy the incidence relation
\begin{equation}
  \label{eq:twistors_incidence}
  \mu_{i\dot{\alpha}}=\lambda_i^{\alpha} y_{i\alpha\dot{\alpha}}.
\end{equation}
Under the usual little group transformation \(\lambda\) and \(\mu\)
scale equally, as can be seen from the incidence relation.
This is consistent with the fact that the \(Z^I\) are homogeneous coordinates in
\(\mathbb{CP}^3\), and a total rescaling of \(Z^I\) describes the same point
in projective momentum-twistor space.

The bracket spinor \(\tilde{\lambda}_i^{\dot{\alpha}}\) 
corresponding to a momentum \(k_i\) can be
obtained via the relation~\cite{Hodges:2009hk},
\begin{equation}
  \label{eq:bracket_spinors_from_twistors}
  \tilde{\lambda}_i^{\dot{\alpha}}=\frac{\spaa{i(i+1)}\mu_{i-1}^{\dot{\alpha}}+
    \spaa{(i+1)(i-1)} \mu_i^{\dot{\alpha}}
    +\spaa{(i-1)i}\mu_{i+1}^{\dot{\alpha}}}
  {\spaa{i(i+1)}\spaa{(i-1)i}}.
\end{equation}
Based on this relation, we can see that 
scaling \(\lambda^\alpha\) and \(\mu^{\dot{\alpha}}\)
leads to the inverse behavior 
for \(\tilde{\lambda}^{\dot{\alpha}}\) as expected.

A convenient way of expressing the momentum twistors \(Z^I_i\) of an
\(n\)-momentum configuration is as a \((4\times N)\)-matrix,
\begin{equation}
  \label{eq:mom_twistor_matrix}
  Z=(Z_1^I Z_2^I\mathellipsis Z_{n-1}^I Z_n^I)=
  \begin{pmatrix}
    \lambda_1^{\alpha}&\lambda_2^{\alpha}&\mathellipsis&\lambda_{n-1}^{\alpha}&\lambda_n^{\alpha}\\
    \mu_1^{\dot{\alpha}}&\mu_2^{\dot{\alpha}}&\mathellipsis&\mu_{n-1}^{\dot{\alpha}}&\mu_n
    ^{\dot{\alpha}}
  \end{pmatrix}.
\end{equation}
In this representation all required spinor products can be obtained by computing
appropriate minors of the matrix \(Z\).

As masslessness and momentum conservation are manifest in kinematics
defined through momentum twistors, any choice of values for the entries of \(Z\)
represents a valid momentum configuration.
This makes them a useful tool for generating (complex)
numeric momentum configurations with
elements in the rational numbers \(\mathbb{Q}\) or a finite field
\(\mathbb{F}_{p}\).

We can use symmetries of our kinematics to reduce the number of independent
entries in \(Z\).
As mentioned before, the \(Z_i\) are homogeneous coordinates, allowing us to fix
one element in each column of \(Z\).
In choosing a frame, spatial rotations, Lorentz 
boosts and shifts fix
an additional \(10\) parameters.
The remaining \((3n-10)\) entries are free, and are sufficient to obtain any
valid configuration of momenta by choosing their values appropriately.
Momentum twistors therefore greatly simplify finding parametrizations of generic
massless kinematics, as any choice of fixing entries leads to a different valid
parametrization.

While spinors and Mandelstam invariants are linked by relations such as
momentum conservation or Schouten identities, the momentum twistor parameters
are independent of one another.
Thus, parameterized momentum twistors allow us to express integral coefficients and
rational parts of amplitudes as multivariate rational functions in the
parameters, while encapsulating the full analytic dependence on the kinematics.
In this representation, simplifications can be easier to carry out as there are no
additional relations, and the simplification of multivariate rational
expressions is a standard---though still challenging---problem.

When moving to momentum-twistor parameters we are removing the explicit
spinorial dependence, and therefore any phase information.
In fact, when we fix \((n+10)\) entries of \(Z\), we make a specific choice for
the phases of the spinors.
When evaluating expressions with non-zero phase-weight, such as helicity
amplitudes, we therefore have to introduce a normalization factor that
cancels the phase dependence, allowing us to restore it when necessary.
In the case of all-plus partial amplitudes a convenient choice of
normalization is assigning a Parke--Taylor factor to every color trace, so
that in general
\begin{multline}
  \TTr(1\mathellipsis i-1)
  \TTr({i}\mathellipsis {j-1})
  \TTr({j}\mathellipsis n)\ 
  \rightarrow \
  \frac{1}{\PTF{1\mathellipsis i-1}}
  \frac{1}{\PTF{i\mathellipsis j-1}}
  \frac{1}{\PTF{j\mathellipsis n}}\,,
\end{multline}
where $\PTF{\mathellipsis}$ was defined in eq.~\eqref{PTFdef}.
As we are usually interested in obtaining a result in terms of spinors
and Mandelstam invariants, we also need to find a solution to the
momentum-twistor parameters in terms of these objects.
Examples of parametrizations for four, five and six momenta together with such
solutions can be found in refs.~\cite{Budge:2020oyl, Badger:2013gxa}.
In the following we give a parametrization of \(n\)-momentum kinematics, that
extends the structure found in these examples.

Splitting the parameters into three sets
$a_p$, $b_q$ and $c_r$, with \(p\in\{1,\mathellipsis ,n-2\}\) and
\(q,r\in\{1,\mathellipsis ,n-4\}\), we choose \(Z\) as follows,
\begin{equation}
  \label{eq:mom_twistor_parametrization_npt}
  Z=
  \begin{pmatrix}
    1&0&y_1&y_2&y_3&\mathellipsis&y_{n-3}&y_{n-2}\\
    0&1&1&1&1&\mathellipsis&1&1\\
    0&0&0&\frac{b_{n-4}}{a_2}&\tilde{b}_{n-5}&\mathellipsis&\tilde{b}_1&1\\
    0&0&1&1&\tilde{c}_{n-4}&\mathellipsis&\tilde{c}_{2}&\tilde{c}_{1}
  \end{pmatrix}.
\end{equation}
The \(y_k\), \(\tilde{b}_k\) and \(\tilde{c}_k\) are defined recursively to be,
\begin{equation}
  \label{eq:npt_mom_twistor_recursive_defs}
  \begin{aligned}
    y_k&=\sum_{i=1}^{k}\prod_{j=1}^{i}\frac{1}{a_j},\\
    \tilde{b}_k&=\tilde{b}_{k-1}+a_{n-k}(\tilde{b}_{k-1}-\tilde{b}_{k-2})+b_k,\\
    \tilde{c}_k&=\tilde{c}_{k-1}+a_{n-k+1}(\tilde{c}_{k-1}-\tilde{c}_{k-2})+
    \frac{b_{k-1}}{b_{n-4}}(c_k-1),\\
  \end{aligned}.
\end{equation}
with $a_{k>(n-2)}=\tilde{b}_{k<0}=\tilde{c}_{k\le 0}=0$ and
$b_{0}=\tilde{b}_{0}=1$.
The associated solution is then
\begin{equation}
  \label{eq:npt_mom_twistor_solution}
  \begin{aligned}
    a_1&=s_{12}&\\
    a_{k>1}&=-\frac{\spaa{k,k+1}\spaa{k+2,1}}{\spaa{1,k}\spaa{k+1,k+2}},\\
    b_{n-4}&=\frac{s_{23}}{s_{12}},\\
    b_k&=\frac{\spab{n-k|n-k+1|2}}{\spab{n-k|1|2}},\\
    c_k&=-\frac{\spab{1|3|n-k+2}}{\spab{1|2|n-k+2}},
  \end{aligned}
\end{equation}
where the spinor labels have to be taken $\mathrm{mod}\ n$ in the solution of $c_k$.

\section{One-Loop Coefficients from \texorpdfstring{$D$}{D}-Dimensional Generalized Unitarity}
\label{GeneralizedUnitarityAppendix}
Here we provide a brief summary of
the one-loop generalized unitarity techniques of
refs.~\cite{Forde:2007mi,Badger:2008cm} that we used in the
computation of the rational parts.
We follow the notation of ref.~\cite{Badger:2008cm}
closely.
\subsection{Boxes}\label{subsec:one-loop_boxes}
\begin{figure}
  \centering
  \includegraphics[width=0.3\textwidth]{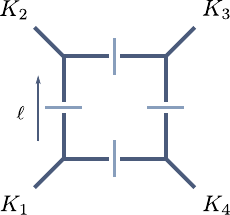}
  \caption{A generic box cut. The momenta $K_1,K_2,K_3,K_{4}$ are assumed to
    be massive. The loop momentum $\ell$ is defined to be flowing from the
    $K_4$ and into the $K_1$ corner.}
  \label{fig:box_cut}
\end{figure}

\FloatBarrier
We start with the coefficient of a generic box integral with external masses
$K_1$, $K_2$, $K_3$, $K_4$.
We define the loop momentum $\ell$ to flow from $K_1$ to $K_2$.
This configuration is shown in \fig{fig:box_cut}.

We define the flattened momenta $K_1^\flat$, $K_2^\flat$ as
\begin{equation}
  K_1^\flat=\frac{\gamma^2 K_1-\gamma K_1^2 K_2}{\gamma^2-K_1^2 K_2^2}, \quad
  K_2^\flat=\frac{\gamma^2 K_4-\gamma K_2^2 K_1}{\gamma^2-K_1^2 K_2^2}\,.
\end{equation}
Here, $\gamma$ is defined by requiring $K_1^\flat$ and $K_2^\flat$ to be
massless,
\begin{align}
  (K_1^\flat)^2&=\frac{\gamma^4 K_1^2+\gamma^2(K_1^2)^2 K_2^2-2\gamma^3
                 K_1^2 (K_1\cdot K_2)}{(\gamma^2-K_1^2 K_2^2)^2}=0\\
               &\Leftrightarrow\gamma^2 K_1^2+K_1^2 K_2^2-2\gamma(K_1\cdot K_2)=0\\
               &\Rightarrow\gamma=(K_1\cdot K_2)\pm \sqrt{(K_1\cdot K_2)^2-K_1^2 K_2^2}.
\end{align}
In case that either $K_1^2=0\ $ or $ K_2^2=0$, the sign is chosen
such that $\gamma$ is non-zero.
Additionally, we have
\begin{equation}
  \gamma=(K_1^\flat+K_2^\flat)^2=2(K_1^\flat\cdot K_2^\flat)
\end{equation}
and we can re-express $K_1$, $K_2$ in terms of the flattened momenta via
\begin{equation}
  \label{eq:D-dim_1L_KfromKflat_box}
  K_1=K_1^\flat+\frac{K_1^2}{\gamma}K_2^\flat,\quad
  K_2=K_2^\flat+\frac{K_2^2}{\gamma}K_1^\flat.
\end{equation}
We use the following 
representation~\cite{Forde:2007mi,Badger:2008cm},
\begin{equation}
  \label{eq:BoxLoopMomentumAnsatz}
  \overline{\ell}^\mu=cK_2^{\flat\mu}+dK_1^{\flat\mu}+\frac{1}{2}
  \left(t\spab{K_2^\flat|\gamma^\mu|K_1^\flat}+b\spab{K_1^\flat|\gamma^\mu|K_2^\flat}\right).
\end{equation}
The on-shell condition $\ell^2=0$ translates into
\begin{align*}
  b=\frac{cd\gamma-\mu^2}{t\gamma}.
\end{align*}
Furthermore, the on-shell conditions
$(\ell-K_2)^2=0$ and $(\ell+K_1)^2=0$ fix the parameters $c$
and $d $ to be
\begin{equation}
  c=-\frac{K_1^2(\gamma+K_2^2)}{\gamma^2-K_2^2 K_1^2},\quad
  d=\frac{K_2^2(\gamma+K_1^2)}{\gamma^2-K_2^2 K_1^2}.
\end{equation}
The last on-shell condition $(\ell-K_2-K_3)^2=0$ provides a quadratic
relation for $t$, with solutions,%
\begin{align}
  \label{eq:box_ell_t_solution}
  t^\pm=\frac{\Delta\pm\sqrt{\Delta^2-4\frac{cd\gamma-\mu^2}{\gamma}
  \trM(K_2^\flat K_3 K_1^\flat K_3)}}{2\spab{K_2^\flat|K_3|K_1^\flat}},\quad
  \Delta=-(2K_3\cdot(K_2+cK_2^\flat+dK_1^\flat)+K_3^2)
\end{align}
For rational terms we only require the leading coefficient in the large
$\mu^2$ expansion.  In computations it is therefore sufficient to use the
leading behavior of the parameterized loop momentum.  Expanding the solutions
$t^{\pm}$ of eq.~\eqref{eq:box_ell_t_solution} around large values of
$\mu^2$, we obtain
\begin{equation}
  t^\pm=\pm
  \sqrt{\frac{\mu^2}{\gamma}}\frac{\spab{K_1^\flat|K_3|K_2^\flat}}{\sqrt{\spab{K_2^\flat|K_3
        K_1^\flat K_3|K_2^\flat}}}+\mathcal{O}\left[\left(\tfrac{1}{\sqrt{\mu^2}}\right)^0\right]
\end{equation}

In the normalization of amplitudes and Feynman integrals of
eqs.\eqref{eq:amplitude_normalization}~and~\eqref{eq:integral_def},
the box $\mu^4$ coefficient is given by
\begin{equation}
  \label{eq:CBox4}
  \CBox_{,[4]}=\frac{1}{2}\sum_{t=t^\pm}\Inf_{\mu^2}\left.\left[A_1^{(0)}A_2^{(0)}A_3^{(0)
      }A_4^{(0)
      }\right]\right\rvert_{\mu^4}
\end{equation}
The operation $\Inf_x[f(x)]$ is defined using the expansion of
$f(x)$ for large values of $x$ as,
\begin{equation}
  \label{eq:inf_residue_decomposition}
  f(x)=\Inf_x[f(x)]+\mathcal{O}\left[\left(\tfrac{1}{x}\right)\right].
\end{equation}
For the following discussion we also define
\begin{equation}
  \label{eq:consecutive_Inf}
  \Inf_{x_1 x_2 \mathellipsis x_n}[f(x)]\equiv\Inf_{x_1}\circ\Inf_{x_2}\circ
  \mathellipsis\circ \Inf_{x_n}[f(x)].
\end{equation}

\subsection{Triangles}\label{subsec:triangles}
\begin{figure}
  \centering
  \includegraphics[width=0.3\textwidth]{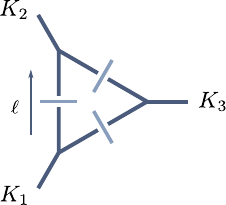}
  \caption{A generic triangle cut. The momenta $K_1,K_2,K_3$ are generally
    massive. The loop momentum $\ell$ is defined to flow out of the $K_3$
    and into the $K_1$ corner.}
  \label{fig:four-d_triangle_cut}
\end{figure}
We now turn to the computation of triangle-integral coefficients.
We again choose a generic representative with all external
momenta $K_1,K_2,K_3$ being massive, with $\ell$ flowing from the $K_1$
to the $K_2$ vertex.
This configuration is shown in Figure~\ref{fig:four-d_triangle_cut}.

We reuse the parametrization of the box loop momentum, relaxing the
on-shell condition that fixes the $t$ parameter.
We again define the massless momenta associated to $K_1$ and
$K_2$,
\begin{equation}
  K_1^\flat=\frac{\gamma^2 K_1-\gamma K_1^2 K_2}{\gamma^2-K_1^2 K_2^2}, \quad
  K_2^\flat=\frac{\gamma^2 K_2-\gamma K_2^2 K_1}{\gamma^2-K_1^2 K_2^2},
\end{equation}
where
\begin{align*}
  \gamma=(K_1\cdot K_2)\pm \sqrt{(K_1\cdot K_2)^2 - K_1^2 K_2^2}.
\end{align*}
Again, when either $K_1^2$ or $K_2^2$ vanishes, we choose the sign
in front of the square-root such that $\gamma$ is non-zero.
The three on-shell conditions
$\ell^2=0$, $(\ell-K_2)^2=0$ and $(\ell+K_1)^2=0$ are then exactly
those of the box case, and are fulfilled by
\begin{equation}
  \label{eq:D-dim_triangle_loop_momentum_parametrization}
  \ell^\mu(t)=cK_2^{\flat\mu}+dK_1^{\flat\mu}+
  \frac{1}{2}\left(t\spab{K_2^\flat|\gamma^\mu|K_1^\flat}+
    \frac{cd\gamma-\mu^2}{t\gamma}\spab{K_1^\flat|\gamma^\mu|K_2^\flat}\right)
\end{equation}
with
\begin{equation}
  c=-\frac{K_1^2(\gamma+K_2^2)}{\gamma^2-K_2^2 K_1^2},\quad
  d=\frac{K_2^2(\gamma+K_1^2)}{\gamma^2-K_2^2 K_1^2}.
\end{equation}
Unlike the case of the box loop momentum, 
$t$ is now a remaining degree of
freedom that is integrated over.

As in the box case, the on-shell conditions yield
two independent solutions.
When neither $K_1^2$ nor $K_2^2$ vanishes these are
given by the two different solutions for $\gamma$.
Should one or both of
the $K_i^2$ vanish, we have to choose the sign in $\gamma$ such that
it is non-zero.
In this case, the second solution can be obtained from the one in
eq.~\eqref{eq:D-dim_triangle_loop_momentum_parametrization} via
spinor conjugation,
\begin{equation}
  \label{eq:D-dim_triangle_loop_momentum_parametrization_cc}
  \ell^{\ast\mu}(t)=cK_2^{\flat\mu}+dK_1^{\flat\mu}+\frac{1}{2}
  \left(t\spba{K_2^\flat|\gamma^\mu|K_1^\flat}+
    \frac{cd\gamma-\mu^2}{t\gamma}\spba{K_1^\flat|\gamma^\mu|K_2^\flat}\right)
\end{equation}
The fact that we have two solutions for the triangle loop momentum
does not mean that imposing an additional on-shell condition for the box
parametrization would lead to four solutions.
Instead, the resulting solutions for $\ell$ would be degenerate, we
would still only end up with two distinct
solutions we found in the discussion of box cuts.

In the parameterization of
eq.~\eqref{eq:D-dim_triangle_loop_momentum_parametrization} the remaining
integrals over positive powers of $t$ vanish, so
that we require only the $t^0$ contributions.
The coefficient of a triangle integral
with a $\mu^2$ numerator insertion is then~\cite{Forde:2007mi,Badger:2008cm}
\begin{equation}
  \label{eq:D-dim_triangle_coefficient}
  \CTriangle_{,[2]}=\frac{1}{2}\sum_{\ellb=\ellb(t),\ellb^{\ast}(t)}
  \Inf_{\mu^2,t}\left.\left[A_1^{(0)}(\ell)A_2^{(0)}(\ell)A_3^{(0)}(\ell)
    \right]\right\rvert_{t^0,\mu^2}.
\end{equation}

\subsection{Bubble}\label{subsec:bubbles} 
\begin{figure}
  \centering
  \includegraphics[width=0.3\textwidth]{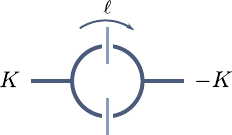}
  \caption{A generic bubble cut.  Assuming the massive momentum $K$ to be
    outgoing, we define the loop momentum $\ell$ to be flowing from the
    $K$ and into the $-K$ vertex.  }
  \label{fig:bubble_cut}
\end{figure}
Finally we explain how to determine the coefficients of 
bubble integrals.  In this case we have only one
external massive momentum $K$, and we define the loop momentum $\ell$ as
shown in Figure~\ref{fig:bubble_cut}.

As in the triangle case, the derivation of the bubble coefficient will depend
on the specific loop momentum parametrization we use to make the on-shell
conditions manifest.  We would like to choose a parametrization similar to the
box and triangle case.  However, in this case we cannot define such a
parametrization entirely in terms of the external kinematics, as we have only
a single momentum $K$.  We therefore need to introduce an arbitrary
reference momentum $\chi$.  Choosing $\chi$ to be lightlike, we define the
flattened version of $K$ to be
\begin{equation}
  K^\flat=K-\frac{K^2}{\gamma}\chi,\quad \text{with}\quad \gamma=2(K \cdot
  \chi)=2(K^\flat \cdot \chi).
\end{equation}
Similarly to the box and triangle cases, we 
use the following representation for the
four-dimensional part of the loop 
momentum~$\ell$~\cite{Forde:2007mi,Badger:2008cm},
\begin{equation}
  \ellb^{\mu}=-K^\mu+y K^{\flat\mu}+d\chi^\mu+\frac{1}{2}
  \left(t\spab{K^\flat|\gamma^\mu|\chi}+b\spab{\chi|\gamma^\mu|K^\flat}\right).
\end{equation}
The on-shell conditions $\ell^2=0$ and $(\ell+K)^2=0$ then fix the
parameters $b$ and $d$ to be,
\begin{equation}
  d=\frac{K^2(1-y)}{\gamma},\hspace{2em} b=\frac{\gamma d y-\mu^2}{ty}
  =\frac{y(1-y)K^2-\mu^2}{t\gamma},
\end{equation}
so that
\begin{equation}
  \label{eq:dim-D_bubble_loop_parametrization}
  \ellb^{\mu}(t,y)=(y-1) K^{\flat\mu}-y\frac{K^2}{\gamma}\chi^\mu+
  \frac{1}{2}\left(t\spab{K^\flat|\gamma^\mu|\chi}+
    \frac{y(1-y)K^2-\mu^2}{t\gamma}\spab{\chi|\gamma^\mu|K^\flat}\right)\,.
\end{equation}

In contrast to the box and triangle coefficients, there are 
two sources
of contributions to the bubble coefficient.  
One is the bubble cut
itself, where two tree amplitudes are evaluated using the parametrization of
eq.~\eqref{eq:dim-D_bubble_loop_parametrization}.  We are left with integrals
over the two remaining parameters $t$ and $y$.  In the chosen
parametrization the integrals over powers of $t$ vanish, 
so that we require only
the parts of the cut of order $t^0$.  The integrals over positive
powers of $y$ do not vanish.  Power counting 
in Yang--Mills theory limits the maximal power of
$y$ to two.  Carrying out the integrals
amounts to expanding the cut for large values of $y$, and replacing $y^0$,
$y^1$ and $y^2$ with~\cite{Kilgore:2007qr,Badger:2008cm}
\begin{equation}
  Y_0=1,\qquad
  Y_1=-\frac{1}{2},\qquad
  Y_2=\frac{1}{3}\left(1-\frac{\mu^2}{s}\right).
\end{equation}

\begin{figure}
  \centering
  \includegraphics[width=0.3\textwidth]{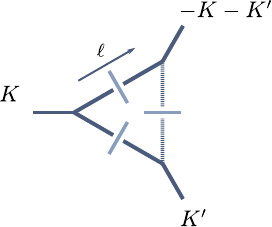}
  \caption{%
    An example of a triangle cut contributing to the bubble
    coefficient.  The momentum $K$ and the solid lines are those belonging to
    the original bubble.  The momentum $K'$ and dashed lines are those of the
    additional on-shell condition.  }
  \label{fig:bubble_triangle_cut}
\end{figure}

The second source of contributions are tensor triangles, which after
Passarino--Veltman reduction also contain terms belonging to the bubble basis
integrals.  Only triangle cuts which have two cuts in common with the bubble
cut will contribute.  As an example, consider the integral in
Figure~\ref{fig:bubble_triangle_cut}.  In addition to the two bubble cut, a
third propagator $(\ell+K+K')^2$ has been cut.  To determine the contribution
to the bubble coefficient we need to use the same parameterization as for the
bubble cut.  The additional on-shell condition leads to two solutions for
$y$~\cite{Kilgore:2007qr,Badger:2008cm}
\begin{equation}
  \begin{aligned}
    y^{\pm}&=\frac{c_1\pm \sqrt{c_1^2+4 c_0 c_2}}{2 c_2}
  \end{aligned}
\end{equation}
where,             %
\begin{equation}
  \begin{aligned}
    c_0 &=  t \left[\gamma(K^{\prime})^2 + 2(K'\cdot \chi) K^2\right]+t^2 \gamma
    \spab{K^\flat|K'|\chi}-\mu^2\spab{\chi|K'|K^\flat}\,,\\
    c_1 & = K^2 \spab{\chi|K'|K^\flat}+t\left[\gamma 2(K'\cdot K^\flat)- K^2 2(K'\cdot
      \chi)\right]\,,\\
    c_2 & = K^2 \spab{\chi|K'|K^\flat}\,.
  \end{aligned}
\end{equation}
Evaluating the triangle cut at $\ellb(t,y^\pm)$, we are left with in
integrals over powers of the remaining parameter $t$.  Only integrals with
$t^{i>0}$ are relevant for the bubble coefficient, as only these are related
to tensor integrals.  Power counting in Yang--Mills
again limits the maximal power which can
appear, this time to $i\le 3$~\cite{Badger:2008cm}.  In contrast to
computation of the triangle coefficient, these integrals no longer vanish, as
their vanishing was a special feature of the triangle parameterization of
eq.~\eqref{eq:D-dim_triangle_loop_momentum_parametrization}.  Additionally, we
do not require the full integral over the $t^i$, but only the bubble
contributions after Passarino--Veltman reduction.  These were determined in
ref.~\cite{Kilgore:2007qr}.  To obtain the contribution of the tensor triangle
to the bubble coefficient, we can follow a simple recipe: we expand the
triangle cut for large values of $t$, remove the constant term, 
and replace
$t$, $t^2$ and $t^3$ with~\cite{Badger:2008cm,Kilgore:2007qr},
\begin{equation}
  \begin{aligned}
    T_1&=-\frac{K^2\spab{\chi | K' | K^\flat}}{2 \gamma \left[(K\cdot K')^2-K^2
        K^{\prime 2}\right]}\,,\\ 
        T_2&=\frac{3 (K^2)^2 \spab{\chi | K' |
        K^\flat}^2}{8 \gamma ^2 \left[(K\cdot K')^2-(K^2) K^2\right]^2}
    \left[(K\cdot K')+K^{\prime 2}\right]\,,\\ 
    T_3&=-\frac{(K^2)^3 \spab{\chi | K' |
        K^\flat}^3}{48 \gamma ^3 \left[(K\cdot K') ^2-K^2 K^{\prime
          2}\right]^3}\\ &\hphantom{=}\times\left((K\cdot K')^2\left(11+16
        \frac{\mu ^2}{K^2}\right)+30 (K\cdot K') K^{\prime 2}+ K^{\prime 2} \left(4
        K^2+15 K^{\prime 2}-16 \mu ^2\right)\right)\,.
  \end{aligned}
  \label{eq:bubble_triangle_integrals}
\end{equation}

Summing over all triangle cuts compatible with the bubble cut,
we obtain the full bubble coefficient via,
\begin{equation}
\label{eq:bubble_coefficient_total}
  \begin{aligned}
    \CBubble_{,[2]}&= \Inf_{\mu^2,t,y}\left.\left[
        A_1^{(0)}A_2^{(0)}\right]\right\rvert_{\mu^2,t^0,y^i\to Y_i}
    +\frac{1}{2}\sum_{\substack{\text{triangle}\\ \text{cuts}}}
    \sum_{y^\pm}\Inf_{\mu^2,t}\left.\left[
        A_1^{(0)}A_2^{(0)}A_3^{(0)}\right]\right\rvert_{\mu^2,t^i\to T_i}.
  \end{aligned}
\end{equation}
\FloatBarrier


\begin{thebibliography}{99}

\bibitem{ManganoParkeReview}
  M.~L.~Mangano and S.~J.~Parke,
  Phys. Rept. \textbf{200} (1991), 301-367
  doi:10.1016/0370-1573(91)90091-Y
  [hep-th/0509223].

\bibitem{Ellis:1985er}
  R.~K.~Ellis and J.~C.~Sexton,
  Nucl. Phys. B \textbf{269}, 445-484 (1986)
  doi:10.1016/0550-3213(86)90232-4

\bibitem{Bern:1993mq}
  Z.~Bern, L.~J.~Dixon and D.~A.~Kosower,
  Phys. Rev. Lett. \textbf{70}, 2677-2680 (1993)
  doi:10.1103/PhysRevLett.70.2677
  [arXiv:hep-ph/9302280 [hep-ph]].

\bibitem{Bern:1993qk}
  Z.~Bern, G.~Chalmers, L.~J.~Dixon and D.~A.~Kosower,
  Phys. Rev. Lett. \textbf{72}, 2134-2137 (1994)
  doi:10.1103/PhysRevLett.72.2134
  [arXiv:hep-ph/9312333 [hep-ph]].

\bibitem{Mahlon:1992fs}
  G.~Mahlon, T.~M.~Yan and C.~Dunn,
  Phys. Rev. D \textbf{48}, 1337-1374 (1993)
  doi:10.1103/PhysRevD.48.1337
  [arXiv:hep-ph/9210212 [hep-ph]].

\bibitem{Mahlon:1993si}
  G.~Mahlon,
  doi:10.1103/PhysRevD.49.4438
  [arXiv:hep-ph/9312276 [hep-ph]].

\bibitem{Bern:1996ja}
  Z.~Bern, L.~J.~Dixon, D.~C.~Dunbar and D.~A.~Kosower,
  Phys.
  Lett.
  B \textbf{394}, 105-115 (1997)
  [arXiv:hep-th/9611127 [hep-th]].

\bibitem{Britto:2020crg}
R.~Britto, G.~R.~Jehu and A.~Orta,
JHEP \textbf{04}, 276 (2021)
doi:10.1007/JHEP04(2021)276
[arXiv:2011.13821 [hep-th]].

\bibitem{AllPlusAndWilsonLoops}
D.~Chicherin and J.~M.~Henn,
[arXiv:2202.05596 [hep-th]];
D.~Chicherin and J.~Henn,
[arXiv:2204.00329 [hep-th]].

\bibitem{Bern:2000dn}
  Z.~Bern, L.~J.~Dixon and D.~A.~Kosower,
  JHEP \textbf{01}, 027 (2000)
  [arXiv:hep-ph/0001001 [hep-ph]].

\bibitem{Badger:2013gxa}
  S.~Badger, H.~Frellesvig and Y.~Zhang,
  JHEP \textbf{12}, 045 (2013)
  [arXiv:1310.1051 [hep-ph]].

\bibitem{TwoLoopAllPlus5ptFull}
S.~Badger, G.~Mogull, A.~Ochirov and D.~O'Connell,
JHEP \textbf{10}, 064 (2015)
doi:10.1007/JHEP10(2015)064
[arXiv:1507.08797 [hep-ph]].

\bibitem{Gehrmann:2015bfy}
  T.~Gehrmann, J.~M.~Henn and N.~A.~Lo Presti,
  Phys. Rev. Lett. \textbf{116}, no.6, 062001 (2016)
  [erratum: Phys. Rev. Lett. \textbf{116}, no.18, 189903 (2016)]
  doi:10.1103/PhysRevLett.116.062001
  [arXiv:1511.05409 [hep-ph]].

\bibitem{Abreu:2017hqn}
S.~Abreu, F.~Febres Cordero, H.~Ita, B.~Page and M.~Zeng,
Phys. Rev. D \textbf{97}, no.11, 116014 (2018)
doi:10.1103/PhysRevD.97.116014
[arXiv:1712.03946 [hep-ph]].

\bibitem{Badger:2019djh}
S.~Badger, D.~Chicherin, T.~Gehrmann, G.~Heinrich, J.~M.~Henn, T.~Peraro, P.~Wasser, Y.~Zhang and S.~Zoia,
Phys. Rev. Lett. \textbf{123}, no.7, 071601 (2019)
doi:10.1103/PhysRevLett.123.071601
[arXiv:1905.03733 [hep-ph]].

\bibitem{Jin:2019nya}
Q.~Jin and H.~Luo,
[arXiv:1910.05889 [hep-ph]].

\bibitem{Caola:2021izf}
F.~Caola, A.~Chakraborty, G.~Gambuti, A.~von Manteuffel and L.~Tancredi,
Phys. Rev. Lett. \textbf{128}, no.21, 212001 (2022)
doi:10.1103/PhysRevLett.128.212001
[arXiv:2112.11097 [hep-ph]].

\bibitem{TwoLoopAllPlusPolylogs}
  D.~C.~Dunbar, G.~R.~Jehu and W.~B.~Perkins,
  Phys. Rev. D \textbf{93}, no.12, 125006 (2016)
  doi:10.1103/PhysRevD.93.125006
  [arXiv:1604.06631 [hep-th]].

\bibitem{Dunbar:2016gjb}
  D.~C.~Dunbar, G.~R.~Jehu and W.~B.~Perkins,
  Phys. Rev. Lett. \textbf{117}, no.6, 061602 (2016)
  doi:10.1103/PhysRevLett.117.061602
  [arXiv:1605.06351 [hep-th]].

\bibitem{Dunbar:2016aux}
  D.~C.~Dunbar and W.~B.~Perkins,
  Phys. Rev. D \textbf{93}, no.8, 085029 (2016)
  doi:10.1103/PhysRevD.93.085029
  [arXiv:1603.07514 [hep-th]].

\bibitem{Dunbar:2016cxp}
  D.~C.~Dunbar, G.~R.~Jehu and W.~B.~Perkins,
  Phys. Rev. D \textbf{93} (2016) no.12, 125006
  doi:10.1103/PhysRevD.93.125006
  [arXiv:1604.06631 [hep-th]].

\bibitem{Dunbar:2017nfy}
  D.~C.~Dunbar, J.~H.~Godwin, G.~R.~Jehu and W.~B.~Perkins,
  Phys. Rev. D \textbf{96}, no.11, 116013 (2017)
  doi:10.1103/PhysRevD.96.116013
  [arXiv:1710.10071 [hep-th]].

\bibitem{Dunbar:2017azf}
  D.~C.~Dunbar, J.~H.~Godwin, G.~R.~Jehu and W.~B.~Perkins,
  PoS \textbf{RADCOR2017}, 026 (2017)
  doi:10.22323/1.290.0026
  [arXiv:1712.05312 [hep-ph]].

\bibitem{Dalgleish:2020mof}
  A.~R.~Dalgleish, D.~C.~Dunbar, W.~B.~Perkins and J.~M.~W.~Strong,
  Phys. Rev. D \textbf{101}, no.7, 076024 (2020)
  doi:10.1103/PhysRevD.101.076024
  [arXiv:2003.00897 [hep-ph]].

\bibitem{Dunbar:2019fcq}
  D.~C.~Dunbar, J.~H.~Godwin, W.~B.~Perkins and J.~M.~W.~Strong,
  Phys. Rev. D \textbf{101}, no.1, 016009 (2020)
  doi:10.1103/PhysRevD.101.016009
  [arXiv:1911.06547 [hep-ph]].

\bibitem{Badger:2016ozq}
  S.~Badger, G.~Mogull and T.~Peraro,
  JHEP \textbf{08}, 063 (2016)
  [arXiv:1606.02244 [hep-ph]].

\bibitem{TwoLoopProjectors}
  D.~A.~Kosower and K.~J.~Larsen,
  Phys. Rev. D \textbf{85}:045017 (2012)
  doi:10.1103/PhysRevD.85.045017
  [arXiv:1108.1180].

\bibitem{BCF}
  R.~Britto, F.~Cachazo and B.~Feng,
  Nucl. Phys. B \textbf{725} (2005), 275-305
  doi:10.1016/j.nuclphysb.2005.07.014
  [arXiv:hep-th/0412103 [hep-th]].

\bibitem{Forde:2007mi}
  D.~Forde,
  Phys.
  Rev.
  D \textbf{75}, 125019 (2007)
  doi:10.1103/PhysRevD.75.125019
  [arXiv:0704.1835 [hep-ph]].

\bibitem{Badger:2008cm}
  S.~D.~Badger,
  JHEP \textbf{01}, 049 (2009)
  [arXiv:0806.4600 [hep-ph]].

\bibitem{Ossola:2006us}
  G.~Ossola, C.~G.~Papadopoulos and R.~Pittau,
  Nucl.
  Phys.
  B \textbf{763}, 147-169 (2007)
  doi:10.1016/j.nuclphysb.2006.11.012
  [arXiv:hep-ph/0609007 [hep-ph]].

\bibitem{Dunbar:2016}
  D.~C.~Dunbar and W.~B.~Perkins,
  Phys.\ Rev.\ D \textbf{93}:085029 (2016)
  [arXiv:1603.07514];\\
  D.~C.~Dunbar, G.~R.~Jehu and W.~B.~Perkins,
  Phys.\ Rev.\ Lett. \textbf{117}:061602 (2016)
  [arXiv:1605.06351];\\
  D.~C.~Dunbar, J.~H.~Godwin, G.~R.~Jehu and W.~B.~Perkins,
  Phys.\ Rev.\ D \textbf{96}:116013 (2017)
  doi:10.1103/PhysRevD.96.116013
  [arXiv:1710.10071].

\bibitem{Dunbar:2020wdh}
  D.~C.~Dunbar, W.~B.~Perkins and J.~M.~W.~Strong,
  Phys. Rev. D \textbf{101}, no.7, 076001 (2020)
  doi:10.1103/PhysRevD.101.076001
  [arXiv:2001.11347 [hep-ph]].

\bibitem{DalgleishDunbar:2020}
  D.~C.~Dunbar, J.~H.~Godwin, W.~B.~Perkins and J.~M.~W.~Strong,
  Phys.\ Rev.\ D \textbf{101}:016009 (2020)
  [arXiv:1911.06547];\\
  A.~R.~Dalgleish, D.~C.~Dunbar, W.~B.~Perkins and J.~M.~W.~Strong,
  Phys.\ Rev.\ D \textbf{101}:076024 (2020)
  [arXiv:2003.00897].

\bibitem{BCFWrecursion}
R.~Britto, F.~Cachazo, B.~Feng and E.~Witten,
Phys. Rev. Lett. \textbf{94}, 181602 (2005)
doi:10.1103/PhysRevLett.94.181602
[arXiv:hep-th/0501052 [hep-th]].

\bibitem{Bern:1990ux}
Z.~Bern and D.~A.~Kosower,
Nucl. Phys. B \textbf{362}, 389-448 (1991)
doi:10.1016/0550-3213(91)90567-H

\bibitem{Giele:1991vf}
  W.~T.~Giele and E.~W.~N.~Glover,
  Phys. Rev. D \textbf{46}, 1980-2010 (1992)
  \url{doi:10.1103/PhysRevD.46.1980}

\bibitem{BDDK}
Z.~Bern, L.~J.~Dixon, D.~C.~Dunbar and D.~A.~Kosower,
Nucl. Phys. B \textbf{425}, 217-260 (1994)
doi:10.1016/0550-3213(94)90179-1
[arXiv:hep-ph/9403226 [hep-ph]];\\
Z.~Bern, L.~J.~Dixon, D.~C.~Dunbar and D.~A.~Kosower,
Nucl. Phys. B \textbf{435}, 59-101 (1995)
doi:10.1016/0550-3213(94)00488-Z
[arXiv:hep-ph/9409265 [hep-ph]].

\bibitem{Kunszt:1994np}
  Z.~Kunszt, A.~Signer and Z.~Trocsanyi,
  Nucl. Phys. B \textbf{420}, 550-564 (1994)
  doi:10.1016/0550-3213(94)90077-9
  [arXiv:hep-ph/9401294 [hep-ph]].

\bibitem{Catani:1996vz}
  S.~Catani and M.~H.~Seymour,
  Nucl. Phys. B \textbf{485}, 291-419 (1997)
  [erratum: Nucl. Phys. B \textbf{510}, 503-504 (1998)]
  doi:10.1016/S0550-3213(96)00589-5
  [arXiv:hep-ph/9605323 [hep-ph]].

\bibitem{Cutkosky:1960sp}
  R.~E.~Cutkosky,
  J. Math. Phys. \textbf{1}, 429-433 (1960)
  doi:10.1063/1.1703676

\bibitem{Glover:2001af}
  E.~W.~N.~Glover, C.~Oleari and M.~E.~Tejeda-Yeomans,
  Nucl.\ Phys\  B \textbf{605}, 467-485 (2001)
  doi:10.1016/S0550-3213(01)00210-3
  [arXiv:hep-ph/0102201 [hep-ph]].

\bibitem{Bern:2002tk}
  Z.~Bern, A.~De Freitas and L.~J.~Dixon,
  JHEP \textbf{03}, 018 (2002)
  \url{doi:10.1088/1126-6708/2002/03/018}
  [arXiv:hep-ph/0201161 [hep-ph]].

\bibitem{Badger:2015lda}
  S.~Badger, G.~Mogull, A.~Ochirov and D.~O'Connell,
  JHEP \textbf{10}, 064 (2015)
  [arXiv:1507.08797 [hep-ph]].

\bibitem{Catani:1998bh}
  S.~Catani,
  Phys. Lett. B \textbf{427}, 161-171 (1998)
  doi:10.1016/S0370-2693(98)00332-3
  [arXiv:hep-ph/9802439 [hep-ph]].

\bibitem{Giele:2008ve}
  W.~T.~Giele, Z.~Kunszt and K.~Melnikov,
  JHEP \textbf{04}, 049 (2008)
  [arXiv:0801.2237 [hep-ph]].

\bibitem{Badger:2018enw}
S.~Badger, C.~Br\o{}nnum-Hansen, H.~B.~Hartanto and T.~Peraro,
JHEP \textbf{01}, 186 (2019)
doi:10.1007/JHEP01(2019)186
[arXiv:1811.11699 [hep-ph]].

\bibitem{Abreu:2017xsl}
S.~Abreu, F.~Febres Cordero, H.~Ita, M.~Jaquier, B.~Page and M.~Zeng,
Phys. Rev. Lett. \textbf{119}, no.14, 142001 (2017)
doi:10.1103/PhysRevLett.119.142001
[arXiv:1703.05273 [hep-ph]].

\bibitem{Abreu:2018jgq}
S.~Abreu, F.~Febres Cordero, H.~Ita, B.~Page and V.~Sotnikov,
JHEP \textbf{11}, 116 (2018)
doi:10.1007/JHEP11(2018)116
[arXiv:1809.09067 [hep-ph]].

\bibitem{Abreu:2018zmy}
S.~Abreu, J.~Dormans, F.~Febres Cordero, H.~Ita and B.~Page,
Phys. Rev. Lett. \textbf{122}, no.8, 082002 (2019)
doi:10.1103/PhysRevLett.122.082002
[arXiv:1812.04586 [hep-ph]].

\bibitem{Abreu:2019rpt}
S.~Abreu, L.~J.~Dixon, E.~Herrmann, B.~Page and M.~Zeng,
JHEP \textbf{03}, 123 (2019)
doi:10.1007/JHEP03(2019)123
[arXiv:1901.08563 [hep-th]].

\bibitem{Abreu:2019odu}
S.~Abreu, J.~Dormans, F.~Febres Cordero, H.~Ita, B.~Page and V.~Sotnikov,
JHEP \textbf{05}, 084 (2019)
doi:10.1007/JHEP05(2019)084
[arXiv:1904.00945 [hep-ph]].

\bibitem{Abreu:2020lyk}
S.~Abreu, F.~Febres Cordero, H.~Ita, M.~Jaquier, B.~Page, M.~S.~Ruf and V.~Sotnikov,
Phys. Rev. Lett. \textbf{124}, no.21, 211601 (2020)
doi:10.1103/PhysRevLett.124.211601
[arXiv:2002.12374 [hep-th]].

\bibitem{Abreu:2020cwb}
S.~Abreu, B.~Page, E.~Pascual and V.~Sotnikov,
JHEP \textbf{01}, 078 (2021)
doi:10.1007/JHEP01(2021)078
[arXiv:2010.15834 [hep-ph]].

\bibitem{Abreu:2021oya}
S.~Abreu, F.~Febres Cordero, H.~Ita, B.~Page and V.~Sotnikov,
JHEP \textbf{07}, 095 (2021)
doi:10.1007/JHEP07(2021)095
[arXiv:2102.13609 [hep-ph]].

\bibitem{AccettulliHuber:2019abj}
  M.~Accettulli Huber, A.~Brandhuber, S.~De Angelis and G.~Travaglini,
  Phys. Rev. D \textbf{101}, no.2, 026004 (2020)
  doi:10.1103/PhysRevD.101.026004
  [arXiv:1910.04772 [hep-th]].
  
\bibitem{KaluzaKlein}
T.~Kaluza,
Sitzungsber. Preuss. Akad. Wiss. Berlin (Math. Phys. ) \textbf{1921}, 966-972 (1921)
doi:10.1142/S0218271818700017
[arXiv:1803.08616 [physics.hist-ph]];
O.~Klein,
Z. Phys. \textbf{37}, 895-906 (1926)
doi:10.1007/BF01397481

\bibitem{Paton:1969je}
  J.~E.~Paton and H.~M.~Chan,
  Nucl. Phys. B \textbf{10}, 516-520 (1969)
  doi:10.1016/0550-3213(69)90038-8

\bibitem{Cheung:2009dc}
  C.~Cheung and D.~O'Connell,
  JHEP \textbf{07}, 075 (2009)
  doi:10.1088/1126-6708/2009/07/075
  [arXiv:0902.0981 [hep-th]].

\bibitem{Bern:2010qa}
  Z.~Bern, J.~J.~Carrasco, T.~Dennen, Y.~t.~Huang and H.~Ita,
  Phys. Rev. D \textbf{83}, 085022 (2011)
  doi:10.1103/PhysRevD.83.085022
  [arXiv:1010.0494 [hep-th]].

\bibitem{Badger:2005zh}
  S.~D.~Badger, E.~W.~N.~Glover, V.~V.~Khoze and P.~Svrcek,
  JHEP \textbf{07}, 025 (2005)
  doi:10.1088/1126-6708/2005/07/025
  [arXiv:hep-th/0504159 [hep-th]].

\bibitem{Forde:2005ue}
  D.~Forde and D.~A.~Kosower,
  Phys.\ Rev\  D \textbf{73}, 065007 (2006)
  doi:10.1103/PhysRevD.73.065007
  [arXiv:hep-th/0507292 [hep-th]].

\bibitem{Rodrigo:2005eu}
  G.~Rodrigo,
  JHEP \textbf{09}, 079 (2005)
  doi:10.1088/1126-6708/2005/09/079
  [arXiv:hep-ph/0508138 [hep-ph]].

\bibitem{Schwinn:2006ca}
  C.~Schwinn and S.~Weinzierl,
  JHEP \textbf{03}, 030 (2006)
  doi:10.1088/1126-6708/2006/03/030
  [arXiv:hep-th/0602012 [hep-th]].

\bibitem{Ferrario:2006np}
  P.~Ferrario, G.~Rodrigo and P.~Talavera,
  Phys.
  Rev.
  Lett. \textbf{96}, 182001 (2006)
  [arXiv:hep-th/0602043 [hep-th]].

\bibitem{Carrasco:2020ywq}
  J.~J.~M.~Carrasco and I.~A.~Vazquez-Holm,
  Phys. Rev. D \textbf{103}, no.4, 045002 (2021)
  doi:10.1103/PhysRevD.103.045002
  [arXiv:2010.13435 [hep-th]].

\bibitem{Britto:2005fq}
  R.~Britto, F.~Cachazo, B.~Feng and E.~Witten,
  Phys. Rev. Lett. \textbf{94}, 181602 (2005)
  doi:10.1103/PhysRevLett.94.181602
  [arXiv:hep-th/0501052 [hep-th]].

\bibitem{Schwinn:2007ee}
  C.~Schwinn and S.~Weinzierl,
  JHEP \textbf{04}, 072 (2007)
  doi:10.1088/1126-6708/2007/04/072
  [arXiv:hep-ph/0703021 [hep-ph]].

\bibitem{Kosower:2004yz}
  D.~A.~Kosower,
  Phys.\ Rev\  D \textbf{71}, 045007 (2005)
  doi:10.1103/PhysRevD.71.045007
  [arXiv:hep-th/0406175 [hep-th]].

\bibitem{Edison:2011ta}
  A.~C.~Edison and S.~G.~Naculich,
  Nucl.\ Phys\  B \textbf{858}, 488-501 (2012)
  \url{doi:10.1016/j.nuclphysb.2012.01.019}
  [arXiv:1111.3821 [hep-th]].

\bibitem{Henn:2014yza}
  J.~M.~Henn and J.~C.~Plefka,
  Lect.\ Notes Phys. \textbf{883}, pp.1-195 (2014)
  doi:10.1007/978-3-642-54022-6

\bibitem{Bern:1992em}
  Z.~Bern, L.~J.~Dixon and D.~A.~Kosower,
  Phys.\ Lett\  B \textbf{302}, 299-308 (1993)
  [erratum: Phys.\ Lett\  B \textbf{318}, 649 (1993)]
  doi:10.1016/0370-2693(93)90400-C
  [arXiv:hep-ph/9212308 [hep-ph]].

\bibitem{Bern:1998sv}
  Z.~Bern, L.~J.~Dixon, M.~Perelstein and J.~S.~Rozowsky,
  Nucl. Phys. B \textbf{546}, 423-479 (1999)
  doi:10.1016/S0550-3213(99)00029-2
  [arXiv:hep-th/9811140 [hep-th]].

\bibitem{Kleiss:1988ne}
  R.~Kleiss and H.~Kuijf,
  Nucl.\ Phys\  B \textbf{312}, 616-644 (1989)
  doi:10.1016/0550-3213(89)90574-9

\bibitem{Hodges:2009hk}
  A.~Hodges,
  JHEP \textbf{05}, 135 (2013)
  doi:10.1007/JHEP05(2013)135
  [arXiv:0905.1473 [hep-th]].

\bibitem{Budge:2020oyl}
  L.~Budge, J.~M.~Campbell, G.~De Laurentis, R.~K.~Ellis and S.~Seth,
  JHEP \textbf{05}, 079 (2020)
  doi:10.1007/JHEP05(2020)079
  [arXiv:2002.04018 [hep-ph]].

\bibitem{Kilgore:2007qr}
  W.~B.~Kilgore,
  [arXiv:0711.5015 [hep-ph]].

\end{thebibliography}
\end{document}